\documentclass[acmsmall, screen]{acmart}

\usepackage{caption, subcaption}
\usepackage[noend]{algpseudocode}
\usepackage{algorithm}
\usepackage{tikz}
\usetikzlibrary{arrows,automata,positioning,backgrounds,calc,patterns}
\usetikzlibrary{arrows.meta}
\usetikzlibrary{decorations.pathreplacing, shapes} %
\usepackage{mathpartir}
\usepackage{framed}
\usepackage{ulem}
\usepackage{adjustbox}
\usepackage{thmtools}
\usepackage{hyperref}
\usepackage{cleveref}
\usepackage{wrapfig}
\usepackage{listings}

\tikzset{transaction state/.style={draw=black!0}}
\tikzset{
	arrow/.pic={\path[tips,every arrow/.try,->,>=#1] (0,0) -- +(.1pt,0);},
	pics/arrow/.default={triangle 90},
	lab dis/.store in=\LabDis,
	lab dis=0.3,
	->-/.style args={at #1 with label #2}{decoration={
			markings,
			mark=at position #1 with {\arrow{>}; \node at (0,\LabDis) {#2};}},postaction={decorate}},
	-<-/.style args={at #1 with label #2}{decoration={
			markings,
			mark=at position #1 with {\arrow{<}; \node at (0,\LabDis)
				{#2};}},postaction={decorate}},
	-*-/.style={decoration={
			markings,
			mark=at position #1 with {\fill (0,0) circle (1.5pt);}},postaction={decorate}}
}

\definecolor{pinegreen}{rgb}{0.0, 0.47, 0.44}
\definecolor{greensilver}{HTML}{669999}
\definecolor{lightGreensilver}{HTML}{a3c2c2}
\definecolor{appendixVersionColor}{HTML}{000000}

\definecolor{noAppendixVersionColor}{HTML}{000000}

\newcommand\appendixver[1]{}

\newcommand{\eg}{e.g., }

\usepackage{amsmath, amsthm, amsfonts, comment, xcolor}

\usepackage{mathtools}

\usepackage[utf8]{inputenc}
\usepackage[T1]{fontenc}
\usepackage{academicons}

\usepackage[noend]{algpseudocode}
\usepackage{algorithm}
\usepackage{tikz}
\usepackage{mathpartir}
\usepackage{framed}
\usepackage{ulem}
\usepackage{adjustbox}
\usepackage{thmtools}
\usepackage{hyperref}
\usepackage{cleveref}
\usepackage{thm-restate}
\usepackage{wrapfig}
\usepackage{listings}
\usepackage{xparse}
\usepackage{soulutf8}

\usepackage[commandnameprefix=ifneeded,authormarkup=none,defaultcolor=newVersionColor]{changes}

\usepackage[framemethod=tikz]{mdframed}

\usepackage[scr=boondoxupr]{mathalpha}

\usetikzlibrary{arrows,automata,positioning,backgrounds,calc,patterns}
\usetikzlibrary{arrows.meta}
\usetikzlibrary{decorations.pathreplacing, shapes} %

\let\so\soSoul

\tikzset{transaction state/.style={draw=black!0}}
\tikzset{
	arrow/.pic={\path[tips,every arrow/.try,->,>=#1] (0,0) -- +(.1pt,0);},
	pics/arrow/.default={triangle 90},
	lab dis/.store in=\LabDis,
	lab dis=0.3,
	->-/.style args={at #1 with label #2}{decoration={
			markings,
			mark=at position #1 with {\arrow{>}; \node at (0,\LabDis) {#2};}},postaction={decorate}},
	-<-/.style args={at #1 with label #2}{decoration={
			markings,
			mark=at position #1 with {\arrow{<}; \node at (0,\LabDis)
				{#2};}},postaction={decorate}},
	-*-/.style={decoration={
			markings,
			mark=at position #1 with {\fill (0,0) circle (1.5pt);}},postaction={decorate}}
}

\newtheorem{theorem}{Theorem}[section]

\newtheorem{proposition}[theorem]{Proposition}
\newtheorem{lemma}[theorem]{Lemma}
\newtheorem{corollary}[theorem]{Corollary}
\newtheorem{example}[theorem]{Example}
\newtheorem{definition}[theorem]{Definition}

\usepackage{mathtools}

\newbool{diffMode}
\setbool{diffMode}{false}

\newbool{appendixMode}
\setbool{appendixMode}{true}

\ifbool{diffMode}{%
\definecolor{newVersionColor}{HTML}{009999}
\definecolor{oldVersionColor}{HTML}{dc2361}
\definecolor{appendixVersionColor}{HTML}{997300}%
\definecolor{noAppendixVersionColor}{HTML}{6600ff}%
\let\oldCref\Cref%
\renewcommand{\Cref}[1]{\mbox{\oldCref{#1}}}%
\let\oldcref\cref%
\renewcommand{\cref}[1]{\mbox{\oldcref{#1}}}%
}{
\colorlet{appendixVersionColor}{black}
\colorlet{noAppendixVersionColor}{black}
\colorlet{newVersionColor}{black}
\colorlet{oldVersionColor}{black}
}%

\definechangesauthor[color=appendixVersionColor]{APP}
\definechangesauthor[color=noAppendixVersionColor]{NOAPP}

\ifbool{appendixMode}{%
}{%
}%

\newcommand{\addedAppendix}[1]{%
    \ifbool{appendixMode}{%
        \added[id=APP]{#1}%
    }{}%
}%
\newcommand{\addedJournal}[1]{%
    \ifbool{appendixMode}{}{%
        \added[id=NOAPP]{#1}%
    }%
}%

\setdeletedmarkup{%
\ifbool{diffMode}{%
\color{oldVersionColor}%
\sout{#1}%
\color{black}%
}%
{%
}%
}%

\NewEnviron{nver}{%
\ifbool{diffMode}{%
    \ignorespaces%
    \leavevmode%
    \color{newVersionColor}%
    \BODY%
    \color{black}%
    \ignorespacesafterend%
    \unskip%
}%
{%
    \ignorespaces%
    \leavevmode%
    \color{black}%
    \BODY%
    \color{black}%
    \ignorespacesafterend%
    \unskip%
}%
}%

\NewEnviron{oldver}{%
\ifbool{diffMode}{%
    \ignorespaces%
    \leavevmode%
    \color{oldVersionColor}%
    \BODY%
    \color{black}%
    \ignorespacesafterend%
    \unskip%
}%
{%
}%
}%

\newcommand{\CE}[1]{\ifbool{diffMode}{{\color{red}{CE: #1}}}{}}
\newcommand{\ERC}[1]{\ifbool{diffMode}{{\color{blue}{ENRIQUE: #1}}}{}}

\definecolor{orcidlogocol}{HTML}{A6CE39}

\newenvironment{sketchproof}{\renewcommand{\proofname}{Proof Sketch}\proof}{\endproof}

\newbool{appendixVerCondition}
\setbool{appendixVerCondition}{true}

\NewEnviron{appendixVer}{%
\ifbool{appendixVerCondition}{%
    \ignorespaces%
    \leavevmode%
    \unskip%
    \color{appendixVersionColor}%
    \BODY%
    \color{black}%
    \ignorespacesafterend%
}{}%
}

\NewEnviron{noappendixVer}{%
\ifbool{appendixVerCondition}{}{%
    \ignorespaces%
    \leavevmode%
    \unskip%
    \color{noAppendixVersionColor}%
    \BODY%
    \color{black}%
    \ignorespacesafterend%
}%
}%

\newcommand{\Sref}[1]{\hyperref[#1]{\S\ref*{#1}}}

\newcommand{\init}{\textup{\textbf{\texttt{init}}}}

\newcommand{\btrue}{\mathsf{true}}
\newcommand{\bfalse}{\mathsf{false}}

\newcommand{\yvar}{{y}}

\newcommand{\hist}{{h}}

\newcommand{\writeVar}[2]{{#1}\ \mathsf{writes}\ {#2}}
\newcommand{\writeVarValue}[3]{{#1}\ \mathsf{writes}\ {#2} \ \mathsf{given} \ {#3}}

\newcommand{\readVar}[2]{{#1}\ \mathsf{reads}\ {#2}}
\renewcommand{\hist}{{h}}

\newcommand{\key}{x}

\newcommand{\tup}[1]{{({#1})}}
\newcommand{\pto}{\rightharpoonup}

\newcommand{\eabort}{\mathsf{abort}}

\renewcommand{\hist}{{h}}

\newcommand{\Vars}{\mathsf{Objs}}

\newcommand{\Vals}{\mathsf{Vals}}
\newcommand{\setRows}{\mathsf{R}}
\newcommand{\mapRows}{\mathsf{U}}
\NewDocumentCommand{\keyof}{m+o}{\mathsf{key}\IfValueT{#1}{(#1)}}

\newcommand{\Rows}{\mathsf{Rows}}

\newcommand{\xvar}{{x}}

\newcommand{\predicate}{\mathsf{p}}

\newcommand{\valuewr}[3][\wro]{\mathtt{value}_{#1}(#2, #3)}

\newcommand{\del}{\dagger}

\newcommand{\choice}[1]{\mathsf{choice}\left(#1\right)}
\newcommand{\historyExecution}[1]{\mathsf{history}(#1)}

\newcommand{\idRelation}{\mathtt{id}}

\newcommand{\length}[1]{\mathsf{len}(#1)}

\definecolor{coColor}{HTML}{b37200}
\definecolor{soColor}{HTML}{800080}
\definecolor{poColor}{HTML}{C41E3A}
\definecolor{wrColor}{HTML}{008000}
\definecolor{rbColor}{HTML}{0026ff}
\definecolor{rwColor}{HTML}{148f83}
\definecolor{arColor}{HTML}{b37200}
\definecolor{srColor}{HTML}{0026ff}

\definecolor{x1Color}{HTML}{003399}
\definecolor{x2Color}{HTML}{ff8000}
\definecolor{x3Color}{HTML}{008040}

\newcommand{\CC}{\textup{\texttt{CC}}}

\newcommand{\SER}{\textup{\texttt{SER}}}
\newcommand{\PC}{\textup{\texttt{PC}}}
\newcommand{\SI}{\textup{\texttt{SI}}}

\newcommand{\SC}{\textup{\texttt{SC}}}
\newcommand{\PSI}{\textup{\texttt{PSI}}}
\newcommand{\RVC}{\textup{\texttt{RVC}}}

\newcommand{\BS}{\textup{\texttt{BS}}}
\newcommand{\CCC}{\textup{\texttt{CCC}}}

\newcommand{\axpre}{\textsf{Prefix}}
\newcommand{\axconf}{\textsf{Conflict}}
\newcommand{\axrvc}{\textsf{Return}$-$\textsf{Value}}

\newcommand{\axcc}{\textsf{Causal}}
\newcommand{\axkbs}{\textsf{k}$-$\textsf{Bounded}}
\newcommand{\axnpsi}{\textsf{n}$-$\textsf{PSI}}

\newcommand{\rbo}{\textcolor{rbColor}{\mathsf{rb}}}
\newcommand{\sro}{\textcolor{srColor}{\mathsf{sr}}}

\newcommand{\so}{\textcolor{soColor}{\mathsf{so}}}

\newcommand{\wro}{\textcolor{wrColor}{\mathsf{wr}}}
\newcommand{\ar}{\textcolor{arColor}{\mathsf{ar}}}

\newcommand{\kvStorage}{\mathsf{k}$-$\mathsf{v}}
\newcommand{\kvmvrStorage}{\mathsf{k}$-$\mathsf{mv}}

\newcommand{\iwritekv}[2]{\mathtt{PUT}(#1, #2)}
\newcommand{\ireadkvName}{\mathtt{GET}}
\newcommand{\ireadkv}[1]{\ireadkvName(#1)}

\newcommand{\counterStorage}{\mathsf{counter}}

\newcommand{\insdelStorage}{\mathsf{ins/del}}

\newcommand{\iinsifabsName}{\mathtt{InsAbs}}
\newcommand{\idelifpresName}{\mathtt{DelPre}}
\newcommand{\iinsifabs}[2]{\iinsifabsName(#1, #2)}
\newcommand{\idelifpres}[1]{\idelifpresName(#1)}

\newcommand{\iincCounter}[1]{\mathtt{inc}(#1)}
\newcommand{\irdCounter}[1]{\mathtt{rd}(#1)}

\newcommand{\icaskvName}{\mathtt{CAS}}
\newcommand{\icaskv}[3]{\icaskvName(#1, #2, #3)}

\newcommand{\faacasStorage}{\mathsf{faacas}}

\newcommand{\ifaakvName}{\mathtt{FAA}}
\newcommand{\ifaakv}[2]{\ifaakvName(#1, #2)}

\newcommand{\icsetz}{\mathtt{TEST\&SET}}
\newcommand{\igetName}{\mathtt{GET}}
\newcommand{\iget}[1]{\igetName(#1)}
\newcommand{\iputName}{\mathtt{PUT}}
\newcommand{\iput}[2]{\iputName(#1, #2)}

\newcommand{\iaddName}{\mathtt{ADD}}
\newcommand{\iadd}[1]{\iaddName(#1)}
\newcommand{\icontainsName}{\mathtt{CONTAINS}}
\newcommand{\icontains}[1]{\icontainsName(#1)}

\newcommand{\simpleSQLStorage}{\mathsf{simple}$-$\mathsf{SQL}}

\newcommand{\transSQLMVRStorage}{\mathsf{SQL}$-$\mathsf{mvr}}

\newcommand{\transaction}[1]{\mathtt{TRANSACTION}(#1)}
\newcommand{\trCode}{\mathsf{body}}

\newcommand{\iabort}{\mathtt{abort}}

\newcommand{\iselect}[1]{\mathtt{SELECT}({#1})}
\newcommand{\iinsert}[1]{\mathtt{INSERT}({#1})}
\newcommand{\ideleteName}{\mathtt{DELETE}}
\newcommand{\idelete}[1]{\ideleteName({#1})}
\newcommand{\iupdate}[2]{\mathtt{UPDATE}({#1},\, {#2})}
\newcommand{\iupsert}[2]{\mathtt{UPSERT}({#1},\, {#2})}
\newcommand{\eselect}{\mathtt{SELECT}}
\newcommand{\einsert}{\mathtt{INSERT}}
\newcommand{\edelete}{\mathtt{DELETE}}
\newcommand{\eupdate}{\mathtt{UPDATE}}
\newcommand{\eupsert}{\mathtt{UPSERT}}
\newcommand{\eend}{\mathtt{end}}

\algrenewcommand\algorithmicindent{1.0em}%
\algnewcommand\algorithmicswitch{\textbf{switch}}
\algnewcommand\algorithmiccase{\textbf{case}}
\algnewcommand\algorithmicassert{\texttt{assert}}
\algnewcommand\Assert[1]{\State \algorithmicassert(#1)}%

\newcommand{\opspec}{\mathsf{OpSpec}}
\newcommand{\cmodel}{\mathsf{CMod}}
\newcommand{\sspec}{\mathsf{Spec}}

\newcommand{\SpecEvents}[1]{\mathsf{Events}[{#1}]}

\newcommand{\Events}[1][]{\mathsf{Events}_{#1}}
\newcommand{\Traces}[1][]{\mathcal{T}_{#1}}
\newcommand{\Contexts}[1][]{\mathsf{Contexts}_{#1}}

\newcommand{\programIotaTag}{\mathsf{p}}

\newcommand{\automataLts}[1][S]{S_{#1}, A_{#1},  s_0^{#1}, \Delta_{#1}} 
\newcommand{\lts}[1][S]{\tup{\automataLts[{#1}]}} 
\newcommand{\program}[1][\mathsf{p}]{\lts[#1]}
\newcommand{\implementation}[1][\mathsf{i}]{\lts[#1]}

\newcommand{\programInstance}[1][E]{{P_{#1}}}
\newcommand{\implementationInstance}[1][E]{I_{#1}}

\newcommand{\implementationInstanceCC}{\implementationInstance}
\newcommand{\implementationCC}{\lts[\mathsf{i}]}

\newcommand{\programInstanceIota}{P}
\newcommand{\programIota}{\lts[\programIotaTag]}

\newcommand{\parallelCompositionInstance}[1][E]{\implementationInstance[{#1}]\customparallel \programInstance[{#1}] }
\newcommand{\parallelCompositionInstanceCC}{\programInstance \customparallel \implementationInstanceCC}
\newcommand{\parallelCompositionInstanceIota}{\programInstanceIota \customparallel \implementationInstance}

\newcommand{\inducedHistory}[2][]{\mathsf{h}_{#1}(#2)}

\newcommand{\stateTrace}[1]{\mathtt{state}(#1)}
\newcommand{\lastTraceReplica}[2][r]{\mathtt{last}_{#1}(#2)}
\mathchardef\mhyphen="2D
\newcommand{\writesSet}[1]{\wro\mhyphen\mathtt{Set}(#1)}
\newcommand{\receiveSet}[1]{\rbo\mhyphen\mathtt{Set}(#1)}
\newcommand{\eventExecuted}[1]{\mathtt{ev}(#1)}

\newcommand{\ereceive}{\mathtt{receive}}
\newcommand{\esend}{\mathtt{send}}

\newcommand{\context}[4]{\mathsf{ctxt}_{#2}(#4, [#3, #1])}

\newcommand{\rspec}[4]{\rspecName(#4)(#2, [#3, #1])}
\newcommand{\rspecContext}[3]{\rspecName(#3)(#1, #2)}
\newcommand{\rspecSimple}[1]{\rspecName(#1)}
\newcommand{\extractspec}[4]{\extractName(#4)(#2, [#3, #1])}
\newcommand{\extractspecContext}[3]{\extractName(#3)(#1, #2)}
\newcommand{\wspec}[4]{\wspecName(#4)(#2, [#3, #1])}
\newcommand{\wspecContext}[3]{\wspecName(#3)(#1, #2)}
\newcommand{\wspecSimple}[2]{\wspecName(#1)(#2)}

\newcommand{\rspecName}{\mathsf{rspec}}
\newcommand{\wspecName}{\mathsf{wspec}}
\newcommand{\extractName}{\mathsf{extract}}

\newcommand{\ReplicaID}[1][]{\mathsf{Reps}_{#1}}

\newcommand{\TracesParallelImplementationCC}{\Traces[\parallelCompositionInstanceCC]}
\newcommand{\TracesIotaParallelImplementation}{\Traces[\parallelCompositionInstanceIota]}

\newcommand{\EventsProgramIota}{\Events[\programIotaTag]}

\newcommand{\initProgramIota}{\init_{\programIotaTag}}
\newcommand{\TracesProgramIota}{\Traces[\programIotaTag]}

\newcommand{\metadataEventNameFunction}{\mathtt{md}}
\newcommand{\metadataEvent}[1]{\metadataEventNameFunction(#1)}

\newcommand{\operationEventNameFunction}{\textup{\texttt{op}}}
\newcommand{\operationEvent}[1]{\operationEventNameFunction(#1)}
\newcommand{\replicaEventNameFunction}{\mathtt{rep}}
\newcommand{\replicaEvent}[1]{\replicaEventNameFunction(#1)}

\newcommand{\outputEventNameFunction}{\mathtt{wval}}
\newcommand{\outputEvent}[1]{\outputEventNameFunction(#1)}
\newcommand{\outputEventObj}[2]{\outputEvent{#1}(#2)}

\newcommand{\operation}{\mathsf{op}}
\newcommand{\replicaID}{\mathsf{r}}
\newcommand{\metadataInstance}{\mathsf{m}}
\newcommand{\executedInstance}{\mathsf{ex}}

\newcommand{\outputInstance}{\mathsf{wval}}

\newcommand{\icas}[3]{\mathtt{CAS}(#1, #2, #3)}

\newcommand{\minRcv}[2]{\mathsf{minRcv}(#1, #2)}
\newcommand{\maxSend}[2]{\mathsf{maxSend}(#1, #2)}
\newcommand{\sendIfDataSend}[2]{\mathsf{sendIfData}(#1, #2)}
\newcommand{\sendAllData}[2]{\mathsf{sendAllData}(#1, #2)}
\newcommand{\maxRcv}[2]{\mathsf{maxRcv}(#1, #2)}

\newcommand{\satRspec}[2]{\mathsf{sat}(#1, #2)}

\newcommand{\simpleConsistency}[1]{\mathsf{simple}(#1)}

\newcommand{\almostSimple}[1]{\mathsf{almost}(#1)}

\newcommand{\normalForm}[2][\opspec]{\mathsf{bn}#2_{#1}}
\newcommand{\generalNormalForm}[2][\opspec]{\mathsf{n}#2_{#1}}

\newcommand{\exec}{\xi}
\newcommand{\event}{\varepsilon}

\newcommand{\execTrace}[1]{\mathtt{exec}(#1)}
\newcommand{\varOf}[1]{\mathsf{obj}({#1})}
\newcommand{\prefixTraceEvent}[2]{\mathsf{prefix}(#1, #2)}

\newcommand{\code}{\mathtt{C}}

\newcommand{\suffixOf}[2]{\mathsf{suff}_{#1}(#2)}

\newcommand{\writeConstraints}[3][\mathsf{v}]{\mathsf{wrCons}^{#1}_{#2}(#3)}
\newcommand{\conflictName}{\mathsf{conflict}}
\newcommand{\writeConflicts}[2][]{\conflictName_{#1}(#2)}
\newcommand{\writeVarExec}[3]{{#1}\ \mathsf{writes}\ {#2} \ \mathsf{in} \ {#3}}
\newcommand{\identifier}{\mathsf{id}}
\newcommand{\identifierEventNameFunction}{\mathtt{id}}
\newcommand{\identifierEvent}[1]{\identifierEventNameFunction(#1)}

\newcommand{\Dom}[1]{\mathsf{Dom}(#1)}

\newcommand{\conflictsOfV}[2]{\mathsf{conflictsOf(#1, #2)}}

\newcommand{\simpleStorageFullName}{{basic}}
\newcommand{\SimpleStorageFullName}{{Basic}}

\newcommand{\simpleConsistencyFullName}{{basic}}
\newcommand{\SimpleConsistencyFullName}{{Basic}}

\newcommand{\valuewrExec}[5][\wro]{\mathtt{value}_{#1}(#2)(#3, [#4, #5])}

\newcommand{\correction}[3]{#1 \overset{#2}{\curlyvee}#3}
\newcommand{\sequence}[1]{\mathsf{seq}(#1)}
\newcommand{\sequenceOrd}[1]{\mathsf{seq}_<(#1)}

\newsavebox{\parallelbox}
\newlength{\parallelheight}

\newcommand{\measureparallelheight}[1]{%
  \sbox{\parallelbox}{$#1\parallel$}%
  \setlength{\parallelheight}{\ht\parallelbox}%
  \addtolength{\parallelheight}{\dp\parallelbox}%
}

\newcommand{\customparallel}{\mathrel{\mathpalette\customparallelaux{}}}%

\newcommand{\customparallelaux}[2]{%
  \measureparallelheight{#1}%
  \vcenter{\hbox{%
    \tikz[baseline={(current bounding box.base)}, x=1ex, y=1ex, yscale=.23]{%
      \pgfmathsetmacro{\parallelht}{\dimexpr\parallelheight\relax/1pt}%
      \draw[line width=0.25mm] (0.5, 0) -- (0.5, \parallelht);
      \pgfmathsetmacro{\marginht}{\parallelht / 7}
      \pgfmathsetmacro{\bottom}{\marginht}
      \pgfmathsetmacro{\top}{\parallelht - \marginht}
      \draw[line width=0.25mm] (0, \bottom) -- (0, \top);
      \draw[opacity=0] (0,0) -- (0,\parallelht);
      \draw[opacity=0] (-0.2,0) -- (0.7,0);
      \draw[opacity=0] (-0.2,\parallelht) -- (0.7,\parallelht);
    }%
  }}%
}%

\algdef{SE}[SWITCH]{Switch}{EndSwitch}[1]{\algorithmicswitch\ #1\ \algorithmicdo}{\algorithmicend\ \algorithmicswitch}%
\algdef{SE}[CASE]{Case}{EndCase}[1]{\algorithmiccase\ #1\ :}{\algorithmicend\ \algorithmiccase}%
\algtext*{EndSwitch}%
\algtext*{EndCase}%
\algtext*{EndFor}%
\algnewcommand\Let{\State\textbf{let} }
\algnewcommand\Break{\State\textbf{break} }
\algnewcommand\InputAlgorithmic{\Statex\textbf{Procedure }}

\algnewcommand\OutputAlgorithmic{\Statex\textbf{Output: }}
\algblockdefx[BlockName]{ForTo1}{EndForTo1}%
[3][i]{\textbf{for} #1 $\gets$ #2 \textbf{to} #3 \textbf{do}}%

\lstdefinelanguage{MyLang}{%
	keywords = { delete, do, each, else, export, finally, for, foreach, function,
		if, in, let, of, return, void, while, with, yield, elements, read, write,
		insert, remove, add, AddItem, DeleteItem, Push, Pop, Enroll, Tweet, Timeline,
		NewsFeed, begin, end, break, throw},
	morecomment = [l]{//},
	morecomment = [s]{/*}{*/},
	morestring  = [b]',
	morestring  = [b]",
	sensitive   = true,
}

\lstdefinelanguage{Java10}{
	language      = Java,
	morekeywords  ={ var },
}

\begin{document}
	\title{Arbitration-Free Consistency is Available (and Vice Versa)}

\author{Hagit Attiya}
\affiliation{
	\institution{Technion --- Israel Institute of Technology}
	\country{Israel}                  %
}
\email{hagit@cs.technion.ac.il}          %
\orcid{0000-0002-8017-6457}

\author{Constantin Enea}
\affiliation{
	\institution{LIX, Ecole Polytechnique, CNRS and Institut Polytechnique de Paris}
	\country{France}                    %
}
\email{cenea@lix.polytechnique.fr}          %
\orcid{0000-0003-2727-8865}

\author{Enrique Román-Calvo}
\affiliation{
	\institution{University of Freiburg}
	\country{Germany}                    %
}
\email{calvo@informatik.uni-freiburg.de}          %
\orcid{0009-0005-7539-2330}

	\begin{abstract}
	
\ifbool{diffMode}{\color{newVersionColor}}{}%
The fundamental tension between \emph{availability}
and \emph{consistency} shapes the design of distributed storage systems.
Classical results capture extreme points of this trade-off: 
the CAP theorem shows that strong models like linearizability 
preclude availability under partitions,
while weak models like causal consistency remain implementable
without coordination.
These theorems apply to simple read-write interfaces,
leaving open a precise explanation of the combinations of
object semantics and consistency models that admit
available implementations.

This paper develops a general semantic framework in which storage specifications combine operation semantics and consistency models. The framework encompasses a broad range of objects (key-value stores, counters, sets, CRDTs, and transactional databases) and consistency models (from causal consistency and sequential consistency to snapshot isolation and transactional and non-transactional SQL).

Within this framework, we prove the \emph{Arbitration-Free Consistency} (AFC) theorem, showing that an object specification within a consistency model admits an available implementation if and only if it is \emph{arbitration-free}, that is, it does not require a total arbitration order to resolve visibility or read dependencies.

The AFC theorem unifies and generalizes previous results, revealing arbitration-freedom as the fundamental property that delineates coordination-free consistency from inherently synchronized behavior.

\ifbool{diffMode}{\color{black}}{}%

\deleted{The ability of distributed storage systems to remain available even in the presence of network failures depends on the consistency guarantees they provide. The CAP theorem states that a distributed storage system cannot deliver both availability and strong consistency, in the presence of network partitions.}

\sloppy
\deleted{
\indent
This paper presents a more precise classification of distributed storage systems, conditioned on the combination of their consistency guarantees and the data type semantics they support, which can be implemented in a highly-available manner. The characterization reveals that the key to availability is the ability to locally determine responses without having to \emph{arbitrate} between several possible dependencies. The results apply to a wide set of consistency models, including prefix consistency, causal consistency, sequential consistency as well as delayed consistency, and to a large collection of objects including counters, last-writer-wins and multi-value key-value stores, key-value stores with compare-and-swap or insert-if-absent/delete-if-present operations, as well as transactional and non-transactional SQL.}

	\end{abstract}

	\maketitle

	\section{Introduction}

Distributed storage systems enable reliable access to objects by replicating them across a wide-area network. Replication is essential for tolerating faults in the system (e.g., machines that crash, network partitions) and for decreasing latency.
In such systems, it is crucial to maintain a trade-off between availability (ensuring prompt access to data) and preserving consistency, even in the presence of communication delays.
The \emph{CAP theorem}~\cite{cap-theorem,Brewer2000} shows that a key-value store
cannot provide strong \emph{Consistency} (atomicity)
while maintaining \emph{Availability}
and tolerating network \emph{Partitions} at the same time.
PACELC~\cite{DBLP:journals/computer/Abadi12,DBLP:journals/sigact/Golab18} refines CAP by adding the case of a
connected network where strong consistency cannot be achieved with low latency.

Many modern storage systems sacrifice strong consistency for availability
(or low latency) and ensure weaker notions of consistency.
There is plethora of weak consistency models~\cite{DBLP:journals/ftpl/Burckhardt14}
(or \emph{isolation levels}~\cite{DBLP:conf/icde/AdyaLO00} in the context of transactions)
that correspond to different trade-offs with respect to availability.
Other modern storage systems relax the semantics of the objects they support,
e.g., multi-value registers, where a get arbitrarily returns a previously stored value.

Given that the guarantees of a storage system are captured through
the subtle combination of its consistency model and its object semantics,
a natural question to consider is:

\begin{quote}
\emph{What class of consistency models support available implementations
of which objects in the presence of network partitions?}
\end{quote}

Previous results provided only partial answers to this question.
The aforementioned CAP theorem only shows a negative result that an \emph{Atomic (Linearizable) Key-Value Store} is not included in this class.
Attiya et al.~\cite{10.1109/TPDS.2016.2556669} identify \replaced{a consistency model, called Observable Causal Consistency (OCC), that is not included in this class; but only for particular objects, Multi-Value Registers. We remark that Causal Consistency, which is strictly weaker than both of them, is in the class.}{a strongest consistency model in this class, called Observable Causal Consistency, but only in the case of particular objects, Multi-Value Registers.}

The goal of this paper is to give a precise answer for the question raised above. To do so, we rely on a very expressive framework for defining consistency models and object semantics that builds on previous work~\cite{DBLP:journals/ftpl/Burckhardt14}. Using this framework, we give a \replaced{tight}{precise} characterization of models and objects that can be expressed within this framework and that support available implementations. Before explaining our characterization, we outline our framework.

\vspace{-1mm}
\paragraph*{A framework for defining storage specifications.}
A storage system is composed of a collection of objects that can be read or modified using a set of operations (the API of the storage).
Specifications are expressed in terms of an abstract model of storage executions, which is defined as a set of binary relations among events—each event corresponding to an invocation of an operation on an object. These relations capture typical control-flow dependencies--such as invocations occurring at the same replica--data-flow dependencies--where certain updates affect the result of a query--and a total order used as a "tie-breaker" to fix an order between concurrent invocations. The latter is called \emph{arbitration order} and it has an important role in our main result.

In a distributed storage system, implementations typically rely on communication protocols to share the effects of invocations among all replicas. They also use specific algorithms to merge the effects received from other replicas into the local replica’s state. As a result, each invocation can be viewed as executing within a specific \emph{context}--that is, the set of prior operations, including those received from remote replicas.

A storage specification defines the expected behavior of the system. It consists of two parts:
\begin{itemize}
	\item a \emph{consistency model}, which restricts the possible contexts in which each invocation may execute.
	\item an \emph{operation specification}, which describes the allowable effects or return values of an invocation, given its context.
\end{itemize}

\replaced{A consistency model consists of a set of \emph{visibility} formulas that say when an invocation belongs to the context of another invocation. This ``being in the context of'' binary relation is defined via combinations of the binary relations mentioned above (by standard composition, union, and transitive closure).}
{A consistency model consists of a set of \emph{visibility} relations, which are defined as combinations of the binary relations mentioned above (by standard composition of relations, union, and transitive closure).}
For instance, a visibility \replaced{formula}{relation} may state that all prior invocations at the same replica should be included in the context. An \emph{operation specification} consists of a set of abstract functions that characterize the read and write behavior of an invocation, in particular, the value written by writes. Note that this value is not always fixed since we allow operations that read and write at the same time, e.g., Compare-and-Swap which writes a given object only if the old value equals some other value given as input. We also allow SQL transactions whose effects are even more complex.

We show that our framework covers many possible storage specifications, including Last-Writer-Wins and Multi-Value Key-Value stores, Key-Value stores with Compare-and-Swap operations, Key-Value stores with counters, as well as transactional and non-transactional SQL stores, and many possible consistency models including Return-Value Consistency, Causal Consistency, Sequential Consistency, and transactional isolation levels like Snapshot Isolation and Serializability.

\paragraph*{The arbitration-free consistency (\emph{AFC}) theorem.}
Our main result states roughly, that a storage system has an available implementation
if and only if the visibility \replaced{formulas}{relations} that define its consistency model \emph{exclude any meaningful use of the total arbitration order.}
Such a consistency model is called \emph{arbitration-free}.
As in previous works, we consider an implementation to be \emph{available} if operations can be answered immediately on every replica (without waiting for messages from other replicas).

The proof of the AFC theorem is quite challenging,
one reason being the very expressive and abstract specification framework that we consider.
Proving that there exist available implementations for arbitration-free consistency models is the easier part since arbitration-freeness implies that the model is weaker than causal consistency and the latter is known to support available implementations~\cite{BailisOverhead,COPS,Eiger,bolton}.
The opposite direction is much more difficult and is described in two stages.

We first consider a basic case, in which operations read and/or write a
single value from/to a single object.
This yields a reasonably simple proof,
while still covering consistency models such as Return-Value Consistency,
Causal Consistency, Prefix Consistency and Sequential Consistency,
and objects such as a key-value store, with ordinary put / get operations
or extended with Fetch-and-Add and Compare-and-Swap operations.

Then, we consider a general class of objects where operations can read and/or write multiple objects at the same time, and reads may compute their return value from multiple updates. In this very generic context, we need to introduce some number of restrictions (assumptions) which are however satisfied by all practical cases that we are aware of (see Section~\ref{sec:generalized-definitions}). This is to exclude pathological cases that arise from starting with a very abstract formal model.

To summarize, we provide the first characterization of distributed storage formal specifications 
that support available implementations which takes into account both consistency constraints 
and the semantics of the implemented objects.
\added{At a high level, the key insight behind our result is that in an asynchronous system, 
where replicas coordinate only through the exchange of messages, 
they can establish at most a causal order between operations.
The arbitration order, in contrast, is total: it compares operations that are concurrent 
and therefore incomparable under causality.
Determining such a total order would require additional synchronization between replicas,
coordination that cannot be achieved in an always-available manner.}

	\section{Motivating Examples}
\label{sec:motivating-example}
\begin{figure}[!t]
\centering
\begin{subfigure}[b]{.24\textwidth}
\centering
\begin{adjustbox}{max width=\textwidth}
{
\begin{tabular}{c||c}
\begin{lstlisting}[xleftmargin=5mm,basicstyle=\ttfamily\small,escapeinside={(*}{*)}, tabsize=1]
PUT((*$x$*),1);
a = GET((*$y$*));
\end{lstlisting} &
\begin{lstlisting}[xleftmargin=5mm,basicstyle=\ttfamily\small,escapeinside={(*}{*)}, tabsize=1]
PUT((*$y$*),2);
b = GET((*$x$*));
\end{lstlisting}
\end{tabular}
}
\end{adjustbox}
\caption{Sequential Consistency and $\iputName$, $\igetName$ operations.}
\label{fig:motivating-example:prog:sc-kv}
\end{subfigure}
\hfill
\begin{subfigure}[b]{.24\textwidth}
\centering
\begin{adjustbox}{max width=\textwidth}
{
\begin{tabular}{c||c}
\begin{lstlisting}[xleftmargin=5mm,basicstyle=\ttfamily\small,escapeinside={(*}{*)}, tabsize=1]
PUT((*$x$*), 1);
PUT((*$x$*), 2);
...
PUT((*$x$*), K);
a = GET((*x*));
\end{lstlisting} &
\begin{lstlisting}[xleftmargin=5mm,basicstyle=\ttfamily\small,escapeinside={(*}{*)}, tabsize=1]
PUT((*$y$*), 1);
PUT((*$y$*), 2);
...
PUT((*$y$*), K);
b = GET((*y*));
\end{lstlisting}
\end{tabular}
}
\end{adjustbox}
\caption{Bounded Staleness and $\iputName$, $\igetName$ operations.}
\label{fig:motivating-example:prog:bs-kv}
\end{subfigure}
\hfill
\begin{subfigure}[b]{.24\textwidth}
\centering
\begin{adjustbox}{max width=\textwidth}
{
\begin{tabular}{c||c}
\begin{lstlisting}[xleftmargin=5mm,basicstyle=\ttfamily\small,escapeinside={(*}{*)}, tabsize=1]
FAA((*$x$*), 1);
\end{lstlisting} &
\begin{lstlisting}[xleftmargin=5mm,basicstyle=\ttfamily\small,escapeinside={(*}{*)}, tabsize=1]
FAA((*$x$*), 2);
\end{lstlisting}
\end{tabular}
}
\end{adjustbox}
\caption{Sequential Consistency and $\ifaakvName$ operations.}
\label{fig:motivating-example:prog:sc-faacas}
\end{subfigure}
\hfill
\begin{subfigure}[b]{.24\textwidth}
\centering
\begin{adjustbox}{max width=\textwidth}
{
\begin{tabular}{c||c}
\begin{lstlisting}[xleftmargin=5mm,basicstyle=\ttfamily\small,escapeinside={(*}{*)}, tabsize=1]
FAA((*$x$*), 1);
FAA((*$y$*), 3);
\end{lstlisting} &
\begin{lstlisting}[xleftmargin=5mm,basicstyle=\ttfamily\small,escapeinside={(*}{*)}, tabsize=1]
FAA((*$y$*), 2);
FAA((*$x$*), 4);
\end{lstlisting}
\end{tabular}
}
\end{adjustbox}
\caption{Prefix Consistency and $\ifaakvName$ operations.}
\label{fig:motivating-example:prog:pc-faacas}
\end{subfigure}
\vspace{-2mm}	
\caption{Different litmus programs with two concurrent sessions showing the absence of available implementations for selected pairs of consistency models and operation specifications.}
\label{fig:motivating-example:prog}

\vspace{-4mm}
\end{figure}

We illustrate the broad applicability of the AFC theorem through various storage specifications, each reflecting different trade-offs between consistency and operation semantics. We argue about the  diversity of reasoning required and motivate the need for a unified framework.

As a starting point, we consider a standard key-value store with $\iputName$ and $\igetName$ operations; $\iput{x}{v}$ writes the value $v$ to object (key) $x$, and $\iget{x}$ reads the latest\footnote{We assume a standard semantics based on the Last-Writer-Wins conflict resolution policy.} value of object $x$. As consistency model, we consider the standard \emph{Sequential Consistency} ($\SC$) whose formalization uses arbitration to postulate an order in which different operations interleave. By the AFC theorem, the latter implies that there exists no available implementation that ensures $\SC$. Intuitively, the proof is based on a \emph{litmus} program like in \Cref{fig:motivating-example:prog:sc-kv}. This program contains two concurrent sessions, each executed at a different replica. Also, $x$ and $y$ are initially $0$.
An $\SC$ available implementation should allow an execution in which, intuitively, the two replicas operate without exchanging any messages, resulting in both final get operations returning 0. However, this outcome violates sequential consistency, as it cannot be produced by any interleaving of the operations—leading to a contradiction.

We remark that this argument proves a version of the CAP theorem that is stronger than the one proved in~\cite{cap-theorem}. The latter proof relies on the \emph{real-time ordering} requirement that is
embedded in linearizability --- a consistency model stronger than sequential consistency (cf.~\cite{DBLP:journals/corr/Kleppmann15}).

Such a proof can be generalized to the case where $\iputName$ / $\igetName$ operations are replaced for instance, by $\iaddName$ / $\icontainsName$ operations on a set, i.e., $\iput{x}{v}$ and $\iget{x}$ in \Cref{fig:motivating-example:prog:sc-kv} are replaced by $\iadd{x}$ and $\icontains{x}$ (and similarly for operations on $y$). As in the previous case, an $\SC$ available implementation should allow an execution without exchange of messages, resulting in both final $\icontainsName$ operations returning $\bfalse$ (the set does not contain the element), which is an $\SC$ violation.

On the other hand, if we consider a weaker consistency model, a straightforward variation of the program in \Cref{fig:motivating-example:prog:sc-kv} can not be used to prove non-existence of available implementations. For instance, consider \emph{Bounded Staleness}~\cite{cosmosdb-consistency} a weakening of $\SC$, which requires that each get operation observes all preceding put operations (on the same object), except possibly the most recent $K-1$, for some fixed value of $K$. The put operations are still required to execute following some fixed arbitration order as in $\SC$ (see \Cref{ssec:conflict-defs} for a precise definition). This weakening for \replaced{$K=2$}{$K=1$} admits an execution of t he program in \Cref{fig:motivating-example:prog:sc-kv} where both final get operations return $0$ (the get operations may miss the only put in the program). Therefore, this program cannot be used to show non-existence of available implementations. Instead, one can use the program given in \Cref{fig:motivating-example:prog:bs-kv}, which contains $K$ put operations in each session. One can follow now the same strategy as above and show that an execution without exchange of messages makes both get operations return $0$, and this violates bounded staleness.

If we weaken consistency even further and consider \emph{Causal Consistency} ($\CC$)~\cite{DBLP:conf/sss/ShapiroPBZ11}, then the AFC theorem will imply existence of available implementations (which is known~\cite{BailisOverhead,COPS,Eiger,bolton}).

Now, if we change the set of operations and consider a storage system with only Fetch-and-Add operations ($\ifaakv{x}{v}$ returns the old value of $x$ and adds $v$, atomically), then a proof for non-existence of $\SC$ available implementations can be done using the program in \Cref{fig:motivating-example:prog:sc-faacas} with only one $\ifaakvName$ in each session. An execution without exchange of messages will imply that both $\ifaakvName$ return the initial value of $x$, and this is a violation of $\SC$.

If we weaken consistency to \emph{Prefix Consistency} ($\PC$) \cite{DBLP:conf/ecoop/BurckhardtLPF15}, then the previous program is not suitable. An execution where both $\ifaakvName$ in \Cref{fig:motivating-example:prog:sc-faacas} return the initial value of $x$ satisfies $\PC$ (see \cref{ssec:consistency} for a formal definition). Instead, we need a litmus program like in \Cref{fig:motivating-example:prog:pc-faacas} which contains two $\ifaakvName$s per session. Here, an execution where all $\ifaakvName$s return an initial value does not satisfy $\PC$. This program can also be used to show the non-existence of available implementations of \emph{Parallel Snapshot Isolation} ($\PSI$)~\cite{DBLP:conf/sosp/SovranPAL11} \added{or \emph{Conflict-preserving Causal Consistency} $(\CCC$), a consistency model defined using the axioms $\axconf$ and $\axcc$ from \cite{DBLP:journals/pacmpl/BiswasE19}}. As a side remark, note that \replaced{$\CCC$}{$\PSI$} is equivalent to $\CC$ for the key-value store with \replaced{$\iputName$}{$\ifaakvName$} and $\igetName$ operations presented at the beginning, and therefore, there exists an available implementation for $\CCC$ in that case.

While these cases follow a broadly similar proof strategy, each demands distinct proof artifacts (such as litmus programs) and tailored reasoning. The AFC theorem unifies these diverse arguments within a common theoretical framework, grounded in a formalization of a wide class of storage specifications encompassing all the examples above.

	\section{Abstracting Storage Executions}
\label{sec:definitions}

We present an abstract model of distributed storage executions that includes the essential components needed to define storage specifications. A \emph{distributed storage} (or simply storage) replicates the state of a set of objects over two or more nodes called \emph{replicas}. We use $\Vars$ to denote the infinite set of objects, ranged over $x, y, z$, and $\ReplicaID$ to denote the set of replica identifiers, ranged over $r$, $r_1$, $r_2$.
Objects are accessed using a set of \emph{operations} which may write or return values in a set $\Vals$. %

An \emph{abstract execution} records operation invocations along with a set of relations that represent control-flow dependencies (two invocations executing on the same replica), and the internal behavior of the storage. The internal behavior includes, broadly, the computation of object states and the return values of invocations, as well as the communication between replicas. The first concerns local computation within each replica, while the second pertains to communication protocols or underlying network assumptions. To distinguish these two aspects, we first introduce the concept of a \emph{history}, which records only the data-flow dependencies relevant to characterizing the local computation. An abstract execution is then defined as an extension of a history, enriched with additional relations that abstract inter-replica communication.

\subsection{Histories}
\label{ssec:histories}

The invocation of an operation on some replica is represented using an \emph{event} \replaced{$e= \tup{\identifier, \replicaID, \operation, \outputInstance, \metadataInstance}$}{$e= \tup{\identifier, \replicaID, \operation, \metadataInstance} $} where $\identifier$ is an event identifier, $\replicaID$ is a replica identifier, $\operation$ is an operation name, \added{$\outputInstance$ is a (partial) mapping that associates an object $x$ with a value $v$ that this event writes to $x$,} and $\metadataInstance$ is additional metadata of the invocation. \deleted{The metadata includes input and return values of the invocation. }We use $\identifierEvent{e}$ $\replicaEvent{e}, \operationEvent{e}$, \added{$\outputEvent{e}$,} and $\metadataEvent{e}$ to denote the event identifier, replica identifier, operation, \added{written value mapping} and metadata of an event $e$, respectively.
We assume that every event $e$ accesses (reads or writes) a fixed finite set of objects denoted as  $\varOf{e}$.
The set of events is denoted by $\Events$. We assume that $\Events$ includes a distinguished type of \emph{initial events} that affect every object, representing the initial state of the storage.

\begin{example}\label{ex:faacas}
As a running example, we consider a Key-value Store with four types of operations: $\iwritekv{x}{v}$ that writes $v$ to object (key) $x$, $\ireadkv{x}$ that reads object $x$, $\ifaakv{x}{v}$ that reads the value $v'$ of object $x$ and writes $v'+v$, and $\icaskv{x}{v}{v'}$, that reads $x$ and writes $v'$ iff the value read is $v$. We use $\faacasStorage$ to refer to this storage (from the Fetch-and-Add and Compare-and-Swap operations). \appendixver{\Cref{ssec:cas-storage} summarizes the full description of $\faacasStorage$.}

\end{example}

A \emph{history} contains a \added{finite} set of events $E$ ordered by a (partial) \emph{session order} $\so$ that relates events on the same replica, and a \emph{write-read} relation $\wro$ (also known as read-from) representing data-flow dependencies between events that update and respectively, read a same object. Histories contain an initial event, $\init$, that precedes every other event in $E$ w.r.t $\so$.
We consider a write-read relation $\wro_{\key}\subseteq \mathcal{P}(E)\times E$ for every object $\key \in \Vars$. The inverse of $\wro_{\key}$ is defined as usual and denoted by $\wro_{\key}^{-1}$. 
We use $\wro : \Vars \to \mathcal{P}(E) \times E$ to denote the mapping associating each object $x$ with $\wro_x$.

For simplicity, we often abuse the notation and extend $\wro_x$ and $\wro$ to pairs of events: \replaced{we say that $(w, r) \in \wro_x$ if $w \in \wro_x^{-1}(r)$, and we say that $(w, r) \in \wro$ if there exists an object $x$ s.t. $(w, r) \in \wro_x$.}{we say that $(w, r) \in \wro_x$ if there exists $\mathsf{W}$ s.t. $w \in \mathsf{W}$ and $(\mathsf{W}, r) \in \wro_x$, and we say that $(w, r) \in \wro$ if there exists an object $x$ s.t. $(w, r) \in \wro_x$.}

\begin{definition}
\label{def:history}
A \emph{history} $\tup{E, \so, \wro}$ is a \added{finite} set of events $E$ along with a strict partial \emph{session order} $\so$, and a 
\emph{write-read} relation $\wro_\key\subseteq \mathcal{P}(E) \times E$ for every $\key \in \Vars$ such that
\begin{itemize}
    \item $E$ contains a single initial event $\init$, which precedes every other event in $E$ w.r.t. $\so$, 
    \item \replaced{$\forall e, e' \in E \setminus \{\init\}$,}{$\forall e, e' \in E$} $\so$ orders $e$ and $e'$ iff $\replicaEvent{e} = \replicaEvent{e'}$,
    \item the inverse of $\wro_{\key}$ is a total function for every $\key \in \Vars$, and
    \item $\so \cup \wro$ is acyclic (here we use the extension of $\wro$ to pairs of events).
\end{itemize}
\end{definition}

\begin{example}
\Cref{fig:example-history-faa} shows two examples of histories of the storage $\faacasStorage$ presented in Example~\ref{ex:faacas}. For readability, we omit replica identifiers from events. The $\wro$ dependencies can be used to explain the ``local'' computation in those invocations as follows: (1) on the left, the $\mathtt{CAS}$ should fail (not write to $x$) because it reads the value written by the $\mathtt{FAA}$ which should be equal to 1 since $\mathtt{FAA}$ reads the initial value, (2) on the right, the $\mathtt{CAS}$ should succeed (write to $x$) because it reads the initial value (the $\mathtt{FAA}$ will concurrently write 1 to $x$).
\end{example}

We say that the event $w$ is \emph{read} by the event $r$ if $\tup{w,r}\in \wro$. 
Since we assumed that $\wro_{\key}^{-1}$ is a total function, we use $\wro_{\key}^{-1}(r)$ to denote the set $\mathsf{W}$ such that $(\mathsf{W},r)\in \wro_{\key}$. We use $\wro_x^{-1}(e) = \emptyset$ to indicate that $e$ does not read $x$ (resp. $\wro_x^{-1}(e) \neq \emptyset$ to indicate that $e$ reads $x$).

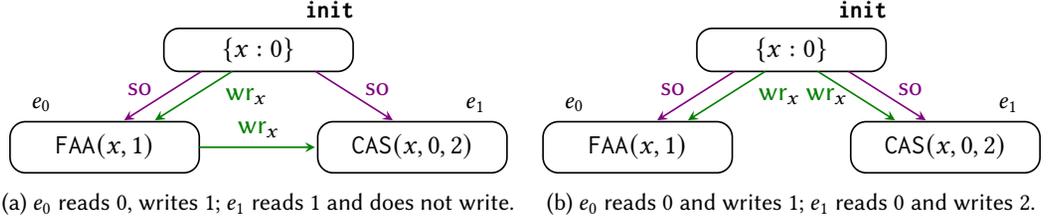
\begin{figure}[t]
\centering

\begin{subfigure}[b]{.49\columnwidth}
\centering
\resizebox{\textwidth}{!}{
\begin{tikzpicture}[>=stealth,shorten >=1pt,auto,node distance=3cm,
    semithick, transform shape,
    B/.style = {%
    decoration={brace, amplitude=1mm,#1},%
    decorate},
    B/.default = ,  %
    ]

    \node[minimum width=7em, draw, rounded corners=2mm,outer sep=0, label={[font=\small]30:$\init$}] (init) at (2, 0) {$\{x: 0\}$};

    \node[minimum width=7em, draw, rounded corners=2mm,outer sep=0, label={[font=\small]150:$e_0$}] (r11) at (0, -1.25) {
        \begin{tabular}{l}
            $\ifaakv{x}{1}$
        \end{tabular}
    };

    \node[minimum width=7em, draw, rounded corners=2mm,outer sep=0, label={[font=\small]30:$e_1$}] (r12) at (4, -1.25) {
        \begin{tabular}{l}
            $\icaskv{x}{0}{2}$
        \end{tabular}
    };

    \path (init) edge[->, soColor, right, transform canvas={xshift=3mm}] node [above, right, shift={(0,0.1)}] {$\so$} (r12);

    \path (init) edge[->, soColor, right, transform canvas={xshift=-3mm}] node [above, left, shift={(0,0.1)}] {$\so$} (r11);

    \path (init) edge[->,  transform canvas={xshift=1mm}, wrColor] node [black, right, shift={(0.3,0)}] {$\wro_{x}$} (r11);

    \path (r11) edge[->, wrColor] node [black, above] {$\wro_{x}$} (r12);

\end{tikzpicture}  
}
\caption{$e_0$ reads $0$, writes $1$; $e_1$ reads $1$ and does not write.}
\label{fig:example-history-faa:sc}
\end{subfigure}
\hfill
\begin{subfigure}[b]{.49\columnwidth}
\centering
\resizebox{\textwidth}{!}{
\begin{tikzpicture}[>=stealth,shorten >=1pt,auto,node distance=3cm,
    semithick, transform shape,
    B/.style = {%
    decoration={brace, amplitude=1mm,#1},%
    decorate},
    B/.default = ,  %
    ]

    \node[minimum width=7em, draw, rounded corners=2mm,outer sep=0, label={[font=\small]30:$\init$}] (init) at (2, 0) {$\{x: 0\}$};

    \node[minimum width=7em, draw, rounded corners=2mm,outer sep=0, label={[font=\small]150:$e_0$}] (r11) at (0, -1.25) {
        \begin{tabular}{l}
            $\ifaakv{x}{1}$
        \end{tabular}
    };

    \node[minimum width=7em, draw, rounded corners=2mm,outer sep=0, label={[font=\small]30:$e_1$}] (r12) at (4, -1.25) {
        \begin{tabular}{l}
            $\icaskv{x}{0}{2}$
        \end{tabular}
    };

    \path (init) edge[->, transform canvas={xshift=-1mm}, wrColor] node [black, left] {$\wro_{x}$} (r12);

    \path (init) edge[->, soColor, right, transform canvas={xshift=3mm}] node [above, right, shift={(0,0.1)}] {$\so$} (r12);

    \path (init) edge[->, soColor, right, transform canvas={xshift=-3mm}] node [above, left, shift={(0,0.1)}] {$\so$} (r11);

    \path (init) edge[->,  transform canvas={xshift=1mm}, wrColor] node [black, right, shift={(0.3,0)}] {$\wro_{x}$} (r11);

\end{tikzpicture}  
}
\caption{$e_0$ reads $0$ and writes $1$; $e_1$ reads $0$ and writes $2$.}
\label{fig:example-history-faa:no-sc}
\end{subfigure}

\caption{Two examples of histories for $\faacasStorage$. Arrows represent $\so$ and $\wro$ relations. The initial event $\init$ defines the initial state where $x$ is $0$. Events $e_0$ and $e_1$ execute a fetch-and-add and compare-and-swap respectively, at different replicas.%
}
\label{fig:example-history-faa}
\vspace{-4mm}
\end{figure}

\subsection{Abstract Executions}
\label{ssec:executions}

An \emph{abstract execution} of a distributed storage is a history with a \added{finite} set of events $E$ along with a relation $\rbo \subseteq E \times E$ called \emph{receive-before}, and a \deleted{well-founded }total order $\ar \subseteq E \times E$ called \emph{arbitration}. These relations are an abstraction of the internal communication behavior, i.e., the propagation of operation invocations between different replicas and conflict-resolution policies. The receive-before relation models information exchange between replicas and intuitively, an event $w$ is received-before an event $e$ on a replica $r$ if $w$ has been propagated to replica $r$ before executing $e$. 
The arbitration order represents a ``last-writer wins'' conflict resolution policy between concurrent events and the order in which events take effect in the storage for ``strong'' consistency models such as Sequential Consistency or Serializability. This order may be ignored by weaker consistency models, where a read is \emph{not} required to read from the latest update that precedes it in arbitration order, or by specific types of storage, e.g., CRDTs (see Section~\ref{sec:generalized-definitions}), where conflict resolution does not rely on the arbitration order. %

\begin{definition}
\label{def:execution}
An \emph{abstract execution} $\exec=\tup{h, \rbo, \ar}$ is a history $h = \tup{E, \so, \wro}$ along with an asymmetric, irreflexive relation \deleted{relation }\emph{receive-before} $\rbo\subseteq E \times E$ and a \deleted{well founded }strict total \emph{arbitration order} $\ar\subseteq E \times E$, such that:
\begin{enumerate}
	\item propagated updates are not ``forgotten'' within the same replica: $\rbo = \rbo ; \so^*$\footnote{\added{The symbol $;$ denotes the usual composition of relations}}, \label{def:execution:rbo-so}
	
	\item events at the same replica or events that are read are necessarily received-before, and $\ar$ is consistent with the receive-before relation: $\so \cup \wro \subseteq \rbo \subseteq \ar$.\label{def:execution:inclusions}
\end{enumerate}
$\exec$ is called an abstract execution of $h$.
\end{definition}

The conditions above are naturally satisfied by storages where replicas execute in a single process, values are not produced ``out of thin air'', \replaced{and the arbitration order is implemented using ``consistent'' timestamps, i.e. timestamps that do not contradict Lamport's clocks~\cite{Lamport1978} or causality. This is the case for implementations where ``ties'' between concurrent operations are solved based on replica IDs (assumed to be totally ordered), or when using timestamps from a (partially-)synchronized clock -- which is most often the case in practice.}{and the arbitration order is implemented using timestamps that are consistent with Lamport's clocks~\cite{Lamport1978} (which is most often the case in practice).}

For an event $e$, we use $e\in \exec$ to denote the fact that $e\in E$.

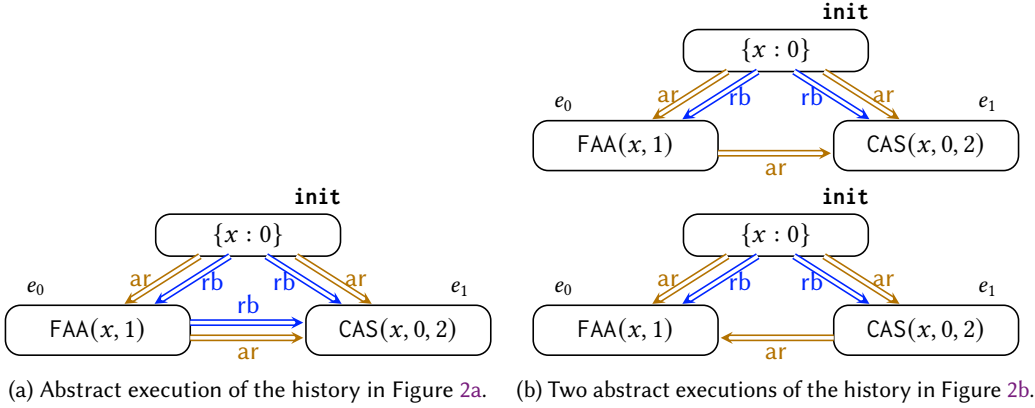
\begin{figure}[t]
\centering

\begin{subfigure}[b]{.495\textwidth}
\centering
\begin{tikzpicture}[>=stealth,shorten >=1pt,auto,node distance=3cm,
    semithick, transform shape,
    B/.style = {%
    decoration={brace, amplitude=1mm,#1},%
    decorate},
    B/.default = ,  %
    ]

    \node[minimum width=7em, draw, rounded corners=2mm,outer sep=0, label={[font=\small]30:$\init$}] (init) at (2, 0) {$\{x: 0\}$};

    \node[minimum width=7em, draw, rounded corners=2mm,outer sep=0, label={[font=\small]150:$e_0$}] (r11) at (0, -1.25) {
        \begin{tabular}{l}
            $\ifaakv{x}{1}$
        \end{tabular}
    };

    \node[minimum width=7em, draw, rounded corners=2mm,outer sep=0, label={[font=\small]30:$e_1$}] (r12) at (4, -1.25) {
        \begin{tabular}{l}
            $\icaskv{x}{0}{2}$
        \end{tabular}
    };

    \path (init) edge[->, double equal sign distance, arColor, right, transform canvas={xshift=-2mm}] node [above, left] {$\ar$} (r11);

    \path (init) edge[->,  transform canvas={xshift=2mm}, double equal sign distance, rbColor] node [black, right] {$\rbo$} (r11);

    \path (init) edge[->, double equal sign distance, arColor, transform canvas={xshift=2mm}] node [black, right] {$\ar$} (r12);

    \path (init) edge[->,  transform canvas={xshift=-2mm}, double equal sign distance, rbColor] node [black, left] {$\rbo$} (r12);

    \path (r11) edge[->, double equal sign distance, arColor, right, transform canvas={yshift=-1mm}] node [black, below] {$\ar$} (r12);

    \path (r11) edge[->,  transform canvas={yshift=1mm}, double equal sign distance, rbColor] node [black, above] {$\rbo$} (r12);

\end{tikzpicture}  
\caption{Abstract execution of the history in \Cref{fig:example-history-faa:sc}.}
\label{fig:example-exec-faa:sc}
\end{subfigure}
\hfill 
\begin{subfigure}[b]{.495\textwidth}
\centering
\begin{minipage}[b]{\textwidth}
\centering
\begin{tikzpicture}[>=stealth,shorten >=1pt,auto,node distance=3cm,
    semithick, transform shape,
    B/.style = {%
    decoration={brace, amplitude=1mm,#1},%
    decorate},
    B/.default = ,  %
    ]

    \node[minimum width=7em, draw, rounded corners=2mm,outer sep=0, label={[font=\small]30:$\init$}] (init) at (2, 0) {$\{x: 0\}$};

    \node[minimum width=7em, draw, rounded corners=2mm,outer sep=0, label={[font=\small]150:$e_0$}] (r11) at (0, -1.25) {
        \begin{tabular}{l}
            $\ifaakv{x}{1}$
        \end{tabular}
    };

    \node[minimum width=7em, draw, rounded corners=2mm,outer sep=0, label={[font=\small]30:$e_1$}] (r12) at (4, -1.25) {
        \begin{tabular}{l}
            $\icaskv{x}{0}{2}$
        \end{tabular}
    };

    \path (init) edge[->, double equal sign distance, arColor, right, transform canvas={xshift=-2mm}] node [above, left] {$\ar$} (r11);

    \path (init) edge[->,  transform canvas={xshift=2mm}, double equal sign distance, rbColor] node [black, right] {$\rbo$} (r11);

    \path (init) edge[->, double equal sign distance, arColor, transform canvas={xshift=2mm}] node [black, right] {$\ar$} (r12);

    \path (init) edge[->,  transform canvas={xshift=-2mm}, double equal sign distance, rbColor] node [black, left] {$\rbo$} (r12);

    \path (r11) edge[->, double equal sign distance, arColor, right, transform canvas={yshift=-1mm}] node [black, below] {$\ar$} (r12);

\end{tikzpicture}  
\end{minipage}
\begin{minipage}[b]{\textwidth}
\centering
\begin{tikzpicture}[>=stealth,shorten >=1pt,auto,node distance=3cm,
    semithick, transform shape,
    B/.style = {%
    decoration={brace, amplitude=1mm,#1},%
    decorate},
    B/.default = ,  %
    ]

    \node[minimum width=7em, draw, rounded corners=2mm,outer sep=0, label={[font=\small]30:$\init$}] (init) at (2, 0) {$\{x: 0\}$};

    \node[minimum width=7em, draw, rounded corners=2mm,outer sep=0, label={[font=\small]150:$e_0$}] (r11) at (0, -1.25) {
        \begin{tabular}{l}
            $\ifaakv{x}{1}$
        \end{tabular}
    };

    \node[minimum width=7em, draw, rounded corners=2mm,outer sep=0, label={[font=\small]30:$e_1$}] (r12) at (4, -1.25) {
        \begin{tabular}{l}
            $\icaskv{x}{0}{2}$
        \end{tabular}
    };

    \path (init) edge[->, double equal sign distance, arColor, right, transform canvas={xshift=-2mm}] node [above, left] {$\ar$} (r11);

    \path (init) edge[->,  transform canvas={xshift=2mm}, double equal sign distance, rbColor] node [black, right] {$\rbo$} (r11);

    \path (init) edge[->, double equal sign distance, arColor, transform canvas={xshift=2mm}] node [black, right] {$\ar$} (r12);

    \path (init) edge[->,  transform canvas={xshift=-2mm}, double equal sign distance, rbColor] node [black, left] {$\rbo$} (r12);

    \path (r12) edge[->, double equal sign distance, arColor, right, transform canvas={yshift=-1mm}] node [black, below] {$\ar$} (r11);

\end{tikzpicture}  
\end{minipage}

\caption{Two abstract executions of the history in \Cref{fig:example-history-faa:no-sc}.}
\label{fig:example-exec-faa:no-sc}
\end{subfigure}

\vspace{-1mm}

\caption{%
Abstract executions of the histories from \Cref{fig:example-history-faa}. Arrows represent $\ar$ and $\rbo$ relations. For readability, we omit the $\so$ and $\wro$ relations. %
The event $e_0$ is received-before executing $e_1$ in \Cref{fig:example-exec-faa:sc} but not in \Cref{fig:example-exec-faa:no-sc}. 
The arbitration relation is the same in both executions.}
\label{fig:example-executions}
\vspace{-4mm}
\end{figure}

\begin{example}
\Cref{fig:example-executions} shows abstract executions for the histories in \Cref{fig:example-history-faa}. In both cases, the receive-before relation includes only the $\wro$ dependencies which is anyway required by definition. Reading a value at some replica $r$ produced by an invocation $e$ at some other replica $r'$ should imply that $e$ propagated to $r$. On the left, the arbitration order includes just the $\wro$ dependencies which already ensure totality. On the right, \replaced{$\mathtt{FAA}$ and $\mathtt{CAS}$ are concurrent, i.e., both invocations were executed before either had a chance to propagate. We present the two possible arbitration orders. This shows that the arbitration order cannot be always determined based on the information exchanged between the replicas, i.e. by the receive-before.}{the $\mathtt{FAA}$ and $\mathtt{CAS}$ are ordered by $\ar$ even if they are ``concurrent'' meaning that both invocations were executed before either had a chance to propagate.}
\end{example}

The concept of abstract execution defined earlier is subsequently used to formalize the specifications of distributed storage systems. We will start with a so-called basic class that concerns ``single-object'' operations.

\section{\SimpleStorageFullName{} Storage Specifications}\label{sec:basic:spec}

We present a first class of storage specifications, called \emph{\simpleStorageFullName{}}, where operations read and/or write a \emph{single value} from/to a \emph{single object} (the operations in Example~\ref{ex:faacas} satisfy this assumption). We will present a more general framework with multi-object operations that read and/or write multiple values or objects in \Cref{sec:generalized-definitions}.

In general, a storage specification has two parts: a \emph{consistency model} characterizing the propagation of invocations between different replicas, and an \emph{operation specification} which defines object states and return values. The definition of consistency models builds on the work of \cite{DBLP:conf/cav/BouajjaniER25,DBLP:journals/pacmpl/BiswasE19} and the definition of operation specifications refines replicated data types as defined in~\cite{DBLP:journals/ftpl/Burckhardt14}. The first two subsections define these concepts for the class of operations mentioned above, and the last subsection formalizes the validity of an abstract execution w.r.t. such storage specifications.

\subsection{\SimpleConsistencyFullName{} Consistency Models}
\label{ssec:consistency}

In general, a consistency model is defined as a non-empty set of \emph{visibility} \replaced{formulas}{relations} that characterize the \emph{context} in which an event (invocation of an operation) is executed (abstractly speaking). The context of an event $e$ at a replica $r$ is defined as the set of events, potentially from other replicas, that propagated to $r$ prior to executing $e$. The notion of validity w.r.t. a consistency model defined later will require that the event $e$ which is read by another event $e'$ is the last in the arbitration order $\ar$ within the context of $e'$. This accurately models the Last-Writer-Wins conflict resolution policy (we consider other conflict resolution policies in \Cref{sec:generalized-definitions}). We define hereafter a class of so-called \simpleConsistencyFullName{} consistency models that will be extended later in \Cref{ssec:conflict-defs}.

Formally, a visibility \replaced{formula $\mathsf{v}$ describes}{relation $\mathsf{v}$ is} a binary relation between events which is parametrized by an object in $\Vars$. This is written as a predicate $\mathsf{v}_\key(e_1,e_2)$ meaning that $\mathsf{v}$ relates $e_1$ to $e_2$ for object $\key$ (explained below). A \emph{consistency model} (criterion) $\cmodel$ is a set of visibility \replaced{formulas}{relations}. 

For a consistency model $\cmodel$ and an abstract execution $\exec$, the context of an event $r$ for object $\key$ is the set of all events $e$ which are related to $r$ by some visibility \replaced{formula}{relation} in $\cmodel$ along with a projection of $\rbo$ and $\ar$ to this set of events, i.e.,
\begin{align}
\context{\cmodel}{\key}{\exec}{r}  = \tup{E_x, \rbo_{E_x\times E_x}, \ar_{E_x \times E_x}} \mbox{ with }
E_x = \{e \in \exec \ | \ \exists \mathsf{v} \in \cmodel. \ \mathsf{v}_\key(e,r)  \}
\label{eq:context} 
\end{align}

We use $\Contexts$ to denote the set of all possible contexts, i.e., tuples $\tup{E, \rbo_{E}, \ar_{E}}$ where $E$ is a finite set of events, $\rbo_E$ is an asymmetric, irreflexive relation over $E$, and $\ar_E$ is a strict total order over $E$, such that $\rbo_E\subseteq \ar_E$.

\emph{\SimpleConsistencyFullName{} visibility \replaced{formulas}{relations}} (used in \simpleConsistencyFullName{} consistency models) have the following form:
\begin{align}
\mathsf{v}_x(\event_0,\event_{n}) \Coloneqq \exists \event_1, \ldots, \event_{n-1}. & 
\bigwedge_{i=1}^{n} (\event_{i-1}, \event_{i}) \in \mathsf{Rel}^{\mathsf{v}}_i \ \land  
\writeVar{\event_0}{x} \land \wro_x^{-1}(\event_{n})\neq  \emptyset %
\label{eq:visibility-criterion} 
\end{align}

where each relation $\mathsf{Rel}^{\mathsf{v}}_i$, \replaced{$1 \leq i \leq n$,}{$1 \leq i \leq n + 1$} is defined by the grammar listed below:
\begin{align} \mathsf{Rel} \Coloneqq \idRelation\  |\  %
\so\  |\ \wro\  |\ \rbo\  |\ \ar\  |\ \mathsf{Rel} \cup \mathsf{Rel}\  |\ \mathsf{Rel} ; \mathsf{Rel}\  |\ \mathsf{Rel}^?\  |\ \mathsf{Rel}^+\  |\ \mathsf{Rel}^*
\label{eq:relations-visibility}
\end{align}

This formula states that $\event_0$ (which is $e$ in Eq.\ref{eq:context}) is connected to $\event_{n}$ (which is $r$ in Eq.\ref{eq:context}) by a path of dependencies that go through some intermediate events $\event_1, \ldots \event_{n-1}$ (all the $\event$ variables are interpreted as events). The constraint $\wro_x^{-1}(\event_{n}) \neq \emptyset $ asks that $\event_{n}$ reads the object $x$. %
Every relation used in the path is a composition of %
$\so, \wro,\rbo$ and $\ar$ via union $\cup$, composition of relations $;$, and 
transitive closure $^+$. $\mathsf{Rel}^?$ is syntactic sugar for $\idRelation \cup \mathsf{Rel}$, and $\mathsf{Rel}^*$ for $\idRelation \cup \mathsf{Rel}^+$. 
Since the grammar includes composition %
the existential quantifiers in Eq.\ref{eq:relations-visibility} do not increase expressivity (one could write $(\event_0,\event_{n + 1})\in \mathsf{Rel}^{\mathsf{v}}_1;\ldots;\mathsf{Rel}^{\mathsf{v}}_n$). These quantifiers are used to simplify proofs in Section~\ref{sec:afc-theorem}. 

\replaced{The predicate $\writeVar{\event_0}{x}$ means that $\event_0$ writes to object $x$, i.e., $\outputEventObj{e}{x} \downarrow$.}
{The predicate $\writeVar{\event_0}{x}$ states that intuitively, $\event_0$ writes to object $x$. We formalize its semantics in Section~\ref{ssec:validity} because it relies on operation specifications defined in Section~\ref{ssec:opspec}.}

We write $\mathsf{v}_x(e_0, \ldots e_n)$ whenever $\mathsf{v}_x(e_0, e_n)$ holds using the events $e_1, \ldots e_{n-1}$ to instantiate the existential quantifiers. The length of $\mathsf{v}_x$, denoted by $\length{\mathsf{v}_x}$, is the number of relations $\mathsf{Rel}^{\mathsf{v}}_i$ used in its definition ($n$ in \Cref{eq:visibility-criterion}).

As mentioned above, a \emph{\simpleConsistencyFullName{} consistency model} is a set of basic visibility \replaced{formulas}{relations}.

\begin{figure*}[t]
\resizebox{\textwidth}{!}{
\footnotesize
\begin{tabular}{|c|c|c|c|}
	\hline & & &\\
	
	\begin{subfigure}[b]{.24\textwidth}
		\centering
		\begin{tikzpicture}[->,>=stealth,shorten >=1pt,auto,node distance=1cm,
			semithick, transform shape]
			\node[transaction state, text=black] at (0,0)       (t_1)           {$e$};
			\node[transaction state] at (2,0)       (t_3)           {$\event_1$};
			\node[transaction state, text=black,label={above:\textcolor{black}{$\writeVar{ }{\xvar}$}}] at (-.5,1.5) (t_2) {$\event_0$};
			\path (t_1) edge[wrColor] node [black] {$\wro_x$} (t_3);
			\path (t_2) edge[bend left] node {$\so \cup \wro$} (t_3);
			\path (t_2) edge[left, dashed, double ,coColor] node {$\ar$} (t_1);
		\end{tikzpicture}
		\vspace{2mm}
		\parbox{\textwidth}{
			
			$\writeVar{\event_0}{x} \ \land \ \wro_x^{-1}(\event_1) \neq \emptyset \ \land$

			\hspace{2mm}$(\event_0, \event_1) \in \so \cup \wro$
			
		}
		
		\caption{$\axrvc$}
		\label{fig:rvc}
	\end{subfigure}
	
	&
	\begin{subfigure}[b]{.24\textwidth}
		\centering
		\begin{tikzpicture}[->,>=stealth,shorten >=1pt,auto,node distance=1cm,
			semithick, transform shape]
			\node[transaction state, text=black] at (0,0)       (t_1)           {$e$};
			\node[transaction state] at (2,0)       (t_3)           {$\event_1$};
			\node[transaction state, text=black,label={above:\textcolor{black}{$\writeVar{ }{\xvar}$}}] at (-.5,1.5) (t_2) {$\event_0$};
			\path (t_1) edge[wrColor] node [black] {$\wro_x$} (t_3);
			\path (t_2) edge[bend left, double equal sign distance, rbColor] node [black] {$\rbo^+$} (t_3);
			\path (t_2) edge[left, dashed, double ,coColor] node {$\ar$} (t_1);
		\end{tikzpicture}
		\vspace{2mm}
		\parbox{\textwidth}{
			
			$\writeVar{\event_0}{x} \ \land \ \wro_x^{-1}(\event_1) \neq \emptyset \ \land$

			\hspace{2mm}$(\event_0, \event_1) \in \rbo^+$
			
		}
		\caption{$\mathsf{Causal}$}
		\label{fig:cc}
		
	\end{subfigure}
	
	&

	\begin{subfigure}[b]{.24\textwidth}
		\centering
		\begin{tikzpicture}[->,>=stealth,shorten >=1pt,auto,node distance=4cm,
			semithick, transform shape]
		\node[transaction state, text=black] at (0,0)       (t_1)           {$e$};
		\node[transaction state] at (2,0)       (t_3)           {$\event_1$};
		\node[transaction state, text=black,label={above:\textcolor{black}{$\writeVar{ }{x}$}}] at (-0.5,1.5) (t_2) {$\event_0$};
		\node[transaction state] at (1.5,1.5) (t_4) {$\bullet$};
		\path (t_1) edge[wrColor] node [black] {$\wro_x$} (t_3);
		\path (t_2) edge[double equal sign distance, coColor] node [black] {$\ar^*$} (t_4);
		\path (t_4) edge[left] node {$(\so \cup \wro)$} (t_3);
		\path (t_2) edge[left, dashed, double equal sign distance, coColor] node {$\ar$} (t_1);
		\end{tikzpicture}
		\vspace{2mm}
		\parbox{\textwidth}{
			
			$\writeVar{\event_0}{x} \ \land \ \wro_x^{-1}(\event_1) \neq \emptyset \ \land$

			\hspace{2mm}$(\event_0, \event_1) \in \ar^* ; (\so \cup \wro)$
			
		}
		
		\caption{$\mathsf{Prefix}$}
		\label{fig:prefix}
	\end{subfigure}

	&

	\begin{subfigure}[b]{.24\textwidth}
		\centering
		\begin{tikzpicture}[->,>=stealth,shorten >=1pt,auto,node distance=1cm,
			semithick, transform shape]
			\node[transaction state, text=black] at (0,0)       (t_1)           {$e$};
			\node[transaction state] at (2,0)       (t_3)           {$\event_1$};
			\node[transaction state, text=black,label={above:\textcolor{black}{$\writeVar{ }{\xvar}$}}] at (-.5,1.5) (t_2) {$\event_0$};
			\path (t_1) edge[wrColor] node [black] {$\wro_x$} (t_3);
			\path (t_2) edge[bend left, coColor, double] node {$\ar$} (t_3);
			\path (t_2) edge[left, dashed, double ,coColor] node {$\ar$} (t_1);
		\end{tikzpicture}
		\vspace{2mm}
		\parbox{\textwidth}{
			
			$\writeVar{\event_0}{x} \ \land \ \wro_x^{-1}(\event_1) \neq \emptyset \ \land$

			\hspace{2mm}$(\event_0, \event_1) \in \ar$

		}
		
		\caption{$\mathsf{SC}$ / $\mathsf{SER}$}
		\label{fig:ser}
	\end{subfigure}

\\\hline
\end{tabular}
}
  \vspace{-2mm}
\caption{Visibility \replaced{formulas}{relations} defining the homonymous consistency models \emph{Return-Value Consistency} ($\RVC$, \Cref{fig:rvc}), %
\emph{Causal Consistency} ($\CC$, \Cref{fig:cc}), \emph{Prefix Consistency} ($\PC$, \Cref{fig:prefix}) and \emph{Sequential Consistency}/\emph{Serializability} ($\SC / \SER$, \Cref{fig:ser}). Solid edges describe the dependencies linking $\event_0$ and $\event_1$. We include the $\wro_x$ edge (and its source $e$) as a visualization of the constraint $\wro_x^{-1}(\event_1) \neq \emptyset$. Dashed $\ar$ edges are not part of the visibility \replaced{formulas}{relations}. These capture the Last-Writer-Wins conflict resolution policy discussed later, requiring that the event $e$ being read succeeds all other events from the context in $\ar$.}
\label{fig:consistency_defs}
\vspace{-3mm}
\end{figure*}
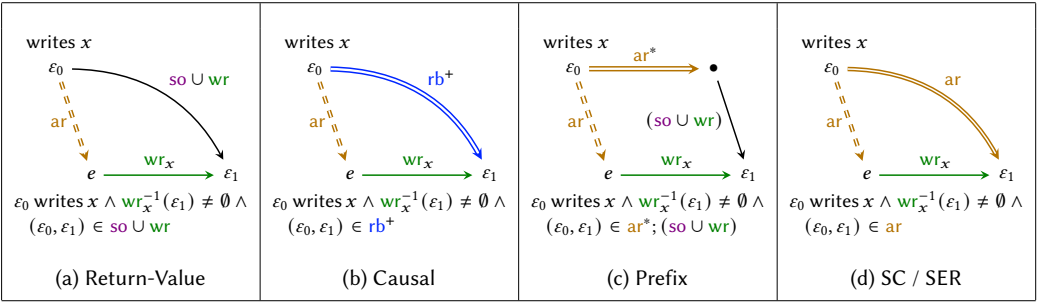

\Cref{fig:consistency_defs} describes several visibility \replaced{formulas}{relations} and \added{their corresponding} consistency models, inspired by Biswas et al.~\cite{DBLP:journals/pacmpl/BiswasE19}. The dashed $\ar$ edges (leading to $e$) should be ignored for now. \replaced{Basic visibility formulas constrain events w.r.t. a single object -- $x$. Later, we will define consistency models whose visibility formulas can impose additional constraints that concern multiple objects.}{Each of these models contains a single visibility relation. We discuss ``non-singleton'' consistency models later in the paper.}

We say that a consistency model $\cmodel_1$ is \emph{weaker than} another consistency model $\cmodel_2$, denoted $\cmodel_1 \preccurlyeq \cmodel_2$ if intuitively, the context of any event w.r.t. $\cmodel_1$ is larger than the context w.r.t. $\cmodel_2$. Formally, $\cmodel_1 \preccurlyeq \cmodel_2$ iff for every abstract execution $\exec$, event $e \in \exec$ and object $x$, $\context{\cmodel_1}{x}{\exec}{e} \subseteq \context{\cmodel_2}{x}{\exec}{e}$ holds.
$\cmodel_1$ and $\cmodel_2$ are \emph{equivalent}, denoted $\cmodel_1 \equiv \cmodel_2$, when $\cmodel_1 \preccurlyeq \cmodel_2$ and $\cmodel_2 \preccurlyeq \cmodel_1$.

We assume that every consistency model $\cmodel$ includes a visibility \replaced{formula}{relation} $\mathsf{v}^{\so}_\key$ (resp. $\mathsf{v}^{\wro}_x$) such that $\so \subseteq \mathsf{v}^{\so}_\key$ (resp. $\wro_x \subseteq \mathsf{v}^{\wro}_\key$) for every object $\key\in\Vars$. The constraint $\so \subseteq \mathsf{v}^{\so}_\key$ corresponds to the so-called "read-my-own-writes" consistency (i.e., an event "observes" every preceding event at the same replica) and $\wro_x \subseteq \mathsf{v}^{\wro}_\key$ is a ``well-formedness'' constraint since visibility \replaced{formulas}{relations} will constrain the write-read relation in a history (see \Cref{def:valid}).

All consistency models in \Cref{fig:consistency_defs} trivially satisfy this constraint as for any abstract execution, $\so \cup \wro \subseteq \rbo \subseteq \ar$. $\RVC$ is the weakest consistency model that our framework can describe.

\subsection{\SimpleStorageFullName{} Operation Specifications}
\label{ssec:opspec}

While visibility \replaced{formulas}{relations} define the context of an invocation in terms of prior invocations, the effect of an invocation is defined using the following semantical functions: $\mathsf{rspec}$ says whether an event reads an object or not, and $\mathsf{wspec}$ defines the value written by the invocation, if any. The written value may depend on the value read by the event in the case of atomic read writes like $\mathtt{FAA}$ and $\mathtt{CAS}$. 
Concerning notations, for a partial function $f:A\pto B$, we use $f(a)\downarrow$ to say that $f$ is defined for $a\in A$, and $f(a)\uparrow$, otherwise.
Similarly, for a predicate $p$ over some set $A$, we use $p(a)\downarrow$ to say that $p$ is true for $a$, and $p(a)\uparrow$, otherwise.

A \emph{\simpleStorageFullName{} read specification} $\mathsf{rspec}$ is a predicate over $\Events$. For example, \Cref{eq:rspec-faa-simple} describes the read specification of $\faacasStorage$. We say that an event $e$ is a \emph{read} event if $\mathsf{rspec}(e) \downarrow$, and in such case, we say that $\readVar{e}{\varOf{e}}$. %
\begin{equation}
\label{eq:rspec-faa-simple}
{\arraycolsep=1.5pt
\rspecSimple{r} = \btrue \text{ iff } \operationEvent{r} = \ireadkvName, \ifaakvName, \icaskvName}
\end{equation}

\sloppy A \emph{\simpleStorageFullName{} write specification} $\mathsf{wspec}$ is a partial function $\mathsf{wspec} : \Events \pto \Vals \pto \Vals$, that associates non-initial events to partial functions that map a read value to a value to be written. %
For example, \Cref{eq:wspec-faa-simple} describes the write specification of $\faacasStorage$.
\begin{equation}
\label{eq:wspec-faa-simple}
\wspecSimple{w}{v} = \left\{ 
    \begin{array}{ll}
        v' & \text{if } w = \iwritekv{x}{v'}\\
        v + v' & \text{if } w = \ifaakv{x}{v'}\\
        v'' & \text{if } w = \icas{x}{v'}{v''} \land v = v'\\
        \mathsf{undefined} & \text{otherwise}
    \end{array} \right.
\end{equation}

For an event $e$, we say that $e$ is a \emph{write} event if $\mathsf{wspec}(e) \downarrow$. We assume that if $\mathsf{wspec}(e) \downarrow$, then the function $\mathsf{wspec}(e): \Vals \pto \Vals$ is defined for at least one value. We say that $\writeVarValue{e}{x}{v}$ if $x = \varOf{e}$ and $\wspecSimple{e}{v}\downarrow$. We assume that every value $v$ can \emph{enable} at least one event to write, i.e., there exists $e\in \Events$ s.t. $\wspecSimple{e}{v} \downarrow$. \added{We also assume that if $e$ is a write event but it is not a read event, e.g., a $\iputName$ invocation, then $\wspecName(e)$ is a total constant function, i.e. $\wspecName(e) : \Vals \to \Vals$ and $\wspecName(e)(v_1)=\wspecName(e)(v_2)$ for all $v_1$, $v_2$.} 
\begin{definition}
\sloppy A \emph{\simpleStorageFullName{} operation specification} is a tuple $\opspec=(E,\rspecName, \wspecName)$ where $E$ is a set of events, such that $\varOf{e}$ is a singleton for every $e\in E$. %
\end{definition}

We use $\SpecEvents{\opspec}$ to refer to the set of events $E$ in an operation specification.

\noindent
\added{\textbf{Operation Closure.}}
\label{ssec:basic-operation-closure}
\replaced{We}{Next, we} define some natural assumptions about \simpleStorageFullName{} operation specifications (it is easy to check that they hold on the $\faacasStorage$ example with the definitions in  \Cref{eq:rspec-faa-simple}  and  \Cref{eq:wspec-faa-simple}).
We assume that $E$ contains at least one read and one write event. We also assume that all objects support a common set of operations with identical read and write behavior, and that these operations can be executed at any replica. Formally, for every event $e\in E$, replica $\replicaID$, identifier $\identifier$ and object $x$ there exists an event $e'\in E$ s.t. $\replicaEvent{e'} = \replicaID$, $\identifierEvent{e'} = \identifier$, $\varOf{e'} = x$, $\rspecSimple{e'} = \rspecSimple{e}$ and $\mathsf{wspec}(e') = \mathsf{wspec}(e)$.

\noindent
\added{\textbf{(Conditional) Read-Write Events.}}
\label{ssec:basic-conditional-read-writes}
We say that $\opspec$ allows read-writes if $E$ contains an event that is a read and a write event at the same time (e.g., $\mathtt{FAA}$ and $\mathtt{CAS}$ invocations); we call such events \emph{read-write} events. If $\opspec$ allows read-writes, then we assume that every value can \emph{enable} some read-write to write, i.e., for every value $v$, $E$ contains a read-write event $e$ s.t. $\wspecSimple{e}{v}\downarrow$. \added{As an example, this condition is not satisfied by a storage with only $\igetName$ and $\icsetz$ operations ($\icsetz$ writes 1 if it reads 0 and nothing otherwise). Indeed, value 1 cannot enable any write.} 

A read-write event is called \emph{unconditional} if for every value $v$, $\wspecSimple{e}{v}\downarrow$ and \emph{conditional} otherwise. For example, a $\mathtt{FAA}$ invocation is unconditional and a $\mathtt{CAS}$ invocation is conditional.
We assume that if $\opspec$ allows conditional writes, then every value $v$ can \emph{disable} some conditional read-write to write, i.e., $E$ contains a conditional read-write event s.t. $\wspecSimple{e}{v}\uparrow$.

\subsection{Validity w.r.t. Basic Storage Specifications}
\label{ssec:validity}

A \emph{\simpleStorageFullName{} storage specification} is a pair $\sspec = (\cmodel, \opspec)$ where $\cmodel$ is a \simpleConsistencyFullName{} consistency model and $\opspec$ is a \simpleStorageFullName{} operation specification. Next, we formalize the validity of an abstract execution w.r.t. a \simpleStorageFullName{} storage specification.

The interpretation of a \simpleConsistencyFullName{} visibility \replaced{formula}{relation} $\mathsf{v}_x(\event_0,\event_{n})$ on an abstract execution $\exec$ is defined as expected. \deleted{We only need to describe the interpretation of the predicate $\writeVar{e}{x}$, which is more involved for conditional writes such as $\icaskvName$ whose write behavior depends on the value they read.}\deleted{To this, we define a function $\valuewr{w}{x}$ which returns the value written by the event $w$ to $x=\varOf{w}$, if any, and it is undefined otherwise:}
\ifbool{diffMode}{%
\begin{equation}
\label{eq:value-wr}
\deleted{
\valuewr{w}{x} = \left\{\begin{array}{ll}
    v_x^0 & \text{if } w \text{ is an initial event} \\
	\wspecSimple{w}{v} &  w \text{ is not an initial event }, \wro_x^{-1}(w) = \emptyset \\
    & \quad \text{and } \wspecSimple{w}{v} \downarrow\text{ for some $v$}\\
    \wspecSimple{w}{v_x^w} & w \text{ is not an initial event }, \{w'\} =  \wro_x^{-1}(w), \\
    & \quad v_x^w = \valuewr{w'}{x}  \text{ and }\wspecSimple{w}{v_x^w} \downarrow\\
    \mathsf{undefined} & \text{otherwise}
	\end{array}\right.}
\end{equation}
}{}%

\deleted{Above, $\valuewr{w}{x}$ is defined if (1) 
$w$ is an initial event and then the output is some fixed initial value $v_x^0$, (2) $w$ is an unconditional write ($w$ is an initial event or $\mathsf{wspec}(w)$ is a constant function) or (3) $w$ is a conditional write and $\wspecSimple{w}{v_x^w}$ is defined for the value $v_x^w$ that $w$ reads. The read value $v_x^w$ is obtained by recursively calling $\mathtt{value}_{\wro}$ on the write read by $w$ (since this can be again a conditional write). Note that this recursion terminates because $\wro$ is an acyclic relation (cf. Definition~\ref{def:history}).} 

\deleted{The predicate $\writeVar{e}{x}$ is interpreted as true iff $\valuewr{e}{x}$ is defined.}

\ifbool{diffMode}{
\begin{example}
\deleted{We compute the function $\mathsf{value}$ for the histories $h_1 = \tup{E^1, \so^1, \wro^1}$ and $h_2 = \tup{E^2, \so^2, \wro^2}$ from Figures\ref{fig:example-history-faa:sc} and \ref{fig:example-history-faa:no-sc} respectively.} 

\deleted{In $h_1$, the initial value written by $\init$ on $x$ is $0$. Hence, $\valuewr[\wro^1]{\init}{x} = 0$. The event $e_0$ reads $x$ from $\init$. Since $e_0$ is an $\ifaakvName$ event, $\valuewr[\wro^1]{e_0}{x} = \wspecSimple{e_0}{0} = 0 + 1 = 1$. The event $e_1$ reads $x$ from $e_0$. Since $e_1$ is a $\icaskvName(x,0,2)$ event, $\valuewr[\wro^1]{e_1}{x}$ is undefined: $\valuewr[\wro^1]{e_0}{x} = 1$ and $\wspecSimple{e_1}{1} \uparrow$. Hence, $e_1$ does not write on $x$.} 

\deleted{In $h_2$, $e_0$ reads the same value from $\init$ and hence, $\valuewr[\wro^2]{e_0}{x} = 1$. However, $e_1$ reads from $\init$ in $h_2$. In this case, as $\valuewr[\wro^2]{\init}{x} = 0$ and $\wspecSimple{e_1}{0} = 2$, $\valuewr[\wro^2]{e_1}{x} = 2$.}
\end{example}
}{}%

\begin{definition}\label{def:valid}
Let $\sspec = (\cmodel, \opspec)$ be a \simpleStorageFullName{} storage specification. %
An abstract execution $\exec = \tup{h, \rbo, \ar}$ of a history $\hist=\tup{E, \so, \wro}$ is \emph{valid} w.r.t. $\sspec$ iff 
\begin{itemize}
\item it contains events from the operation specification, i.e., $E \subseteq \SpecEvents{\opspec}$,

\item \added{the write-read dependencies of each event $e \in E$ for object $x$ satisfy the following:}
\begin{itemize}
    \item \added{if $e$ reads object $x$, i.e. $\rspecSimple{e} \downarrow$ and $x \in \varOf{e}$, $e$ reads from the write event in its context that is maximal w.r.t. the arbitration order: $\wro_x^{-1}(e) = \{w_x^e\}$,}
    \item \added{if $e$ does not read object $x$, i.e. $\rspecSimple{e} \uparrow$ or $x \not\in \varOf{e}$,  then $\wro_x^{-1}(e) = \emptyset$.}
\end{itemize}
\item \added{the value written by each event $e \in E$ to object $x$ is consistent with $\wspecName$:}
\begin{itemize}
    \item \added{if $e$ reads object $x$, i.e. $\rspecSimple{e} \downarrow$ and $x \in \varOf{e}$, then it writes based on the value read: $\outputEventObj{e}{x} = \wspecSimple{e}{\outputEventObj{w_x^e}{x}}$}\footnote{\added{Since $\outputEventNameFunction$ and $\wspecName$ are partial functions, the equality also means that the left side is defined iff the right side is defined.}},
    \item \added{if $e$ does not read object $x$, i.e. $\rspecSimple{e} \uparrow$ or $x \not\in \varOf{e}$, then $\outputEventObj{e}{x} = \wspecName(e)(\_)$}\footnote{\added{$\_$ represents any value in $\Vals$.}},
\end{itemize}
\added{where $w_x^e = \max_{\ar} \context{\cmodel}{x}{\exec}{e}$.}\deleted{for every event $r\in E$ that accesses object $x$, if $r$ is a read event, then it reads from the write event in its context that is maximal w.r.t. the arbitration order, and otherwise, it does not read from any event. Formally, if $\rspecSimple{r} \downarrow$ then $\wro_x^{-1}(r) = \{\max_{\ar} \context{\cmodel}{x}{\exec}{r}\}$, and $\rspecSimple{r} \uparrow$ implies $\wro_x^{-1}(r) = \emptyset$.}

\end{itemize}
A history $\hist$ is valid w.r.t. $\sspec$ iff there exists an abstract execution of $\hist$ which is valid w.r.t. $\sspec$.
\label{def:consistency}
\end{definition}

Recall that the $\mathtt{value}$ function, and \replaced{implicitly}{implicitely}, the operation specification, are used to interpret the visibility \replaced{formulas}{relations} of $\cmodel$ and thus define invocation contexts.

\begin{example}
The abstract executions described in \Cref{fig:example-executions} are both valid w.r.t. $(\CC, \faacasStorage)$ as every event which is read is also received-before ($\wro \subseteq \rbo$). However, only \Cref{fig:example-exec-faa:sc} is valid w.r.t. $(\SC, \faacasStorage)$. In \Cref{fig:example-exec-faa:sc}, $e_1$ reads from the writing event that precedes it w.r.t. $\ar$. On the other hand, in \Cref{fig:example-exec-faa:no-sc}, $e_1$ reads $x$ from $\init$ and not from $e_0$ which is its maximal visible event w.r.t. $\ar$ that writes $x$. 
Moreover, by the symmetry between $e_0$ and $e_1$, it can be proven that any abstract execution of such history is not valid w.r.t. $(\SC, \faacasStorage)$. 
\end{example}

	\section{
Programs and Storage Implementations}
\label{sec:lts}

We model programs accessing a storage and storage implementations using \emph{Labeled Transition Systems (LTSs)}. Their interaction via invocations of operations will be defined as the usual parallel composition of LTSs. 
We also present the notions of availability and validity of a storage implementation, key to the AFC theorem.

\subsection{Labeled Transition Systems}

An LTS $L = (S, A, s_0, \Delta)$ is a tuple formed of a (possibly infinite) set of \emph{states} $S$, a set of \emph{actions} $A$, an \emph{initial state} $s_0 \in S$ and a (partial) \emph{transition function} $\Delta : S \times A \pto S$. An \emph{execution} of $L$ is an alternating sequence of states and actions $\rho = s_0, a_0,s_1,a_{1},s_2,\ldots$ such that $\Delta(s_i, a_i)=s_{i+1}$ for each $i$. A state $s$ is \emph{reachable} if there exists an execution ending in $s$. 
A \emph{trace} of an execution $\rho$ is the projection of $\rho$ over actions (the maximum subsequence of $\rho$ formed of actions). %
The final state of a finite trace $t$, denoted by $\stateTrace{t}$, is the last state of $\rho$. The set of all traces of $L$ is denoted by $\Traces[L]$.
An LTS is \emph{finite} if all its traces are finite.
For any finite trace $t$ and action $a$, $\Delta(t, a)$ is defined as $\Delta(\stateTrace{t}, a)$. If $\Delta(t, a) \downarrow$, then $t \oplus a$ is defined by appending $a$ to $t$.

Let $L_1 = (S_1, A_1, s_0^1, \Delta_1)$ and $L_2 = (S_2, A_2, s_0^2, \Delta_2)$ be two LTSs. We define a parallel composition operator between $L_1$ and $L_2$ that is parametrized by a partial function $\pi: A_1 \pto A_2$. This function allow us to define a relationship between a subset of $A_1$ and a subset of $A_2$, called \emph{synchronized actions} of $L_1$ and $L_2$. The set of actions $a \in A_1$ for which $\pi(a)$ is not defined (resp. actions $a \in A_2$ for which $\pi^{-1}(a)$ is not defined) are the \emph{local actions} of $L_1$ (resp. $L_2$). \added{Without loss of generality, we assume that the set of local actions of $L_1$ and $L_2$ are disjoint.}

The parallel composition of $L_1$ and $L_2$ w.r.t. $\pi$ is the LTS $L_1 \customparallel_{\pi} L_2  = (S, A, s_0, \Delta)$ where $S = S_1 \times S_2$, $A = A_1 \cup A_2$, $s_0 = (s_0^1, s_0^2)$, and $\Delta$ is defined as follows:
\begin{align*}
{\arraycolsep=0pt
\begin{array}{lll}
   \Delta((s_1, s_2), a) & \Coloneqq 
   & \left\{ 
   {\arraycolsep=2pt
   \begin{array}{ll}
   (\Delta(s_1, a), \Delta(s_2, \pi(a))) & \text{if } a \in A_1, \pi(a) \downarrow, \Delta(s_1, a) \downarrow, 
    \text{ and }  \Delta(s_2, \pi(a)) \downarrow\\[.5mm]
   (\Delta(s_1, a), s_2) & \text{if }  a \in A_1, \pi(a) \uparrow,\text{ and }\Delta(s_1, a) \downarrow \\[.5mm]
   (s_1, \Delta(s_2, a)) & \text{if } a \in A_2, \pi^{-1}(a) \uparrow,\text{ and } \Delta(s_2, a) \downarrow\\[.5mm]
   \mathsf{undefined} & \text{otherwise}
   \end{array}} \right.
\end{array}
}
\end{align*}
\added{(note the asymmetry due to using the function $\pi$). Whenever there is no ambiguity w.r.t. $\pi$ we simply write} $L_1 \customparallel L_2$.

\subsection{Programs and Storage Implementations}

Let $E$ be a set of events.
A \emph{program} over $E$ is an LTS $\programInstance=\program$ such that $E \subseteq A_\mathsf{p}$. Intuitively, this LTS models all possible interleavings between invocations on different replicas. Actions in $A_\mathsf{p}\setminus E$ represent computation steps performed by the program locally, before or after invoking operations on the storage. Also, to simplify the technical exposition, we do not consider separate transitions for calling and returning from a storage operation. Intuitively, the transitions labeled by events occur at the return time.

A \emph{storage implementation} over $E$ is an LTS $\implementationInstance=\implementation$ such that $A_\mathsf{i}$ contains (1) an arbitrary set of local actions (representing computation/communication steps internal to the storage), and (2) pairs of events in $E$ and their read-dependencies, i.e., pairs $(e, m)$ where $e \in E$ and $m : \Vars \pto \mathcal{P}(E)$. %
Intuitively, $m$ represents the \deleted{(possibly undefined) }write-read dependencies of $e$. We also assume that each action includes an identifier, denoted by $\identifierEvent{a}$, so that along an execution every action occurs only once. For any action $a = (e, m)$, $\eventExecuted{a}$ and $\writesSet{a}$ denote the event $e$ and the write-read dependencies $m$ respectively. Also, $\operationEvent{a}= \operationEvent{e}$ is the operation type of $a$. To model communication, we assume that $A_\mathsf{i}$ includes two types of local actions, $\esend$ actions for sending a message (from one replica to another) and $\ereceive$ to receive a message. %

The formalization of send/receive actions is straightforward and we omit it. We will say that a send action \emph{matches} a receive action if they concern precisely the same message (messages are associated with unique identifiers).
For any $\esend$, resp., $\ereceive$, action $a$ at some replica $r$, $\receiveSet{a}$ denotes the set of events that $r$ sends in this message, resp., that $r$   receives in this  message. We assume that if a trace $t$ contains any such action, for every event $e \in \receiveSet{a}$ there must exist an action $(e, \_)$ preceding $a$ in $t$. As expected, if $a_s$ and $a_r$ match, then $\receiveSet{a_s} = \receiveSet{a_r}$.  %

For any action $a\in A_\mathsf{p}\cup A_\mathsf{i}$, $\replicaEvent{a}$ denotes the replica executing $a$.

The interaction between a storage implementation $\implementationInstance$  and a program $\programInstance$ is defined as their \added{asymmetric} parallel composition w.r.t. a partial function $\pi:A_\mathsf{i}\pto A_\mathsf{p}$ which is defined only for actions of the form $(e, m)$ (as described above) by $\pi(e,m)=e$. The program and the storage implementation synchronize on events representing operation invocations. It is denoted by $\parallelCompositionInstance$. By definition, traces of $\parallelCompositionInstance$ include actions of the form $(e, m)$ (coming from $A_\mathsf{i}$), and local actions of $\programInstance$ or $\implementationInstance$.

Traces of $\implementationInstance$ (or $\parallelCompositionInstance$) induce histories and abstract executions. The \emph{induced history} of a trace $t$ of $\implementationInstance$ (or $\parallelCompositionInstance$) is the history $h = \tup{E^t, \so^t, \wro^t}$ where $E^t$ is the set events $e$ such that some action $a_e=(e,m)$ occurs in $t$, $\so^t$ orders events from the same replica as they occur in $t$, and for every object $x$ and event $e$, $(\wro^t_x)^{-1}(e) = \mathsf{W}$ iff $\writesSet{a_e} = (x, \mathsf{W})$ ($a_e$ is the action that contains $e$). We implicitly assume that for any event $e \in E$ different from $\init$, $(\init, e) \in \so^t$. We use $\inducedHistory{t}$ to denote the induced history of a trace $t$.

The \emph{induced receive-before} of a trace $t$ of $\implementationInstance$ (or $\parallelCompositionInstance$) is the relation $\rbo^t$ over events induced by the matching relation between sends and receives: $(e, e') \in \rbo^t$ iff $(e, e') \in \so^t$ or there exists matching $\esend$ and $\ereceive$ actions, $a_s$,  $a_r$ and a synchronized action $a = (e', \_)$ s.t. %
$\replicaEvent{a_r}=\replicaEvent{a}$, $a_r$ occurs before $a$ in $t$, and $e \in \receiveSet{a_s}$ (which coincides with $\receiveSet{a_r}$).%

\replaced{
A trace $t$ of $\implementationInstance$ also induces a set of abstract executions of the form $\exec=\tup{\inducedHistory{t}, \rbo^t, \ar^t}$
where $\ar^t$ is any total order between the events in $\exec$ that is consistent with $\rbo^t$, i.e., $\rbo^t\subseteq \ar^t$ (to satisfy the requirements in \Cref{def:execution}).}
{The induced abstract execution of a trace $t$ of $\implementationInstance$ (or $\parallelCompositionInstance$), denoted by $\execTrace{t}$, is the set of tuples $\execTrace{t} = \tup{\inducedHistory{t}, \rbo^t, \ar^t}$ where $\rbo^t$ is the induced receive before of $t$ and $\ar^t$ is the order between events as they occur in $t$.
We only consider storage implementations s.t. for every trace $t$, $\execTrace{t}$ satisfies the requirements in \Cref{def:execution}.}

\subsection{Availability and Validity of a Storage Implementation}

We say that a storage implementation $\implementationInstance$ is \emph{available} if, intuitively, every execution of $\implementationInstance$ terminates when interacting with a finite program $\programInstance$ \added{(executing a single synchronized action does not make a replica enter an infinite loop of local steps)}, and no invocation is delayed due to a replica waiting for messages.

We say that a replica $r \in \ReplicaID$ is \emph{waiting} in a trace $t$ of some composition $\parallelCompositionInstance$ if
\begin{itemize}
	\item the program can execute some action at replica $r$: there is an action $a\in A_\mathsf{p}$ s.t. $\replicaEvent{a} = r$ and $\Delta_{\programInstance}(t', a) \downarrow$; where $t'$ is obtained from $t$ by removing all local actions of $\implementationInstance$ and replacing every action $(e',m)$ with $e'$, and
	\item the only actions of replica $r$ that the parallel composition can execute are $\ereceive$ actions: for every action $a \in A_\mathsf{p}\cup A_\mathsf{i}$ s.t. $a$ is not a $\ereceive$ action and $\replicaEvent{a} = r$, $\Delta_{\parallelCompositionInstance}(t, a) \uparrow$\added{.}
\end{itemize}
Note that the latter implies that the action $a$ that $\programInstance$ can execute after $t'$ is necessarily an event in $E$ (otherwise, $a$ is a local action of $\programInstance$ and the parallel composition could execute it).

\begin{definition}
\label{def:consistently-available}
An implementation $\implementationInstance$ is \emph{available} if the following hold:
\begin{itemize}
\item for every finite program $\programInstance$, the composition $\parallelCompositionInstance$ is also finite, and
\item for every program $\programInstance$ and every trace $t$ of $\parallelCompositionInstance$, there is no replica waiting in $t$.
\end{itemize}
\end{definition}

Given a storage specification $\sspec$ over a set of events $E$, a storage implementation $\implementationInstance$ is \emph{valid w.r.t. $\sspec$} if 
\replaced{every trace $t$ induces some abstract execution which}
{the induced abstract execution of every trace $t$}
is valid w.r.t. $\sspec$.%
An implementation valid w.r.t. $\sspec$ is simply called a \emph{$\sspec$-implementation} %
(or implementation of $\sspec$).

	\section{The Basic Arbitration-Free Consistency Theorem}
\label{sec:afc-theorem}

We present a simpler instance of our main result (the AFC theorem) for \simpleStorageFullName{} storage specifications.

To simplify the statement of the theorem, we define a normal form for \simpleConsistencyFullName{} consistency models w.r.t. a \simpleStorageFullName{} operation specification $\opspec$. A visibility \replaced{formula}{relation} is called \emph{simple} if it does \emph{not} use composition operators between relations, i.e., the grammar in \Cref{eq:relations-visibility} is replaced by:
$\mathsf{Rel} \Coloneqq \ \so\  |\ \wro\  |\ \rbo\  |\ \ar$. Also, a visibility \replaced{formula}{relation} $\mathsf{v}$ from a consistency model $\cmodel$ is called \emph{vacuous} w.r.t. $\opspec$ iff for every abstract execution $\exec$, $\exec$ is valid w.r.t. $(\cmodel, \opspec)$ iff $\exec$ is valid w.r.t. $(\cmodel\setminus\{\mathsf{v}\}, \opspec)$. 
For example, if $\mathsf{Rel}_i^\mathsf{v}$ and $\mathsf{Rel}_{i+1}^\mathsf{v}$ in \Cref{eq:visibility-criterion} are $\wro$ (for some $i$), then any instance of $\event_i$ must be an invocation of a read-write that both reads and writes. If the operation specification does not include read-writes (e.g., a key-value store with only \texttt{PUT} and \texttt{GET} operations), such visibility \replaced{formulas}{relations} are vacuous.

\begin{definition}
\label{def:normal-form}
A \simpleConsistencyFullName{} consistency model $\cmodel$ is called in \emph{normal form w.r.t. a basic operation specification $\opspec$} if it contains only simple visibility \replaced{formulas}{relations} and no visibility \replaced{formula}{relation} from $\cmodel$ is vacuous w.r.t. $\opspec$. 
\end{definition}

A normal form of a \simpleConsistencyFullName{} consistency model $\cmodel$ w.r.t. $\opspec$ is any \simpleConsistencyFullName{} consistency model $\cmodel'$ in normal form, such that for every abstract execution $\exec$, $\exec$ is valid w.r.t. $(\cmodel, \opspec)$ iff $\exec$ is valid w.r.t. $(\cmodel', \opspec)$. \addedAppendix{\Cref{app:finite-normal-consistency}}\addedJournal{\textit{the arXiv}} shows that every \simpleConsistencyFullName{} consistency model $\cmodel$ has a normal form. A normal form can be obtained by replacing each visibility \replaced{formula}{relation} $\mathsf{v}$ with an equivalent \added{(possibly infinite)} set of simple visibility \replaced{formulas}{relations} $S_\mathsf{v}$. Each set $S_\mathsf{v}$ is obtained by recursively decomposing the union, composition and transitive closure operators in each relation $\mathsf{Rel}^\mathsf{v}$ (see \Cref{eq:visibility-criterion}).

A visibility \replaced{formula}{relation} is called \emph{arbitration-free} if its definition does not use the arbitration relation $\ar$, i.e. the grammar in \Cref{eq:relations-visibility} omits the $\ar$ relation. For example, in \Cref{fig:consistency_defs}, $\RVC$ and $\CC$ are arbitration-free while $\PC$ and $\SC$ are not. 

\begin{definition}\label{def:arbitration-free-model}
A consistency model is called \emph{arbitration-free} w.r.t. an operation specification $\opspec$ if the visibility \replaced{formulas}{relations} contained in some normal form w.r.t. $\opspec$ are arbitration-free.
\end{definition}

\added{Defining arbitration-free via a normal form removes ``redundant'' occurrences of the arbitration-order, i.e. visibility relations that employ $\ar$ but are vacuous w.r.t. $\opspec$.}
\addedAppendix{\Cref{app:finite-normal-consistency}}\addedJournal{\textit{the arXiv}} also shows that for every \simpleConsistencyFullName{} consistency model $\cmodel$, if some normal form consists of arbitration-free visibility \replaced{formulas}{relations}, then this holds for any other normal form (this is actually proved for the more general class of consistency models defined in Section~\ref{ssec:conflict-defs}).

\begin{theorem}[Basic Arbitration-Free Consistency (AFC$_0$)]
\label{th:characterization-cons-available-lww}
Let $\sspec=(\cmodel, \opspec)$ be a \simpleStorageFullName{} storage specification. %
The following statements are equivalent:
\begin{enumerate}
    \item $\cmodel$ is arbitration-free w.r.t. $\opspec$,
    \item there exists an available $\sspec$-implementation. %
\end{enumerate}
\end{theorem}

In the following, we present a summary for the proof of AFC$_0$, which contains a series of lemmas. We refer the reader to \addedAppendix{\Cref{app:main-theorem-extra-proofs}}\addedJournal{\textit{the arXiv}} for a detailed proof. \replaced{\Cref{lemma:normal-form:consistency-stronger-than-criterion,lemma:saturable:cc-strongest,lemma:cons-available-lww:implementation-cc} show that if $\cmodel$ is arbitration-free then there exists an available $\sspec$-implementation, whereas \Cref{lemma:cons-available-lww:no-implementation-with-ar} is used to show the converse.}{The first lemma below proves the simpler direction.}

\subsection{Arbitration-Freeness Implies Availability}
\label{ssec:afc012}

\added{Assume that $\cmodel$ is arbitration-free w.r.t. $\opspec$. We first show that $\cmodel$ is weaker than $\CC$.}

\begin{restatable}{lemma}{weakerCC}
\label{lemma:saturable:cc-strongest}
\sloppy
\added{Let $\sspec = (\cmodel, \opspec)$ be a \simpleStorageFullName{} storage specification. If $\cmodel$ is arbitration-free w.r.t. $\opspec$, then $\cmodel$ is weaker than $\CC$.}
\end{restatable}

\begin{sketchproof}
\added{If $\cmodel$ is arbitration-free, then every simple visibility formula $\mathsf{v}$ in a normal form of $\cmodel$ does not use $\ar$, i.e. it only uses $\so, \wro$ and $\rbo$. By \Cref{def:execution}, $\so \cup \wro \subseteq \rbo$ in any abstract execution $\exec$. Hence, for every object $x$, $\context{x}{\cmodel}{\exec}{r} \subseteq \context{x}{\CC}{\exec}{r}$, i.e. $\cmodel \preccurlyeq \CC$.}
\end{sketchproof}

\added{\Cref{lemma:normal-form:consistency-stronger-than-criterion} below implies that if a consistency model $\cmodel$ is weaker than $\CC$, then any available $(\CC, \opspec)$-implementation is also an available $(\cmodel, \opspec)$-implementation.}

\begin{restatable}{lemma}{strongerThan}
\label{lemma:normal-form:consistency-stronger-than-criterion}
\added{Let $\opspec$ be a \simpleStorageFullName{} operation specification, and let $\cmodel_1, \cmodel_2$ be a pair of \simpleConsistencyFullName{} consistency models s.t. $\cmodel_2$ is weaker than $\cmodel_1$.
Any abstract execution valid w.r.t. $(\cmodel_2, \opspec)$ is also valid w.r.t. $(\cmodel_1, \opspec)$.}
\end{restatable}

\added{\Cref{lemma:cons-available-lww:implementation-cc} shows that there exists an available $(\CC, \opspec)$-implementation, which concludes the proof of this direction.}

\begin{restatable}{lemma}{satConsAvailableLWW}%
\label{lemma:cons-available-lww:1-2}%
\label{lemma:cons-available-lww:implementation-cc}%
\replaced{Let $\opspec$ be a \simpleStorageFullName{} operation specification. There exists an available $(\CC, \opspec)$-implementation.}{Let $\sspec = (\cmodel, \opspec)$ be a \simpleStorageFullName{} storage specification. If $\cmodel$ is arbitration-free w.r.t. $\opspec$, then there exists an available $\sspec$-implementation.}%
\end{restatable}%

\begin{sketchproof}
\sloppy
\replaced{We define an available storage implementation of $(\CC, \opspec)$}{We show that if $\cmodel$ is arbitration-free w.r.t. $\opspec$, then $\cmodel$ is weaker than $\CC$. This implies that any available storage implementation for $(\CC, \opspec)$ is also an available $\sspec$-implementation. We complete the proof by defining an available storage implementation of $(\CC, \opspec)$,} which is an abstraction of existing $\CC$ implementations~\cite{BailisOverhead,COPS,Eiger,bolton}.

The storage implementation $\implementationInstanceCC$ describes a transition function associating events with the write-read relation obtained by computing the maximum writing event on its causal past (i.e. all write events that are already received in its replica). Each replica $r$ maintains the causal past as follows: (1) every event invoked at $r$ is added to $r$'s causal past, (2) after every invocation, $r$ broadcasts a message to all other replicas that contains its causal past, (3) whenever a replica $r'$ receives this message, it adds the included causal past to its own. Sent messages are not required to be received before executing an invocation. The latter implies trivially that $\implementationInstanceCC$ is an available storage implementation. The validity w.r.t. $(\CC, \opspec)$ follows easily from the ``transitive'' communication of causal pasts between replicas.
\end{sketchproof}

\subsection{Availability Implies Arbitration-Freeness}
\label{ssec:afc021}

\added{
We prove the contrapositive: if $\cmodel$ is not arbitration-free, then no available $\sspec$-implementation exists. Indeed, if $\cmodel$ is not arbitration-free, every normal form $\cmodel'$ of $\cmodel$ contains a simple visibility formula involving $\ar$ (see \Cref{def:arbitration-free-model}). By \Cref{lemma:cons-available-lww:3-1}, such a formula precludes the existence of an available $(\cmodel', \opspec)$-implementation. Consequently, there is no available $(\cmodel, \opspec)$-implementation, since any such implementation would also be an available $(\cmodel', \opspec)$-implementation -- this is an easy observation as $\cmodel$ is equivalent to $\cmodel'$.}

\begin{restatable}{lemma}{notSatnotConsAvailableLWW}%
\label{lemma:cons-available-lww:3-1}%
\label{lemma:cons-available-lww:no-implementation-with-ar}
\replaced{Let $\sspec = (\cmodel, \opspec)$ be a \simpleStorageFullName{} storage specification. Assume that $\cmodel$ contains a simple visibility formula $\mathsf{v}$ which is non-vacuous w.r.t. $\opspec$, such that for some $i, 0 \leq i \leq \length{\mathsf{v}}$, $\mathsf{Rel}_i^\mathsf{v} = \ar$. Then, there is no available $(\cmodel, \opspec)$-implementation.}{Let $\sspec = (\cmodel, \opspec)$ be a \simpleStorageFullName{} storage specification. If there exists an available $\sspec$-implementation, then $\cmodel$ is arbitration-free w.r.t. $\opspec$.}%
\end{restatable}%

\begin{sketchproof}

We assume by contradiction that there is an available implementation $\implementationInstance$ of $\sspec$. \deleted{but $\cmodel$ is not arbitration-free w.r.t. $\opspec$}. We use the \replaced{visibility formula $\mathsf{v}$}{latter fact} to construct a specific program, which by the assumption, admits a trace (in the composition with this implementation) that contains no $\ereceive$ action. We show that \replaced{any}{the} abstract execution induced by this trace, which is admissible by any available implementation of $\sspec$, is not valid w.r.t. $\sspec$.

\deleted{The key idea is observing that since $\cmodel$ is not arbitration-free, then there exists a non-vacuous visibility relation $\mathsf{v}$ in a normal form of $\cmodel$ that is not arbitration-free. Recall that $\mathsf{v}$ is a formula defined as in Equation~\ref{eq:visibility-criterion}. Since $\mathsf{v}$ is simple but not arbitration-free, there exists an index $i$ such that $\mathsf{Rel}_{i}^\mathsf{v} = \ar$. Let $d_\mathsf{v}$ be the largest such index $i$ (last occurrence of $\ar$).}%

\replaced{The program $\programInstanceIota$ we construct}{We construct a program $\programInstanceIota$ that} generalizes the litmus programs presented in \Cref{fig:motivating-example:prog}. $\programInstanceIota$ uses two replicas $r_0, r_1$, two distinguished objects $x_0, x_1$ and a collection of events $e_{i}^{x_l}, 0 \leq i \leq n, l \in \{0,1\}$. The events are used to ``encode'' two instances $\mathsf{v}_{x_0}$ and $\mathsf{v}_{x_1}$ of the visibility \replaced{formula}{relation}.

\begin{figure}[t]
\centering
\centering
\resizebox{.8\textwidth}{!}{
    \begin{tikzpicture}[>=stealth,shorten >=1pt,auto,node distance=3cm, every node/.style={scale=1.3},
        semithick, transform shape,
        B/.style = {%
        decoration={brace, amplitude=1mm,#1},%
        decorate},
        B/.default = ,  %
        cross/.style={cross out, draw, 
         minimum size=2*(#1-\pgflinewidth), 
         inner sep=0pt, outer sep=0pt}
        ]

        \node[font=\large, minimum width=2em, draw, rounded corners=2mm,outer sep=0] (init) at (2, -1) {$\init$};

        \node[font=\large, minimum width=2em, minimum height=1em, draw, rounded corners=2mm,outer sep=0, label={[font=\large]180:$e_0^{x_0}$}] (t11) at (0, -2) {
            $\ldots$
        };

        \node[font=\large, minimum width=2em, minimum height=1em, rounded corners=2mm,outer sep=0,] (dots1) at (0, -3.75) {
            $\ldots$
        };

        \node[font=\large, minimum width=2em, minimum height=1em, draw, rounded corners=2mm,outer sep=0, label={[font=\large]180:$e_{d_\mathsf{v} -1}^{x_0}$}] (tk1) at (0, -5.5) {
            $\ldots$
        };

        \node[font=\large, minimum width=2em, minimum height=1em, draw, rounded corners=2mm,outer sep=0, label={[font=\large]180:$e_{d_\mathsf{v} }^{x_1}$}] (tk11) at (0, -7.25) {
            $\ldots$
        };

        \node[font=\large, minimum width=2em, minimum height=1em, rounded corners=2mm,outer sep=0,] (dots21) at (0, -9) {
            $\ldots$
        };

        \node[font=\large, minimum width=2em, minimum height=1em, draw, rounded corners=2mm,outer sep=0, label={[font=\large]180:$e_{n}^{x_1}$}] (tn1) at (0, -10.75) {
            $\ldots$
        };

        \node[font=\large, minimum width=2em, minimum height=1em, draw, rounded corners=2mm,outer sep=0, label={[font=\large]0:$e_0^{x_1}$}] (t12) at (4, -2) {
            $\ldots$
        };

        \node[font=\large, minimum width=2em, minimum height=1em, rounded corners=2mm,outer sep=0,] (dots2) at (4, -3.75) {
            $\ldots$
        };

        \node[font=\large, minimum width=2em, minimum height=1em, draw, rounded corners=2mm,outer sep=0, label={[font=\large]0:$e_{d_\mathsf{v} -1}^{x_1}$}] (tk2) at (4, -5.5) {
            $\ldots$
        };

        \node[font=\large, minimum width=2em, minimum height=1em, draw, rounded corners=2mm,outer sep=0, label={[font=\large]0:$e_{d_\mathsf{v} }^{x_0}$}] (tk12) at (4, -7.25) {
            $\ldots$
        };

        \node[font=\large, minimum width=2em, minimum height=1em, rounded corners=2mm,outer sep=0,] (dots22) at (4, -9) {
            $\ldots$
        };

        \node[font=\large, minimum width=2em, minimum height=1em, draw, rounded corners=2mm,outer sep=0, label={[font=\large]0:$e_{n}^{x_0}$}] (tn2) at (4, -10.75) {
            $\ldots$
        };

        \node[font=\large, minimum width=2em, minimum height=1em, rounded corners=2mm,outer sep=0] (label-x0) at (-4, -2) {
            \added{$\textsf{writes} \ x_0$}
        };

        \node[font=\large, minimum width=2em, minimum height=1em, rounded corners=2mm,outer sep=0] (label-no-x0) at (9, -5.5) {
            \added{$\textsf{do not write} \ x_0$}
        };

        \node[font=\large, minimum width=2em, minimum height=1em, rounded corners=2mm,outer sep=0] (label-r-x0) at (9, -10.75) {
            \added{$\textsf{reads} \ x_0$}
        };

        \path (init) edge[->, soColor, right] node [above] {$\so$} (t11);
        \path (init) edge[->, soColor, right] node [above] {$\so$} (t12); %

        \path (label-x0.east) edge[->, thick, greensilver, right] node {} ([xshift=-4.5em] t11);
    
        \coordinate (middleright) at ($(label-no-x0.west)!0.5!([xshift=4.5em] tk2)$) {};

        \path (label-no-x0.west) edge[->,thick, greensilver, right] node {} ([xshift=4.5em] tk2);

        \draw[thick, ->,  -latex, greensilver, rounded corners=5mm] (label-no-x0.west) to ([xshift=6.5em] t12) to ([xshift=4.5em] t12);

        \draw[thick, ->,  -latex, greensilver, rounded corners=5mm] (label-no-x0.west) to ([xshift=6.5em] tk12) to ([xshift=4.5em] tk12);

        \draw[thick, ->, -latex, greensilver, rounded corners=5mm] (label-no-x0.west) to ([xshift=6.5em] dots2) to ([xshift=4.5em] dots2);

        \draw[thick, ->, -latex, greensilver, rounded corners=5mm] (label-no-x0.west) to ([xshift=6.5em] dots22) to ([xshift=4.5em] dots22);
        
        \path (label-r-x0.west) edge[->, thick, greensilver, right] node {} ([xshift=4.5em] tn2);

        \path (t11) edge[->, transform canvas={xshift=-1mm}, soColor, right] node [left] {$\so$} (dots1);
        \path (t11) edge[->, transform canvas={xshift=1mm}, black, right] node [right] {$\mathsf{Rel}_1^\mathsf{v} $} (dots1);

        \path (dots1) edge[->, transform canvas={xshift=-1mm}, soColor, right] node [left] {$\so$} (tk1);
        \path (dots1) edge[->, transform canvas={xshift=1mm}, black, right] node [right] {$\mathsf{Rel}_{d_\mathsf{v} -1}^\mathsf{v} $} (tk1);

        \path (tk1) edge[->, soColor, right] node [left] {$\so$} (tk11);
        \path (tk11) edge[->, transform canvas={xshift=-1mm}, soColor, right] node [left] {$\so$} (dots21);
        \path (tk11) edge[->, transform canvas={xshift=1mm}, black, right] node [right] {$\mathsf{Rel}_{d_\mathsf{v} +1 }^\mathsf{v} $} (dots21);

        \path (dots21) edge[->, transform canvas={xshift=-1mm}, soColor, right] node [left] {$\so$} (tn1);
        \path (dots21) edge[->, transform canvas={xshift=1mm}, black, right] node [right] {$\mathsf{Rel}_{n}^\mathsf{v} $} (tn1);

        \path (t12) edge[->, transform canvas={xshift=-1mm}, soColor, right] node [left] {$\so$} (dots2);
        \path (t12) edge[->, transform canvas={xshift=1mm}, black, right] node [right] {$\mathsf{Rel}_1^\mathsf{v} $} (dots2);

        \path (dots2) edge[->, transform canvas={xshift=-1mm}, soColor, right] node [left] {$\so$} (tk2);
        \path (dots2) edge[->, transform canvas={xshift=1mm}, black, right] node [right] {$\mathsf{Rel}_{d_\mathsf{v} -1}^\mathsf{v} $} (tk2);

        \path (tk2) edge[->,soColor, right] node [left] {$\so$} (tk12);
        \path (tk12) edge[->, transform canvas={xshift=-1mm}, soColor, right] node [left] {$\so$} (dots22);
        \path (tk12) edge[->, transform canvas={xshift=1mm}, black, right] node [right] {$\mathsf{Rel}_{d_\mathsf{v} +1 }^\mathsf{v} $} (dots22);

        \path (dots22) edge[->, transform canvas={xshift=-1mm}, soColor, right] node [left] {$\so$} (tn2);
        \path (dots22) edge[->, transform canvas={xshift=1mm}, black, right] node [right] {$\mathsf{Rel}_{n}^\mathsf{v} $} (tn2);

        \path (tk1) edge[->, double equal sign distance, arColor, right] node [above, shift={(-0.3,0)}] {$\ar$} (tk2);
        \path (tk1) edge[->, double equal sign distance, arColor, right] node [above, shift={(-0.7,-0.2)}] {$\ar$} (tk12);

        \node [above, shift={(0.3,0)}] (rbolabel) at (2, -8.75) {$\rbo$};

        \draw[->, double equal sign distance, rbColor,rounded corners,] (t11.east) .. controls ($(t11)!1!(t12) $) and ($(tn1)!0!(tn2) $) .. (tn2.west);

        \draw (1.93,-7.2) node[font=\large, cross=8pt,red,line width=1.5pt] {};

        \path[thick, lightGreensilver, double distance=10\pgflinewidth, opacity=.25, line width=10] (t11.south) edge[-] (tk1.north);
        \path[thick, lightGreensilver, double distance=10\pgflinewidth, opacity=.25, line width=10] (tk12.south) edge[-] (tn2.north);
        \path[thick, lightGreensilver, double distance=10\pgflinewidth, opacity=.25, line width=10] (tk1) edge[-] (tk12);

    \end{tikzpicture}  
}

\vspace{-1mm}
\caption{Abstract execution of a trace without $\ereceive$ actions for the visibility \replaced{formula}{relation} $\mathsf{v}$. If $i \neq d_\mathsf{v} $, $(e_{i-1}^{x_l}, e_i^{x_l}) \in \mathsf{Rel}_i^\mathsf{v} $ holds because the two events are executed at the same replica (recall that $\so\subseteq\rbo\subseteq\ar$). If $(e_{d_\mathsf{v} -1}^{x_0}, e_{d_\mathsf{v} -1}^{x_1}) \in \ar$, then since $\so\subseteq\ar$ and $\ar$ is transitive, we get that $(e_{d_\mathsf{v} -1}^{x_0}, e_{d_\mathsf{v} }^{x_0}) \in \mathsf{Rel}_{d_\mathsf{v} }^\mathsf{v} = \ar$; and therefore, that $\mathsf{v}_{x_0}(e_0^{x_0}, e_{n}^{x_0})$ holds. However, in the absence of receives, $(e_0^{x_0}, e_{n}^{x_0}) \not\in \rbo$. %
}
\vspace{-3mm}
\label{fig:diagram-theorem-proof}
\end{figure}
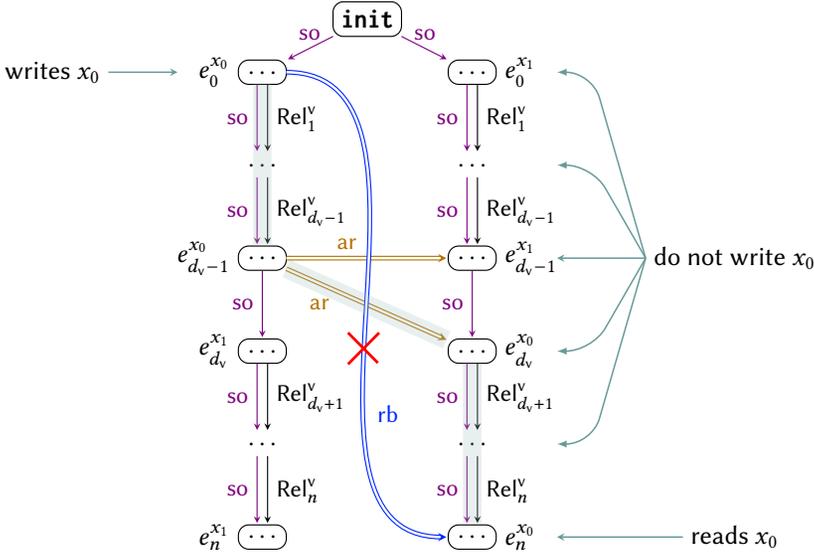

\added{Let $d_\mathsf{v}$ be the largest index $i$ s.t. $\mathsf{Rel}_i^\mathsf{v} = \ar$ (last occurrence of $\ar$).} Then, $\mathsf{v}$ is formed of two parts: the path of dependencies from $\event_0$ to $\event_{d_\mathsf{v}}$ which is not arbitration-free, and the suffix from $\event_{d_\mathsf{v}}$ up to $\event_{\length{\mathsf{v}}}$, the arbitration-free part. Thus, $\mathsf{v}$ is of the form:
\begin{align*}
\mathsf{v}_x(\event_0,\event_{n}) \Coloneqq \exists \event_1, \ldots, \event_{n-1}. &
\bigwedge_{i=1}^{n} (\event_{i-1}, \event_{i}) \in \mathsf{Rel}^{\mathsf{v}}_i \ \land
\writeVar{\event_0}{x} \land \wro_x^{-1}(\event_{n})\neq  \emptyset %
\end{align*}
where
$n = \length{\mathsf{v}}$, %
$\mathsf{Rel}_{i}^\mathsf{v} \in \{ \so,\wro, \rbo, \ar\}$ for $i < d_v$, $\mathsf{Rel}_{d_\mathsf{v}}^\mathsf{v} = \ar$, and $\mathsf{Rel}_{i}^\mathsf{v} \in \{\so,\wro, \rbo\}$ for $i > d_\mathsf{v}$.

Replica $r_l$ executes first events $e_i^{x_l}$ with $i < d_\mathsf{v}$ and then, events $e_i^{x_{1-l}}$ with $i\geq d_\mathsf{v}$ -- the replica $r_l$ executes the non arbitration-free part of $\mathsf{v}$ for object $x_l$ and the arbitration-free suffix of $\mathsf{v}$ for $x_{1-l}$. All events in replica $r_l$ access (read and/or write) object $x_l$ except for $e_{n}^{x_l}$ which reads $x_{1-l}$. For ensuring that $\mathsf{v}_x(e_0^{x_l}, \ldots e_{n}^{x_l})$ holds in an induced abstract execution of a trace without $\ereceive$ actions, we require that if $\mathsf{Rel}_i^{\mathsf{v}} = \wro$, then $e_{i-1}^{x_l}$ is a write event and $e_i^{x_l}$ is a read event. \Cref{fig:diagram-theorem-proof} exhibits a diagram of such execution.

\begin{example}
We illustrate the construction for Prefix Consistency (\PC) and a Key-Value store with $\iputName$ and $\igetName$ operations (their specification is defined in \Cref{ssec:opspec}). \PC{} can be defined as the following set of \emph{simple} visibility \replaced{formulas}{relations} (obtained from \deleted{the axiom }$\mathsf{Prefix}$ in \Cref{fig:prefix}):
\begin{equation}
\label{eq:prefix-simple}
\begin{array}{ll}
\mathsf{v}^1_x(\event_0, \event_1) \Coloneqq  & \writeVar{\event_0}{x} \ \land \ \wro_x^{-1}(\event_1) \neq \emptyset \ \land \ (\event_0, \event_1) \in \so \\
\mathsf{v}^2_x(\event_0, \event_1) \Coloneqq  & \writeVar{\event_0}{x} \ \land \ \wro_x^{-1}(\event_1) \neq \emptyset \ \land \ (\event_0, \event_1) \in \wro  \\
\mathsf{v}^3_x(\event_0, \event_2) \Coloneqq  & \exists \event_1.\ \writeVar{\event_0}{x} \ \land \ \wro_x^{-1}(\event_2) \neq \emptyset \ \land \ (\event_0, \event_1) \in \ar \ \land \ (\event_1, \event_2) \in \so \\
\mathsf{v}^4_x(\event_0, \event_2) \Coloneqq  & \exists \event_1.\ \writeVar{\event_0}{x} \ \land \ \wro_x^{-1}(\event_2) \neq \emptyset \ \land \ (\event_0, \event_1) \in \ar \ \land \ (\event_1, \event_2) \in \wro
\end{array}
\end{equation}
Observe that $\mathsf{v}^4_x$ is vacuous w.r.t. the specification of $\iputName$ and $\igetName$ since it implies that $\event_2$ reads from multiple events, and $\iputName$ and $\igetName$ read a single object at a time. Thus, the normal form of $\PC$ w.r.t. the specification of $\iputName$ and $\igetName$ contains only the first three visibility \replaced{formulas}{relations} above.

The only visibility \replaced{formula}{relation} which is not arbitration-free is $\mathsf{v}^3_x$. We have that the index $d_\mathsf{v} = 1$ and we consider the following types of events:
\begin{align*}
e_0^{x_0}: \iput{x_0}{\_}, e_1^{x_0}: \iput{x_1}{\_}, e_2^{x_0}: \iget{x_0} \\
e_0^{x_1}: \iput{x_1}{\_}, e_1^{x_1}: \iput{x_0}{\_}, e_2^{x_1}: \iget{x_1}
\end{align*}
Replica $r_0$ executes $e_0^{x_0}$ and then $e_1^{x_1}$ and $e_2^{x_1}$. Replica $r_1$ executes $e_0^{x_1}$ and then $e_1^{x_0}$ and $e_2^{x_0}$.

\end{example}

Given such a program $\programInstanceIota$, the proof proceeds as follows:
\begin{enumerate}
    \item There exists a finite trace $t$ of $\parallelCompositionInstanceIota$ that contains no receive action\addedAppendix{ (\Cref{lemma:program-a-la-decker:trace-without-receive})}: Since $\implementationInstance$ is available, it can always delay receiving messages, and execute other actions instead. Then, as $\programInstanceIota$ is a finite program, such an execution must be finite. \label{lemma:cons-available-lww:3-1:1}

    \item The trace $t$ induces a history $h_\mathsf{v} = \tup{E, \so, \wro}$ and an abstract execution $\exec_\mathsf{v} = \tup{h, \rbo, \ar}$ where $\rbo = \so$ \added{($\ar$ is arbitrary as long as $\rbo\subseteq\ar$)}. As $\implementationInstance$ is valid w.r.t. $\sspec$, $\exec_\mathsf{v}$ is valid w.r.t. $\sspec$. Next, we prove that since $\rbo = \so$, events in $\exec_\mathsf{v}$ read the latest value w.r.t. $\so$ written on their associated object in $\exec_\mathsf{v}$\addedAppendix{ (\Cref{lemma:program-a-la-decker-lww:reads})}. In particular, we deduce that all traces of $\programInstanceIota$ without $\ereceive$ events induce the same history and therefore, the induced history does not change when the induced arbitration order changes. \label{lemma:cons-available-lww:3-1:2}

    \item \sloppy Since $\ar$ is a total order, either $(e_{d_\mathsf{v}-1}^{x_0}, e_{d_\mathsf{v} - 1}^{x_{1}}) \in \ar$ or $(e_{d_\mathsf{v}-1}^{x_1}, e_{d_\mathsf{v} - 1}^{x_{0}}) \in \ar$. W.l.o.g., assume that $(e_{d_\mathsf{v}-1}^{x_0}, e_{d_\mathsf{v} - 1}^{x_{1}}) \in \ar$. \addedAppendix{By \Cref{lemma:program-a-la-decker:f0-visible}}\addedJournal{Then}, we deduce that $e_0^{x_0} \in \context{\cmodel}{x_0}{\exec_\mathsf{v}}{e_{n}^{x_0}}$. The proof is explained in \Cref{fig:diagram-theorem-proof}: if $(e_{d_\mathsf{v}-1}^{x_0}, e_{d_\mathsf{v} - 1}^{x_{1}}) \in \ar$, then all events $e_i^{x_0}$ form a path in such way that $\mathsf{v}_{x_0}(e_0^{x_0}, \ldots e_{n}^{x_0})$ holds in $\exec_\mathsf{v}$. \label{lemma:cons-available-lww:3-1:3}

    \item Since $e_{n}^{x_0}$ is the only event at $r_1$ that reads or writes $x_0$ and events in $\exec_\mathsf{v}$ read the latests values w.r.t. $\so$ in $\exec_\mathsf{v}$, we deduce that $e_{n}^{x_0}$ reads $x_0$ from $\init$. However, as $e_0^{x_0} \in \context{\cmodel}{x_0}{\exec_\mathsf{v}}{e_{n}^{x_0}}$ and $\init$ precedes $e_0^{x_0}$ in arbitration order, we deduce that $e_{n}^{x_0}$ does not read the latest value w.r.t. $\ar$, i.e. $\rspecSimple{e_{n}^{x_0}} \downarrow$ but $\wro_{x_0}^{-1}(e_{n}^{x_0}) \neq \{\max_{\ar} \context{\cmodel}{x_0}{\exec_\mathsf{v}}{e_{n}^{x_0}}\}$. Therefore, $\exec_\mathsf{v}$ is not valid w.r.t. $\sspec$ (see \Cref{def:valid}). This contradicts the hypothesis that $\implementationInstance$ is an implementation of $\sspec$. \qedhere \label{lemma:cons-available-lww:3-1:4}

\end{enumerate}
\end{sketchproof}

The corollary below is a direct consequence of \replaced{\Cref{th:characterization-cons-available-lww} and \Cref{lemma:saturable:cc-strongest}.}{and from the observation that any consistency model $\cmodel$ which is arbitration-free w.r.t. some $\opspec$ is necessarily weaker than $\CC$ (as established in an intermediate step of the proof of Lemma~\ref{lemma:cons-available-lww:1-2}).}

\begin{corollary}
\label{corollary:cc-strongest-lww}
Let $\opspec$ be a \simpleStorageFullName{} operation specification.
The strongest consistency model $\cmodel$ for which $(\cmodel, \opspec)$ admits an available implementation is $\CC$.
\end{corollary}

	\section{Generalized Distributed Storage Specifications}
\label{sec:generalized-definitions}

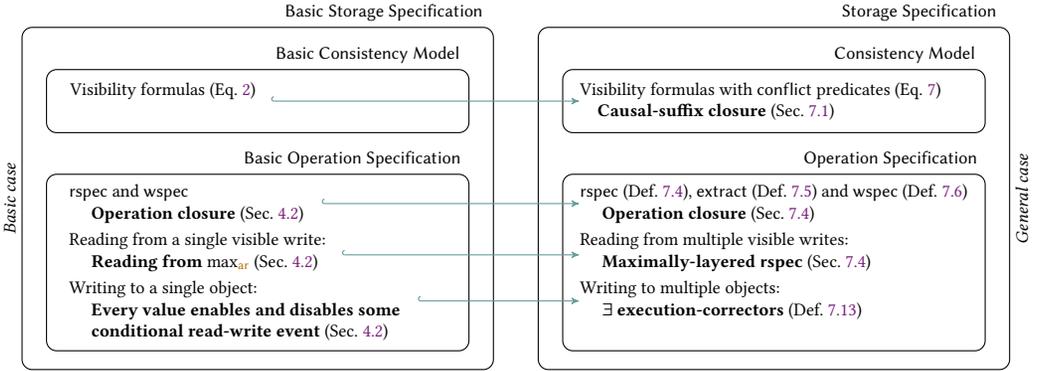
\begin{figure}[t]
\centering
\resizebox{\textwidth}{!}{
    \begin{tikzpicture}[>=stealth',shorten >=1pt,auto,node distance=3cm,
        hookarrow/.style={{Hooks[right]}->},
        semithick, transform shape,
        B/.style = {%
        decoration={brace, amplitude=1mm,#1},%
        decorate},
        B/.default = ,  %
        ]

        \node[minimum width=26.5em, minimum height=19em, outer sep=0,rounded corners=2mm, draw, label={[font=\normalsize]83:\sf{Basic Storage Specification}}] (sspecBasicBox) at (0, -2.3) {
        
        };

        \node[minimum width=23.75em, minimum height=9.8em, outer sep=0,rounded corners=2mm, draw, label={[font=\normalsize,xshift=-1.4em]87:\sf{Basic Operation Specification}}] (opspecBasicBox) at (0, -3.55) {
        
        };

        \node[minimum width=19.5em, outer sep=0] (opspecBasic) at (0, -3.55) {
            \begin{tabular}{l}
                $\rspecName$ and $\wspecName$ \\
                \hspace{1em} \textbf{Operation closure} (Sec. \ref{ssec:basic-operation-closure}) \\[1mm]
                Reading from a single visible write: \\
                \hspace{1em} \textbf{Reading from $\max_{\ar}$} (Sec. \ref{def:valid}) \\[1mm]
                Writing to a single object: \\
                \hspace{1em} \textbf{Every value enables and disables some}\\ 
                \hspace{1em} \textbf{conditional read-write event} (Sec. \ref{ssec:basic-conditional-read-writes}) \hspace{3em}$ $
            \end{tabular}
        };

        \node[outer sep=0] (opClosureBasic) at (0.5, -0.4) {
            
        };           

        \node[outer sep=0] (maxArBasic) at (1.5, -2.40) {
           
        };

        \node[outer sep=0] (singleObjectBasic) at (1.9, -3.40) {
           
        };

        \node[outer sep=0] (wellformedBasic) at (3.4, -4.30) {
           
        };

        \node[minimum width=19.5em, outer sep=0] (consistencyBasic) at (0, -0.4) {
             \begin{tabular}{l}
                Visibility formulas (Eq. \ref{eq:visibility-criterion})\hspace{10.4em} $ $\\
                $ $
            \end{tabular}
            
        };

        \node[minimum width=23.75em, minimum height=3.5em, outer sep=0,rounded corners=2mm, draw, label={[font=\normalsize]70:\sf{Basic Consistency Model}}] (consBasicBox) at (0, -0.4) {
        
        };

        \node[minimum width=26.5em, minimum height=19em, outer sep=0,rounded corners=2mm, draw, label={[font=\normalsize]70:\sf{Storage Specification}}] (sspecGeneralBox) at (10.2,-2.3) {
        
        };

        \node[minimum width=23.75em, minimum height=9.8em, outer sep=0,rounded corners=2mm, draw, label={[font=\normalsize]75:\sf{Operation Specification}}] (opspecGeneralBox) at (10.2, -3.55) {
        
        };

        \node[minimum width=19em, outer sep=0] (opspecGeneral) at (10.2, -3.55) {
            \begin{tabular}{l}
                $\rspecName$ (Def. \ref{def:general-read-spec}), $\extractName$ (Def. \ref{def:extract-spec}) and $\wspecName$ (Def. \ref{def:write-spec}) \\
                \hspace{1em} \textbf{Operation closure} (Sec. \ref{ssec:general-operation-closure}) \\[1mm]
                Reading from multiple visible writes: \\
                \hspace{1em} \textbf{Maximally-layered \textsf{rspec}} (Sec. \ref{ssec:maximally-layered}) \\[1mm]
                Writing to multiple objects: \\
                \hspace{1em} \textbf{$\exists$ execution-correctors} (Def. \ref{def:execution-corrector}) \\
                $ $\hspace{1em}$ $ 
            \end{tabular}
        };

        \node[outer sep=0] (opClosureGeneral) at (6.5, -0.4) {
            
        };           

        \node[outer sep=0] (maxArGeneral) at (6.5, -2.40) {
           
        };

        \node[outer sep=0] (correctorsGeneral) at (6.5, -3.40) {
            
        };

        \node[outer sep=0] (wellformedGeneral) at (6.5, -4.30) {
            
        };

        \node[minimum width=23.75em, minimum height=3.5em, outer sep=0,rounded corners=2mm, draw, label={[font=\normalsize]30:\sf{Consistency Model}}] (consGeneralBox) at (10.2, -0.4) {
        
        };
        \node[minimum width=19.5em, outer sep=0] (consistencyGeneral) at (10.2, -0.4) {
            \begin{tabular}{l}
                Visibility formulas with $\conflictName$ predicates (Eq. \ref{eq:general-visibility-criterion})\\
                \hspace{1em}\textbf{Causal-suffix closure} (Sec. \ref{ssec:causal-suffix-closure}) \hspace{7.25em}$ $
            \end{tabular}
        };

        \node[outer sep=0, rotate=90] (basicCase) at (-4.9,-2.3) {
           \textit{Basic case}
        };

        \node[outer sep=0, rotate=90] (basicCase) at (15.1,-2.3) {
           \textit{General case}
        };

        \path ([xshift=-1em]opClosureBasic.east) edge[hookarrow, greensilver] (opClosureGeneral.west);
        
        \path ([xshift=-1em]maxArBasic.east) edge[hookarrow, greensilver] (maxArGeneral.west);

        \path ([xshift=-1em]singleObjectBasic.east) edge[hookarrow, greensilver] (correctorsGeneral.west);

        \path ([xshift=-1em]wellformedBasic.east) edge[hookarrow, greensilver] (wellformedGeneral.west);

    \end{tikzpicture}  
}
\vspace{-4mm}
\caption{\added{Conceptual map relating \simpleStorageFullName{} and generalized storage specifications (\Cref{sec:basic:spec,sec:generalized-definitions}). Storage specifications are composed of consistency models and operation specifications. Assumptions are written in bold text. Arrows denote how definitions/assumptions translate from the basic case to the general case.}
}
\label{fig:conceptal-map}
\vspace{-2mm}
\end{figure}

We describe a generalization of the basic storage specifications from \Cref{sec:basic:spec} along three dimensions: a larger class of consistency models, multi-object operations, and more general read behaviors. To rule out anomalous behaviors in this generalization, we introduce a set of additional assumptions. \added{\Cref{fig:conceptal-map} summarizes the structure of storage specifications and the relationship between basic and generalized specifications in terms of assumptions.}

\subsection{Consistency Models}
\label{ssec:conflict-defs}

The set of \simpleConsistencyFullName{} consistency models (\Cref{ssec:consistency}) does \emph{not} include (parallel) snapshot isolation, and the version of $k$-bounded staleness considered in Section~\ref{sec:motivating-example}. 
\added{Snapshot Isolation, $k$-Bounded Staleness and Parallel Snapshot Isolation are defined, respectively, using the visibility formulas $\axconf$ (\Cref{fig:conflict}), $\axkbs$ (\Cref{fig:kbs})}\footnote{Our version of $k$-Bounded Staleness corresponds to the $(k, T)$-Bounded Staleness with $T = \infty$ as defined in~\cite{cosmosdb-consistency}.}\replaced{, and $\axnpsi$ (\Cref{fig:npsi}).}{Snapshot isolation (and the parallel version) are defined using the axiom $\axconf$ (\Cref{fig:conflict}), where event $\event_0$ is visible to $\event_2$ whenever some intermediate event $\event_1$ writes on a same object as $\event_2$. The version of $k$-bounded Staleness is defined using the axiom $\axkbs$ (\Cref{fig:kbs}), where event $\event_0$ is visible to $\event_k$ whenever there are some intermediate events $\event_1, \ldots \event_{k-1}$ that write on object $x$.}
To include such consistency models in our formalization, we extend the syntax of visibility \replaced{formulas}{relations} so that the intermediate events can be further constrained via the $\mathsf{wrCons}$ formula:
\begin{align}
\mathsf{v}_x(\event_0,\event_{n}) \Coloneqq \exists \event_1, \ldots, \event_{n-1}. &
\bigwedge_{i=1}^{n} (\event_{i-1}, \event_{i}) \in \mathsf{Rel}^{\mathsf{v}}_i \land \wro_x^{-1}(\event_{n}) \neq \emptyset 
\land \writeConstraints{x}{\event_0, \ldots \event_{n}} \label{eq:general-visibility-criterion}
\end{align}

\sloppy The formula $\writeConstraints{x}{\event_0, \ldots \event_{n}}$ is a conjunction of predicates $\writeConflicts{E}$ and $\writeConflicts[x]{E}$  with $E \subseteq \{\event_0, \ldots \event_{n}\}$.
The predicate $\writeConflicts{E}$ (resp., $\writeConflicts[x]{E}$) means that \emph{all} the events in $E$ write on some object $y$ (resp., the object $x$).
Since we want to preserve the constraint $\writeVar{\event_0}{x}$ from \simpleConsistencyFullName{} visibility \replaced{formulas}{relations}, we require that there exists a set $E \subseteq \{\event_0, \ldots \event_{n}\}$ s.t. $\event_0 \in E$ and $\writeConflicts[x]{E}$ is included in $\writeConstraints{x}{\event_0, \ldots \event_{n}}$ ($E$ can be the singleton $\event_0$). 
The interpretation of a conflict predicate in an abstract execution $\exec$ is done as expected: a predicate $\writeConflicts{E}$ (resp., $\writeConflicts[x]{E}$) holds iff there exists an object $y$ s.t. for every $e \in E$, $\writeVarExec{e}{y}{\exec}$ (resp. $\writeVarExec{e}{x}{\exec}$). \replaced{As before, the predicate $\writeVar{\event}{y}$ is true iff $\outputEventObj{e}{y} \downarrow$.}
{The interpretation of $\mathsf{writes}$ will be defined later in \Cref{ssec:general-object-specs}.}

From this point on, a consistency model is defined as a set of visibility \replaced{formulas}{relations}, as in \Cref{eq:general-visibility-criterion}.

\begin{figure}[t]
\resizebox{\columnwidth}{!}{
\footnotesize
\begin{tabular}{|c|c|c|}
	\hline & & \\
	
	\begin{subfigure}[b]{.32\textwidth}
		\centering
		\begin{tikzpicture}[->,>=stealth,shorten >=1pt,auto,node distance=4cm,
			semithick, transform shape]
			\node[transaction state, text=black] at (0,0)       (t_1)           {$e$};
			\node[transaction state, label={[xshift=2]right:$\writeVar{ }{\yvar}$}] at (2,0)       (t_3)           {$\event_2$};
			\node[transaction state, text=black,label={above:$\writeVar{}{x}$}] at (-.5,1.5) (t_2) {$\event_0$};
			\node[transaction state, label={[yshift=-2]above:{$\writeVar{}{\yvar}$}}] at (1.5,1.5) (t_4) {$\event_1$};
			\path (t_1) edge[wrColor] node {$\wro_x$} (t_3);
			\path (t_2) edge [double equal sign distance, coColor] node {$\ar^*$} (t_4);
			\path (t_4) edge [double equal sign distance, coColor] node {$\ar$} (t_3);
			\path (t_2) edge[left, dashed, double equal sign distance, coColor] node {$\ar$} (t_1);
		\end{tikzpicture}

		\vspace{2mm}
		\parbox{\textwidth}{

			$ $

			$\exists \event_1. 
			(\event_0, \event_1) \in \ar^* \land (\event_1, \event_2) \in \ar  \ \land$ 
			
			\hspace{1mm}$\wro_x^{-1}(\event_1) \neq \emptyset \ \land \  \writeVar{\event_0}{x} \ \land $

			\hspace{1mm}$\writeConflicts{\event_1, \event_2}$

		}
		
		\caption{$\mathsf{Conflict}$}
		\label{fig:conflict}
	\end{subfigure}
	&
	\begin{subfigure}[b]{.32\textwidth}
		\centering
		\begin{tikzpicture}[->,>=stealth,shorten >=1pt,auto,node distance=1cm,
			semithick, transform shape]
			\node[transaction state, text=black] at (0,0)       (t_e)           {$e$};
			\node[transaction state, text=black,label={above:\textcolor{black}{$\writeVar{ }{\xvar}$}}] at (-.5,1.5) (t_0) {$\event_0$};
			\node[transaction state, label={above:{$\writeVar{}{\xvar}$}}] at (1,1.5) (t_1) {$\event_1$};
			\node[transaction state, label={[yshift=0]right:{$\writeVar{}{\xvar}$}}] at (2,1) (t_k1) {$\event_{k-1}$};
			\node[transaction state] at (2,0)       (t_k)           {$\event_k$};

			\path (t_e) edge[wrColor] node {$\wro_x$} (t_k);
			\path (t_0) edge[double equal sign distance, arColor] node {$\ar$} (t_1);
			\path (t_1) edge[bend left=20, double equal sign distance, arColor] node {$\ldots$} (t_k1);
			\path (t_k1) edge[double equal sign distance, arColor] node {$\ar$} (t_k);
			\path (t_0) edge[left, dashed, double ,coColor] node {$\ar$} (t_e);
		\end{tikzpicture}

		\vspace{2mm}
		\parbox{\textwidth}{
			$ $

			$ $ 

			$\exists \event_1 \ldots \event_{k-1}. \ \bigwedge_{i=1}^{k}(\event_{i-1}, \event_i) \in \ar \ \land $ 

			\hspace{1mm}$\wro_x^{-1}(\event_1) \neq \emptyset \land \writeConflicts[x]{\event_0, \ldots \event_{k-1}}$
			
		}
		\caption{$\axkbs$}
		\label{fig:kbs}
	\end{subfigure}
	&
	\begin{subfigure}[b]{.32\textwidth}
		\centering
		\resizebox{\textwidth}{!}{
		\begin{tikzpicture}[->,>=stealth,shorten >=1pt,auto,node distance=1cm,
			semithick, transform shape]
			\node[transaction state, text=black] at (-1,0)       (t_e)           {$e$};
			\node[transaction state,label={[yshift=2,xshift=-4]above:{$\writeVar{ }{\xvar}$}}] at (-1.5,2) (t_0) {$\event_0$};
			\node[transaction state, label={[yshift=1.1,xshift=6]above:{$\writeVar{}{\xvar_1}$}}] at (-0.5,2) (t_1) {$\event_1$};
			\node[transaction state, label={[yshift=1.1,xshift=15]above:{$\writeVar{}{\xvar_1}$}}] at (.5,2) (t_2) {$\event_2$};
			\node[transaction state, label={[yshift=6,xshift=-8]right:{$\writeVar{}{\xvar_n}$}}] at (1.5,1.5) (t_nm1) {$\event_{2 \cdot n-1}$};
			\node[transaction state, label={[yshift=0,xshift=-2]right:{$\writeVar{}{\xvar_n}$}}] at (1.5,.75) (t_n) {$\event_{2 \cdot n}$};
			\node[transaction state] at (1.5,0) (t_n1) {$\event_{2 \cdot n + 1}$};

			\path (t_e) edge[wrColor] node {$\wro_x$} (t_n1);
			\path (t_0) edge[double equal sign distance, rbColor] node {$\rbo^*$} (t_1);
			\path (t_1) edge[double equal sign distance, arColor] node {$\ar$} (t_2);
			\path (t_2) edge[black, dashed] node {$\ldots$} (t_nm1);
			\path (t_nm1) edge[double equal sign distance, arColor] node {$\ar$} (t_n);
			\path (t_n) edge[double equal sign distance, rbColor] node {$\rbo^*$} (t_n1);
			\path (t_0) edge[left, dashed, double ,coColor] node {$\ar$} (t_e);
		\end{tikzpicture}
		}

		\vspace{2mm}
		\parbox{\textwidth}{
			\added{$\exists \event_1 \ldots \event_{2 \cdot n }. \bigwedge_{i=0}^{n}(\event_{2\cdot i}, \event_{2\cdot i +1}) \in \rbo^* \land $} 

			\hspace{1mm} \added{$\bigwedge_{i=1}^{n}(\event_{2\cdot i - 1}, \event_{2\cdot i }) \in \ar \ \land$}

			\hspace{1mm} \added{$\wro_x^{-1}(\event_1) \neq \emptyset \land \ \writeVar{\event_0}{x}$}

			\hspace{1mm} \added{$\bigwedge_{i=1}^{n}\writeConflicts{\event_{2 \cdot i - 1}, \event_{2 \cdot i}}$}
		}
		\caption{$\axnpsi$}
		\label{fig:npsi}
	\end{subfigure}
	
\\\hline
\end{tabular}
}
\vspace{-2mm}
\caption{$\axconf$, $\axkbs$ \added{and $\axnpsi$} visibility \replaced{formulas}{relations} used to define \emph{Snapshot Isolation} ($\SI$), \emph{Bounded Staleness} ($\BS$) and \emph{Parallel Snapshot Isolation} ($\PSI$). $\SI$ is defined by $\axpre$ (\Cref{fig:prefix}) and $\axconf$, $\BS$ is defined by $\axkbs$ and $\axrvc$ (\Cref{fig:rvc}), and $\PSI$ is defined by $\axcc$ (Figure~\ref{fig:cc}) and \replaced{the set of visibility formulas $\{\axnpsi \ | \ n\geq 1\}$}{$\axconf$}.}
\label{fig:consistency_conflict_def}
\vspace{-2mm}
\end{figure}
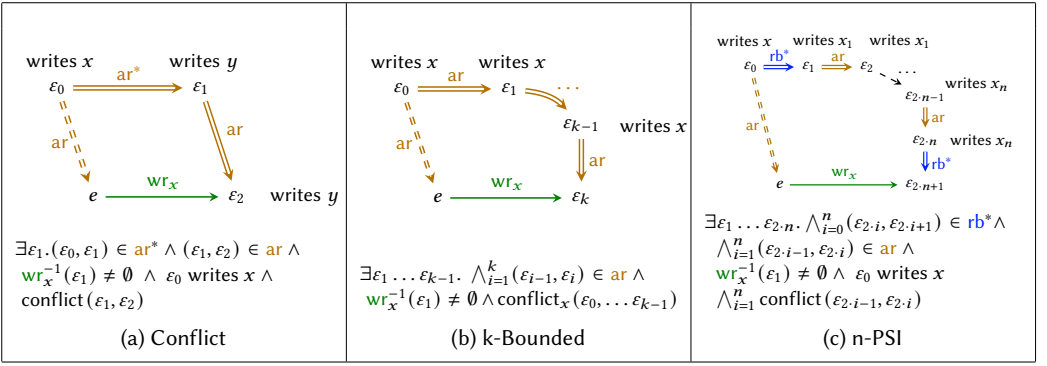

\smallskip
\noindent
\textbf{Normal Form.} We generalize the normal form of a consistency model to take into account conflict predicates. A consistency model in normal form only contains visibility \replaced{formulas}{relations} that are simple, non-vacuous and ``conflict-maximal''. 
A \emph{conflict-strengthening} of a visibility \replaced{formula}{relation} $\mathsf{v}$ is a visibility \replaced{formula}{relation} $\mathsf{v}'$ obtained from $\mathsf{v}$ by (1) replacing some occurrence of $\writeConflicts{E}$ (resp., $\writeConflicts[x]{E}$) with $\writeConflicts{E'}$ (resp., $\writeConflicts[x]{E'}$) where $E'$ is a strict superset of $E$ or (2) removing predicate $\writeConflicts{E}$ if $\writeConflicts[x]{E}$ also belongs to $\mathsf{v}$. A visibility \replaced{formula}{relation} $\mathsf{v}$ is \emph{conflict-maximal} w.r.t. $\opspec$ iff there is no conflict-strengthening $\mathsf{v}'$ such that for every execution $\exec$ over events in $\SpecEvents{\opspec}$, object $x$, and events $e_0, \ldots e_{\length{\mathsf{v}}}$, if $\mathsf{v}_x(e_0, \ldots e_{\length{\mathsf{v}}})$ holds in $\exec$, then $\mathsf{v}'_x(e_0, \ldots e_{\length{\mathsf{v}}})$ holds in $\exec$ as well. A consistency model $\cmodel$ is \emph{conflict-maximal} w.r.t. $\opspec$ iff all its visibility \replaced{formulas}{relations} are conflict-maximal w.r.t. $\opspec$.

For example, if $\mathsf{Rel}_i^\mathsf{v} = \wro$, any instance of $\event_i$ must write on some object $y$. In conflict-maximal visibility \replaced{formulas}{relations}, \replaced{this fact is represented with a conflict predicate}{there is a conflict predicate} ($\writeConflicts{E}$ or $\writeConflicts[x]{E}$) s.t. $\event_{i-1} \in E$. If $\opspec$ requires that every event reading $y$ also writes on $y$, then in a conflict-maximal visibility \replaced{formula}{relation}, both $\event_{i-1}, \event_i$ belong to $E$. In general, if in any abstract execution, the events instantiating $\event_{i_1}, \ldots, \event_{i_j}$ from $\mathsf{v}_x$ always conflict (resp. they always write $x$), then the visibility \replaced{formula}{relation} $\mathsf{v}$ must contain the predicate $\writeConflicts{\event_{i_1}, \ldots, \event_{i_j}}$ (resp. $\writeConflicts[x]{\event_{i_1}, \ldots, \event_{i_j}}$).

\begin{definition}
A consistency model $\cmodel$ is called in \emph{normal form w.r.t. a operation specification $\opspec$} if it contains only simple, conflict-maximal visibility \replaced{formulas}{relations} and no visibility \replaced{formula}{relation} from $\cmodel$ is vacuous w.r.t. $\opspec$.
\end{definition}

Under some operation specifications, consistency models can be equivalent due to conflict predicates. For example, in a storage with only $\ifaakvName$ operations, \deleted{both $\PSI$, }$\SI$ and $\SER$ are equivalent due to the $\axconf$ visibility \replaced{formula}{relation}: in this specification, every event is both a read and a write event and so any event reading $x$ conflicts with an event writing $x$. \deleted{On the other hand, in a storage with only $\iputName$ and $\igetName$ operations, $\PSI$ and $\CC$ are equivalent: the $\axconf$ visibility {relation} is vacuous as there are no events that can both read and write at the same time.}

Similarly to \Cref{sec:afc-theorem}, we say that a consistency model $\cmodel$ is \emph{arbitration-free} w.r.t. an operation specification $\opspec$ if there exists a consistency model in general normal form w.r.t. $\opspec$ that is equivalent to $\cmodel$ and whose visibility \replaced{formulas}{relations} are arbitration-free. \addedAppendix{\Cref{app:finite-normal-consistency}}\addedJournal{\textit{the arXiv}} demonstrates the existence of a normal form and shows that it is not possible for two normal forms to differ solely in that one includes only arbitration-free visibility \replaced{formulas}{relations} while the other does not. This result confirms that arbitration-freedom is not a property of the chosen normal form, but rather an inherent characteristic of the definitions of $\cmodel$ and $\opspec$.

\smallskip
\noindent
\textbf{Causal Suffix Closure.}
\label{ssec:causal-suffix-closure}
\added{We introduce an assumption about consistency models which is used in the proof of the AFC theorem in order to find counterexamples to availability that involve only two replicas. This assumption is satisfied by all practical cases that we are aware of (see \Cref{ex:causal_suffix_models}).}

Therefore, we assume that every normal form $\cmodel$ of a consistency model is \emph{closed under causal suffixes}, i.e., for every visibility \replaced{formula}{relation} $\mathsf{v}_\key\in \cmodel$, $\cmodel$ contains every arbitration-free ``suffix'' of $\mathsf{v}_\key$ that starts with an event writing $x$. Thinking about a visibility \replaced{formula}{relation} $\mathsf{v}$ as a path of dependencies (between the pairs $(\event_{i-1}, \event_{i})$), a suffix of $\mathsf{v}$ is a suffix of that path. For example, the visibility \replaced{formulas}{relations} $s$ and $s'$ described in \Cref{eq:counterexample-suffix-closed:suffix1} and \Cref{eq:counterexample-suffix-closed:suffix2} are suffixes of the visibility \replaced{formula}{relation} in \Cref{eq:counterexample-suffix-closed:initial}. 
\begin{align}
& \mathsf{v}_x(\event_0, \event_3) = \exists \event_1, \event_2. (\event_0, \event_1) \in \rbo \land (\event_1, \event_2) \in \ar \land (\event_2, \event_3) \in \so \land \wro_x^{-1}(\event_3) \neq \emptyset \land \writeConflicts[x]{\event_0, \event_1, \event_2} \label{eq:counterexample-suffix-closed:initial} \\
& s_x(\event_1, \event_3) = \exists \event_2. (\event_1, \event_2) \in \ar \ \land \ (\event_2, \event_3) \in \so \ \land\ \wro_x^{-1}(\event_3) \neq \emptyset  \land \ \writeConflicts[x]{\event_1, \event_2}\label{eq:counterexample-suffix-closed:suffix1} \\[.5mm]
& s'_x(\event_2, \event_3) = (\event_2, \event_3) \in \so \ \land\ \wro_x^{-1}(\event_3) \neq \emptyset  \land \ \writeConflicts[x]{\event_2} \label{eq:counterexample-suffix-closed:suffix2}
\end{align}

Formally, let $\mathsf{v}_x$ be a visibility \replaced{formula}{relation} defined as in \Cref{eq:general-visibility-criterion}. Let $\writeConflicts[x]{\mathsf{v}_x}$ be the union of the sets $E$ such that $\writeConflicts[x]{E}$ occurs in the definition of $\mathsf{v}_x$. For any variable $\event_k\in \writeConflicts[x]{\mathsf{v}_x}$, the $\event_k$-suffix of $\mathsf{v}_x$ is the \replaced{formula}{relation} \added{obtained by (1) removing the quantifiers for the first $k$ quantified events, $e_1 \ldots e_{k}$, and (2) removing all occurrences of the (now) free variables $e_0, \ldots e_{k-1}$, i.e.:}
\begin{align*}
\suffixOf{x}{\mathsf{v}_x,k}(\event_{k},\event_{n}) \Coloneqq  \exists \event_{k+1}, \ldots, \event_{n-1}. 
\bigwedge_{i=k+1}^{n} (\event_{i-1}, \event_{i}) \in \mathsf{Rel}^{\mathsf{v}}_i  \land  
\wro_x^{-1}(\event_{n}) \neq \emptyset \land \writeConstraints{x}{\event_{k}, \ldots \event_{n}} 
\end{align*}
where 
$\writeConstraints{x}{\event_{k}, \ldots \event_{n}}$ is obtained from $\writeConstraints{x}{\event_{0}, \ldots \event_{n}}$ by projecting all the conflict predicates over the set of events $E_k = \{\event_{k},\ldots,\event_{n}\}$, i.e., a predicate $\writeConflicts{E}$ (resp. $\writeConflicts[x]{E}$) occurs in $\writeConstraints{x}{\event_{0}, \ldots \event_{n}}$ iff $\writeConflicts{E\cap E_k}$ (resp. $\writeConflicts[x]{E\cap E_k}$) occurs in $\writeConstraints{x}{\event_{k}, \ldots \event_{n}}$.%

We refer to arbitration-free suffixes as \emph{causal}, since the remaining dependencies intuitively reflect broader notions of causality. 
The intuition behind this notion of closure is that the context of an invocation should be upward-closed with respect to causality—meaning that if an update \added{(writing $x$)} is included, then any later updates  \added{(writing $x$)} along the dependency path defined by the visibility \replaced{formula}{relation} that lie in its causal past must also be included.

We say that a visibility \replaced{formula}{relation} $\mathsf{v}'$ \emph{subsumes} a visibility \replaced{formula}{relation} $\mathsf{v}$ of the same length if for every $i, 1 \leq i \leq \length{v}$, $\mathsf{Rel}_i^{v'}$ is stronger or equal than $\mathsf{Rel}_i^\mathsf{v}$. We say that $\rbo$ is stronger than $\so$ and $\wro$, and $\ar$ is stronger than $\rbo, \so$ and $\wro$. The extension of ``being stronger'' to any relation $\mathsf{Rel}$ described using \Cref{eq:relations-visibility} is done as expected, as all our operators are positive (there are no negations).

\begin{definition}
A consistency model $\cmodel$ is \emph{closed under causal suffixes} if for every $\mathsf{v}_x\in \cmodel$ and $\event_k\in \writeConflicts[x]{\mathsf{v}_x}$, $\cmodel$ includes some visibility \replaced{formula}{relation} $\mathsf{v}'$ that subsumes every arbitration-free suffix of $v$.%
\end{definition}

\begin{example}\label{ex:causal_suffix_models}
A consistency model containing the visibility \replaced{formula}{relation} $\mathsf{v}$ in \Cref{eq:counterexample-suffix-closed:initial} must also contain the visibility \replaced{formula}{relation} $s'$ in order to be closed under causal suffixes. Note that $s$ uses arbitration and it is not required to be included. 

Any \simpleConsistencyFullName{} consistency model is closed under causal suffixes because every \simpleConsistencyFullName{} visibility formula has no proper arbitration-free suffix. Indeed, $\writeConflicts[x]{\mathsf{v}_x}$ contains just the first event $\event_0$ (assuming that $\writeVar{\event_0}{x}$ is rewritten as $\writeConflicts[x]{\{\event_0\}}$).
The models described in \Cref{fig:consistency_defs,fig:consistency_conflict_def} are trivially closed under causal suffixes because their visibility \replaced{formulas}{relations} have no arbitration-free suffixes.
\end{example}

\subsection{Operation Specifications}
\label{ssec:general-object-specs}

We generalize operation specifications to allow operations to access (read or write) multiple objects, and to support read values that are not limited to the inputs of individual write operations. For example, this includes multi-value reads that return all concurrently written values for an object, or counter reads that return an aggregated value computed from all observed increments.

The generalized reading behavior is modeled using two functions $\mathsf{rspec}$ and $\mathsf{extract}$ described hereafter. We also introduce a generalized $\mathsf{wspec}$ function. Therefore, $\mathsf{rspec}$ selects from a given context the events (updates) which are relevant for a reading invocation, $\mathsf{extract}$ defines the value read by an invocation, if any (based on the output of $\mathsf{rspec}$), and $\mathsf{wspec}$ defines the value written by an invocation, if any (to model conditional read-writes, this is based on the output of $\mathsf{extract}$).

\begin{definition}
\label{def:general-read-spec}
A \emph{read specification} $\mathsf{rspec}: \Events \to \Vars \to \Contexts \to \mathcal{P}(\Events)$ is a function such that for every object $x$, context $c = \tup{E, \rbo, \ar}$ and event $e$:
\begin{enumerate}
    \item well-formedness: $\rspecContext{x}{c}{e} \subseteq E$, and if $e$ is an initial event, $\rspecContext{x}{c}{e} = \emptyset$, and\label{def:general-read-spec:well-formedness}

    \item unconditional reading: if $\rspecContext{\key}{c}{e} \neq \emptyset$ for some context $c$, then for every non-empty context $c'$, $\rspecContext{\key}{c'}{e} \neq \emptyset$ \label{def:general-read-spec:unconditional-read}
\end{enumerate}
\end{definition}

\Cref{eq:rspec-faa-lww,eq:rspec-mvr,eq:rspec-counter} describe the read specifications of $\faacasStorage$, a key-value store $\kvmvrStorage$ %
with $\iput{x}{v}$ and multi-value $\iget{x}$ operations\addedAppendix{ (\Cref{ssec:mvr-kv-storage})},
and a collection of distributed counters $\counterStorage$ with $\iincCounter{x}$ and $\irdCounter{x}$ operations\addedAppendix{ (\Cref{ssec:counter})}. Concerning the relationship to \emph{\simpleStorageFullName{}} read specifications, note that the $\faacasStorage$ specification in \Cref{eq:rspec-faa-simple} was simpler because the constraint from \Cref{eq:rspec-faa-lww} was imposed in the notion of validity for abstraction executions (\Cref{def:valid}). For multi-value reads (\Cref{eq:rspec-mvr}), the read specification selects the maximal elements in the receive-before relation (which models causality), and for a counter (\Cref{eq:rspec-counter}), it returns all events in the context.
\begin{equation}
\label{eq:rspec-faa-lww}
{\arraycolsep=1.5pt
\rspecContext{x}{c}{r} =  \left\{ \begin{array}{ll}
    \{\max_{\ar} E\}, & \text{if } r \in \{\ireadkv{x}, \ifaakv{x}{v}, \icaskv{x}{v}{v'}\} \text{ and } c = \tup{E, \rbo, \ar} \\
    \ \ \emptyset, & \text{otherwise}
\end{array} \right.
}
\end{equation}

\begin{equation}
\label{eq:rspec-mvr}
{\arraycolsep=2pt
\rspecContext{x}{c}{r} =  \left\{ \begin{array}{ll}
\max_{\rbo}E, &  \text{if } r  = \iget{x} \text{ and } c = \tup{E, \rbo, \ar} \\
\ \ \emptyset, & \text{otherwise}
\end{array} \right.
}
\end{equation}
    
\begin{equation}
\label{eq:rspec-counter}
\rspecContext{x}{c}{r} =  \left\{ \begin{array}{ll}
    E, & \text{if } r = \irdCounter{x} \text{ and } c = \tup{E, \rbo, \ar} \\
    \ \ \emptyset, & \text{otherwise}
\end{array} \right.
\end{equation}

The extract specification below computes the value returned from an object $\key$ based on the set of invocations writing $\key$ returned by the read specification which are paired with values they write (this will become clearer when defining the application of these functions on an abstract execution). 

\begin{definition}
\label{def:extract-spec}
An \emph{extract specification} $\mathsf{extract}: \Events \to \Vars \to \mathcal{P}(\Events \times \Vals) \to \Vals$, such that $\mathsf{extract}(\init)$ is defined for every initial event $\init$.
\end{definition}

\Cref{eq:extract-faa-lww} \replaced{describes the extract specification of $\faacasStorage$:}{describes the extract specifications of both $\faacasStorage$ and $\kvmvrStorage$:} the value extracted for $\mathtt{GET}$, $\mathtt{FAA}$ and $\mathtt{CAS}$ coincides with the value written by some previous $\mathtt{PUT}$/$\mathtt{FAA}$/$\mathtt{CAS}$ operation. \Cref{eq:extract-kv-mvr} \added{describes the extract specification of $\kvmvrStorage$: the value extracted for $\mathtt{GET}$ is the set of values written by some previous $\iputName$.} In the case of $\counterStorage$, \Cref{eq:extract-counter}, the value extracted for $\mathtt{rd}$ returns the number of increment invocations in the input, which equals $|R|$ minus one for the initial event $\init$ which is always included in $R$ (since it is $\so$ before all other events). %
\begin{equation}
\label{eq:extract-faa-lww}
{\arraycolsep=1pt
\begin{array}{lll}
\extractspecContext{x}{R}{r} = \left\{ 
    \begin{array}{ll}
        v & \text{if } r \in \{\ireadkv{x}, \ifaakv{x}{v'}, \icaskv{x}{v'}{v''}\}\text{ and } R = \{(w, v)\} \\
        \mathsf{undefined} & \text{otherwise}
    \end{array} \right. \\
\end{array}}
\end{equation}

\begin{equation}
\label{eq:extract-kv-mvr}
\begin{array}{lll}
\extractspecContext{x}{R}{r} = \left\{ 
    \begin{array}{ll}
        \{v \ | \ (\_, v)\in R\} & \text{if } r = \ireadkv{x}\\
        \mathsf{undefined} & \text{otherwise}
    \end{array} \right. \\
\end{array}
\end{equation}
    
\begin{equation}
\label{eq:extract-counter}
\extractspecContext{x}{R}{r} = \left\{ 
    \begin{array}{ll}
        |R| - 1& \text{if } r = \irdCounter{x} \\
        \mathsf{undefined} & \text{otherwise}
    \end{array} \right.
\end{equation}

Finally, the write specification computes the value written by an invocation to an object $x$, based on the values it reads. This makes it possible to model atomic read-writes, e.g., a compare-and-swap, which may write or not depending on what they read, or the value they write may change depending on what they read, e.g., a Fetch-and-Add. %

\begin{definition}
\label{def:write-spec}
A \emph{write specification} $\mathsf{wspec} : \Events \to \Vars \to \Vals \to \Vals$ is a function such that $\mathsf{wspec}(\init)$ is defined for every initial event $\init$.
\end{definition}

The write specification of $\faacasStorage$ and $\kvmvrStorage$, \Cref{eq:wspec-faa-lww}, describes that its write operations are $\mathtt{PUT}$, $\mathtt{FAA}$ and $\mathtt{CAS}$. $\mathtt{PUT}$ and $\mathtt{FAA}$ unconditionally writes on $x$ while $\mathtt{CAS}$ does it depending on the read-and-extracted value of $x$; where $x$ is the only object accessed by the invocation. In the case of $\counterStorage$, \Cref{eq:wspec-counter}, only the operation $\iincCounter{x}$ writes, writing a dummy value $1$ just to indicate that the write has taken place.
\vspace{-3mm}
\begin{equation}
\label{eq:wspec-faa-lww}
\wspecContext{x}{v}{w} = \left\{ 
    \begin{array}{ll}
        v' & \text{if } w = \iwritekv{x}{v'}\\
        v + v' & \text{if } w = \ifaakv{x}{v'}\\
        v'' & \text{if } w = \icas{x}{v'}{v''} \land v = v'\\
        \mathsf{undefined} & \text{otherwise}
    \end{array} \right.
\end{equation}

\begin{equation}
\label{eq:wspec-kv-mvr}
\wspecContext{x}{v}{w} = \left\{ 
    \begin{array}{ll}
        v' & \text{if } w = \iwritekv{x}{v'}\\
        \mathsf{undefined} & \text{otherwise}
    \end{array} \right.
\end{equation}

\begin{equation}
\label{eq:wspec-counter}
\wspecContext{x}{\_}{w} = \left\{ 
    \begin{array}{ll}
        1 & \text{if } w = \iincCounter{x}\\
        \mathsf{undefined} & \text{otherwise}
    \end{array} \right.
\end{equation}

\begin{definition}
\sloppy An \emph{operation specification} is a tuple $\opspec=(E,\rspecName, \extractName, \wspecName)$ where $E$ is a set of events. %
$\SpecEvents{\opspec}$ refers to the set of events $E$ in an operation specification.
\end{definition}

\addedAppendix{\Cref{app:examples}}\addedJournal{\textit{the arXiv}} contains more examples of operation specifications, including SQL statements.

\subsection{Validity w.r.t. Storage Specifications}

We extend the notion of validity for abstract executions to (general) storage specifications, in a way that is similar to the case of \simpleStorageFullName{} storage specifications (\Cref{ssec:validity}). \deleted{We similarly define a $\mathsf{value}$ function which returns the value written by an event $w$ to an object $x$ from an abstract execution $\exec$ under a consistency model $\cmodel$, if any.}
\ifbool{diffMode}{%
\begin{equation}%
\label{eq:value-wr-general}%
\deleted{%
\valuewrExec{w}{x}{\exec}{\cmodel} = \left\{\begin{array}{ll}%
    v_x^0 & \text{if } w \text{ is an initial event} \\
    \wspec{\cmodel}{x}{\exec}{w} & \text{otherwise}
	\end{array}\right.%
}%
\end{equation}%
}{%
}%
\deleted{where $v_x^0$ is some fixed initial value.In the above, we use an extension of $\mathsf{wspec}$ to abstract executions whose definition (below) relies on the extended versions of $\mathsf{rspec}$ and $\mathsf{extract}$ for abstract executions:}%
\ifbool{diffMode}{%
\color{oldVersionColor}%
\begin{align*}%
&\rspec{\cmodel}{x}{\exec}{e} = \rspecContext{x}{\context{\cmodel}{x}{\exec}{e}}{e} \\%
&\extractspec{\cmodel}{x}{\exec}{e} = \mathsf{extract}(e)\left(x, \left\{ (e',v) \left| \begin{array}{l}%
    e' \in \rspec{\cmodel}{x}{\exec}{e} \ \land \\%
    \ v= \wspec{\cmodel}{x}{\exec}{e'}%
\end{array} \right.%
\right\}\right)\\%
&\wspec{\cmodel}{x}{\exec}{e} = \wspecContext{x}{\extractspec{\cmodel}{x}{\exec}{e}}{e}%
\end{align*}%
\color{black}%
}%
{%
}%

\deleted{The mutual recursion between the functions $\wspec{\cmodel}{x}{\exec}{e}$ and $\extractspec{\cmodel}{x}{\exec}{e}$ terminates because $\rspec{\cmodel}{x}{\exec}{e} \subseteq \context{\cmodel}{x}{\exec}{e}$ (see \mbox{\Cref{def:general-read-spec}}), every event $e'$ in the context $\context{\cmodel}{x}{\exec}{e} = \tup{E, \rbo, \ar}$ precedes $e$ in $\ar$, the order $\ar$ is well-founded (with some initial event as minimal event) and $\mathsf{extract}$ is defined for all initial events. If some of the applications of $\mathsf{extract}$ or $\mathsf{wspec}$ are undefined, then the extensions are undefined as well.}%
\deleted{
The predicate $\writeVar{e}{x}$ is interpreted as true iff $\valuewrExec{w}{x}{\exec}{\cmodel}$ is defined.
\Cref{def:general-valid} generalizes the notion of validity for \simpleStorageFullName{} storage specifications (\Cref{def:valid}).} 
\added{We use the extension of $\rspecName, \extractName$, and $\wspecName$ to abstract executions defined below:}
\begin{align*}%
&\added{\rspec{\cmodel}{x}{\exec}{e} = \rspecContext{x}{\context{\cmodel}{x}{\exec}{e}}{e}} \\%
&\added{\extractspec{\cmodel}{x}{\exec}{e} = \mathsf{extract}(e)\left(x, \left\{ (e',\outputEventObj{e}{x}) \left| \ e' \in \rspec{\cmodel}{x}{\exec}{e} \right. \right\}\right)}\\%
&\added{\wspec{\cmodel}{x}{\exec}{e} = \wspecContext{x}{\extractspec{\cmodel}{x}{\exec}{e}}{e}}%
\end{align*}%

\begin{definition}
\label{def:general-valid}
Let $\sspec = (\cmodel, \opspec)$ be a storage specification. %
An abstract execution $\exec = \tup{h, \rbo, \ar}$ of a history $\hist=\tup{E, \so, \wro}$ is \emph{valid} w.r.t. $\sspec$ iff 
\begin{itemize}
\item $\exec$ contains events from the operation specification, i.e., $E \subseteq \SpecEvents{\opspec}$, 
\item for every event $r\in E$, $\wro_x^{-1}(r) = \rspec{\cmodel}{x}{\exec}{r}$, and
\item \added{the value written by each event $e \in E$ to object $x$ is consistent with $\wspecName$, i.e., $\outputEventObj{e}{x} = \wspec{\cmodel}{x}{\exec}{e}$.}
\end{itemize}
A history $\hist$ is valid w.r.t. $\sspec$ iff there exists an abstract execution of $\hist$ which is valid w.r.t. $\sspec$.
\end{definition}

Observe that \Cref{def:general-valid} coincides with \Cref{def:valid} for storage systems that also admit \simpleStorageFullName{} storage specifications, e.g., $\faacasStorage$.

\subsection{Assumptions About Operation Specifications}
\label{ssec:assumptions-generalized-specs}

To avoid pathological behaviors in the generalization of specifications, we make several assumptions.

\smallskip
\noindent
\textbf{Maximally-Layered Read Specifications.}
\label{ssec:maximally-layered}
For any \simpleStorageFullName{} operation specification $\opspec$, the validity of an abstract execution w.r.t. a stronger consistency model (and $\opspec$) implies validity w.r.t. a weaker one \added{(see \Cref{lemma:normal-form:consistency-stronger-than-criterion})}. \replaced{In general, this}{This} is not true for operation specifications as described in this section (see \Cref{ex:rspec:counter-maxl}). Therefore, we introduce an assumption about read specifications, called \emph{maximally-layered}, which ensures that this property remains true.

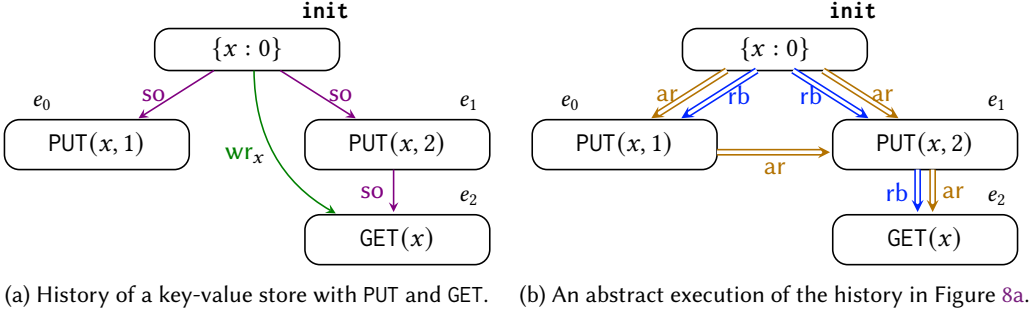
\begin{figure}[t]
\centering
\begin{subfigure}[b]{.495\textwidth}
\centering
\begin{tikzpicture}[>=stealth,shorten >=1pt,auto,node distance=3cm,
    semithick, transform shape,
    B/.style = {%
    decoration={brace, amplitude=1mm,#1},%
    decorate},
    B/.default = ,  %
    ]

    \node[minimum width=7em, draw, rounded corners=2mm,outer sep=0, label={[font=\small]25:$\init$}] (init) at (2, 0) {$\{x: 0\}$};

    \node[minimum width=7em, draw, rounded corners=2mm,outer sep=0, label={[font=\small]145:$e_0$}] (r11) at (0, -1.25) {
        \begin{tabular}{l}
            $\iput{x}{1}$
        \end{tabular}
    };

    \node[minimum width=7em, draw, rounded corners=2mm,outer sep=0, label={[font=\small]25:$e_1$}] (r12) at (4, -1.25) {
        \begin{tabular}{l}
            $\iput{x}{2}$
        \end{tabular}
    };

     \node[minimum width=7em, draw, rounded corners=2mm,outer sep=0, label={[font=\small]25:$e_2$}] (r22) at (4, -2.5) {
        \begin{tabular}{l}
            $\iget{x}$
        \end{tabular}
    };

    \path (init) edge[->, soColor, right] node [above, left] {$\so$} (r11);

    \path (init) edge[->, soColor, right] node [black, right] {$\so$} (r12);

    \path (r12) edge[->, soColor, right, transform canvas={xshift=-.5mm}] node [black, left] {$\so$} (r22);

    \path (init) edge[->, bend right, wrColor, right, transform canvas={xshift=.5mm}] node [black, left] {$\wro_x$} (r22);

\end{tikzpicture}  
\caption{\added{History of a key-value store with $\iputName$ and $\igetName$.}}
\label{fig:counterexample-maxl:history}
\end{subfigure}
\hfill
\begin{subfigure}[b]{.495\textwidth}
\centering
\begin{tikzpicture}[>=stealth,shorten >=1pt,auto,node distance=3cm,
    semithick, transform shape,
    B/.style = {%
    decoration={brace, amplitude=1mm,#1},%
    decorate},
    B/.default = ,  %
    ]

    \node[minimum width=7em, draw, rounded corners=2mm,outer sep=0, label={[font=\small]25:$\init$}] (init) at (2, 0) {$\{x: 0\}$};

    \node[minimum width=7em, draw, rounded corners=2mm,outer sep=0, label={[font=\small]145:$e_0$}] (r11) at (0, -1.25) {
        \begin{tabular}{l}
            $\iput{x}{1}$
        \end{tabular}
    };

    \node[minimum width=7em, draw, rounded corners=2mm,outer sep=0, label={[font=\small]25:$e_1$}] (r12) at (4, -1.25) {
        \begin{tabular}{l}
            $\iput{x}{2}$
        \end{tabular}
    };

     \node[minimum width=7em, draw, rounded corners=2mm,outer sep=0, label={[font=\small]25:$e_2$}] (r22) at (4, -2.5) {
        \begin{tabular}{l}
            $\iget{x}$
        \end{tabular}
    };

    \path (init) edge[->, double equal sign distance, arColor, right, transform canvas={xshift=-2mm}] node [above, left] {$\ar$} (r11);

    \path (init) edge[->,  transform canvas={xshift=2mm}, double equal sign distance, rbColor] node [black, right] {$\rbo$} (r11);

    \path (init) edge[->, double equal sign distance, arColor, transform canvas={xshift=2mm}] node [black, right] {$\ar$} (r12);

    \path (init) edge[->,  transform canvas={xshift=-2mm}, double equal sign distance, rbColor] node [black, left] {$\rbo$} (r12);

    \path (r11) edge[->, double equal sign distance, arColor, right, transform canvas={yshift=-1mm}] node [black, below] {$\ar$} (r12);

    \path (r12) edge[->, rbColor, right, double equal sign distance, transform canvas={xshift=-1mm}] node [black, left] {$\rbo$} (r22);

    \path (r12) edge[->, arColor, right, double equal sign distance, , transform canvas={xshift=1mm}] node [black, right] {$\ar$} (r22);

\end{tikzpicture}  
\caption{\added{An abstract execution of the history in \Cref{fig:counterexample-maxl:history}.}}
\label{fig:counterexample-maxl:execution}
\end{subfigure}

\vspace{-1mm}
\caption{\added{A history and an abstract execution of the operation specification in \Cref{ex:rspec:counter-maxl}. For readability, we omit the $\so$ and $\wro$ relations from the abstract execution. Events $e_1$ and $e_2$ are executed in the same replica, different from $e_0$'s replica.}}
\label{fig:counterexample-maxl}
\vspace{-4mm}
\end{figure}

\begin{example}
\label{ex:rspec:counter-maxl}
\added{Let $\opspec = \tup{E, \rspecName,\extractName, \wspecName}$ be an operation specification of a key-value store with $\igetName$ and $\iputName$ operations whose read specification is given by \Cref{eq:counterexample-maximally-layered}.
\begin{equation}
\label{eq:counterexample-maximally-layered}
	\rspecContext{x}{c}{e} = \left\{ \begin{array}{ll}
		\{\max_{\ar} E\} & \text{if } \nexists e' \in E \text{ s.t. } \replicaEvent{e} \neq \replicaEvent{e'} \text{ and } c = \tup{E, \rbo, \ar}\\
		\init & \text{otherwise}
	\end{array}\right.
\end{equation}
We compare the validity of the abstract execution $\exec$ depicted in \Cref{fig:counterexample-maxl} w.r.t. $\SC$ and $\CC$ (observe that $\CC \preccurlyeq \SC$). Under $\SC$ both $e_0$ and $e_1$ are visible to $e_2$, which implies $\rspec{\SC}{x}{\exec}{e_2} = \{\init\}$. Therefore, $\exec$ is valid w.r.t. $\SC$. However, under $\CC$, only $e_1$ is visible to $e_2$, which implies $\rspec{\CC}{x}{\exec}{e_2} = \{e_1\}$, and therefore, $\exec$ is not valid w.r.t. $\CC$. 
}	
\end{example}

Let $\leq$ be a partial order over a set $A$. A chain of $\leq$ is a subset of $A$ which is totally ordered w.r.t. $\leq$. The \emph{layer} of an element $a\in A$ is the size of the largest chain of $\leq$ which includes $a$ but no elements smaller than $a$, and a maximal element. For instance, the layer of a maximal element is $1$ (the aforementioned largest chain includes just the element itself), the level of a strict predecessor of a maximal element is \replaced{$2$}{$1$}, and so on. A subset $B\subseteq A$ is called \emph{$k$-maximally layered} w.r.t. $\leq$ if $B$ is the set of all elements in $A$ of layer $k'\leq k$. %
When $\leq$ is also a total order, the notion of maximally layered is equivalent to being upward closed w.r.t. $\leq$. Otherwise, it is equivalent to being upward closed w.r.t. every total extension of $\leq$.

A read specification $\mathsf{rspec}$ is \emph{$k$-maximally layered} w.r.t. $\ar$ (resp. $\rbo^+$) if for every object $x$, context $c  = \{E, \rbo, \ar\}$, and event $e$, either $\rspecContext{x}{c}{e}= \emptyset$ or $\rspecContext{x}{c}{e}$ is $k$-maximally layered w.r.t. $\ar$ (resp. $\rbo^+$). The \emph{layer bound} of $\mathsf{rspec}$ is defined as $k$. To cover cases where there is no such $k$, we say that a read specification $\mathsf{rspec}$ is \emph{$\infty$-maximally layered} if for every $x$, context $c  = \{E, \rbo, \ar\}$, and event $e$, either $\rspecContext{x}{c}{e}= \emptyset$ or $E$; and we say that the layer bound is $\infty$. When the layer bound and the partial order ($\ar$ or $\rbo^+$) are irrelevant, we simply say that $\mathsf{rspec}$ is \emph{maximally layered}.

\begin{example}\label{ex:rspec:maxl}
For example, $\faacasStorage$ is $1$-maximally layered w.r.t. $\ar$, $\kvmvrStorage$ is $1$-maximally layered w.r.t. $\rbo^+$ and $\counterStorage$ is $\infty$-maximally layered. 
\replaced{On the other hand, the read specification in \Cref{ex:rspec:counter-maxl} is not maximally layered since it can sometimes return $\init$ from a non-empty context.}
{On the other hand, a read specification which returns the minimal event w.r.t. $\ar$ is not maximally-layered. Such a read specification always returns the initial event, and hence, \emph{all consistency models are equivalent}. Thus, an available implementation of $(\CC, \opspec)$, where $\opspec$ includes this read specification, is also an implementation of $(\SER, \opspec)$, and no separation can be made between consistency models w.r.t. availability.}
\vspace{-1mm}
\end{example}

\begin{restatable}{lemma}{generalizationStrongerThan}
\label{lemma:general-normal-form:consistency-stronger-than-criterion}
\added{Let $\opspec$ be a maximall-layered operation specification and let $\cmodel_1, \cmodel_2$ be a pair of consistency models such that $\cmodel_2$ is stronger than $\cmodel_1$.
Any abstract execution valid w.r.t. $(\cmodel_2, \opspec)$ is also valid w.r.t. $(\cmodel_1, \opspec)$.}
\end{restatable}

\noindent
\textbf{Operation Closure.}
\label{ssec:general-operation-closure}
As in \Cref{ssec:opspec}, we assume that $\opspec$ contains at least a read and a write event. Also, we assume that all objects support a common set of operations with identical read and write behavior, and that these operations can be executed at any replica. Formally, for every event $e\in E$, replica $\replicaID$, and identifier $\identifier$, there exists an event $e'$ s.t. 
$\replicaEvent{e'} = \replicaID$, $\identifierEvent{e'} = \identifier$, $\varOf{e'} = \varOf{e}$, $\rspecSimple{e'} = \rspecSimple{e}$, $\mathsf{extract}(e')=\mathsf{extract}(e)$, and $\mathsf{wspec}(e') = \mathsf{wspec}(e)$.

We also assume that operations apply uniformly to any set of objects. To formalize this assumption, we define a notion of  \emph{domain} for an operation specification $\opspec$ which is any set of objects $D$ s.t. there is an event $e\in \SpecEvents{\opspec}$ for which $\varOf{e} = D$. We assume that domains are ``symmetric'', i.e. if $D$ is a domain for $\opspec$, then for every pair of objects $x \in D$ and $y \in \Vars \setminus D$, the set $D' = D \setminus \{x\} \cup \{y\}$ is also a domain for $\opspec$. If $\opspec$ allows single-object read/write/read-write events (defined as in \Cref{ssec:opspec}), we assume that for every object $x$, there exists a read/write/read-write event whose domain is $\{x\}$. Also, we assume that if $\opspec$ allows a \emph{multi-object} read/write/read-write event $e$ such that $\varOf{e}$ is a finite set of size at least $2$, then for every non-empty finite set $D \subseteq \Vars$, $D$ is a domain of a read/write/read-write event in $\opspec$.

\vspace{.5mm}
\noindent
\textbf{Correctors.}
In addition, we assume that if $\opspec$ permits conditional read-write events--which write to a set of objects $X$ based on values they read (possibly from other objects) in some context--then any execution can be extended with some conditional read-write event $e$ that writes to every object in $X$, modulo a so-called correction defined below. This property is only relevant for events with $|\varOf{e}|>1$ (and therefore, irrelevant for basic storage specifications). \added{Our proof will rely on the existence of such extensions.}

\begin{example}
\label{ex:corrector-intuition}
To provide some intuition about \added{the need for} corrections, consider \replaced{a specification formed of prefix consistency ($\PC$) and  an operation specification with two multi-object operations, $\iinsifabsName$ and $\idelifpresName$, under Last-Writer-Wins (LWW) conflict resolution (i.e., the read specification selects the maximal invocation from the context w.r.t. $\ar$)\addedAppendix{ (see \Cref{ssec:mvr-insdel-storage})}. $\iinsifabs{X}{v}$ checks for every object $x \in X$ if it is present, and inserts it with value $v$ if not, and $\idelifpres{X}$ deletes every object $x \in X$ as long as it was present.}{a storage system with operations $\mathtt{INSERT}$/$\mathtt{DELETE}$ that can insert/delete an arbitrary set of objects. We assume that an $\mathtt{INSERT}$, resp., $\mathtt{DELETE}$, operation \emph{writes} if and only if the object is absent, resp., present.}Assume an abstract execution $\exec$, \deleted{a consistency model $\cmodel$, }and an event $e$ from $\exec$ whose context implies that $x$ is absent and $y$ is present. If $e$ is an invocation of $\iinsifabsName$ (resp., $\idelifpresName$), then it can not write both objects since $x$ is absent and $y$ is present.
\end{example}
We introduce the notion of \emph{corrector}, a set of auxiliary events that modify the context, ensuring the existence of an event that can write to both objects. For instance, \added{in the scenario presented in \Cref{ex:corrector-intuition},} if $e$ is an invocation of $\iinsifabs{\{x, y\}}{1}$, the corrector will add a $\idelifpres{\{y\}}$ invocation in its context, so both objects are absent.

We start by defining some notations. Let $\sspec = (\cmodel, \opspec)$ be a storage specification, $\exec = \tup{h, \rbo, \ar}$ an abstract execution of a history $h = \tup{E, \so, \wro}$, and $e \in E$ an event. A \emph{correction} of $e$ in $\exec$ with an event $a$, denoted by $\correction{\exec}{a}{e}$, is an abstract execution $\exec' = \tup{h', \rbo', \ar'}$ associated to a history $h' = \tup{E \cup \{a\}, \so', \wro'}$ obtained by adding $a$ as the immediate $\rbo$-predecessor and $\ar$-predecessor of $e$. If $\replicaEvent{e} = \replicaEvent{a}$, then $a$ is also the immediate $\so$-predecessor of $e$. The write-read dependencies ($\wro^{-1}$) of every event in $\exec$ remain the same. %
Multiple corrections exist because the write-read and receive-before dependencies of $a$ are not constrained. This allows flexibility on correcting $\exec$ while preserving validity w.r.t. $\sspec$. %

The correction of $\exec$ with a sequence of events $\vec{\mathsf{s}} = (a_1, a_2,\ldots)$, denoted by $\correction{\exec}{\vec{\mathsf{s}}}{e}$, is defined as expected, by iteratively correcting $\exec$ with all events in $\vec{\mathsf{s}}$ in the order defined by $\vec{\mathsf{s}}$. Therefore, if $e'$ is the immediate $\ar$-predecessor of $e$ in $\exec$, the $\ar$ order in $\correction{\exec}{\vec{\mathsf{s}}}{e}$ will have $a_1, a_2, \ldots$ inserted in between $e'$ and $e$ (in this order). Similarly for $\rbo$ and possibly for $\so$.
 
For a (partial) mapping $f : A \to B$ and a total order $<$ over $A$, the sequence of elements in $B$ mapped by $f$ and ordered according to $<$ is denoted by $\sequenceOrd{f}$. Formally, $\sequenceOrd{f} = (f(a_1), f(a_2), \ldots)$ such that $f(a_i)\downarrow$ and $a_i < a_{i+1}$ for all $i$. We omit the subscript $<$ when it is understood from the context.

Also, if $\exec$ is an abstract execution, then $\exec \oplus e$ is an abstract execution obtained from $\exec$ by appending $e$ to $\exec$ as the last event w.r.t. $\ar$.

\noindent
\textbf{Corrector Assumption.}
If $\opspec$ allows conditional read-writes, then we assume that for every domain $D$, $W \subseteq D$, $x\in \Vars$ s.t. $x\in W$ if $W \neq \emptyset$, and abstract execution $\exec$, there exists 
\begin{enumerate}
	\item a conditional read-write $e$ with $\varOf{e} = D$ which is not contained in $\exec$, and
	\item a partial mapping $a : D\setminus \{x\} \to \Events$ called \emph{execution-corrector} for event $e$ in an abstract execution $\exec \oplus e$.
\end{enumerate}

We define execution-correctors as follows.

\vspace{-1mm}
\begin{definition}
\label{def:execution-corrector} 
Let $\sspec = (\cmodel, \opspec)$ be a storage specification, $\exec$ an abstract execution, 
$e$ a conditional read-write event from $\exec$
with $\varOf{e} = D$, $W \subseteq D$ a set of objects, and $x \in D$ an object s.t. $x \in W$ if $W \neq \emptyset$. Also, let $<$ be a fixed total order on the set of objects. An \emph{execution-corrector} for $\tup{e, W, x,\exec}$ is a partial mapping $a : D\setminus \{x\} \to \Events$ such that %
if 
\begin{align*}
\
\exec' = \correction{\exec}{\sequence{a}}{e}\text{ and }\exec' \restriction y =  (\correction{\exec}{\sequence{a \restriction y}}{e} )\setminus \{e\} \text{ where } a \restriction y = a \restriction_{\{z \in \Dom{a} \ | \ z \leq y\}},
\end{align*}
then the following hold:
\begin{enumerate}

\item for every $y \in  D\setminus \{x\}$, if $a(y)$ is defined and the correction up to $a(y)$ is valid w.r.t. $\sspec$, then $a(y)$ writes only $y$ in the correction: if $a(y) \downarrow$ and $\exec' \restriction y$ is valid w.r.t. $\sspec$, then for every object $z \in \Vars$, $\wspec{\cmodel}{z}{\exec' \restriction y}{a(y)}\downarrow$ iff $z = y$, and\label{def:execution-corrector:1-context-corrector-writes}

\item for every $y \in  D$, if the correction is valid w.r.t. $\sspec$, then $e$ reads $y$ and additionally, $e$ writes $y$ only if $y \in W$, i.e., $\rspec{\cmodel}{y}{\exec'}{e} \neq \emptyset$ and $\wspec{\cmodel}{y}{\exec'}{e} \downarrow$ iff $y \in W$.
\label{def:execution-corrector:3-e-writes-x} 
\label{def:execution-corrector:4-e-writes-w} 
\label{def:execution-corrector:5-e-no-write} 
\end{enumerate}
\end{definition}
\begin{example}
We illustrate execution-correctors for the \replaced{storage specification presented in \Cref{ex:corrector-intuition}, with $\iinsifabsName$ and $\idelifpresName$ as operations and $\PC$ as consistency model.}{$\insdelStorage$ operation specification (\Cref{ssec:mvr-insdel-storage}) and prefix consistency ($\PC$). $\insdelStorage$ uses two multi-object operations, $\iinsifabsName$ and $\idelifpresName$, under LWW conflict resolution. $\iinsifabs{X}{v}$ checks for every object $x \in X$ if it is present, and inserts it with value $v$ if not, and $\idelifpres{X}$ deletes every object $x \in X$ as long as it was present.}

Let $\exec$ be an abstract execution, $e$ a $\idelifpres{D}$ event from $\exec$, $W \subseteq D$ a non-empty set of objects and $x \in W$. For every object $y$, let $w_y$ be the last event from the ``read'' context of $e$ w.r.t. $\PC$ which writes $y$ (by read context we mean the set of events selected by $\mathsf{rspec}$ from the context). In the following we assume that $w_x$ is an insert event. Note that if $w_x$ is a delete event, then there exists no execution-corrector for $e$ (intuitively, the correction concerns objects different from $x$, and $\idelifpres{D}$ will not delete an object which is already deleted).

An execution-corrector for $(e, W, x, \exec)$ is the mapping $a: D \setminus \{x\} \to \Events$ defined below. The mapping $a$ observes the update on $y$ made by $w_y$, and overwrites it when necessary. Thus, when $e$ reads $y$, $y$ is present iff $y \in W$.
\begin{equation}
a(y) = \left\{ 
	\begin{array}{ll}
		\iinsifabs{\{y\}}{v} & \text{if } y \in W \text{ and } w_y \text{ deletes } y \text{ in } \exec \\
		\idelifpres{\{y\}} & \text{if } y \not\in W \text{ and } w_y \text{ inserts } y \text{ in } \exec \\
		\mathsf{undefined} & \text{otherwise}
	\end{array} 
\right.
\end{equation}
\end{example}%

Observe that requiring that $a$ is defined for all objects in $D$ is too strict: if the read specification has a layer-bound of $1$ and the events read a single object (as $\faacasStorage$), any correction will change the entire context read by $e$.

	\section{The Arbitration-Free Consistency Theorem}
\label{sec:generalized-afc}

We now present our main result in its most general form, which extends Theorem~\ref{th:characterization-cons-available-lww}.

\begin{theorem}[Arbitration-Free Consistency (AFC)]
\label{th:characterization-cons-available}
Let $\sspec = (\cmodel, \opspec)$ be a storage specification. The following statements are equivalent:
\begin{enumerate}
    \item $\cmodel$ is arbitration-free w.r.t. $\opspec$,

    \item there exists an available $\opspec$-implementation.%

\end{enumerate}
\end{theorem}

\sloppy The proof of (1) $\Rightarrow$ (2) is very similar to that \replaced{in \Cref{th:characterization-cons-available-lww} (see \Cref{ssec:afc012}). The only difference is replacing \Cref{lemma:normal-form:consistency-stronger-than-criterion} with \Cref{lemma:general-normal-form:consistency-stronger-than-criterion} where we use the maximally-layered assumption of read specifications. }{of \Cref{lemma:cons-available-lww:1-2}. We rely on maximal-layerness to show that if $\cmodel$ is arbitration-free w.r.t. $\opspec$, then it is weaker than $\CC$.}%
For the reverse, we follow the \replaced{reasoning explained in the beginning of \Cref{ssec:afc021} to reduce to consistency models in normal form. \Cref{lemma:cons-available:3-1} extends the arguments in \Cref{lemma:cons-available-lww:3-1} to generalized storage specifications.}{same reasoning by contradiction as in \Cref{lemma:cons-available-lww:3-1}.}

\begin{restatable}{lemma}{notSatnotConsAvailable}%
\label{lemma:cons-available:3-1}%
\label{lemma:cons-available:no-implementation-with-ar}
\added{Let $\sspec = (\cmodel, \opspec)$ be a storage specification. Assume that $\cmodel$ contains a simple visibility formula $\mathsf{v}$ which is non-vacuous w.r.t. $\opspec$, such that for some $i, 0 \leq i \leq \length{\mathsf{v}}$, $\mathsf{Rel}_i^\mathsf{v} = \ar$. Then, there is no available $(\cmodel, \opspec)$-implementation.}
\end{restatable}%
\begin{sketchproof}
\replaced{As in \Cref{lemma:cons-available-lww:3-1}, we}{We} assume by contradiction that there is an available implementation $\implementationInstance$ of $\sspec$. \deleted{$\cmodel$ is not arbitration-free w.r.t. $\opspec$.} We use \replaced{the visibility formula $\mathsf{v}$}{the latter} to construct a specific program, which by the assumption, admits a trace (in the composition with this implementation) that contains no $\ereceive$ action. We show that \replaced{any}{the} abstract execution induced by this trace, which is admissible by any available implementation of $\sspec$, is not valid w.r.t. $\sspec$. This contradicts the hypothesis.

\deleted{As in \mbox{\Cref{lemma:cons-available-lww:3-1}}, since $\cmodel$ is not arbitration-free w.r.t. $\opspec$, there exists a non-vacuous visibility relation $\mathsf{v}$ in a general normal form $\generalNormalForm{\cmodel}$ of $\cmodel$ that is not arbitration-free. Recall that $\mathsf{v}$ is a formula as in \mbox{\Cref{eq:general-visibility-criterion}}. Since $\mathsf{v}$ is simple but not arbitration-free, there exists an index $i$ such that $\mathsf{Rel}_{i}^\mathsf{v} = \ar$. Let $d_\mathsf{v}$ be the largest such index $i$ (last occurrence of $\ar$).}\added{Let $d_\mathsf{v}$ be the largest index $i$ s.t. $\mathsf{Rel}_i^\mathsf{v} = \ar$ (last occurrence of $\ar$).} Then, $\mathsf{v}$ is formed of two parts: the path of dependencies from $\event_0$ to $\event_{d_\mathsf{v}}$ which is not arbitration-free, and the suffix from $\event_{d_\mathsf{v}}$ up to $\event_{\length{\mathsf{v}}}$, the arbitration-free part.

\replaced{The program $\programInstanceIota$ that we construct}{We construct a program $\programInstanceIota$ that} uses two replicas $r_0, r_1$, two objects $x_0, x_1$ and a collection of events $e_{i}^{x_l}, 0 \leq i \leq \length{\mathsf{v}}, l \in \{0,1\}$. The events are used to ``encode'' two instances of $\mathsf{v}_{x_0}$ and $\mathsf{v}_{x_1}$. Replica $r_l$ executes first events $e_i^{x_l}$ with $i < d_\mathsf{v}$ and then, events $e_i^{x_{1-l}}$ with $i\geq d_\mathsf{v}$ -- the replica $r_l$ executes the non arbitration-free part of $\mathsf{v}$ for object $x_l$ and the arbitration-free suffix of $\mathsf{v}$ for $x_{1-l}$. For every $l$, the event $e_{\length{\mathsf{v}}}^{x_l}$ reads $x_{1-l}$.

For ensuring that $\mathsf{v}_x(e_0^{x_l}, \ldots e_{n}^{x_l})$ holds in an induced abstract execution of a trace without $\ereceive$ actions, we require that if $\mathsf{Rel}_i^{\mathsf{v}} = \wro$, then $e_{i-1}^{x_l}$ is a write event and $e_i^{x_l}$ is a read event. For ensuring that $\writeConstraints{x}{e_0, \ldots e_{\length{\mathsf{v}}}}$ holds in such an abstract execution, for each set $E \in \mathcal{P}(\event_0, \ldots e_{\length{\mathsf{v}}})$ s.t. $\writeConflicts{E}$ occurs in $\mathsf{v}$, we consider a distinct object $y_E$, which is also distinct from $x_0$ and $x_1$. These objects represent each conflict in $\mathsf{v}$ in a distinct manner. Then, we require that events $e_i^{x_l}$ write to object $y_E$ iff $\event_i \in E$ and to object $x_l$ iff $\event_i$ belongs to the set $E_x$ s.t. $\writeConflicts[x]{E_x}$ occurs in $\mathsf{v}$ (since $\mathsf{v}$ is conflict-maximal, there is only one occurrence of a $\mathsf{conflict}_x$ predicate). In the case $e_i^{x_l}$ is a conditional read-write, we add a set of events $A_i^{x_l}$ that form an execution-corrector  so $\writeConflicts[x]{e_0^{x_l}, \ldots e_{\length{\mathsf{v}}}^{x_l}}$ holds in an abstract execution of a trace without $\ereceive$ actions. These additional events do not write on objects $x_0$ or $x_1$.

\Cref{fig:diagram-theorem-proof-general} exhibits a diagram of the abstract execution of the program.

\input{figures-tex/program-a-la-decker-general.tex}

The rest of the proof, which proceeds as follows, is a generalization of the proof of \Cref{lemma:cons-available-lww:3-1} which takes into considerations the assumptions we make about storage specifications:
\begin{enumerate}
    \item There exists a finite trace $t$ of $\parallelCompositionInstanceIota$ that contains no receive action\addedAppendix{ (\Cref{lemma:program-a-la-decker:trace-without-receive})}.
    
    \item The trace $t$ induces a history $h_\mathsf{v} = \tup{E, \so, \wro}$ and an abstract execution $\exec_\mathsf{v} = \tup{h, \rbo, \ar}$ where $\rbo = \so$. As $\implementationInstance$ is valid w.r.t. $\sspec$, $\exec_\mathsf{v}$ is valid w.r.t. $\sspec$. Next, we prove that since $\rbo = \so$, events in $\exec_\mathsf{v}$ read the latests value w.r.t. $\so$ for any object. In particular, we deduce that $\exec_v$ is valid w.r.t. $(\CC, \opspec)$\addedAppendix{ (\Cref{corollary:afc-theorem:validity-exec-trace})}.
    
    \item \sloppy Since $\ar$ is a total order, either $(e_{d_\mathsf{v}-1}^{x_0}, e_{d_\mathsf{v} - 1}^{x_{1}}) \in \ar$ or $(e_{d_\mathsf{v}-1}^{x_1}, e_{d_\mathsf{v} - 1}^{x_{0}}) \in \ar$. W.l.o.g., assume that $(e_{d_\mathsf{v}-1}^{x_0}, e_{d_\mathsf{v} - 1}^{x_{1}}) \in \ar$. \addedAppendix{By \Cref{proposition:afc-theorem:visibility}}\addedJournal{Then}, we deduce that $e_0^{x_0} \in \context{\cmodel}{x_0}{\exec_\mathsf{v}}{e_{\length{\mathsf{v}}}^{x_0}}$. The proof is explained in \Cref{fig:diagram-theorem-proof-general}: if $(e_{d_\mathsf{v}-1}^{x_0}, e_{d_\mathsf{v} - 1}^{x_{1}}) \in \ar$, then all events $e_i^{x_0}$ form a path in such way that $\mathsf{v}_{x_0}(e_0^{x_0}, \ldots e_{\length{\mathsf{v}}}^{x_0})$ holds in $\exec_\mathsf{v}$. If some event $e_i^{x_l}$ is a conditional read-write event, the predicate $\writeConflicts[x]{e_0^{x_0}, \ldots e_{\length{\mathsf{v}}}^{x_0}}$ holds in $\exec_{\mathsf{v}}$ thanks to the corrector events $A_i^{x_l}$.%
    
    \item As $e_0^{x_0} \in \context{\cmodel}{x_0}{\exec_\mathsf{v}}{e_{\length{v}}^{x_0}}$ but $(e_0^{x_0}, e_{\length{\mathsf{v}}}^{x_0}) \not\in \rbo$ (no message is received), we deduce \addedAppendix{in \Cref{proposition:general-normal-form:stratified-ar}}that $\opspec$ is layered w.r.t. $\ar$. By contrapositive, if $\opspec$ would be layered w.r.t. $\rbo$, as $e_0^{x_0} \in \context{\cmodel}{x_0}{\exec_\mathsf{v}}{e_{\length{\mathsf{v}}}^{x_0}}$, there would exist an event $e$ s.t. $(e_0^{x_0}, e) \in \rbo$ and $e \in \rspec{\cmodel}{x_0}{\exec_\mathsf{v}}{e_{\length{\mathsf{v}}}^{x_0}}$. However, as $\rbo = \so$, $\replicaEvent{e_0^{x_0}} = \replicaEvent{e} = \replicaEvent{e_{\length{\mathsf{v}}}^{x_0}}$ which is false because $\replicaEvent{e_0^{x_0}} = r_0$ and $\replicaEvent{e_{\length{\mathsf{v}}}^{x_0}} = r_1$.
    
    \item Since $\rspecName$ is maximally layered, we can show that the layer bound of $\rspecName$ is smaller than or equal to the number of arbitration-free suffixes of $\mathsf{v}$\addedAppendix{ (\Cref{proposition:general-normal-form:k-suffixes})}. %
    Observe that an event writes $x_0$ only if it is $\init$ or is an event $e_i^{x_l}$ s.t. $\event_i \in E_x$ and $l = 0$. Any such index $i$ corresponds to a suffix of $\mathsf{v}$. By causal suffix closure, for any arbitration-free suffix $v'$ of $v$ there is a visibility \replaced{formula}{relation} that subsumes $v'$ in $\generalNormalForm{\cmodel}$. As $d_\mathsf{v}$ is the maximum index for which $\mathsf{Rel}_i^\mathsf{v} = \ar$, the number of events writing $x_0$ in replica $r_1$ distinct from $\init$ coincide with the number of arbitration-free suffixes of $\mathsf{v}$. Hence, as $\rspecName$ is layered w.r.t. $\ar$, if its layer bound would be greater than the number of arbitration-free suffixes, $e_{\length{\mathsf{v}}}^{x_0}$ would necessarily read $x_0$ from $\init$ (other events writing $x_0$ are in replica $r_0$ and $e_{\length{\mathsf{v}}}$ only reads from events in $r_1$). However, as $\rspecName$ is maximally-layered and $e_0^{x_0}$ succeeds $\init$ w.r.t. $\ar$ and $\rbo^+$, we would conclude that $e_{\length{\mathsf{v}}}^{x_0}$ would also read $x_0$ from $e_0^{x_0}$. However, this is impossible as $\wro \subseteq \rbo = \so$ but $e_0^{x_0}$ is in replica $r_0$ and $e_{\length{\mathsf{v}}}^{x_0}$ is in replica $r_1$.
    
    \item Lastly, we show \addedAppendix{in \Cref{proposition:general-normal-form:layer-bound-implies-v-vacuous} }that if the layer bound of $\rspecName$ is smaller than or equal to the number of arbitration-free suffixes of $v$, then $v$ is vacuous w.r.t. $\opspec$, which contradicts the fact that $\mathsf{v}$ is a visibility \replaced{formula}{relation} from the normal form $\generalNormalForm{\cmodel}$.    
    \qedhere
\end{enumerate}
\end{sketchproof}

\Cref{corollary:cc-strongest} is an immediate consequence of \Cref{th:characterization-cons-available} \added{and \Cref{lemma:saturable:cc-strongest}}.

\begin{corollary}
\label{corollary:cc-strongest}
Let $\opspec$ be an operation specification.
The strongest consistency model $\cmodel$ for which $(\cmodel, \opspec)$ admits an available implementation is $\CC$.
\end{corollary}

	\section{Related Work and Discussion}
\label{sec:related-work}

The CAP conjecture~\cite{Brewer2000} claims that a distributed key-value store cannot be both consistent, available and tolerate partitions. The proof of the \emph{CAP theorem}~\cite{cap-theorem}, uses a so-called \emph{split brain} behavior, where two sets of replicas are isolated from each other, and a \emph{get} (read) operation misses the result of an \emph{earlier} \emph{set} (write) operation (which completes before the get starts). We remark that our proof in \cref{sec:motivating-example} actually extends the proof of the CAP theorem so it holds without the \replaced{real}{read}-time requirement used in the original proof~\cite{cap-theorem}.

As pointed by some critiques of the CAP theorem (\eg~\cite{DBLP:journals/corr/Kleppmann15}), the proof equates consistency with atomicity of read / write variables. Moreover, network partitioning is a stand-in for end-to-end delays in geo-distributed systems. The PACELC (\emph{if Partition then Availability or Consistency, Else Latency or Consistency}) theorem~\cite{DBLP:journals/computer/Abadi12} (see~\cite{DBLP:journals/sigact/Golab18}) captures these observations; its proof extends results proved for \emph{sequential consistency}~\cite{MISC:tr/princeton/Lipton88,DBLP:journals/tocs/AttiyaW94}. These results are also proved for read / write variables, capturing key-value stores. In the executions we construct, messages between replicas are delayed, in a way that corresponds to split brain behavior, and our emphasis is on constructing the right interaction sequences. We believe this behavior can be used to extend the AFC theorem so it talks about latency, rather than availability, for the same interaction sequences.

The CALM (\emph{consistency as logical monotonicity})
conjecture~\cite{Hellerstein-CALM} relates monotonicity of queries to lack of coordination.
Informally, it states that a query has a coordination-free execution strategy if and only if it is monotonic.
In order to make this statement more concrete, it is necessary do define what coordination freedom means. In their proof of the CALM theorem, Ameloot et al.~\cite{AmelootNvB-CALM} equate coordination-freedom with the ability of clients to produce an output even when there is no communication between replicas. The proof relies on a split brain behavior, somewhat similar to the one used in the CAP theorem~\cite{cap-theorem}. Extensions of this theorem~\cite{AmelootKNZ2015} equip replicas with knowledge of the data distribution.
The CALM theorem is motivated, in part, by Bloom~\cite{DBLP:conf/cidr/AlvaroCHM11}, a programming language that encourages order-insensitive programming. The applications they present are to key-value stores and to a shopping cart, essentially, a counter.
Later work extends the CALM approach to a programming environment for composing small \emph{lattices}~\cite{DBLP:conf/cidr/AlvaroCHM11},
and relates it to CRDTs~\cite{LaddadPMCCH2022CRDT}.

One key challenge in deriving our result is considering abstract, generic consistency models, while prior work considers specific models. The other challenge is to allow their composition with abstract, generic shared objects, while prior work mostly consider key-value stores. On the possibility side, this is facilitated by the relating arbitration-freeness to causality; the necessity side relies on finding carefully-designed client interactions that ``stress'' dependencies between replicas.

Defining available implementations for causal consistency has been considered in several works~\cite{BailisOverhead,COPS,Eiger,bolton}. The work of \citeauthor*{10.1109/TPDS.2016.2556669}~\cite{10.1109/TPDS.2016.2556669} \replaced{and \citeauthor*{mahajan2011consistency}~\cite{mahajan2011consistency} show that, in the case of multi-value registers, consistency models stronger than causal consistency cannot support available implementations. In~\cite{10.1109/TPDS.2016.2556669} the condition is \emph{observable causal consistency} (OCC) whereas in \cite{mahajan2011consistency} the condition is \emph{real-time causal consistency} (RTC). The definition of both OCC and RTC are specific to multi-value registers, and the impossibility result depends on several restrictions that we do not consider. Both papers make some (nontrivial, but different) assumptions about the implementations. Furthermore, both of them do not truly prove a tight result: while both \cite{10.1109/TPDS.2016.2556669,mahajan2011consistency} prove the positive result for CC, in \cite{10.1109/TPDS.2016.2556669}, the impossibility is proved for OCC, and in~\cite{mahajan2011consistency} it is for RTC (both stronger than CC).}{shows that for multi-value registers, a condition stronger than causal consistency, called \emph{observable causal consistency} (OCC), cannot support an available implementation. The definition of OCC is specific to multi-value registers, and the impossibility result depends on several restrictions that we do not consider.} Besides handling a more general class of operations, the AFC theorem is a strengthening of their results, as it applies to causal consistency and is therefore tight.

Our specification framework builds on previous work~\cite{DBLP:journals/ftpl/Burckhardt14,DBLP:journals/pacmpl/BiswasE19,DBLP:conf/popl/BurckhardtGYZ14,DBLP:conf/concur/Cerone0G15}. Similarly to Burckhardt et al.~\cite{DBLP:conf/popl/BurckhardtGYZ14,DBLP:journals/ftpl/Burckhardt14}, storage system specifications decouple consistency from the object semantics. We re-use the same ideas of defining consistency using visibility \replaced{formulas}{relations}, contexts, and an arbitration relation. Our object semantics is split into several semantical functions ($\mathsf{rspec}$, $\mathsf{extract}$, and $\mathsf{wspec}$) in order to be more general (modeling transactions), and be able to express ``normal'' constrains. The extension to transaction isolation levels is similar to \citeauthor*{DBLP:conf/concur/Cerone0G15}~\cite{DBLP:conf/concur/Cerone0G15} and \citeauthor*{DBLP:journals/pacmpl/BiswasE19}~\cite{DBLP:journals/pacmpl/BiswasE19}. 

\added{The works of \cite{DBLP:journals/pacmpl/BiswasE19,DBLP:conf/cav/BouajjaniER25} study the complexity of checking consistency under different scenarios. There is no apparent relation between the complexity of checking consistency and the existence of available implementations: the AFC theorem shows that Read Committed admits available implementations but Sequential Consistency does not whereas \cite{DBLP:conf/cav/BouajjaniER25} shows that checking consistency of an SQL history under Read Committed (equivalent to Return-Value) or Sequential Consistency is NP-complete (inclusion in NP is trivial for any model).}

	\bibliographystyle{ACM-Reference-Format}

    \bibliography{bibliography/main,bibliography/dblp,bibliography/acmart,bibliography/misc}


\begin{thebibliography}{30}


\ifx \showCODEN    \undefined \def \showCODEN     #1{\unskip}     \fi
\ifx \showDOI      \undefined \def \showDOI       #1{#1}\fi
\ifx \showISBNx    \undefined \def \showISBNx     #1{\unskip}     \fi
\ifx \showISBNxiii \undefined \def \showISBNxiii  #1{\unskip}     \fi
\ifx \showISSN     \undefined \def \showISSN      #1{\unskip}     \fi
\ifx \showLCCN     \undefined \def \showLCCN      #1{\unskip}     \fi
\ifx \shownote     \undefined \def \shownote      #1{#1}          \fi
\ifx \showarticletitle \undefined \def \showarticletitle #1{#1}   \fi
\ifx \showURL      \undefined \def \showURL       {\relax}        \fi
\providecommand\bibfield[2]{#2}
\providecommand\bibinfo[2]{#2}
\providecommand\natexlab[1]{#1}
\providecommand\showeprint[2][]{arXiv:#2}

\bibitem[Abadi(2012)]%
        {DBLP:journals/computer/Abadi12}
\bibfield{author}{\bibinfo{person}{Daniel Abadi}.}
  \bibinfo{year}{2012}\natexlab{}.
\newblock \showarticletitle{Consistency Tradeoffs in Modern Distributed
  Database System Design: {CAP} is Only Part of the Story}.
\newblock \bibinfo{journal}{\emph{Computer}} \bibinfo{volume}{45},
  \bibinfo{number}{2} (\bibinfo{year}{2012}), \bibinfo{pages}{37--42}.
\newblock
\urldef\tempurl%
\url{https://doi.org/10.1109/MC.2012.33}
\showDOI{\tempurl}


\bibitem[Adya(1999)]%
        {adya-thesis}
\bibfield{author}{\bibinfo{person}{A. Adya}.} \bibinfo{year}{1999}\natexlab{}.
\newblock \emph{\bibinfo{title}{Weak Consistency: A Generalized Theory and
  Optimistic Implementations for Distributed Transactions}}.
\newblock \bibinfo{thesistype}{Ph.\,D. Dissertation}.
\newblock


\bibitem[Adya et~al\mbox{.}(2000)]%
        {DBLP:conf/icde/AdyaLO00}
\bibfield{author}{\bibinfo{person}{Atul Adya}, \bibinfo{person}{Barbara
  Liskov}, {and} \bibinfo{person}{Patrick~E. O'Neil}.}
  \bibinfo{year}{2000}\natexlab{}.
\newblock \showarticletitle{Generalized Isolation Level Definitions}. In
  \bibinfo{booktitle}{\emph{Proceedings of the 16th International Conference on
  Data Engineering, San Diego, California, USA, February 28 - March 3, 2000}},
  \bibfield{editor}{\bibinfo{person}{David~B. Lomet} {and}
  \bibinfo{person}{Gerhard Weikum}} (Eds.). \bibinfo{publisher}{{IEEE} Computer
  Society}, \bibinfo{pages}{67--78}.
\newblock
\urldef\tempurl%
\url{https://doi.org/10.1109/ICDE.2000.839388}
\showDOI{\tempurl}


\bibitem[Alvaro et~al\mbox{.}(2011)]%
        {DBLP:conf/cidr/AlvaroCHM11}
\bibfield{author}{\bibinfo{person}{Peter Alvaro}, \bibinfo{person}{Neil
  Conway}, \bibinfo{person}{Joseph~M. Hellerstein}, {and}
  \bibinfo{person}{William~R. Marczak}.} \bibinfo{year}{2011}\natexlab{}.
\newblock \showarticletitle{Consistency Analysis in Bloom: a {CALM} and
  Collected Approach}. In \bibinfo{booktitle}{\emph{Fifth Biennial Conference
  on Innovative Data Systems Research, {CIDR} 2011, Asilomar, CA, USA, January
  9-12, 2011, Online Proceedings}}. \bibinfo{publisher}{www.cidrdb.org},
  \bibinfo{pages}{249--260}.
\newblock
\urldef\tempurl%
\url{http://cidrdb.org/cidr2011/Papers/CIDR11\_Paper35.pdf}
\showURL{%
\tempurl}


\bibitem[Ameloot et~al\mbox{.}(2015)]%
        {AmelootKNZ2015}
\bibfield{author}{\bibinfo{person}{Tom~J. Ameloot}, \bibinfo{person}{Bas
  Ketsman}, \bibinfo{person}{Frank Neven}, {and} \bibinfo{person}{Daniel
  Zinn}.} \bibinfo{year}{2015}\natexlab{}.
\newblock \showarticletitle{Weaker Forms of Monotonicity for Declarative
  Networking: A More Fine-Grained Answer to the CALM-Conjecture}.
\newblock \bibinfo{journal}{\emph{ACM Trans. Database Syst.}}
  \bibinfo{volume}{40}, \bibinfo{number}{4}, Article \bibinfo{articleno}{21}
  (\bibinfo{date}{Dec.} \bibinfo{year}{2015}), \bibinfo{numpages}{45}~pages.
\newblock
\urldef\tempurl%
\url{https://doi.org/10.1145/2809784}
\showDOI{\tempurl}


\bibitem[Ameloot et~al\mbox{.}(2013)]%
        {AmelootNvB-CALM}
\bibfield{author}{\bibinfo{person}{Tom~J. Ameloot}, \bibinfo{person}{Frank
  Neven}, {and} \bibinfo{person}{Jan Van Den~Bussche}.}
  \bibinfo{year}{2013}\natexlab{}.
\newblock \showarticletitle{Relational transducers for declarative networking}.
\newblock \bibinfo{journal}{\emph{J. ACM}} \bibinfo{volume}{60},
  \bibinfo{number}{2}, Article \bibinfo{articleno}{15} (\bibinfo{date}{May}
  \bibinfo{year}{2013}), \bibinfo{numpages}{38}~pages.
\newblock
\urldef\tempurl%
\url{https://doi.org/10.1145/2450142.2450151}
\showDOI{\tempurl}


\bibitem[Attiya et~al\mbox{.}(2017)]%
        {10.1109/TPDS.2016.2556669}
\bibfield{author}{\bibinfo{person}{Hagit Attiya}, \bibinfo{person}{Faith
  Ellen}, {and} \bibinfo{person}{Adam Morrison}.}
  \bibinfo{year}{2017}\natexlab{}.
\newblock \showarticletitle{Limitations of Highly-Available
  Eventually-Consistent Data Stores}.
\newblock \bibinfo{journal}{\emph{IEEE Trans. Parallel Distrib. Syst.}}
  \bibinfo{volume}{28}, \bibinfo{number}{1} (\bibinfo{date}{Jan.}
  \bibinfo{year}{2017}), \bibinfo{pages}{141–155}.
\newblock
\showISSN{1045-9219}
\urldef\tempurl%
\url{https://doi.org/10.1109/TPDS.2016.2556669}
\showDOI{\tempurl}


\bibitem[Attiya and Welch(1994)]%
        {DBLP:journals/tocs/AttiyaW94}
\bibfield{author}{\bibinfo{person}{Hagit Attiya} {and}
  \bibinfo{person}{Jennifer~L. Welch}.} \bibinfo{year}{1994}\natexlab{}.
\newblock \showarticletitle{Sequential Consistency versus Linearizability}.
\newblock \bibinfo{journal}{\emph{{ACM} Trans. Comput. Syst.}}
  \bibinfo{volume}{12}, \bibinfo{number}{2} (\bibinfo{year}{1994}),
  \bibinfo{pages}{91--122}.
\newblock
\urldef\tempurl%
\url{https://doi.org/10.1145/176575.176576}
\showDOI{\tempurl}


\bibitem[Bailis et~al\mbox{.}(2012)]%
        {BailisOverhead}
\bibfield{author}{\bibinfo{person}{Peter Bailis}, \bibinfo{person}{Alan
  Fekete}, \bibinfo{person}{Ali Ghodsi}, \bibinfo{person}{Joseph~M.
  Hellerstein}, {and} \bibinfo{person}{Ion Stoica}.}
  \bibinfo{year}{2012}\natexlab{}.
\newblock \showarticletitle{The Potential Dangers of Causal Consistency and an
  Explicit Solution}. In \bibinfo{booktitle}{\emph{Proceedings of the 3rd ACM
  Symposium on Cloud Computing}}.
\newblock
\urldef\tempurl%
\url{http://doi.acm.org/10.1145/2391229.2391251}
\showURL{%
\tempurl}


\bibitem[Bailis et~al\mbox{.}(2013)]%
        {bolton}
\bibfield{author}{\bibinfo{person}{Peter Bailis}, \bibinfo{person}{Ali Ghodsi},
  \bibinfo{person}{Joseph~M. Hellerstein}, {and} \bibinfo{person}{Ion Stoica}.}
  \bibinfo{year}{2013}\natexlab{}.
\newblock \showarticletitle{{Bolt-on Causal Consistency}}. In
  \bibinfo{booktitle}{\emph{Proceedings of the 2013 ACM SIGMOD International
  Conference on Management of Data}}. \bibinfo{pages}{761--772}.
\newblock
\urldef\tempurl%
\url{http://doi.acm.org/10.1145/2463676.2465279}
\showURL{%
\tempurl}


\bibitem[Biswas and Enea(2019)]%
        {DBLP:journals/pacmpl/BiswasE19}
\bibfield{author}{\bibinfo{person}{Ranadeep Biswas} {and}
  \bibinfo{person}{Constantin Enea}.} \bibinfo{year}{2019}\natexlab{}.
\newblock \showarticletitle{On the complexity of checking transactional
  consistency}.
\newblock \bibinfo{journal}{\emph{Proc. {ACM} Program. Lang.}}
  \bibinfo{volume}{3}, \bibinfo{number}{{OOPSLA}} (\bibinfo{year}{2019}),
  \bibinfo{pages}{165:1--165:28}.
\newblock
\urldef\tempurl%
\url{https://doi.org/10.1145/3360591}
\showDOI{\tempurl}


\bibitem[Bouajjani et~al\mbox{.}(2025)]%
        {DBLP:conf/cav/BouajjaniER25}
\bibfield{author}{\bibinfo{person}{Ahmed Bouajjani},
  \bibinfo{person}{Constantin Enea}, {and} \bibinfo{person}{Enrique
  Rom{\'{a}}n{-}Calvo}.} \bibinfo{year}{2025}\natexlab{}.
\newblock \showarticletitle{On the Complexity of Checking Mixed Isolation
  Levels for {SQL} Transactions}. In \bibinfo{booktitle}{\emph{Computer Aided
  Verification - 37th International Conference, {CAV} 2025, Zagreb, Croatia,
  July 23-25, 2025, Proceedings, Part {IV}}} \emph{(\bibinfo{series}{Lecture
  Notes in Computer Science}, Vol.~\bibinfo{volume}{15934})},
  \bibfield{editor}{\bibinfo{person}{Ruzica Piskac} {and}
  \bibinfo{person}{Zvonimir Rakamaric}} (Eds.). \bibinfo{publisher}{Springer},
  \bibinfo{pages}{315--337}.
\newblock
\urldef\tempurl%
\url{https://doi.org/10.1007/978-3-031-98685-7\_15}
\showDOI{\tempurl}


\bibitem[Brewer(2000)]%
        {Brewer2000}
\bibfield{author}{\bibinfo{person}{Eric~A. Brewer}.}
  \bibinfo{year}{2000}\natexlab{}.
\newblock \showarticletitle{Towards robust distributed systems (Invited Talk)}.
  In \bibinfo{booktitle}{\emph{Proceedings of the Nineteenth Annual ACM
  Symposium on Principles of Distributed Computing}}. \bibinfo{address}{New
  York, NY, USA}.
\newblock
\showISBNx{1581131836}
\urldef\tempurl%
\url{https://doi.org/10.1145/343477.343502}
\showURL{%
\tempurl}


\bibitem[Burckhardt(2014)]%
        {DBLP:journals/ftpl/Burckhardt14}
\bibfield{author}{\bibinfo{person}{Sebastian Burckhardt}.}
  \bibinfo{year}{2014}\natexlab{}.
\newblock \showarticletitle{Principles of Eventual Consistency}.
\newblock \bibinfo{journal}{\emph{Found. Trends Program. Lang.}}
  \bibinfo{volume}{1}, \bibinfo{number}{1-2} (\bibinfo{year}{2014}),
  \bibinfo{pages}{1--150}.
\newblock
\urldef\tempurl%
\url{https://doi.org/10.1561/2500000011}
\showDOI{\tempurl}


\bibitem[Burckhardt et~al\mbox{.}(2014)]%
        {DBLP:conf/popl/BurckhardtGYZ14}
\bibfield{author}{\bibinfo{person}{Sebastian Burckhardt},
  \bibinfo{person}{Alexey Gotsman}, \bibinfo{person}{Hongseok Yang}, {and}
  \bibinfo{person}{Marek Zawirski}.} \bibinfo{year}{2014}\natexlab{}.
\newblock \showarticletitle{Replicated data types: specification, verification,
  optimality}. In \bibinfo{booktitle}{\emph{41st Symposium on Principles of
  Programming Languages, {POPL}}}. \bibinfo{publisher}{{ACM}},
  \bibinfo{pages}{271--284}.
\newblock
\urldef\tempurl%
\url{https://doi.org/10.1145/2535838.2535848}
\showURL{%
\tempurl}


\bibitem[Burckhardt et~al\mbox{.}(2015)]%
        {DBLP:conf/ecoop/BurckhardtLPF15}
\bibfield{author}{\bibinfo{person}{Sebastian Burckhardt}, \bibinfo{person}{Daan
  Leijen}, \bibinfo{person}{Jonathan Protzenko}, {and} \bibinfo{person}{Manuel
  F{\"{a}}hndrich}.} \bibinfo{year}{2015}\natexlab{}.
\newblock \showarticletitle{Global Sequence Protocol: {A} Robust Abstraction
  for Replicated Shared State}. In \bibinfo{booktitle}{\emph{29th European
  Conference on Object-Oriented Programming, {ECOOP} 2015, July 5-10, 2015,
  Prague, Czech Republic}} \emph{(\bibinfo{series}{LIPIcs},
  Vol.~\bibinfo{volume}{37})}, \bibfield{editor}{\bibinfo{person}{John~Tang
  Boyland}} (Ed.). \bibinfo{publisher}{Schloss Dagstuhl - Leibniz-Zentrum
  f{\"{u}}r Informatik}, \bibinfo{pages}{568--590}.
\newblock
\urldef\tempurl%
\url{https://doi.org/10.4230/LIPICS.ECOOP.2015.568}
\showDOI{\tempurl}


\bibitem[Cerone et~al\mbox{.}(2015)]%
        {DBLP:conf/concur/Cerone0G15}
\bibfield{author}{\bibinfo{person}{Andrea Cerone}, \bibinfo{person}{Giovanni
  Bernardi}, {and} \bibinfo{person}{Alexey Gotsman}.}
  \bibinfo{year}{2015}\natexlab{}.
\newblock \showarticletitle{A Framework for Transactional Consistency Models
  with Atomic Visibility}. In \bibinfo{booktitle}{\emph{26th International
  Conference on Concurrency Theory, {CONCUR}}}. \bibinfo{pages}{58--71}.
\newblock
\urldef\tempurl%
\url{https://doi.org/10.4230/LIPICS.CONCUR.2015.58}
\showDOI{\tempurl}


\bibitem[Gilbert and Lynch(2002)]%
        {cap-theorem}
\bibfield{author}{\bibinfo{person}{Seth Gilbert} {and} \bibinfo{person}{Nancy
  Lynch}.} \bibinfo{year}{2002}\natexlab{}.
\newblock \showarticletitle{Brewer's Conjecture and the Feasibility of
  Consistent, Available, Partition-Tolerant Web Services}.
\newblock \bibinfo{journal}{\emph{SIGACT News}} \bibinfo{volume}{33},
  \bibinfo{number}{2} (\bibinfo{date}{June} \bibinfo{year}{2002}),
  \bibinfo{pages}{51--59}.
\newblock
\showISSN{0163-5700}
\urldef\tempurl%
\url{https://doi.org/10.1145/564585.564601}
\showDOI{\tempurl}


\bibitem[Golab(2018)]%
        {DBLP:journals/sigact/Golab18}
\bibfield{author}{\bibinfo{person}{Wojciech~M. Golab}.}
  \bibinfo{year}{2018}\natexlab{}.
\newblock \showarticletitle{Proving {PACELC}}.
\newblock \bibinfo{journal}{\emph{{SIGACT} News}} \bibinfo{volume}{49},
  \bibinfo{number}{1} (\bibinfo{year}{2018}), \bibinfo{pages}{73--81}.
\newblock
\urldef\tempurl%
\url{https://doi.org/10.1145/3197406.3197420}
\showDOI{\tempurl}


\bibitem[Hellerstein(2010)]%
        {Hellerstein-CALM}
\bibfield{author}{\bibinfo{person}{Joseph~M. Hellerstein}.}
  \bibinfo{year}{2010}\natexlab{}.
\newblock \showarticletitle{The declarative imperative: {Experiences} and
  conjectures in distributed logic}.
\newblock \bibinfo{journal}{\emph{SIGMOD Rec.}} \bibinfo{volume}{39},
  \bibinfo{number}{1} (\bibinfo{date}{Sept.} \bibinfo{year}{2010}),
  \bibinfo{pages}{5–19}.
\newblock
\urldef\tempurl%
\url{https://doi.org/10.1145/1860702.1860704}
\showDOI{\tempurl}


\bibitem[Kleppmann(2015)]%
        {DBLP:journals/corr/Kleppmann15}
\bibfield{author}{\bibinfo{person}{Martin Kleppmann}.}
  \bibinfo{year}{2015}\natexlab{}.
\newblock \showarticletitle{A Critique of the {CAP} Theorem}.
\newblock \bibinfo{journal}{\emph{CoRR}}  \bibinfo{volume}{abs/1509.05393}
  (\bibinfo{year}{2015}).
\newblock
\showeprint[arXiv]{1509.05393}
\urldef\tempurl%
\url{http://arxiv.org/abs/1509.05393}
\showURL{%
\tempurl}


\bibitem[Laddad et~al\mbox{.}(2022)]%
        {LaddadPMCCH2022CRDT}
\bibfield{author}{\bibinfo{person}{Shadaj Laddad}, \bibinfo{person}{Conor
  Power}, \bibinfo{person}{Mae Milano}, \bibinfo{person}{Alvin Cheung},
  \bibinfo{person}{Natacha Crooks}, {and} \bibinfo{person}{Joseph~M.
  Hellerstein}.} \bibinfo{year}{2022}\natexlab{}.
\newblock \showarticletitle{Keep {CALM} and {CRDT} On}.
\newblock \bibinfo{journal}{\emph{Proc. {VLDB} Endow.}} \bibinfo{volume}{16},
  \bibinfo{number}{4} (\bibinfo{year}{2022}), \bibinfo{pages}{856--863}.
\newblock
\urldef\tempurl%
\url{https://doi.org/10.14778/3574245.3574268}
\showDOI{\tempurl}


\bibitem[Lamport(1978)]%
        {Lamport1978}
\bibfield{author}{\bibinfo{person}{Leslie Lamport}.}
  \bibinfo{year}{1978}\natexlab{}.
\newblock \showarticletitle{Time, Clocks, and the Ordering of Events in a
  Distributed System}.
\newblock \bibinfo{journal}{\emph{Commun. {ACM}}} \bibinfo{volume}{21},
  \bibinfo{number}{7} (\bibinfo{year}{1978}), \bibinfo{pages}{558--565}.
\newblock
\urldef\tempurl%
\url{https://doi.org/10.1145/359545.359563}
\showDOI{\tempurl}


\bibitem[Lipton and Sandberg(1988)]%
        {MISC:tr/princeton/Lipton88}
\bibfield{author}{\bibinfo{person}{Richard~J Lipton} {and}
  \bibinfo{person}{Jonathan~S Sandberg}.} \bibinfo{year}{1988}\natexlab{}.
\newblock \bibinfo{booktitle}{\emph{{PRAM}: A scalable shared memory}}.
\newblock \bibinfo{type}{{T}echnical {R}eport} TR-180-88.
  \bibinfo{institution}{Princeton University, Department of Computer Science}.
\newblock


\bibitem[Lloyd et~al\mbox{.}(2011)]%
        {COPS}
\bibfield{author}{\bibinfo{person}{Wyatt Lloyd}, \bibinfo{person}{Michael~J.
  Freedman}, \bibinfo{person}{Michael Kaminsky}, {and}
  \bibinfo{person}{David~G. Andersen}.} \bibinfo{year}{2011}\natexlab{}.
\newblock \showarticletitle{{Don't Settle for Eventual: Scalable Causal
  Consistency for Wide-area Storage with COPS}}. In
  \bibinfo{booktitle}{\emph{Proceedings of the 23rd ACM Symposium on Operating
  Systems Principles}}. \bibinfo{pages}{401--416}.
\newblock
\urldef\tempurl%
\url{http://doi.acm.org/10.1145/2043556.2043593}
\showURL{%
\tempurl}


\bibitem[Lloyd et~al\mbox{.}(2013)]%
        {Eiger}
\bibfield{author}{\bibinfo{person}{Wyatt Lloyd}, \bibinfo{person}{Michael~J.
  Freedman}, \bibinfo{person}{Michael Kaminsky}, {and}
  \bibinfo{person}{David~G. Andersen}.} \bibinfo{year}{2013}\natexlab{}.
\newblock \showarticletitle{Stronger Semantics for Low-Latency Geo-Replicated
  Storage}. In \bibinfo{booktitle}{\emph{Proceedings of the 10th {USENIX}
  Symposium on Networked Systems Design and Implementation, {NSDI} 2013,
  Lombard, IL, USA, April 2-5, 2013}}, \bibfield{editor}{\bibinfo{person}{Nick
  Feamster} {and} \bibinfo{person}{Jeffrey~C. Mogul}} (Eds.).
  \bibinfo{publisher}{{USENIX} Association}, \bibinfo{pages}{313--328}.
\newblock
\urldef\tempurl%
\url{https://www.usenix.org/conference/nsdi13/technical-sessions/presentation/lloyd}
\showURL{%
\tempurl}


\bibitem[Mahajan et~al\mbox{.}(2011)]%
        {mahajan2011consistency}
\bibfield{author}{\bibinfo{person}{Prince Mahajan}, \bibinfo{person}{Lorenzo
  Alvisi}, \bibinfo{person}{Mike Dahlin}, {et~al\mbox{.}}}
  \bibinfo{year}{2011}\natexlab{}.
\newblock \showarticletitle{Consistency, availability, and convergence}.
\newblock \bibinfo{journal}{\emph{University of Texas at Austin Tech Report}}
  \bibinfo{volume}{11} (\bibinfo{year}{2011}), \bibinfo{pages}{158}.
\newblock


\bibitem[Microsoft(2022)]%
        {cosmosdb-consistency}
\bibfield{author}{\bibinfo{person}{Microsoft}.}
  \bibinfo{year}{2022}\natexlab{}.
\newblock \bibinfo{booktitle}{\emph{Consistency Levels in Azure Cosmos DB}}.
\newblock
\newblock
\shownote{\url{https://docs.microsoft.com/en-us/azure/cosmos-db/consistency-levels}}.


\bibitem[Shapiro et~al\mbox{.}(2011)]%
        {DBLP:conf/sss/ShapiroPBZ11}
\bibfield{author}{\bibinfo{person}{Marc Shapiro}, \bibinfo{person}{Nuno~M.
  Pregui{\c{c}}a}, \bibinfo{person}{Carlos Baquero}, {and}
  \bibinfo{person}{Marek Zawirski}.} \bibinfo{year}{2011}\natexlab{}.
\newblock \showarticletitle{Conflict-Free Replicated Data Types}. In
  \bibinfo{booktitle}{\emph{Stabilization, Safety, and Security of Distributed
  Systems {SSS}}}, Vol.~\bibinfo{volume}{6976}. \bibinfo{publisher}{Springer},
  \bibinfo{pages}{386--400}.
\newblock
\urldef\tempurl%
\url{https://doi.org/10.1007/978-3-642-24550-3\_29}
\showDOI{\tempurl}


\bibitem[Sovran et~al\mbox{.}(2011)]%
        {DBLP:conf/sosp/SovranPAL11}
\bibfield{author}{\bibinfo{person}{Yair Sovran}, \bibinfo{person}{Russell
  Power}, \bibinfo{person}{Marcos~K. Aguilera}, {and} \bibinfo{person}{Jinyang
  Li}.} \bibinfo{year}{2011}\natexlab{}.
\newblock \showarticletitle{Transactional storage for geo-replicated systems}.
  In \bibinfo{booktitle}{\emph{Proceedings of the 23rd {ACM} Symposium on
  Operating Systems Principles 2011, {SOSP} 2011, Cascais, Portugal, October
  23-26, 2011}}, \bibfield{editor}{\bibinfo{person}{Ted Wobber} {and}
  \bibinfo{person}{Peter Druschel}} (Eds.). \bibinfo{publisher}{{ACM}},
  \bibinfo{pages}{385--400}.
\newblock
\urldef\tempurl%
\url{https://doi.org/10.1145/2043556.2043592}
\showDOI{\tempurl}


\end{thebibliography}

	\ifbool{appendixMode}{
		\newpage%
		\appendix%

		\section{Examples of Operation Specifications}
\label{app:examples}

We present here several well-know operation specifications. 

\subsection{Key-Value Store with Fetch-And-Add and Compare-And-Swap Operations}
\label{ssec:cas-storage}

The Key-Value Store with Fetch-And-Add and Compare-And-Swap ($\faacasStorage$) is an operation specification with four operations, $\iwritekv{x}{v}$, that puts value $v$ to object $x$, $\ireadkv{x}$ that reads object $x$, $\ifaakv{x}{v}$ that reads the value $v'$ of object $x$ and writes $v'+v$, and $\icaskv{x}{v}{v'}$, that reads $x$ and writes $v'$ iff the value read is $v$.

The following equations, corresponding to \Cref{eq:rspec-faa-lww,eq:extract-faa-lww,eq:wspec-faa-lww}, describe the operation specification of $\faacasStorage$.
\begin{equation}
\rspecContext{x}{c}{r} =  \left\{ \begin{array}{ll}
    \{\max_{\ar} E\} & \text{if } r \in \{\ireadkv{x}, \ifaakv{x}{v}, \icas{x}{v'}{v''}\} \text{ and } c = \tup{E, \ar, \rbo} \\
    \emptyset & \text{otherwise}
\end{array} \right.
\end{equation}
\begin{equation}
\extractspecContext{x}{R}{r} = \left\{ 
    \begin{array}{ll}
        v & \text{if } r \in \{\ireadkv{x}, \ifaakv{x}{v}, \icas{x}{v}{v'}\} \text{ and } R = \{(w, v)\} \\
        \mathsf{undefined} & \text{otherwise}
    \end{array} \right.
\end{equation}
\begin{equation}
\wspecContext{x}{v}{w} = \left\{ 
    \begin{array}{ll}
        v' & \text{if } w = \iwritekv{x}{v'}\\
        v + v' & \text{if } w = \ifaakv{x}{v'}\\
        v'' & \text{if } w = \icas{x}{v'}{v''} \land v = v'\\
        \mathsf{undefined} & \text{otherwise}
    \end{array} \right.
\end{equation}

The $\faacasStorage$ is maximally layered w.r.t. $\ar$, with $1$ as its layer bound. As $\icaskvName$ is a single-object conditional read-write operation, it trivially allows execution-correctors.

\subsection{Key-Value Multi-Value Store}
\label{ssec:mvr-kv-storage}

The Key-Value Multi-Value Store ($\kvmvrStorage$) \cite{DBLP:conf/popl/BurckhardtGYZ14,10.1109/TPDS.2016.2556669}
is an operation specification with two operations, $\ireadkv{x}$, reading multiple concurrent values on a single object $x$, and $\iwritekv{x}{v}$, writing on a single object $x$ the value $v$.

The following equations, corresponding to \Cref{eq:rspec-mvr,eq:extract-faa-lww,eq:wspec-faa-lww}, describe the operation specification of $\kvmvrStorage$.
\begin{equation}
\rspecContext{x}{c}{r} =  \left\{ \begin{array}{ll}
    \{\max_{\rbo} E\} & \text{if } r = \ireadkv{x} \text{ and } c = \tup{E, \ar, \rbo} \\
    \emptyset & \text{otherwise}
\end{array} \right.
\end{equation}
\begin{equation}
\extractspecContext{x}{R}{r} = \left\{ 
\begin{array}{ll}
    \{v \ | \ (\_, v)\in R\} & \text{if } r = \ireadkv{x}\\
    \mathsf{undefined} & \text{otherwise}
\end{array} \right.
\end{equation}
\begin{equation}
\wspecContext{x}{\_}{w} = \left\{ 
    \begin{array}{ll}
        v & \text{if } w = \iwritekv{x}{v}\\
        \mathsf{undefined} & \text{otherwise}
    \end{array} \right.
\end{equation}

The $\kvStorage$ is maximally layered w.r.t. $\rbo^+$, with $1$ as its layer bound. %

\subsection{Distributed Counter}
\label{ssec:counter}

The distributed counter ($\counterStorage$) \cite{DBLP:conf/popl/BurckhardtGYZ14} is an operation specification with two operations, $\iincCounter{x}$, incrementing the value of $x$ by $1$, and $\irdCounter{x}$, reading the amount of increments of $x$.

The following equations, corresponding to \Cref{eq:rspec-counter,eq:extract-counter,eq:wspec-counter}, describe the operation specification of $\counterStorage$.
\begin{equation}
\rspecContext{x}{c}{r} =  \left\{ \begin{array}{ll}
    E & \text{if } r = \irdCounter{x} \text{ and } c = \tup{E, \ar, \rbo} \\
    \emptyset & \text{otherwise}
\end{array} \right.
\end{equation}

\begin{equation}
\extractspecContext{x}{R}{r} = \left\{ 
    \begin{array}{ll}
        |R| - 1& \text{if } r = \irdCounter{x} \\
        \mathsf{undefined} & \text{otherwise}
    \end{array} \right.
\end{equation}

\begin{equation}
\wspecContext{x}{\_}{w} = \left\{ 
    \begin{array}{ll}
        1 & \text{if } w = \iincCounter{x}\\
        \mathsf{undefined} & \text{otherwise}
    \end{array} \right.
\end{equation}

The $\counterStorage$ is maximally layered w.r.t. $\ar$, with $\infty$ as its layer bound. %

\subsection{Insert/Delete Last-Write-Wins}
\label{ssec:mvr-insdel-storage}

The Insert/Delete Last-Write-Wins ($\insdelStorage$) is an operation specification with two multi-object operations. $\iinsifabs{X}{v}$ checks for every object $x \in X$ if it is present, and inserts it with value $v$ if not, and $\idelifpres{X}$ deletes every object $x \in X$ as long as it was already present.

Its operation specification is described as follows:
\begin{equation}
\rspecContext{x}{c}{r} =  \left\{ \begin{array}{ll}
    \{\max_{\ar}E\}& \text{if } r \in \left\{
        \begin{array}{l}
            \iinsifabs{X}{v}, \idelifpres{X}
    \end{array} \right\}, x \in X \text{ and } c = \tup{E, \ar, \rbo} \\
    \emptyset & \text{otherwise}
\end{array} \right.
\end{equation}
\begin{equation}
\extractspecContext{x}{R}{r} =
\left\{
    \begin{array}{ll}
        v & \text{if } w \in \{\iinsifabs{X}{\_}, \idelifpres{X}\}, x \in X \text{ and } R = \{(\_, v)\} \\
        \mathsf{undefined} & \text{otherwise}
    \end{array}  \\ \right.
\end{equation}
\begin{equation}
\wspecContext{x}{v}{w} = \left\{ 
    \begin{array}{ll}
        v' & \text{if } w = \iinsifabs{X}{v'} \land v = \del \\
        \del & \text{if } w = \idelifpres{X} \land v \neq \del\\
        \mathsf{undefined} & \text{otherwise}
    \end{array} \right.
\end{equation}
where $\del$ is a special value representing absence. We assume that $\iinsifabs{X}{\del}$ is not defined.

The $\insdelStorage$ is maximally layered w.r.t. $\ar$, with $1$ as its layer bound. $\insdelStorage$ allows execution-correctors: let $\cmodel$ be a consistency model, $\exec$ be an abstract execution, $D$ be a domain, $W \subseteq D$ be a set of objects and $x$ be an object s.t. $x \in W$ if $W \neq \emptyset$. 

Let be $v$ the value that event $e$ reads in $\exec \oplus e$. If $v = \del$, we select $e = \iinsifabs{D}{\_}$ while otherwise, $e = \idelifpres{D}$. The mapping $a$ below is an execution-corrector for $(e, W, x, \exec)$:

\begin{equation}
a(y) = \left\{ 
	\begin{array}{ll}
		\iinsifabs{\{y\}}{v'} & \text{if } y \in W \land v_y = \del \neq v, \text{ or } y \not\in W \land v_y = \del = v\\
		\idelifpres{\{y\}} & \text{if } y \in W \land v_y \neq \del = v, \text{ or } y \not\in W \land v_y \neq \del \neq v\\
		\mathsf{undefined} & \text{otherwise}
	\end{array} 
\right.
\end{equation}
where $v_y = \wspec{\cmodel}{y}{\exec}{e_p}$ and $e_p$ is the maximal event w.r.t. $\so$ on the same replica as $e$.

\subsection{Non-Transactional SQL with Last-Writer-Wins Store}
\label{ssec:non-trans-sql-lww}

The Non-Transactional SQL with Last-Writer-Wins Store ($\simpleSQLStorage$) is an operation specification modelling SQL-like databases \cite{adya-thesis}. Each object represents a row identifier and the set of values is defined abstractly as $\Rows$. $\Rows$ contain a special value denoted $\del$, different from $\bot$, indicating that the row is deleted. 

This operation specification employs four operations: $\einsert$, $\eselect$, $\eupsert$ and $\edelete$. Each operation has a finite set of objects $D$ as domain. $\iinsert{\setRows}$ inserts in the database each row $r$ on an object $d \in D$ using the mapping $\setRows : D \to \Rows$. $\iselect{\predicate}$ selects the rows on the storage satisfying the predicate $\predicate : D \times \Rows \to \{\bfalse,\btrue\}$. $\iupsert{\predicate}{\mapRows}$ updates the rows that satisfy $\predicate$ using the mapping $\mapRows : D \times \Rows \to \Rows$, inserting them if they are absent. Finally, $\idelete{\predicate}$, deletes the objects satisfying the predicate (i.e. replaces its row by $\del$). We assume that in for any predicate $\predicate$ and object $x$, $\predicate(x, \del) = \bfalse$. %

\begin{equation}
\rspecContext{x}{c}{r} =  \left\{ \begin{array}{ll}
    \{\max_{\ar}E\} & \text{if }  r \in \{
        \iselect{\predicate},  \iupsert{\predicate}{\mapRows}, \idelete{\predicate}\}  \text{ and } c = \tup{E, \ar, \rbo}   \\
    \emptyset & \text{otherwise}
\end{array} \right.
\end{equation}

\begin{equation}
\extractspecContext{x}{R}{r} = \left\{ 
    \begin{array}{ll}
        v & 
        \begin{array}{l}
            \text{if }  r \in \{\iselect{\predicate},  \iupsert{\predicate}{\mapRows}, \idelete{\predicate}\}, \\
            \quad R = \{(w, v)\} \text{ and } \predicate_x(v)
        \end{array}   \\
        \mathsf{undefined} & \text{otherwise}
    \end{array} \right.
\end{equation}

\begin{equation}
\wspecContext{x}{v}{w} = \left\{ 
    \begin{array}{ll}
        \setRows(x) & \text{if } w = \iinsert{\setRows}\\
        \mapRows_x(v) & \text{if } w = \iupsert{\predicate}{\mapRows} \\
        \del & \text{if } w = \idelete{\predicate} \land  v \not\in \{\bot, \del\}\\
        \mathsf{undefined} & \text{otherwise}
    \end{array} \right.
\end{equation}

The $\simpleSQLStorage$ is maximally layered w.r.t. $\ar$, with $1$ as its layer bound. $\simpleSQLStorage$ allows execution-correctors: let $\cmodel$ be a consistency model, $\exec$ be an abstract execution, $D$ be a domain, $W \subseteq D$ be a set of objects and $x$ be an object s.t. $x \in W$ if $W \neq \emptyset$. 

Let be $v$ the value that event $e$ reads in $\exec \oplus e$. We select the event $e = \iupsert{\predicate_{D,W}}{\mapRows_D}$, where $\predicate_{D,W}$ and $\mapRows_D$ are defined below. 
\begin{equation}
\begin{array}{ll}
    \predicate_{D,W}(d, r) = \left\{
    \begin{array}{ll}
         \btrue & \text{if } d \in W \\
         \bfalse & \text{if } d \in D \setminus W \\
         \mathsf{undefined} & \text{otherwise}
    \end{array}\right.\\
    \mapRows_D(d, r) = \left\{
    \begin{array}{ll}
        r & \text{if } d \in D \\
        \mathsf{undefined} & \text{otherwise}
    \end{array}\right.\\
\end{array} \nonumber
\end{equation}

For such event, we define the execution-corrector $a: D \setminus \{x\} \to \Events$ as the totally-undefined mapping, i.e. the function that no object $y \in D$ is associated with some event.

\subsection{Transactional SQL Multi-Value Store}
\label{ssec:trans-sql-mvr}

The Transactional SQL Multi-Value Store ($\transSQLMVRStorage$) is an operation specification modelling SQL-like databases using \emph{transactions}. Each object represents a row identifier and the set of values, $\Rows$, is defined as in \Cref{ssec:non-trans-sql-lww}.

Transactions are blocks of simple instructions that are executed sequentially. Transactions start its execution by selecting a \emph{snapshot} of the database (i.e. a mapping associating each object a constant value) from which operations can read. Each instruction may execute a writing operation, but its effect it is only viewed internally. After their completion, the writing effects of the transaction can be seen by other transactions; giving the impression of atomicity.

We model the store with the aid of a unique operation, $\transaction{\trCode}$ that reads the snapshot of the database and then executes the instructions declared in $\code$. $\code$ is defined as a sequence of five type of operations: $\einsert$, $\eselect$, $\eupdate$ and $\edelete$. Each operation has a finite set of objects $D$ as domain. $\iinsert{\setRows}$ inserts in the database each row $r$ on an object $d \in D$ using the mapping $\setRows : D \to \Rows$. $\iselect{\predicate}$ selects the rows on the storage satisfying the predicate $\predicate : D \times \Rows \to \{\bfalse,\btrue\}$. $\iupdate{\predicate}{\mapRows}$ updates the rows that satisfy $\predicate$ using the mapping $\mapRows : D \times \Rows \to \Rows$. Finally, $\idelete{\predicate}$, deletes the objects satisfying the predicate (i.e. replaces its row by $\del$). $\eabort$ represents states declared by the user where the transaction should not execute any more instructions and any declared write should be aborted. 
We assume that in for any predicate $\predicate$ and object $x$, $\predicate(x, \del) = \bfalse$.

We model snapshots as mappings $\Vars \to \Vals$. Unlike in \Cref{ssec:non-trans-sql-lww}, $\transSQLMVRStorage$ requires that local effects of SQL-like instructions are only seen internally, during the execution of the transaction. Such effects are modelled in \Cref{eq:execute-transaction} as a recursive function that simulates the transaction execution w.r.t. a concrete object. The function $\mathsf{exe}$ executes one instruction at a time, and it stops whenever all instructions are executed, indicating that the execution was correct, or halting it midway in case some abortion occurred (modelled with the constant value $\bot$).

\begin{equation}
\label{eq:execute-transaction}
\mathsf{exe}_x(\trCode, \sigma) = 
\left\{\begin{array}{ll}
    \mathsf{exe}_x(\mathsf{body}', \sigma') & \text{if } \trCode = e ; \mathsf{body}', \sigma' = \mathsf{exI}_x(e, \sigma) \text{ and } \sigma' \neq (\bot, \bfalse) \\
    \sigma & \text{if } \trCode = \emptyset \\
    (\bot, \bfalse) &  \text{otherwise}
\end{array} \right.
\end{equation}

The behavior of each instruction is modelled in \Cref{eq:execute-one-instruction}, updating the snapshot in object $x$ in a similar way as $\mathsf{wspec}$ does in \Cref{ssec:non-trans-sql-lww}, and indicating if the event $e$ indeed wrote object $x$.

\begin{equation}
\label{eq:execute-one-instruction}
\begin{array}{lll}
\mathsf{exI}_x(e, (\sigma, \mathsf{w})) & = & 
\left\{\begin{array}{ll}
    (\sigma, \mathsf{w}) & \text{if } e = \iselect{\predicate} \\
    (\sigma, \mathsf{w}) & \text{if } e = \idelete{\predicate} \land  \lnot \predicate_x(\sigma)\\
    (\del, \btrue) & \text{if } e = \idelete{\predicate} \land  \predicate_x(\sigma)\\
    (\sigma, \mathsf{w}) & \text{if } e = \iupdate{\predicate}{\mapRows} \text{ and either } \lnot \predicate_x(\sigma) \text{ or } U_x(\sigma) \uparrow \\
    (\mapRows_x(\sigma), \btrue) & \text{if } e = \iupdate{\predicate}{\mapRows}, \predicate_x(\sigma) \land U_x(\sigma) \uparrow \\
    (\sigma, \mathsf{w}) & \text{if } e = \iinsert{\setRows} \land \setRows(x) \uparrow\\
    (\setRows(x), \btrue) & \text{if } e = \iinsert{\setRows} \land \setRows(x) \downarrow\\
    (\bot, \bfalse) & \text{if } e = \iabort
\end{array} \right.
\end{array}
\end{equation}

The operation specifications of $\transSQLMVRStorage$ are an adaptation of those of $\kvmvrStorage$: 

\begin{equation}
\rspecContext{x}{c}{r} =  \left\{ \begin{array}{ll}
    \{\max_{\rbo}E \}& \text{if }  r = \transaction{\trCode} \text{ and } c = \tup{E, \ar, \rbo} \\
    \emptyset & \text{otherwise}
\end{array} \right.
\end{equation}

\begin{equation}
\extractspecContext{x}{R}{r} = \left\{ 
        \begin{array}{ll}
            \sigma' & \text{if }  r = \transaction{\trCode}, \sigma = \{(v, \bfalse) \ | \ (w, v) \in R\} \\
            & \quad \text{and }\sigma' = \mathsf{exe}_x(\trCode, \sigma)
        \end{array} \right. \\
\end{equation}

\begin{equation}
\wspecContext{x}{\sigma}{w} = \left\{ 
    \begin{array}{ll}
        v & \text{if }  r = \transaction{\trCode}, \text{ and } \sigma = (v, \btrue) \\
        \mathsf{undefined} & \text{otherwise}
    \end{array} \right. \\
\end{equation}

The $\transSQLMVRStorage$ operation specification is maximally layered w.r.t. $\rbo^+$, with $1$ as its layer bound. $\transSQLMVRStorage$ allows execution-correctors: let $\cmodel$ be a consistency model, $\exec$ be an abstract execution, $D$ be a domain, $W \subseteq D$ be a set of objects and $x$ be an object s.t. $x \in W$ if $W \neq \emptyset$. 

We define $e = \transaction{\iselect{\predicate_D} ; \iinsert{\setRows_W}}$, where $\predicate_W$ and $\mapRows_D$ are defined below.  
\begin{equation}
\begin{array}{ll}
    \predicate_D(d, r) = \left\{
    \begin{array}{ll}
        \btrue & \text{if } d \in D \\
        \mathsf{undefined} & \text{otherwise}
    \end{array}\right.\\
    \setRows_W(d) = \left\{
    \begin{array}{ll}
        \_ & \text{if } d \in D \\
        \mathsf{undefined} & \text{otherwise}
    \end{array}\right.\\
\end{array} \nonumber
\end{equation}
where $\_$ indicates some arbitrary unspecified value.

For such event, we define the execution-corrector $a: D \setminus \{x\} \to \Events$ as the totally-undefined mapping, i.e. the function that no object $y \in D$ is associated with some event.

\newpage

\section{Normal Form of a Consistency Model w.r.t. an Operation Specification}
\label{app:finite-normal-consistency}

In this section, we prove the existence of a consistency model in normal form equivalent to a given one (\Cref{th:general-normal-form:existence}), and we show as well that arbitration-freeness is well-defined  (\Cref{th:general-normal-from:arbitration-well-defined}), i.e. that either all its normal forms are arbitration-free or none.

For compare consistency models when restricted to an operation specification $\opspec$, we introduce the notion of $\opspec$-equivalence. Two consistency models $\cmodel_1$, $\cmodel_2$ are \emph{$\opspec$-equivalent}, denoted $\cmodel_1 \equiv_{\opspec} \cmodel_2$, if for every abstract execution of $\opspec$, $\exec$, $\exec$ is valid w.r.t. $(\cmodel_1, \opspec)$ iff $\exec$ is valid w.r.t. $(\cmodel_2, \opspec)$. In particular, if $\cmodel_1$ and $\cmodel_2$ are equivalent, they are also $\opspec$-equivalent. 
The converse is not true: vacuous visibility \replaced{formulas}{relations} under an operation specification $\opspec$ may not be vacuous for every possible operation specification.

\subsection{Existence of a Normal Form of a Consistency Model}
\label{app:ssec:existence-general-normal-form}

\Cref{th:general-normal-form:existence} states the existence of a normal form of a consistency model w.r.t. $\opspec$.

\begin{restatable}{theorem}{generalNormalFormTheorem}
\label{th:general-normal-form:existence}
Let $\opspec$ be an operation specification. For every consistency model $\cmodel$, there exists a consistency model that is in normal form w.r.t. $\opspec$ and that is $\opspec$-equivalent to $\cmodel$.
\end{restatable}

The proof of such result is divided in three parts, proving the existence of a consistency model with only simple visibility \replaced{formulas}{relations} (\Cref{lemma:normal-form:simple-equivalent-consistency}), proving that such model can be refined for removing vacuous visibility \replaced{formulas}{relations} (\Cref{lemma:general-normal-form:non-vacuous}) and finally, showing that conflict-maximality can be assumed without loss of generality (\Cref{lemma:general-normal-form:conflict-maximal}). \deleted{The proof is shown for a particular class of operation specifications, called \emph{monotonic} operation specifications. Lemma~\ref{lemma:maximal-layered-implies-monotonic} shows that the operation specifications we consider are monotonic.}

\noindent
\textbf{Monotonicity}

\replaced{Maximally-layered operation specifications are \emph{monotonic}.}{\emph{Monotonicity} is a useful property of maximally-layered operation specifications.} Intuitively, an operation specification is \emph{monotonic} if (1) the values that are not read under a consistency model $\cmodel_1$ should be also not read under a stronger model $\cmodel_2$, and (2) whenever some values are read under a consistency model $\cmodel_1$ but not under a stronger one $\cmodel_2$, some other values must be read under $\cmodel_2$ which were not visible under $\cmodel_1$. 

\begin{definition}
\label{def:cons-monotonic}
Let $\opspec=(E,\mathsf{rspec}, \mathsf{extract}, \mathsf{wspec})$ be an operation specification. $\opspec$ is called \emph{monotonic} if for every pair of consistency models $\cmodel_1, \cmodel_2$, $\cmodel_1 \preccurlyeq \cmodel_2$, abstract execution $\exec$, event $r \in \exec$, and object $x$ the following hold:
\begin{enumerate}
    
    \item \sloppy $\rspec{\cmodel_2}{x}{\exec}{r} \subseteq \rspec{\cmodel_1}{x}{\exec}{r} \cup (\context{\cmodel_2}{x}{\exec}{r} \setminus \context{\cmodel_1}{x}{\exec}{r})$. \label{def:cons-monotonic:1}
    
    \item if $\rspec{\cmodel_1}{x}{\exec}{r}\setminus \rspec{\cmodel_2}{x}{\exec}{r} \neq \emptyset$, then $\rspec{\cmodel_2}{x}{\exec}{r} \setminus \context{\cmodel_1}{x}{\exec}{r} \neq \emptyset$ \label{def:cons-monotonic:2}   
\end{enumerate} 
\end{definition}

\deleted{A particular case of monotonic operation specifications are maximally-layered operation specifications.}

\begin{lemma}
\label{lemma:maximal-layered-implies-monotonic}
A maximally-layered operation specification is monotonic.
\end{lemma}

\begin{proof}
Let $\opspec$ be a maximally-layered operation specification, $\cmodel_1, \cmodel_2$ be two consistency models s.t. $\cmodel_1 \preccurlyeq \cmodel_2$, $\exec$ be an abstract execution, $r$ be an event in $\exec$ and $x$ be an object. Observe that by the unconditional read property of $\opspec$ (Property~\ref{def:general-read-spec:unconditional-read} of \Cref{def:general-read-spec}), we can assume w.l.o.g. that $r$ is a read event.

On one hand, we observe that if the layer bound of $\opspec$ is $\infty$, $\opspec$ is trivially monotonic: as $r$ is a read event and the layer bound of $\opspec$ is $\infty$, $\rspec{\cmodel_2}{x}{\exec}{r} = \context{\cmodel_2}{x}{\exec}{r}$ and $\rspec{\cmodel_1}{x}{\exec}{r} = \context{\cmodel_1}{x}{\exec}{r}$. Using the fact that $\context{\cmodel_1}{x}{\exec}{r} \subseteq \context{\cmodel_2}{x}{\exec}{r}$, is easy to see that Properties~\ref{def:cons-monotonic:1} and \ref{def:cons-monotonic:2} hold in this case.

On the other hand, if the layer bound of $\opspec$, $k$, is finite, let $\mathsf{R}$ be the relation for which $\opspec$ is $k$-maximally layered. For proving Property~\ref{def:cons-monotonic:1} of \Cref{def:cons-monotonic}, let us partition $\context{\cmodel_2}{x}{\exec}{r}$ in the three disjoint sets $C_1$, $C_2$ and $C_3$ described in \Cref{eq:partition-contexts}.

\begin{equation}
\label{eq:partition-contexts}
\begin{array}{ll}
    C_1 \coloneqq & \rspec{\cmodel_1}{x}{\exec}{r} \\
    C_2 \coloneqq & \context{\cmodel_1}{x}{\exec}{r} \setminus \rspec{\cmodel_1}{x}{\exec}{r}  \\
    C_3 \coloneqq & \context{\cmodel_2}{x}{\exec}{r} \setminus \context{\cmodel_1}{x}{\exec}{r} \\
\end{array}
\end{equation}

We note that by Property~\ref{def:general-read-spec:well-formedness} of \Cref{def:general-read-spec}, $\rspec{\cmodel_1}{x}{\exec}{r} \subseteq \context{\cmodel_1}{x}{\exec}{r}$. As $\cmodel_1 \preccurlyeq \cmodel_2$, we deduce that $\rspec{\cmodel_1}{x}{\exec}{r} \subseteq \context{\cmodel_2}{x}{\exec}{r}$; so $\{C_1, C_2, C_3\}$ is indeed a partition of $\context{\cmodel_2}{x}{\exec}{r} $. Observe that showing Property~\ref{def:cons-monotonic:1} of \Cref{def:cons-monotonic} is equivalent to show that $\rspec{\cmodel_2}{x}{\exec}{r} \subseteq C_1 \cup C_3$. By Property~\ref{def:general-read-spec:well-formedness} of \Cref{def:general-read-spec}, $\rspec{\cmodel_2}{x}{\exec}{r} \subseteq \context{\cmodel_2}{x}{\exec}{r} = C_1 \cup C_2 \cup C_3$. We conclude the result by showing that $C_2 \cap \rspec{\cmodel_2}{x}{\exec}{r} = \emptyset$.

For showing it, we observe that the layer of an event $w$ in $\context{\cmodel_1}{x}{\exec}{r}$ is less or equal than the layer of $w$ in $\context{\cmodel_2}{x}{\exec}{r}$: as $\context{\cmodel_1}{x}{\exec}{r} \subseteq \context{\cmodel_2}{x}{\exec}{r}$, every chain of events in $\context{\cmodel_1}{x}{\exec}{r}$ containing $w$ and ordered w.r.t. $\mathsf{R}$ belongs to $\context{\cmodel_2}{x}{\exec}{r}$. Thus, as $\opspec$ is maximally layered, an event $w$ in $C_2$ does not belong to $\rspec{\cmodel_2}{x}{\exec}{r}$: if $w \in C_2$, its layer in $\context{\cmodel_1}{x}{\exec}{r}$ is greater than $k$; so it is also greater than $k$ in $\context{\cmodel_2}{x}{\exec}{r}$. Hence, as $\opspec$ has $k$ as layer bound, $w \not\in \rspec{\cmodel_2}{x}{\exec}{r}$.

For proving Property~\ref{def:cons-monotonic:2}, we observe that if there exists an event $w \in \rspec{\cmodel_1}{x}{\exec}{r}\setminus \rspec{\cmodel_2}{x}{\exec}{r}$, then the layer of $w$ in $\context{\cmodel_2}{x}{\exec}{r}$ is greater than $k$. Let $k'$ be the layer of $w$ and let $\{e_i\}_{i=1}^{k'}$ be a chain of $\mathsf{R}$ of length $k'$ s.t. $e_{k'} = w$. As the layer of $w$ in $\context{\cmodel_1}{x}{\exec}{r}$ is $k$ and $\mathsf{R}$ is a partial order, there exists an event $e_i, 1 \leq i \leq k$ s.t. $e_i \in C_3$. We observe that as the layer of $w$ is $k$, the layer of event $e_i$ is $i$. Hence, as $\mathsf{rspec}$ is $k$-maximally layered, we conclude that $e_i \in \rspec{\cmodel_2}{x}{\exec}{r} \setminus \context{\cmodel_1}{x}{\exec}{r}$.
\end{proof}

\Cref{lemma:general-normal-form:consistency-stronger-than-criterion} shows that\replaced{ for maximally-layered operation specifications}{, under monotonic operation specifications}, ensuring a strong consistency criteria is enough for ensuring a weaker one. \added{The proof relies on the fact that maximally-layered operation specifications are monotonic (\Cref{lemma:maximal-layered-implies-monotonic}).}

\generalizationStrongerThan*

\begin{proof}
Let $h = \tup{E, \so, \wro}$ be a history and let $\cmodel_1$ and $\cmodel_2$ be two consistency models s.t. $\cmodel_1 \preccurlyeq \cmodel_2$. Let also $\exec = \tup{h, \rbo, \ar}$ be an abstract execution that witness the validity of $h$ w.r.t. $(\cmodel_2, \opspec)$.
To prove that $\exec$ also witnesses $h$'s validity w.r.t. $(\cmodel_1, \opspec)$, by \Cref{def:general-valid}, \replaced{it suffices to}{we} prove that for every event $r \in h$ and object $x$, $\wro_x^{-1}(r) = \rspec{\cmodel_1}{x}{\exec}{r}$.%

\begin{itemize}
    \item \sloppy \underline{$\wro_x^{-1}(r) \subseteq \rspec{\cmodel_1}{x}{\exec}{r}$:} Let $w$ be a write event in $\wro_x^{-1}(r)$. As $(w, r) \in \wro_x$, $w \in \context{\cmodel_1}{x}{\exec}{r}$. Moreover, as $\exec$ witnesses $h$'s validity w.r.t. $\cmodel_2$, $\wro_x^{-1}(r) = \rspec{\cmodel_2}{x}{\exec}{r}$. Hence, as $w \in \rspec{\cmodel_2}{x}{\exec}{r} \cap \context{\cmodel_1}{x}{\exec}{r}$, by Property~\ref{def:cons-monotonic:1} of \Cref{def:cons-monotonic}, $w \in \rspec{\cmodel_1}{x}{\exec}{r}$.

    \item \sloppy \underline{$\wro_x^{-1}(r) \supseteq \rspec{\cmodel_1}{x}{\exec}{r}$:} Let $w \in\rspec{\cmodel_1}{x}{\exec}{r}$ s.t. $w \not\in \rspec{\cmodel_2}{x}{\exec}{r}$. By property \ref{def:cons-monotonic:2} from \Cref{def:cons-monotonic}, there exists $w' \in \rspec{\cmodel_2}{x}{\exec}{r}$ s.t. $w' \not\in \context{\cmodel_1}{x}{\exec}{r}$. However, as $\rspec{\cmodel_2}{x}{\exec}{r} = \wro_x^{-1}(r) \subseteq \context{\cmodel_1}{x}{\exec}{r}$, this is impossible. Therefore, $\rspec{\cmodel_1}{x}{\exec}{r} \subseteq \rspec{\cmodel_2}{x}{\exec}{r} = \wro_x^{-1}(r)$. \qedhere
\end{itemize}
\end{proof}

\added{An immediate consequence of \Cref{lemma:general-normal-form:consistency-stronger-than-criterion} is the following result.}

\strongerThan*

\noindent
\textbf{Simple Form}

For proving \Cref{th:general-normal-form:existence}, we first prove the existence of a consistency model in simple form (i.e. \added{a consistency model with} all its visibility \replaced{formulas}{relations} are simple) that is equivalent to $\cmodel$.

\begin{restatable}{lemma}{simpleForm}
\label{lemma:normal-form:simple-equivalent-consistency}
For any consistency model $\cmodel$, there exists a consistency model in simple form that is equivalent to $\cmodel$.
\end{restatable}

Intuitively, the proof of \Cref{lemma:normal-form:simple-equivalent-consistency} is as follows: we first unfold union and transitive closure operators, and then trim $\idRelation$ and compositional operators to obtain a consistency model in simple form.
As an intermediate step, we define the consistency model obtained after unfolding union and transitive closure operators. Such consistency model is the \emph{almost simple form} of $\cmodel$, $\almostSimple{\cmodel}$, and it is described as the union of the \emph{almost simple form} of each of its visibility \replaced{formulas}{relations}, i.e. $\almostSimple{\cmodel} = \bigcup_{v \in \cmodel} \almostSimple{v}$. A visibility \replaced{formula}{relation} $a$ belongs to the almost simple form of a visibility \replaced{formula}{relation} $v$, $a \in \almostSimple{v}$ if (1) $\length{v} = \length{a}$ and (2) for every $i, 1 \leq i \leq \length{v}$, $\mathsf{Rel}_i^a \in \sigma(\mathsf{Rel}_i^\mathsf{v})$; where $\sigma(\mathsf{Rel}i^v)$ is the set of relations described as follows:

\begin{equation}
\label{eq:sigma-function-relation}
\sigma(\mathsf{R}) = \left\{ 
    \begin{array}{ll}
        \{\mathsf{R}\} & \text{if } \mathsf{R} = \idRelation, \so, \wro, \rbo \text{ or } \ar \\
        \sigma(\mathsf{S}) \cup \sigma(\mathsf{T}) & \text{if } \mathsf{R} = \mathsf{S} \cup \mathsf{T} \\
        \sigma(\mathsf{S}) ; \sigma(\mathsf{T})& \text{if } \mathsf{R} = \mathsf{S};\mathsf{T} \\
        \bigcup_{k \in \mathbb{N} \land k \geq 1}\sigma(\mathsf{S})^k & \text{if } \mathsf{R} = \mathsf{S}^+
    \end{array}\right.
\end{equation}
where the composition of two sets of relations $A, B$ is defined as $A ; B \Coloneqq \{a;b \ | \ a \in A, b \in B\} $.

We prove that $\cmodel$ and $\almostSimple{\cmodel}$ are equivalent.

\begin{proposition}
\label{proposition:simple-form:almost-simple-coincide}
For any consistency model $\cmodel$, $\cmodel$ and $\almostSimple{\cmodel}$ are equivalent.
\end{proposition}

\begin{proof}
For proving the result, we show that for any abstract execution $\exec$, object $x$ and event $r$, $\context{\cmodel}{x}{\exec}{r} = \context{\almostSimple{\cmodel}}{x}{\exec}{r}$. In particular, it suffices to prove that for every visibility \replaced{formula}{relation} $v \in \cmodel$ and event $w$, $v_x(w, r)$ holds in $\exec$ iff there exists a visibility \replaced{formula}{relation} $a \in \almostSimple{v}$ s.t. $a_x(w, r)$ holds in $\exec$. Observe that for every $a \in \almostSimple{v}$, $\length{v} = \length{a}$; so we reduce the proof to show that for every pair of events $e, e'$, $(e, e') \in \mathsf{Rel}_i^\mathsf{v}$ iff there exists $\mathsf{R}' \in \sigma( \mathsf{Rel}_i^\mathsf{v})$ s.t. $(e, e') \in \mathsf{R}'$. 

In the following, we prove that for every relation $\mathsf{R}$ over pair of events obtained by the grammar described in \Cref{eq:relations-visibility}, the following holds: $(e, e') \in \mathsf{R}$ iff there exists $\mathsf{R'} \in \sigma(R)$ s.t. $(e, e') \in \mathsf{R}'$. We show the result by induction on the depth of $\mathsf{R}$\footnote{By depth of $\mathsf{R}$ we mean the depth of the tree obtained by deriving $\mathsf{R}$ using \Cref{eq:relations-visibility}.}. The base case, when the depth of $\mathsf{R}$ is $0$, refers to the case $\mathsf{R} = \idRelation, \so, \wro, \rbo, \ar$. In such case, the result immediately holds by the definition of $\sigma(\mathsf{R})$.

Let us assume that for any relation of depth at most $n$ the result holds, and let us prove that for relations of depth $n+1$. Three alternatives arise:
\begin{itemize}
    \item If $\mathsf{R} = \mathsf{S} \cup \mathsf{T}$, $(e, e') \in \mathsf{R} $ if and only if $(e, e') \in \mathsf{S} \cup \mathsf{T}$. By induction hypothesis on both $\mathsf{S}$ and $\mathsf{T}$, $(e, e') \in \mathsf{S} \cup \mathsf{T}$ iff there exists $\mathsf{R}' \in \sigma(\mathsf{S}) \cup \sigma(\mathsf{T})$ s.t. $(e, e') \in \mathsf{R}'$. Finally, by \Cref{eq:sigma-function-relation}, we conclude that there exists $\mathsf{R}' \in \sigma(\mathsf{S}) \cup \sigma(\mathsf{T})$ s.t. $(e, e') \in \mathsf{R}'$ if and only if  there exists $\mathsf{R}' \in \sigma(\mathsf{R})$ s.t. $(e, e') \in \mathsf{R}'$.
    
    \item If $\mathsf{R} = \mathsf{S} ; \mathsf{T}$, $(e, e') \in \mathsf{R} $ if and only if $(e, e') \in \mathsf{S} ; \mathsf{T}$. By the definition of composition, $(e, e') \in \mathsf{S} ; \mathsf{T}$ iff there exists $e''$ s.t. $(e, e'') \in \mathsf{S}$ and $(e'', e') \in \mathsf{T}$. By induction hypothesis on both $\mathsf{S}$ and $\mathsf{T}$, there exists $e''$ s.t. $(e, e'') \in \mathsf{S}$ and $(e'', e') \in \mathsf{T}$ iff there exists $e''$ and relations $\mathsf{S}' \in \sigma(\mathsf{S}), \mathsf{T}' \in \sigma(\mathsf{T})$ $e''$ s.t. $(e, e'') \in \mathsf{S}'$ and $(e'', e') \in \mathsf{T}'$. By the definition of $\sigma(S) ; \sigma(T)$, we observe that there exists $e''$ and relations $\mathsf{S}' \in \sigma(\mathsf{S}), \mathsf{T}' \in \sigma(\mathsf{T})$ $e''$ s.t. $(e, e'') \in \mathsf{S}'$ and $(e'', e') \in \mathsf{T}'$ iff there exists relation $\mathsf{R}' \in \sigma(\mathsf{S}; \mathsf{T})$ s.t. $(e, e') \in \mathsf{R}'$. Finally, by \Cref{eq:sigma-function-relation}, we conclude that there exists relation $\mathsf{R}' \in \sigma(\mathsf{S}; \mathsf{T})$ s.t. $(e, e') \in \mathsf{R}'$ if and only if there exists $\mathsf{R}' \in   \sigma(\mathsf{R})$ s.t. $(e, e') \in \mathsf{R}' $. 
    
    \item If $\mathsf{R} = \mathsf{S}^+$, $(e, e') \in \mathsf{R} $ if and only if there exists $k \in \mathbb{N}^+$ s.t. $(e, e') \in \mathsf{S}^k$. By the previous point, there exists $k \in \mathbb{N}^+$ s.t. $(e, e') \in \mathsf{S}^k$ if and only if there exists $k \in \mathbb{N}^+$ and relation $\mathsf{S}' \in \sigma(\mathsf{S})^k $ s.t. $(e, e') \in \mathsf{S}'$. Finally, by \Cref{eq:sigma-function-relation}, we conclude that there exists $k \in \mathbb{N}^+$ and relation $\mathsf{S}' \in \sigma(\mathsf{S})^k$ s.t. $(e, e') \in \sigma(\mathsf{S})^k$ if and only if there exists relation $\mathsf{R}' \in \sigma(\mathsf{R})$ s.t. $(e, e') \in \mathsf{R}'$. 
\end{itemize}
\end{proof}

Obtaining a consistency model in simple form from a consistency model in almost simple form is straightforward: every visibility \replaced{formula}{relation} is transformed by splitting composed relations into simpler subrelations and omitting $\idRelation$ by merging two existentially quantified events. \Cref{lemma:normal-form:simple-equivalent-consistency} formally describes such procedure.

\simpleForm*

\begin{proof}
We construct a consistency model, $\simpleConsistency{\cmodel}$, that is in simple form and it is equivalent to $\cmodel$. The model is formally defined as follows: 

\begin{equation}
\simpleConsistency{\cmodel} = \{\simpleConsistency{a} \ | \ a \in \almostSimple{\cmodel}\}
\end{equation}
where $\simpleConsistency{a}$ is the \emph{simple visibility \replaced{formula}{relation}} of $a$. 

The simple visibility \replaced{formula}{relation} of a visibility \replaced{formula}{relation} in almost form $a$ is the visibility \replaced{formula}{relation} $f$ obtained by supressing $\idRelation$ and compositional operators. Formally, $f$ is the visibility \replaced{formula}{relation} s.t. (1) $\length{f} = \sum_{i=1}^{\length{a}}\mathsf{count}(\mathsf{Rel}_i^a)$ and (2) for every $i, 1 \leq i \leq \length{f}$, $\mathsf{Rel}_i^f = \mathsf{rel}(\mathsf{Rel}_j^a, i-k_j)$; where $j$ is the maximum index s.t. $k_j < i$ and $k_j = \sum_{l=1}^{j} \mathsf{count}(\mathsf{Rel}_l^a)$, and $\mathsf{count}$ and $\mathsf{rel}$ are the functions described in \Cref{eq:count-function-relation} and \Cref{eq:relation-function-relation} respectively.

The function $\mathsf{count}$ counts the number of additional quantifiers the correspondant simple form requires:
\begin{equation}
\label{eq:count-function-relation}
\mathsf{count}(\mathsf{R}) = \left\{ 
    \begin{array}{ll}
        0 & \text{if } \mathsf{R} = \idRelation \\
        1 & \text{if } \mathsf{R} = \so, \wro, \rbo \text{ or } \ar \\
        \mathsf{count}(\mathsf{S}) + \mathsf{count}(\mathsf{T}) & \text{if } \mathsf{R} = \mathsf{S} ; \mathsf{T} \\
    \end{array}\right.
\end{equation}

Also, the function $\mathsf{rel}$, given a relation using compositional operator and an index $i$, returns the $i$-th component:

\begin{equation}
\label{eq:relation-function-relation}
\mathsf{rel}(\mathsf{R}, i) = \left\{ 
    \begin{array}{ll}
        \mathsf{R} & \text{if } \mathsf{R} = \so, \wro, \rbo \text{ or } \ar \\
        \mathsf{rel}(\mathsf{S}, i) & \text{if } i \leq \mathsf{count}(\mathsf{S})\\
        \mathsf{rel}(\mathsf{T}, i - \mathsf{count}(\mathsf{S})) & \text{otherwise}
    \end{array}\right.
\end{equation}

By construction, $\simpleConsistency{\cmodel}$ is in simple form. Clearly, $\simpleConsistency{\cmodel}$ is equivalent to $\almostSimple{\cmodel}$. Then, thanks to \Cref{proposition:simple-form:almost-simple-coincide}, we conclude that $\simpleConsistency{\cmodel}$ is equivalent to $\cmodel$.
\end{proof}

\noindent
\textbf{Removing Vacuous Visibility \replaced{Formulas}{Relations}}

After proving the existence of a consistency model $\cmodel$ in simple form equivalent to a given one, we show how to transform it for obtaining an equivalent consistency model $\cmodel$ without vacuous visibility \replaced{formulas}{relations} (\Cref{lemma:general-normal-form:non-vacuous}). We say that any such consistency model is in \emph{basic normal form}, extending \Cref{def:normal-form} to any consistency model whose visibility \replaced{formulas}{relations} are described using \Cref{eq:general-visibility-criterion}.

The following result, key to prove \Cref{lemma:general-normal-form:non-vacuous}, it is a simple consequence of \Cref{def:cons-monotonic} \added{and \Cref{lemma:maximal-layered-implies-monotonic}}.

\begin{proposition}
\label{proposition:monotonic:write-read-not-visible}
Let $\opspec$ be a maximally-layered operation specification, and let $\cmodel_1, \cmodel_2$ be two consistency models s.t. $\cmodel_1 \not\equiv_{\opspec} \cmodel_2$ but $\cmodel_1 \preccurlyeq \cmodel_2$. There exists an abstract execution $\exec$ valid w.r.t. $\cmodel_1$, an object $x$ and events $w, r$ s.t. $w \in \rspec{\cmodel_2}{x}{\exec}{r} \setminus \context{\cmodel_1}{x}{\exec}{r}$.
\end{proposition}

\begin{proof}
\sloppy First of all, as $\cmodel_1 \not\equiv_{\opspec} \cmodel_2$ but $\cmodel_1 \preccurlyeq \cmodel_2$, by \Cref{lemma:general-normal-form:consistency-stronger-than-criterion}, there exists an abstract execution $\exec$ valid w.r.t. $\cmodel_1$, an object $x$ and an event $r$ s.t. $\rspec{\cmodel_2}{x}{\exec}{r} \neq \rspec{\cmodel_1}{x}{\exec}{r}$. Thus, either $\rspec{\cmodel_2}{x}{\exec}{r} \setminus \rspec{\cmodel_1}{x}{\exec}{r} \neq \emptyset$ or $\rspec{\cmodel_1}{x}{\exec}{r} \setminus \rspec{\cmodel_2}{x}{\exec}{r} \neq \emptyset$. 

On one hand, if $\rspec{\cmodel_2}{x}{\exec}{r} \setminus \rspec{\cmodel_1}{x}{\exec}{r} \neq \emptyset$, by Property~\ref{def:cons-monotonic:1} of \Cref{def:cons-monotonic}, then $\rspec{\cmodel_2}{x}{\exec}{r} \setminus \context{\cmodel_1}{x}{\exec}{r} \neq \emptyset$. On the other hand, if $\rspec{\cmodel_1}{x}{\exec}{r} \setminus \rspec{\cmodel_2}{x}{\exec}{r} \neq \emptyset$, by Property~\ref{def:cons-monotonic:2} of \Cref{def:cons-monotonic}, $\rspec{\cmodel_2}{x}{\exec}{r} \setminus \context{\cmodel_1}{x}{\exec}{r} \neq \emptyset$.
\end{proof}

\begin{lemma}
\label{lemma:general-normal-form:non-vacuous}
Let $\opspec$ be an operation specification. For every consistency model $\cmodel$ in simple form, there exists a $\opspec$-equivalent consistency model, $\normalForm{\cmodel}$, that is in basic normal form w.r.t. $\opspec$.
\end{lemma}

\begin{proof}
To prove the result, we construct a consistency model in basic normal form w.r.t. $\opspec$, $\normalForm{\cmodel}$, that is $\opspec$-equivalent to $\cmodel$. Without loss of generality we can assume that $\cmodel$ is ordered. Let $\alpha$ be an ordinal of cardinality $|\cmodel|$. We denote by $v^i, 0 \leq i < \alpha$ to the $i$-th visibility \replaced{formula}{relation} in $\cmodel$\footnote{Without loss of generality, we can assume that limit ordinals in $\alpha$ are not associated to a visibility \replaced{formula}{relation}.}.

We construct a sequence of nested consistency models $\cmodel_{k}, 0 \leq k \leq \alpha$ s.t. (1) $\cmodel_k$ is $\opspec$-equivalent to $\cmodel$, (2) $\cmodel_k$ is more succinct than $\cmodel_i$ (i.e., for every $i < k$, $v^i \in \cmodel_k$ iff $v^i \in \cmodel_i$ and for every $i > k$, $v^i \in \cmodel_k$), and (3) the first $k$ visibility \replaced{formulas}{relations} of $\cmodel_k$ are simple and non-vacuous w.r.t. $(\cmodel_k, \opspec)$ (i.e., for every $i, 0 \leq i < k$, if $v^i \in \cmodel_k$, then $\cmodel_k \setminus \{v^i\} \not\equiv_{\opspec} \cmodel$).%

We construct such sequence using transfinite induction. The base case, $k = 0$, corresponds to $\cmodel_0 = \cmodel$, which trivially satisfies (1), (2) and (3). For the successor case, let us assume that the property holds for the consistency model $\cmodel_{k}$, and let us prove it for $\cmodel_{k+1}$. If $\cmodel_{k} \setminus \{v^k\} \equiv_{\opspec} \cmodel$, we denote $\cmodel_{k+1}$ as $\cmodel_{k} \setminus \{v^k\}$; and otherwise, $\cmodel_{k+1} = \cmodel_{k}$. 

Clearly, by construction of $\cmodel_{k+1}$, (1) and (2) immediately hold. For proving (3), we observe that if $v^i \in \cmodel_{k+1}$, $v^i \in \cmodel_i$. In such case, $\cmodel_i \setminus \{v^i\} \not\equiv_{\opspec} \cmodel$. Hence, by \Cref{lemma:general-normal-form:consistency-stronger-than-criterion}, there exists an abstract execution valid w.r.t. $(\cmodel_i \setminus \{v^i\}, \opspec)$ that is not valid w.r.t. $(\cmodel, \opspec)$. As $\cmodel_{k+1} \subseteq \cmodel_i$, $\cmodel_{k+1} \setminus \{v^i\} \subseteq \cmodel_i \setminus \{v^i\}$ and hence, $\cmodel_{k+1} \setminus \{v^i\} \preccurlyeq \cmodel_i \setminus \{v^i\}$. Therefore, by \Cref{lemma:general-normal-form:consistency-stronger-than-criterion}, $\exec$ is valid w.r.t. $(\cmodel_{k+1} \setminus \{v^i\}, \opspec)$. Thus, as $\exec$ is not valid w.r.t. $(\cmodel, \opspec)$, $\cmodel_{k+1} \setminus \{v^i\} \not\equiv_{\opspec} \cmodel$; so we conclude (3).

For the limit case, we define $\cmodel_k$ as the intersection of all consistency models $\cmodel_i, i < k$. We observe that in this case, (2) immediately holds by construction of $\cmodel_k$.

For proving (3) we observe that $v^i \in \cmodel_k$ iff $v^i \in \cmodel_i$. In such case, $\cmodel_i \setminus \{v^i\} \not\equiv_{\opspec} \cmodel$; so by \Cref{lemma:general-normal-form:consistency-stronger-than-criterion}, there exists an abstract execution $\exec$ valid w.r.t. $(\cmodel_i \setminus \{v^i\}, \opspec)$ that is not valid w.r.t. $(\cmodel, \opspec)$. Similarly to the inductive case, we deduce using \Cref{lemma:general-normal-form:consistency-stronger-than-criterion} that $\exec$ is valid w.r.t. $(\cmodel_k \setminus \{v^i\}, \opspec)$. Therefore, we conclude that $\cmodel_i \setminus \{v^i\} \not\equiv_{\opspec} \cmodel$.

For proving (1), we reason by contradiction, assuming that $\cmodel_k \not\equiv_{\opspec} \cmodel$ and reaching a contradiction. In such case, by \Cref{lemma:general-normal-form:consistency-stronger-than-criterion} there exists an abstract execution $\exec = \tup{h, \rbo, \ar}$ valid w.r.t. $(\cmodel_k, \opspec)$ that is not valid w.r.t. $(\cmodel, \opspec)$.
W.l.o.g., we can assume that $\exec$ is minimal w.r.t. the number of events in it; and let $\length{\exec}$ the number of events in such execution.

For each event $r \in \exec$, we define an ordinal $i(r), i(r) < k$ associated to every visibility \replaced{formula}{relation} $v^i, i < k$ that can be applied on $\exec$. First, we note that for every pair of events, $e, e'$ and object $x$, if a visibility \replaced{formula}{relation} $v_x(e, e')$ holds in $\exec$, $\length{v} \leq \length{\exec}$. Observe that there exists finite number of visibility \replaced{formulas}{relations} $v$ in $\cmodel$ with at most length $\length{\exec}$: on one hand, for each $j, 1 \leq j \leq \length{v}$, $\mathsf{Rel}_j^v$ is either $\so, \wro, \rbo$ or $\ar$. On the other hand, $\mathsf{wrCons}$ is defined as a conjunction of predicates from a finite set. Thus, the number of possible visibility \replaced{formulas}{relations} $v$ of length $\length{v} \leq \length{\exec}$ is finite. Let $i_r$ be the biggest index of a visibility \replaced{formula}{relation} $v^i \in \cmodel$ s.t. $\length{v^i} \leq \length{\exec}$ and $i < k$; and let $i(r) = i_r + 1$. Observe that $k$ is a limit ordinal, $i(r) < k$.

Let $x$ be an object and $r$ be an event in $\exec$. We show that $\context{\cmodel_k}{x}{\exec}{r} = \context{\cmodel_{i(r)}}{x}{\exec}{r}$. As $\cmodel_k \subseteq \cmodel_{i(r)}$, $\context{\cmodel_k}{x}{\exec}{r} \subseteq \context{\cmodel_{i(r)}}{x}{\exec}{r}$. For showing $\context{\cmodel_{i(r)}}{x}{\exec}{r} \subseteq \context{\cmodel_k}{x}{\exec}{r}$, let $w \in \context{\cmodel_{i(r)}}{x}{\exec}{r} $. In such case, there exists a visibility \replaced{formula}{relation} $v^i$ s.t. $v^i(w, r)$ holds in $\exec$. If $i > k$, by (2) $v^i \in \cmodel_k$. Otherwise, $i < i(r)$, so by (2), $v^i \in \cmodel_i$. Observe that in this case, applying the induction hypothesis (2) on every consistency model $\cmodel_j, j < k$, $v^i \in \cmodel_j$, we deduce that $v^i \in \cmodel_k$. Either way, we deduce that $w \in \context{\cmodel_k}{x}{\exec}{r}$. In conclusion, $\context{\cmodel_k}{x}{\exec}{r} = \context{\cmodel_{i(r)}}{x}{\exec}{r}$.

We conclude a contradiction by showing that $\exec$ is valid w.r.t. $(\cmodel, \opspec)$; which by assumption it is not. Let $e$ be the last event w.r.t. $\ar$ in $\exec$. For reaching such contradiction, as $i(e) < k$ and $\cmodel_{i(e)} \equiv_{\opspec} \cmodel$, it suffices to show that $\exec$ is valid w.r.t. $(\cmodel_{i(e)}, \opspec)$. We show that $\wro_x^{-1}(e') = \rspec{\cmodel_{i(e)}}{x}{\exec}{e'}$. 

On one hand, if $e' = e$, we note that $\context{\cmodel_k}{x}{\exec}{e} = \context{\cmodel_{i(e)}}{x}{\exec}{e}$. As $\exec$ is valid w.r.t. $(\cmodel_k, \opspec)$, we conclude that $\rspec{\cmodel_{i(e)}}{x}{\exec}{e} = \wro_x^{-1}(e)$. 

On the other hand, if $e' \neq e$, let $\exec'$ be the execution obtained by removing $e$ from $\exec$. By the minimality of $\exec$, $\exec'$ is valid w.r.t. $(\cmodel, \opspec)$. By induction hypothesis (1), $\cmodel \equiv_{\opspec} \cmodel_{i(e)}$. Hence, $\exec'$ is valid w.r.t. $(\cmodel_{i(e)}, \opspec)$. We thus deduce that $\wro_x^{-1}(e') =  \rspec{\cmodel_{i(e)}}{x}{\exec}{e'}$. In conclusion, $\cmodel_k$ satisfies (1) and thus, the inductive step.

Finally, we define $\normalForm{\cmodel} = \cmodel_{\alpha}$. As $\cmodel_{\alpha}$ satisfies (1) and (3), it is a consistency model $\opspec$-equivalent to $\cmodel$ composed of finite, non-vacuous w.r.t. $(\cmodel_{\alpha}, \opspec)$ visibility \replaced{formulas}{relations}; so we conclude that it is a consistency model in basic normal form. 

\end{proof}

\textbf{Conflict-Strengthening a Consistency Model}

\begin{lemma}
\label{lemma:general-normal-form:conflict-maximal}
Let $\opspec$ be an operation specification. For every consistency model $\cmodel$ in basic normal form w.r.t. $\opspec$ there exists a $\opspec$-equivalent consistency model that is in normal form.
\end{lemma}

\begin{proof}
We transform $\cmodel$ to define $\generalNormalForm{\cmodel}$, a consistency model in normal form that is $\opspec$-equivalent to $\cmodel$.

For every visibility \replaced{formula}{relation} $v \in \cmodel$, we define $v'$ as the visibility \replaced{formula}{relation} that only differs with $v$ on its conflict predicate. More specifically, we require that for every set $E \in \mathcal{P}(\event_0, \ldots \event_{\length{v}})$, we require that $\writeConflicts{E} \in v'$ (resp. $\writeConflicts[x]{E} \in v'$) iff (1) for every abstract execution $\exec$, every object $x$ and every collection of events $e_0, \ldots e_{\length{v}}$ s.t. $v_x(e_0, \ldots e_{\length{v}})$ holds in $\exec_v$, there exists an object $y \neq x$ s.t. if $\event_i \in E, 0 \leq i \leq \length{v}$, then $\wspec{\cmodel}{y}{\exec}{e_i} \downarrow$ (resp. $\wspec{\cmodel}{x}{\exec}{e_i} \downarrow$) and (2) there is no strict superset of $E$ satisfying (1). We define $\generalNormalForm{\cmodel}$ as the set containing all such visibility \replaced{formulas}{relations}.
For conclude the result, we first prove that $\generalNormalForm{\cmodel} \equiv_{\opspec} \cmodel$ for then deduce that $\generalNormalForm{\cmodel}$ is indeed a consistency model in normal form.

We show that $\generalNormalForm{\cmodel} \equiv_{\opspec} \cmodel$. On one hand, as every visibility \replaced{formula}{relation} $v'$ enforces more conflicts than $v$, $\generalNormalForm{\cmodel} \preccurlyeq \cmodel$. On the other hand, by the definition of $v'$, for every abstract execution $\exec$, object $x$ and events $w,r$, if $v'_x(w, r)$ holds in $\exec$, $v_x(w, r)$ also holds in $\exec$. Altogether, we conclude that $\generalNormalForm{\cmodel} \equiv_{\opspec} \cmodel$. %

To show that $\generalNormalForm{\cmodel}$ is a consistency model in normal form, we observe that by construction, every visibility \replaced{formula}{relation} $v \in \generalNormalForm{\cmodel}$ is in simple form and it is conflict-maximal w.r.t. $\opspec$. Hence, it suffices to prove that every visibility \replaced{formula}{relation} $v \in \generalNormalForm{\cmodel}$ is non-vacuous w.r.t. $\generalNormalForm{\cmodel}$.

Let $v'$ be a visibility \replaced{formula}{relation} of $\generalNormalForm{\cmodel}$. Observe that by construction of $\generalNormalForm{\cmodel}$, $\generalNormalForm{\cmodel} \setminus \{v'\} \equiv \cmodel \setminus \{v\}$. Hence, as $\cmodel \setminus \{v\} \not\equiv_{\opspec} \cmodel$, we deduce that $\generalNormalForm{\cmodel} \setminus \{v'\} \equiv \cmodel \setminus \{v\} \not\equiv_{\opspec} \cmodel \equiv \generalNormalForm{\cmodel}$. In other words, $v'$ is non-vacuous w.r.t. $(\generalNormalForm{\cmodel}, \opspec)$.

\end{proof}

\subsection{Arbitration-Free Well-Formedness}
\label{app:ssec:arbitration-free-general-normal-form}

As described in \Cref{ssec:conflict-defs}, a consistency model is arbitration-free if a $\opspec$-equivalent consistency model in normal form is arbitration-free. In \Cref{th:general-normal-from:arbitration-well-defined}, we present a result that states that arbitration-free is well-defined, as either every $\opspec$-equivalent consistency model in normal form are arbitration-free or none.

Regarding notations, for a visibility \replaced{formula}{relation} $v$ and $i, 0 \leq i \leq \length{v}$ we denote hereinafter $\conflictsOfV{v}{i} \in \mathcal{P}(\mathcal{P}(\event_0, \ldots \event_{\length{v}}))$ to the sets of conflicts of $\event_i$ in $v$, i.e. $E \in \conflictsOfV{v}{i}$ iff $\event_i \in E$ and $\writeConflicts{E} \in v$. %

\begin{theorem}
\label{th:general-normal-from:arbitration-well-defined}
Let $\opspec = (E, \rspecName, \extractName, \wspecName)$ be an operation specification and let $\cmodel$ be a consistency model. For every pair of consistency models in normal form $n_1, n_2$ that are $\opspec$-equivalent to $\cmodel$, $n_1$ is arbitration-free iff $n_2$ is arbitration-free.
\end{theorem}

\begin{proof}
We prove the result by contradiction, assuming that there exists two consistency models $n_1, n_2$ in normal form, $\opspec$-equivalent to $\cmodel$, but one of them arbitration-free and the other one no. W.l.o.g., we can assume that $n_1$ is arbitration-free and $n_2$ is not. On one hand, as $n_2$ is not arbitration-free w.r.t. $\opspec$, there exists a visibility \replaced{formula}{relation} $v \in n_2$ s.t. $v$ is not arbitration-free. We construct an abstract execution that is valid w.r.t. $(n_1, \opspec)$ but not valid w.r.t. $(n_2, \opspec)$ using $v$, reaching a contradiction.

First of all, observe that by \Cref{lemma:saturable:cc-strongest}, $n_1$ is weaker than $\CC$. The abstract execution we construct contains a collection events $e_0, \ldots e_{\length{v}}$%
s.t. $\exec$ is valid w.r.t. $(\CC, \opspec)$ and $v_x(e_0, \ldots e_{\length{v}})$ holds on it; for some object $x$.

Let $x$ be an object. For each set $E \in \mathcal{P}(\event_0, \ldots e_{\length{v}})$ we consider a distinct object $y_E$, also distinct from $x$. These objects represents each different conflict in $v$ in an explicit manner.

We denote by $E_x \in \mathcal{P}(\event_0, \ldots e_{\length{v}})$ to the set s.t. $\writeConflicts[x]{E_x} \in v$. Also, for every $i, 0 \leq i \leq \length{v}$, we denote by $X_i$ to the set containing objects $y_E$ (resp. $x$) iff $E \in \conflictsOfV{v}{i}$ (resp. $E_x \in \conflictsOfV{v}{i}$). We denote by $X$ to the union of sets $X_i, 0 \leq i \leq \length{v}$.

For obtaining $\exec$, we construct a sequence of executions $\exec^i, 0 \leq i \leq \length{v}$ inductively, starting from an initial event $\init$, and incorporating at each time a new event $e_i$. We use the notation $h^{-1}$ and $\exec^{-1}$ to describe the history and abstract execution containing only $\init$ respectively. We use the convention $e_{-1} = \init$, $\conflictsOfV{v}{-1} = \Vars$ and $\tilde{x}_{-1} = o_{-1} = x$ (the usage of such conventions will be clearer later).

For the inductive step, we assume that the abstract execution $\exec^{i-1} = \tup{h^{i-1}, \rbo^{i-1}, \ar^{-1}}$ associated to the history $h^{i-1} = \tup{E^{i-1}, \so^{i-1}, \wro^{i-1}}$ contains events $e_{-1} \ldots e_{i-1}$ and is well-defined (satisfies \Cref{def:execution}) and we construct the history $h^i$ and the abstract execution $\exec^i$. First of all, we impose the constraint that if $i > 0$, then $r_i = r_{i-1}$ iff $\mathsf{Rel}_i^\mathsf{v} = \so$, and otherwise $r_i \neq r_j, 0 \leq j < i$. 

Also, we define a pair of special objects, $\tilde{x}_i$ and $o_i$. The purpose of object $\tilde{x}_i$ is control the number of events in $\exec$ that write object $x$. \Cref{eq:x-i} describes $\tilde{x}_i$; where $\mathsf{choice}$ is a function that deterministically chooses an element from a non-empty set. The object $o_i$ is an object different from objects $x, y_E, E \in \mathcal{P}(\event_0, \ldots \event_{\length{v}})$ and $o_j, -1 \leq j < i$ that we use for ensuring that if $\mathsf{Rel}_i^\mathsf{v} = \wro$, then $(e_{i-1}, e_i) \in \wro$. 

\begin{equation}
\label{eq:x-i}
    \tilde{x}_i = \left\{ \begin{array}{ll}
        \tilde{x}_{i-1} & \text{if } X_i = \emptyset \\
        x & \text{if } X_i \neq \emptyset \text{ and } x \in X_i \\
        \choice{X_i} & \text{if } X_i \neq \emptyset \text{ and } x \not\in X_i
    \end{array}\right.
\end{equation}

We select a domain $D_i$, a set of objects $W_i, W_i \subseteq D_i$ that event $e_i$ must write, and a set of objects $C_i \subseteq D_i$ whose value needs to be corrected for $e_i$ in $\exec_{i+1}$ -- in the sense of \Cref{def:execution-corrector}. We distinguishing between several cases:

\begin{itemize}
    \item \underline{$i = 0$ or $0 < i \leq \length{v}$ and $\mathsf{Rel}_i^\mathsf{v} \neq \wro$ and $\conflictsOfV{v}{i} \neq \emptyset$}: In this case, we select $e_i$ to be a write event. If $\opspec$ only allows single-object atomic read-write events, we define $D_i = X_i$; while if not, we consider a domain containing $o_{i-1}, o_i$, every object in $X_i$ but no object from $X \setminus X_i$ nor objects $o_j, 0 \leq j < \length{v}, j \neq i-1,i$. Observe that by \Cref{proposition:general-normal-form:arbitration-well-defined:domains}, such domain always exist on $\opspec$. 
    
    If there is an unconditional write event whose domain is $D_i$, we define $W_i = D_i$. Otherwise, we define $W_i = X_i \cup \{o_{i}\}$.%

    \item \underline{$0 < i \leq \length{v}$, $\mathsf{Rel}_i^\mathsf{v} = \wro$ and $\conflictsOfV{v}{i} \neq \emptyset$}: In this case, by \Cref{proposition:general-normal-form:arbitration-well-defined:events-well-defined}, $\opspec$ allows atomic read-write events. If $\opspec$ only allows single-object atomic read-write events, we define $D_i = X_i$; while if not, we consider a domain containing $o_{i-1}, o_i$, every object in $X_i$ but no object from $X \setminus X_i$ nor objects $o_j, 0 \leq j < \length{v}, j \neq i-1,i$. Observe that by \Cref{proposition:general-normal-form:arbitration-well-defined:domains}, such domain always exist on $\opspec$. 
    
    Similarly to the previous case, if there is an unconditional atomic read-write event whose domain is $D_i$, we define $W_i = D_i$. Otherwise, we define $W_i = X_i \cup \{o_{i}\}$.%

    \item \underline{$0 < i \leq \length{v}$ and $\conflictsOfV{v}{i} = \emptyset$}: In this case, by \Cref{proposition:general-normal-form:arbitration-well-defined:events-well-defined}, $\opspec$ allows events that do not unconditionally write. %
    If $\opspec$ allows read events that are not write events, we select $D_i$ to be the domain of any such event and $W_i = \emptyset$. Otherwise, $\opspec$ must allow conditional write events; so we select $D_i$ to be the domain of any such event, $W_i = \emptyset$. %
    Observe that in this case, thanks to the assumptions on $\opspec$ (see \Cref{ssec:assumptions-generalized-specs}), we can assume without loss of generality that whenever $o_{i-1} \in D_{i-1}$, $o_{i-1} \in D_i$ as well; while otherwise, that $\tilde{x}_{i-1} \in D_i$.

\end{itemize}

Finally we describe the event $e_i$ thanks to the sets $D_i$ and $W_i$. If $W_i = D_i$ and $\mathsf{Rel}_i^\mathsf{v} = \wro$, we select an unconditional atomic read-write event whose domain is $D_i$. If $W_i = D_i$ and $\mathsf{Rel}_i^\mathsf{v} \neq \wro$, we select an unconditional write event whose domain is $D_i$. If $W_i = \emptyset$ and $\opspec$ allows read events that are not write events, we select a read event whose domain is $D_i$. Finally, if that is not the case, we select a conditional write event $e_i$ s.t. $\varOf{e_i} = D_i$ and s.t. an execution-corrector exists for $(e_i, W_i, \tilde{x}_i, \exec^{i-1} \oplus e_i)$. Such event always exists by the assumptions on operation specifications (\Cref{ssec:assumptions-generalized-specs}). W.l.o.g. we can assume that $e_i$ happens on replica $r_i$.

For concluding the description of $h^i = \tup{E_i, \so^i, \wro^i}$ and $\exec^i= \tup{h^i, \rbo^i, \ar^i}$, we use an auxiliary history and abstract execution, $h^i_0 = \tup{E^i_0, \so^i_0, \wro^i_0}$ and $\exec^i_0= \tup{h^i_0, \rbo^i_0, \ar^i_0}$ respectively. For describing the write-read dependencies of $e_i$ in $\exec_i^0$, we define the context mapping $c^i: \Vars \to \Contexts$, associating each object $y$ to the context $c^i(y)$ described in \Cref{eq:general-normal-form:arbitration-context-def}.

\begin{equation}
\label{eq:general-normal-form:arbitration-context-def}
c^i(y) = (F^i(y), \rbo^{i-1}_{\restriction F^i(y) \times F^i(y)}, \ar^{i-1}_{\restriction F^i(y) \times F^i(y)})
\end{equation}
where $F^i(y)$ is the mapping associating each object $y$ with the set of events described below:

\begin{equation}
F^i(y) = \left\{ \begin{array}{ll}
    \{\init\} & \text{if } i = 0 \text{ or if } 0 < i \leq \length{v} \land \mathsf{Rel}_i = \ar \\
    \left\{e \in E^{i-1} \ \left| \ 
    \begin{array}{l}
    \wspec{\CC}{y}{\exec^{i-1}}{e}\downarrow \text{and} \\
    \ (e, e_{i-1}) \in (\rbo^{i-1})^* 
    \end{array}\right.\right\} & \text{otherwise} \\
\end{array}\right. \nonumber
\end{equation}

Then, we define $\exec^i_0$ as the abstract execution of the history $h^i_0 = \tup{E^i_0, \so^i_0, \wro^i_0}$ obtained by appending $e_i$ to $h^i_0$ and $\exec^i_0$ as follows: $E^i_0$ contains $E^{i-1}$ and event $e_i$. First of all, we require that the relations $\so^i_0$, $\wro^i_0$, $\rbo^i_0$ and $\ar^i_0$ contain $\so^{i-1}$, $\wro^{i-1}$, $\rbo^{i-1}$ and $\ar^{i-1}$ respectively. With respect to event $e_i$, we impose that $e_i$ is the maximal event w.r.t. $\so^i_0$ among those on the same replica. Also, $e_i$ is maximal w.r.t. $\wro$ as we define that for every object $z$, ${\wro^i_0}_z^{-1}(e_i) = \rspecContext{z}{c_i(z)}{e_i}$. For describing $\rbo^i_0$, we require that for every event $e$ s.t. $(e, e_i) \in \so^i_0$, $(e, e_i) \in \rbo^i$. Also, if $\mathsf{Rel}_i^\mathsf{v} = \rbo$, we impose that $(e_{i-1}, e_i) \in \rbo^i_0$. Finally, we require that for every pair of events $e, e' \in E^{i-1}$ s.t. $(e, e') \in \rbo^{i-1}$ and $(e', e_i) \in \so^i_0$, $(e, e_i) \in \rbo^i_0$. With respect to $\ar^i_0$, we impose that $e_i$ is the maximum event w.r.t. $\ar$ in $\exec^i_0$.

We use $\exec^i_0$ to construct $\exec^i$. If event $e_i$ is not a conditional write event, $\exec^i = \exec^i_0$. Otherwise, if event $e_i$ is a conditional write event, given $W_i$ and object $\tilde{x}_i$, we select an execution-corrector for $e_i$ w.r.t. $(\CC, \opspec)$ and $a_i$. W.l.o.g., we assume that every event mapped by $a_i$ happens on replica $r_i$. Observe that by the choice of sets $D_i$ and $W_i$, and thanks to the assumptions on storages (see \Cref{ssec:assumptions-generalized-specs}), such event(s) are always well-defined.

In addition, we denote by $C_i$ to the set of objects we need to correct for $e_i$. More specifically, if $e_i$ is a conditional write-read, we denote by $C_i$ to the set of objects $y$ s.t. $a_i(y)$ is defined, i.e. $C_i = \{y \in \Vars \ | \ a_i(y) \downarrow\}$. In the case $e_i$ is not a conditional write-read, we use the convention $C_i = \emptyset$. The set of events in $\exec^i$ is the following: $E^i  = E^{i-1} \cup \{e_i\} \bigcup_{y \in C_i \setminus \{o_{i-1}\}} a_i(y)$. Observe that by the choice of $C_i$, the set $E^i$ is well-defined.

Concerning notations, we use $c \oplus a$ to denote the context obtained by appending $a$ to the context $c = \{E, \rbo, \ar\}$ as the $\rbo$-maximum and $\ar$-maximum event.

From $\exec^i_0$, we define $\exec^i = \correction{\exec^i_0}{\sequence{a_i}}{e_i}$ as the corrected execution of $\exec$ and $e_i$ with events $a_i$. For describing $\exec^i$, we consider $<$ to be a well-founded order over $\Vars$. $\exec^i$ satisfies the following:

\begin{itemize}
    \item \underline{$\so^i$}: Let $y \in C_i$. We require that for every event $e \in E^{i-1}$, $(e, a_i(y)) \in \so^i$ iff $\replicaEvent{e} = r_i, 0 \leq j < i$. We also require that $(\init, a_i(y)) \in \so^i$ and $(a_i(y), e_i) \in \so^i$. Finally, we require that for every objects $y' \in C_i, y' < y$, $(a_i(y'), a_i(y)) \in \so^i$.
    
    \item \sloppy \underline{$\wro^i$}: Let $y$ be an object in $C_i$. For every object $z$, if $z \in C_i$ and $z < y$, we require that $(\wro^i_z)^{-1}(a_i(y)) = \rspecContext{z}{c^i(z) \oplus a_i(z)}{a_i(y)}$; while otherwise, we require that $(\wro^i_z)^{-1}(a_i(y)) = \rspecContext{z}{c^i(z)}{a_i(y)}$. We also require that for every object $z$, if $z \in C_i$, then $(\wro^i_z)^{-1}(e_i) = \rspecContext{z}{c^i(z) \oplus a_i(z)}{e_i}$, while otherwise, $(\wro^i_z)^{-1}(e_i) = \rspecContext{z}{c^i(z)}{e_i}$.

    \item \underline{$\rbo^i$}: Let $y \in C_i$. We require that for every object $y \in  C_i$ and event $e$ s.t. $(e, a_i(y)) \in \so^i \cup \wro^i$, $(e, a_i(y)) \in \rbo^i$. Also, if $\mathsf{Rel}_i^\mathsf{v} = \rbo$, we impose that $(e_{i-1}, a_i(y)) \in \rbo^i$. Finally, we require that for every pair of events $e, e' \in E^{i-1}$ s.t. $(e, e') \in \rbo^{i-1}$ and $(e', a_i(y)) \in \so^i$, $(e, a_i(y)) \in \rbo^i$.

    \item \underline{$\ar^i$}: We impose that for every event $e \in E^{i-1}$, $(e, a_i(y)) \in \ar^i, y \in  C_i$. We also require that for every pair of objects $y_1, y_2 \in  C_i$ s.t. $y_1, y_2$, $(a_i(y_1), a_i(y_2)) \in \ar^i$.
\end{itemize}

We then define $h^i = \tup{E^i, \so^i, \wro^i}$ and $\exec^i = \tup{h^i, \rbo^i, \ar^i}$. Observe that by construction of $h^i$ and $\exec^i$, they satisfy \Cref{def:history,def:execution} respectively; so they are a history and an abstract execution respectively. In particular, observe that $\exec^i$ is a correction of the abstract execution $\exec^{i-1}_0$ with events $a_i$.

Finally, we define $h = \tup{E, \so, \wro}$ and $\exec = \tup{h, \rbo, \ar}$ as, respectively, the history $h^{\length{v}}$ and the abstract execution $\exec^{\length{v}}$. We prove that $\exec$ is the abstract execution we were looking for. 

First, we show that $\exec$ is valid w.r.t. $n_2$: as $\exec$ is valid w.r.t. $(\CC, \opspec)$ (\Cref{corollary:general-normal-form:arbitration-well-defined:validity}), so by \Cref{lemma:saturable:cc-strongest}, it is valid w.r.t. $(n_1, \opspec)$. As $n_1 \equiv_{\opspec} n_2$, $\exec$ is valid w.r.t. $(n_2, \opspec)$. Next, we deduce in \Cref{proposition:general-normal-form:stratified-ar} that $\opspec$ is maximally layered w.r.t. $\ar$. For proving such result, we rely on \Cref{proposition:general-normal-form:arbitration-well-defined:visibility,proposition:general-normal-form:arbitration-well-defined:ei-en-rbo}. Finally, we conclude in \Cref{proposition:general-normal-form:k-suffixes} that the layer bound of $\rspecName$ is bounded by the number of arbitration-free suffixes of $v$. However, this implies that $v$ is vacuous w.r.t. $n_2$ (\Cref{proposition:general-normal-form:layer-bound-implies-v-vacuous}); which is impossible by the choice of $v$. The contradiction arises from assuming that $n_1$ is arbitration-free but $n_2$ is not; so we conclude the result.
\end{proof}

\begin{proposition}
\label{proposition:general-normal-form:arbitration-well-defined:domains}
Let $\opspec$ be a storage that allows multi-object write (resp. read-write) events whose domain is not $\Vars$. Then, for every pair of finite disjoint sets $F_1, F_2$ there exists a domain  $D$ in $\opspec$ s.t. $F_1 \subseteq D$ but $F_2 \cap D = \emptyset$.
\end{proposition}

\begin{proof}
The result is immediate as $F_1$ is finite. Hence, by the assumptions on operation specifications (\Cref{ssec:assumptions-generalized-specs}), $F_1$ is a domain on $\opspec$.
\end{proof}

\begin{proposition}
\label{proposition:general-normal-form:arbitration-well-defined:events-well-defined}
Let $v$ be a visibility \replaced{formula}{relation} and  $i, 0 < i \leq \length{v}$. If $\conflictsOfV{v}{i} \neq \emptyset$ and $\mathsf{Rel}_i^\mathsf{v} = \wro$, $\opspec$ allows read-write events. If $\conflictsOfV{v}{i} = \emptyset$ allows events that do not unconditionally write.
\end{proposition}

\begin{proof}
Observe that as $v$ is non-vacuous w.r.t. $(\cmodel, \opspec)$, $\cmodel \setminus \{v\} \not\equiv_{\opspec} \cmodel$. By \Cref{proposition:monotonic:write-read-not-visible}, there exists an execution $\overline{\exec}$ valid w.r.t. $\cmodel \setminus \{v\}$, an object $z$ and events $f_0, \ldots f_{\length{v}}$ s.t. $v_z(f_0, \ldots f_{\length{v}})$ holds in $\overline{\exec}$. 

On one hand, if $\conflictsOfV{v}{i} \neq \emptyset$ and $\mathsf{Rel}_i^\mathsf{v} = \wro$, as $\overline{\exec}$ is valid w.r.t. $\cmodel \setminus \{v\}$, there exists $z$ s.t. $\rspec{\cmodel \setminus \{v\}}{z}{\overline{\exec}}{f_i} \neq \emptyset$. Also, $\conflictsOfV{v}{i} \neq \emptyset$ iff $f_i$ writes on some object $z'$. Hence, $f_i$ is a read-write event.

On the other hand, if $\conflictsOfV{v}{i} = \emptyset$, as $v$ is conflict-maximal w.r.t. $\opspec$, event $f_i$ does not necessarily write any object. Thus, $\opspec$ allows events that do not unconditionally write.
\end{proof}

\begin{proposition}
\label{proposition:general-normal-form:arbitration-well-defined:contexts}
The abstract execution $\exec$ described in \Cref{th:general-normal-from:arbitration-well-defined} satisfies that for every $i, 0 \leq i \leq \length{v}$:
\begin{enumerate}

    \item For every object $y \in C_i$, the following conditions hold:
     \begin{enumerate}
        \item For every object $z \in\Vars$, if $z \in C_i$ and $z < y$, $G(a_i(y), z) = F^i(z) \cup \{a_i(z)\}$, while otherwise, $G(a_i(y), z) = F^i(z)$.
        
        \item The execution $\exec^i \restriction y$ is valid w.r.t. $(\CC, \opspec)$.
    \end{enumerate}

    \item For the event $e_i$, the following conditions hold:
    \begin{enumerate}
        \item  For every object $z$, if $z \in C_i$, $G(e_i, z) = F^i(z) \cup \{a_i(z)\}$, while otherwise $G(e_i, z) = F^i(z)$.
        \item The execution $\exec^i$ is valid w.r.t. $(\CC, \opspec)$.
    \end{enumerate}

\end{enumerate}
where $\context{\CC}{z}{\exec}{e} = \tup{G(e, z), \rbo_{\restriction G(e, z) \times G(e, z)}, \ar_{\restriction G(e, z) \times G(e, z)}}$.
\end{proposition}

\begin{proof}
We prove the result by induction. In particular, we show that for every $i, -1 \leq i \leq \length{v}$ and object $y$, either (0) $i = -1$ or (1) and (2) hold. The base case, $i = -1$, is immediate as (0) holds; so let us suppose that the result holds for every $j, -1 \leq j < i$, and let us prove it for $i$.

For proving the inductive step, we first prove (1) and then (2). As both (1) and (2) have an identical proof (observe that the role of object $y$ in the former is just to declare that event $a_i(y)$ is well-defined), we present only the proof of (1).

We show (1) by transfinite induction. Let $\alpha$ be an ordinal of cardinality $|\Vars|$. For every $k, 0 \leq k \leq \alpha$, we denote by $V_k$ to the set containing the first $k$ elements in $\Vars$ according to $<$. We show that (1) holds for every $y \in V_k \cap C_i$.

The base, $V_0$ is immediate as $V_0 = \emptyset$. We thus focus on the successor case (i.e., showing that if (1) holds for every object $y \in V_k \cap C_i$ it also holds for $V_{k+1}$), as the limit case is immediate: if $k$ is a limit ordinal, $V_k = \bigcup_{i, i < k} V_i$; so (1) immediately holds. For showing that (1) holds for every object $y \in V_{k+1} \cap C_i$, as by induction hypothesis it holds for every object $y \in V_k \cap C_i$, it suffices to show it for the only object $y \in V_{k+1} \setminus V_i$. W.l.o.g., we can assume that $y \in C_i$; as otherwise the result is immediate. 

We first prove (1a) and then we show (1b). Let $z \in \Vars$ be an object. Two cases arise depending on $\mathsf{Rel}_i^\mathsf{v}$.

On one hand, if $i = 0$ or $i > 0 \land \mathsf{Rel}_i^\mathsf{v} = \ar$, $F^i(z) = \{\init\}$. As $\init \in G(a_i(y), z)$, it suffices to show that the only non-initial event in $E$ in $G(a_i(y), z)$ is $a_i(z)$ (whenever $z \in C_i$ and $z < y$). Observe that an event $e$ belongs to $G(a_i(y), z)$ if $\wspec{\CC}{z}{\exec}{e} \downarrow$ and $(e, a_i(y)) \in \rbo^+$. As $a_i(y) \in E^i$, by construction of $\exec$, $e$ must belong to $E^i$, $\wspec{\CC}{z}{\exec^i}{e} \downarrow$ and $(e, a_i(y)) \in (\rbo^i)^+$. 

Observe that as either $i = 0$ or $0 < i \leq \length{v} \land \mathsf{Rel}_i^\mathsf{v} = \ar$, by definition of $\rbo^i$, $e \not\in E^{i-1}$. Thus, $e$ must be an event in $E^i \setminus E^{i-1}$. Observe that by construction of $\exec$, as $(e, a_i(y)) \in (\rbo^i)^+$, such event must be an event $a_i(w), w \in C_i, w < y$. As $\exec^i = \correction{\exec^i_0}{a_i}{e_i}$, by induction hypothesis (1b), we deduce that $\exec^i \restriction w$ is valid w.r.t. $(\CC, \opspec)$. Hence, as $\wspec{\CC}{z}{\exec^i}{a_i(w)} \downarrow$, we deduce thanks to Property~\ref{def:execution-corrector:1-context-corrector-writes} of \Cref{def:execution-corrector} that $z = w$ -- so $z \in C_i$ and $z < y$.

On the other hand, if $0 < i \leq \length{v} \land \mathsf{Rel}_i^\mathsf{v} \neq \ar$, two sub-cases arise: $z \in C_i, z < y$ or not. Both cases are identical, so we present the former, i.e., if $z \in C_i, z < y$, then $F^i(z) \cup \{a_i(z)\} = G(a_i(y), z)$.

For proving that $F^i(z) \cup \{a_i(z)\} \subseteq G(a_i(y), z)$, we split the proof in two blocks: showing that $F^i(z) \subseteq G(a_i(y), z)$ and showing that $a_i(z) \in G(a_i(y), z)$.

For showing that $F^i(z) \subseteq G(a_i(y), z)$, let $e$ be an event in $F^i(z)$. In such case, to $e \in E^{i-1}$, $\wspec{\CC}{z}{\exec^i}{e} \downarrow$ and $(e, e_{i-1}) \in (\rbo^i)^*$. By the construction of $\exec$, it is easy to see that any such event belongs to $E^i$, $\wspec{\CC}{z}{\exec}{e} \downarrow$ and $(e, e_{i-1}) \in \rbo^*$. As $\mathsf{Rel}_i^\mathsf{v} \neq \ar$, we deduce that $(e_{i-1}, a_i(y)) \in \rbo^i \subseteq \rbo$. Hence, $(e, a_i(y)) \in \rbo^+$; so $e \in G(a_i(y), z)$. This show that $F^i(z) \subseteq G(a_i(y), z)$.

For showing that $a_i(z) \in G(a_i(y), z)$, we observe that $\exec^i = \correction{\exec^i_0}{a_i}{e_i}$. As $z < y$, by induction hypothesis (1b), $\exec^i \restriction z$ is valid w.r.t. $(\CC, \opspec)$. Thus, by Property~\ref{def:execution-corrector:1-context-corrector-writes} of \Cref{def:execution-corrector}, $\wspec{\CC}{z}{\exec^i}{a_i(z)} \downarrow$. Hence, $\wspec{\CC}{z}{\exec}{a_i(z)} \downarrow$. As $z < y$, $(a_i(z), a_i(y)) \in \so^i \subseteq \so$; so we conclude that $a_i(z) \in G(a_i(y), z)$.

We conclude the proof of the inductive step of (1a) by showing the converse i.e. $F^i(z) \cup \{a_i(z)\} \supseteq G(a_i(y), z)$. Let $e \in G(a_i(y), z)$. First of all, by the definition of $\axcc$ visibility \replaced{formula}{relation} (see \Cref{fig:cc}), $e \in G(a_i(y), z)$ iff $\wspec{\CC}{z}{\exec}{e} \downarrow$ and $(e, a_i(y)) \in \rbo^+$. Observe that if $(e, a_i(y)) \in \rbo^+$, by construction of $\exec$, such event must belong to $E^i$, $\wspec{\CC}{z}{\exec^i}{e} \downarrow$ and $(e, a_i(y)) \in (\rbo^i)^+$. We prove that if $e \in E^{i-1}$ then $e \in F^i(z)$, while otherwise, if $e \in E^i \setminus E^{i-1}$, then $e = a_i(z)$.

If $e \in E^{i-1}$, as $\wspec{\CC}{z}{\exec^i}{e} \downarrow$, $\wspec{\CC}{z}{\exec^{i-1}}{e} \downarrow$. Also, as $\mathsf{Rel}_i^\mathsf{v} \neq \ar$ and $(e, a_i(y)) \in (\rbo^{i})^+$, we deduce that $(e, e_{i-1}) \in (\rbo^{i-1})^*$. In other words, $e \in F^i(z)$. 

Otherwise, if $e \in E^i \setminus E^{i-1}$, we note that by construction of $\exec$, the only events in $E^i \setminus E^{i-1}$ s.t. $(e, a_i(y)) \in (\rbo^i)^+$ are events $a_i(w), w \in C_i, w < y$. As $\exec^i = \correction{\exec^i_0}{\sequence{a_i}}{e_i}$ and $z < y$, $\exec^i \restriction z$ is valid w.r.t. $(\CC, \opspec)$. Thus, as $\wspec{\CC}{z}{\exec^i}{e} \downarrow$, by Property~\ref{def:execution-corrector:1-context-corrector-writes} of \Cref{def:execution-corrector} we conclude that $e = a_i(z)$.

For concluding the inductive step, we show that (1b) holds. This is immediate by the definition of $\wro^i$: for every event $e \in \exec^{i} \restriction y$, by induction hypothesis (1a) or (2a) -- depending on whether $e = e_j$ or $a_j(w)$, where $0 \leq j \leq i, w \in C_i$ -- $(\wro^i)_z^{-1}(e) = \rspec{z}{\CC}{\exec^i \restriction y}{e}$. Thus, $\exec^i \restriction y$ is valid w.r.t. $(\CC, \opspec)$.

\end{proof}

A consequence of \Cref{proposition:general-normal-form:arbitration-well-defined:contexts} is the following result.

\begin{corollary}
\label{corollary:general-normal-form:arbitration-well-defined:validity}
The abstract execution $\exec$ described in \Cref{th:general-normal-from:arbitration-well-defined} is valid w.r.t. $(\CC, \opspec)$. 
\end{corollary}

\begin{proposition}
\label{proposition:general-normal-form:arbitration-well-defined:visibility}
The predicate $v_{x_0}(e_0, \ldots e_{\length{v}})$ holds in the abstract execution $\exec$ described in \Cref{th:general-normal-from:arbitration-well-defined}.
\end{proposition}

\begin{proof}
The proof is a simple consequence of $\exec$'s construction. To show that $v_{x_0}(e_0, \ldots e_{\length{v}})$ holds in $\exec$, we first show that for every $i, 1 \leq i \leq \length{v}$, $(e_{i-1}, e_i) \in \mathsf{Rel}_i^\mathsf{v}$ and to then prove that $\writeConstraints{x}{e_0, \ldots e_{\length{v}}}$ holds in $\exec$.

We prove that for every $i, 1 \leq i \leq \length{v}$, $(e_{i-1}, e_i) \in \mathsf{Rel}_i^\mathsf{v}$. Four cases arise depending on $\mathsf{Rel}_i^\mathsf{v}$.

\begin{itemize}
    \item \underline{$\mathsf{Rel}_i^\mathsf{v} = \so$}: In this case, by construction of events $e_{i-1}, e_i$, we know that $r_i = r_{i-1}$. Hence, $(e_{i-1}, e_i) \in \so^i \subseteq \so$.
    
    \item \underline{$\mathsf{Rel}_i^\mathsf{v} = \wro$}: In this case, we first show that there is an object $y \in D_i \cap W_{i-1}   \setminus C_i$, and then show that $(e_{i-1}, e_i) \in \wro_y$. For showing the first part, we distinguish between cases depending on whether $o_{i-1} \in D_i$ or not. 
    \begin{itemize}
        \item \underline{$o_{i-1} \in D_i$}: In this sub-case, we show that $y = o_{i-1}$. On one hand, if $\conflictsOfV{v}{i} = \emptyset$, by the choice of event $e_i$, $o_{i-1} \in D_{i-1} \setminus C_i$. On the other hand, if $\conflictsOfV{v}{i} \neq \emptyset$, as $o_{i-1} \in D_{i}$, we deduce that $\opspec$ allows multi-object read-write events. Observe that as $v$ is conflict-maximal w.r.t. $\opspec$, $\conflictsOfV{v}{i-1} \neq \emptyset$. Hence, as $\opspec$ allows multi-object read-write events, we deduce that $o_{i-1} \in D_{i-1} \setminus C_i$. In both cases, as $\conflictsOfV{v}{i-1} \neq \emptyset$ and $o_{i-1} \in D_{i-1}$, by the choice of $W_{i-1}$, we conclude that $o_{i-1} \in W_{i-1}$.

        \item \underline{$o_{i-1} \not\in D_i$}: In this case, we show that $y = \tilde{x}_i$. On one hand, if $\conflictsOfV{v}{i} = \emptyset$, $X_i = \emptyset$; so by the choice of $\tilde{x}_i$ (see \Cref{eq:x-i}), $\tilde{x}_i = \tilde{x}_{i-1}$. By the choice of $D_i$, $\tilde{x}_{i-1} \in D_i \setminus C_i$. Moreover, as $v$ is conflict-maximal w.r.t. $\opspec$, $\conflictsOfV{v}{i-1} \neq \emptyset$; so $\tilde{x}_{i-1} \in X_{i-1}$. By the choice of event $e_{i-1}$, $X_{i-1} \subseteq W_{i-1}$. Altogether, we conclude that $\tilde{x}_i \in W_{i-1}$. 
        
        On the other hand, if $\conflictsOfV{v}{i} \neq \emptyset$, we note that $\tilde{x}_i \in D_i \setminus C_i$. As $o_{i-1} \not\in D_{i}$, we deduce that $\opspec$ only allows single-object read-write events. Thus, $D_i = \{\tilde{x}_i\}$. As $v$ is conflict-maximal w.r.t. $\opspec$, we deduce that $X_i \subseteq X_{i-1}$. As by the choice of $e_{i-1}$, $X_{i-1} \subseteq W_{i-1}$, we conclude that $\tilde{x}_i \in W_{i-1}$.

    \end{itemize}

    We prove now that $(e_{i-1}, e_i) \in \wro_{y}$. First, we show that $\writeVarExec{e_{i-1}}{y}{\exec}$. On one hand, if $e_{i-1}$ is an unconditional write event, $\wspecContext{y}{c^i(y)}{e_{i-1}} \downarrow$. On the other hand, if $e_{i-1}$ is a conditional write event, as $\exec$ is valid w.r.t. $(\CC, \opspec)$ (\Cref{corollary:general-normal-form:arbitration-well-defined:validity}) and $y \in W_i$, by Property~\ref{def:execution-corrector:4-e-writes-w} of \Cref{def:execution-corrector}, we deduce that $\wspecContext{y}{c^i(y)}{e_{i-1}} \downarrow$. Then, as $\mathsf{Rel}_i^\mathsf{v} = \wro$, $e_{i-1} \in F^i(y)$. Observe that by construction of $\exec$, $e_{i-1}$ is the $\ar$-maximum event in $c^i(y)$. We note that as $y \not\in C_i$, by \Cref{proposition:general-normal-form:arbitration-well-defined:contexts}, $F^i(y) = G(e_i, y)$. To sum up, $e_{i-1}$ is the $\ar$-maximum event in $\context{\CC}{y}{\exec}{e_i}$. As $\rspecName$ is maximally layered, we deduce that $e_{i-1} \in \rspec{\CC}{y}{\exec}{e_i}$. Finally, as $\exec$ is valid w.r.t. $\CC$ (\Cref{corollary:general-normal-form:arbitration-well-defined:validity}), we conclude that $(e_{i-1}, e_i) \in \wro_y$.

    \item \underline{$\mathsf{Rel}_i^\mathsf{v} = \rbo$}: In this case, we explicitly stated that $(e_{i-1}, e_i) \in \rbo^i \subseteq \rbo$ during the construction of $\exec$.

    \item \underline{$\mathsf{Rel}_i^\mathsf{v} = \ar$}: Similarly, by definition of $\ar^i$, we know that $(e_{i-1}, e_i) \in \ar^i \subseteq \ar$.

\end{itemize}

For showing that show that $\writeConstraints{x}{e_0, \ldots e_{\length{v}}}$, we show that for every $i, 0 \leq i \leq \length{v}$ and every set $E \in \conflictsOfV{v}{i}$, the event $e_i$ writes on object $y_E$\footnote{For simplifying the proof, we abuse of notation and say that $y_E = x$ if $E = E_x$. Observe that $v$ is conflict-maximal w.r.t. $\opspec$, either $\writeConflicts[x]{E_x}$ or $\writeConflicts{E_x}$ do not belong to $v$.}. If $e_i$ is an unconditional write, by the choice of $e_i$, it writes on every object in $D_i$. As $y_E \in D_i$, we conclude that $e_i$ writes on $y_E$. Otherwise, if $e_i$ is a conditional write, we observe that $y_E \in W_i$. Hence, as $\exec^i = \correction{\exec^i_0}{\sequence{a_i}}{e_i}$ and $\exec^i$ is valid w.r.t. $(\CC, \opspec)$ (\Cref{proposition:general-normal-form:arbitration-well-defined:contexts}), we deduce using Property~\ref{def:execution-corrector:4-e-writes-w} of \Cref{def:execution-corrector} that $\wspec{\CC}{y_E}{\exec^i}{e_i} \downarrow$. By construction of $\exec$, we conclude that $\wspec{\CC}{y_E}{\exec}{e_i} \downarrow$.
\end{proof}

\begin{proposition}
\label{proposition:general-normal-form:arbitration-well-defined:ei-en-rbo}
Let $\exec$ be the abstract execution described in \Cref{th:general-normal-from:arbitration-well-defined}. For every $i, 0 \leq i < \length{v}$, if the $\event_i$ suffix of $v$ is non-arbitration-free, then $(e_i, e_{\length{v}}) \not\in \rbo^+$.
\end{proposition}

\begin{proof}
The proof is just an observation about the construction of $\exec$: for every $j, 0 < j \leq \length{v}$, $(e_{j-1}, e_j) \in \rbo$ iff $\mathsf{Rel}_i^\mathsf{v} \neq \ar$. Hence, $(e_i, e_{\length{v}} \in \rbo^+)$ iff for every $j, i < j \leq \length{v}$, $\mathsf{Rel}_j \neq \ar$. In particular, if the $\event_i$ suffix of $v$ is non-arbitration-free, then $(e_i, e_{\length{v}}) \not\in \rbo^+$.
\end{proof}

\begin{proposition}
\label{proposition:general-normal-form:stratified-ar}
Let $\opspec$ be a storage, $\cmodel$ be a consistency model in normal form w.r.t. $\opspec$ and $v$ be a visibility \replaced{formula}{relation} in $\cmodel$. If there exists an abstract execution $\exec = \tup{h, \rbo, \ar}$ valid w.r.t. $\cmodel$, an object $x$ and events $w, r$ s.t. $v_x(w, r)$ holds in $\exec$ but $(w, r) \not\in (\rbo)^+$, then $\opspec$ is maximally layered w.r.t. $\ar$.
\end{proposition}

\begin{proof}
First of all, as $v_x(w, r)$ holds in $\exec$, $w \in \context{\cmodel}{x}{\exec}{r}$. If $\opspec$ would be maximally layered w.r.t. $(\rbo)^+$, $\rspec{\cmodel}{x}{\exec}{r}$ contains at least the first layer of $\context{\cmodel}{x}{\exec}{r}$ w.r.t. $\rbo$. Hence, there would exist an event $w'$ s.t. $w' \in \rspec{\cmodel}{x}{\exec}{r}$ and $(w, w') \in (\rbo)^+$. As $\exec$ is valid w.r.t. $\cmodel$, we deduce that $(w', r) \in \wro$. By \Cref{def:execution}, we deduce that $(w', r) \in \rbo$. However, this implies that $(w, w') \in \rbo^+$; which contradicts the assumptions. Hence, $\opspec$ must be maximally layered w.r.t. $\ar$.
\end{proof}

\begin{proposition}
\label{proposition:general-normal-form:k-suffixes}
Let $\opspec$ be a storage maximally layered w.r.t. $\ar$, $\cmodel$ be a consistency model in normal form w.r.t. $\opspec$ and $v$ be a non-arbitration free visibility \replaced{formula}{relation} in $\cmodel$. Let us suppose that there exists an abstract execution $\exec = \tup{h, \rbo, \ar}$ valid w.r.t. $\cmodel$, an object $x$ and events $e_0, \ldots e_{\length{v}}$ satisfying the following:
\begin{enumerate}
    \item for every non-initial event $e$ in $\exec$, if $e \not\in \{e_i \ | \ 0 \leq i \leq \length{v}\}$, then $e$ does not write on $x$ in $\exec$, \label{proposition:general-normal-form:k-suffixes:1-no-write}
    
    \item $v_x(e_0, \ldots e_{\length{v}})$ holds in $\exec$, and \label{proposition:general-normal-form:k-suffixes:2-vis}
    
    \item for every non-arbitration-free $\event_k$-suffix of $v$, $(e_k, e_{\length{v}}) \not\in \rbo^+$. \label{proposition:general-normal-form:k-suffixes:3-rbo}
\end{enumerate}

\noindent
In such case, the layer bound of $\opspec$ is bounded by the number of arbitration-free suffixes of $v$.
\end{proposition}

\begin{proof}
We reason by contradiction, assuming that $k$ is bigger than the number of saturable suffixes of $v$. We first show that $\rspec{\cmodel}{x}{\exec}{e_{\length{v}}}$ contains less than $k$ events in $\{e_i \ | \ 0 \leq i < \length{v}\}$, for then deduce that $\init \in \rspec{\cmodel}{x}{\exec}{e_{\length{v}}}$. After that, we reach a contradiction by showing that $e_0 \in \rspec{\cmodel}{x}{\exec}{e_{\length{v}}}$ but $(e_0, e_{\length{v}}) \not\in \wro$; which contradicts that $\exec$ is valid w.r.t. $\cmodel$.

We first show that $\rspec{\cmodel}{x}{\exec}{e_{\length{v}}}$ contains less than $k$ events in $\{e_i \ | \ 0 \leq i < \length{v}\}$. As $v$ contains less than $k$ saturable suffixes, by the Assumption~\ref{proposition:general-normal-form:k-suffixes:3-rbo}, there is less than $k$ events in $\{e_i \ | \ 0 \leq i < \length{v}\}$ that write on $x$ in $\exec$ and that succeed $e_{\length{v}}$ w.r.t. $\rbo^+$. As $\wro \subseteq \rbo$ (see \Cref{def:execution}), we deduce that $\wro_{x}^{-1}(e_{\length{v}})$ contains less than $k$ events in $\{e_i \ | \ 0 \leq i < \length{v}\}$. As $\exec$ is valid w.r.t. $\cmodel$, $\wro_x^{-1}(e_{\length{v}}) = \rspec{\cmodel}{x}{\exec}{e_{\length{v}}}$; so we prove the first part.

For showing that $\init \in \rspec{\cmodel}{x}{\exec}{e_{\length{v}}}$, we observe that by the Assumption~\ref{proposition:general-normal-form:k-suffixes:1-no-write}, no other non-initial event in $\exec$ writes on $x$ in $\exec$. Hence, $\rspec{\cmodel}{x}{\exec}{e_{\length{v}}}$ contain less than $k$ non-initial events. As $\init \in \context{\cmodel}{x}{\exec}{e_{\length{v}}}$, and $\opspec$ is maximally layered with layer bound $k$, we conclude that $\init \in \rspec{\cmodel}{x}{\exec}{e_{\length{v}}}$.

For proving that $e_0 \in \rspec{\cmodel}{x}{\exec}{e_{\length{v}}}$ but $(e_0, e_{\length{v}}) \not\in \wro$, we observe that by the Assumption~\ref{proposition:general-normal-form:k-suffixes:2-vis}, $e_0 \in \context{\cmodel}{x}{\exec}{e_{\length{v}}}$. As $\opspec$ is maximally layered w.r.t. $\ar$, $\init \in \rspec{\cmodel}{x}{\exec}{e_{\length{v}}}$, $(\init, e_0) \in \ar$ and $e_0 \in \context{\cmodel}{x}{\exec}{e_{\length{v}}}$; we conclude that $e_0 \in \rspec{\cmodel}{x}{\exec}{e_{\length{v}}}$. 

For reaching a contradiction, we observe that $v$ is non-arbitration-free. Hence, by the Assumption~\ref{proposition:general-normal-form:k-suffixes:3-rbo}, $(e_0, e_{\length{v}}) \not\in \rbo$. Once again, as $\wro \subseteq \rbo$ (see \Cref{def:execution}), we deduce that $e_0 \not\in \wro_x^{-1}(e_{\length{v}})$. However, as $e_0 \in \rspec{\cmodel}{x}{\exec}{e_{\length{v}}}$, we conclude that $\exec$ is not valid w.r.t. $\cmodel$; which is contradicts the hypothesis. Thus, the layer bound of $\opspec$ is bounded by the number of arbitration-free suffixes of $v$.
\end{proof}

\begin{proposition}
\label{proposition:general-normal-form:layer-bound-implies-v-vacuous}
Let $\opspec = (E, \rspecName, \extractName, \wspecName)$ be an operation specification maximally layered w.r.t. $\ar$, $\cmodel$ be a consistency model in normal form w.r.t. $\opspec$ and $v$ be a simple, conflict-maximal w.r.t. $\opspec$, non-arbitration-free visibility \replaced{formula}{relation}. If the layer bound of $\rspecName$ is smaller or equal by the number of arbitration-free suffixes of $v$, then $v \not\in\cmodel$.
\end{proposition}

\begin{proof}
Let $v$ be a simple, conflict-maximal w.r.t. $\opspec$, non-arbitration-free visibility \replaced{formula}{relation}. We show that $v$ is vacuous w.r.t. $\cmodel$; so $v \not\in \cmodel$.

We reason by contradiction, assuming that $v$ is non-vacuous w.r.t. $\cmodel$. In such case, $\cmodel \setminus \{v\} \not\equiv_{\opspec} \cmodel$ but $\cmodel \setminus \{v\} \preccurlyeq \cmodel$. By \Cref{proposition:monotonic:write-read-not-visible}, there exists an abstract execution on $\opspec$, and object $x$, and events $w, r$ s.t. $\rspec{\cmodel}{x}{\exec}{r} \setminus \context{\cmodel \setminus \{v\}}{x}{\exec}{r}$. 

We observe that by Property~\ref{def:general-read-spec:unconditional-read} of \Cref{def:general-read-spec}, $w \in \context{\cmodel}{x}{\exec}{r}$. Hence, as $w \in \context{\cmodel}{x}{\exec}{r} \setminus \context{\cmodel \setminus \{v\}}{x}{\exec}{r}$, we deduce that $v_x(w, r)$ holds in $\exec$. As $v$ is simple, there exist events $e_0, \ldots e_{\length{v}}$ s.t. $e_0 = w$, $e_{\length{v}} = r$ and $v_x(e_0, \ldots e_{\length{v}})$ holds in $\exec$. 

First of all, as $\rspecName$ is maximally layered w.r.t. $\ar$ and $e_0 \in \rspec{\cmodel}{e_0}{\exec}{e_{\length{v}}}$, every event in $\{e_0, \ldots e_{\length{v}}\}$ that writes $x$ is also in $\rspec{\cmodel}{e_0}{\exec}{e_{\length{v}}}$. As $v$ is conflict-maximal w.r.t. $\opspec$, at least $|E_x|$ events write on $x$; where $E_x \in \mathcal{P}(\event_0, \ldots e_{\length{v}})$ s.t. $\writeConflicts[x]{E_x} \in v$. Observe that for every event $e_i$ s.t. $\event_i \in E_x$ and $\suffixOf{x}{\mathsf{v}_x,i}$ is arbitration-free, as $\cmodel$ is closed under causal suffixes, there exists a visibility \replaced{formula}{relation} $v' \in \cmodel$ s.t. $v'_x(e_i, e_{\length{v}})$. Thus, $|E_x| \geq \mathsf{af}(v)$, where $\mathsf{af}(v)$ is the number of arbitration-free suffixes of $v$. Moreover, as $e_0 \in \rspec{\cmodel}{e_0}{\exec}{e_{\length{v}}}$, and $v$ is not arbitration-free, $|E_x| > \mathsf{af}(v)$. However, as the layer bound of $\rspecName$, $k$, is smaller or equal than the number of arbitration-free suffixes of $v$, the number of events read by $f_{\length{v}}$ is at most $\mathsf{af}(v)$. Hence, $|E_x| \leq \mathsf{af}(v)$, which contradicts that $|E_x| > \mathsf{af}(v)$. We reach a contradiction; so the initial hypothesis, that $v$ is non-vacuous w.r.t. $\cmodel$, is false. Thus, $v \not\in \cmodel$.
\end{proof}

\newpage

\section{Proof of the Basic Arbitration-Free Consistency Theorem \deleted{(\Cref{th:characterization-cons-available-lww})}}
\label{app:main-theorem-extra-proofs}

\added{Let in the folloeing $\sspec = (\cmodel, \opspec)$ be a \simpleStorageFullName{} storage specification. We show that there exists an available $\sspec$-implementation iff $\cmodel$ is arbitration-free w.r.t. $\opspec$.}

\subsection{\replaced{Arbitration-Freeness Implies Availability}{Proof of Lemma~\ref{lemma:cons-available-lww:1-2}}}
\label{ssec:cons-available-lww:1-2}

\added{As discussed in \Cref{sec:afc-theorem}, the proof of such result is decomposed in three steps:} 
\begin{enumerate}
    \item \added{We show that arbitration-free consistency models w.r.t. $\opspec$ are weaker than $\CC$ (\Cref{lemma:saturable:cc-strongest}).}
    \item \added{We deduce that available $(\CC, \opspec)$-implementations are also available $(\cmodel, \opspec)$-implementations as an immediate consequence of \Cref{lemma:normal-form:consistency-stronger-than-criterion}.}
    \item \added{We prove that there exists available $(\CC, \opspec)$-implementations (\Cref{lemma:cons-available-lww:1-2}).}
\end{enumerate}

\weakerCC*

\begin{proof}
For showing that $\cmodel$ is weaker than $\CC$, let $h = \tup{E, \so, \wro}$ be a history and $\exec = \tup{h, \rbo, \ar}$ be an abstract execution of $h$ valid w.r.t. $\sspec$. Let $n$ be a consistency model in normal form that is $\opspec$-equivalent to $\cmodel$. By \Cref{th:general-normal-form:existence}, such model always exists. As $\cmodel$ is arbitration-free, every visibility \replaced{formula}{relation} $v \in n$ is arbitration-free. We conclude the result by showing that $n \preccurlyeq \CC$, i.e. showing that for every object $x$ and every pair of distinct events $e, e' \in E$, if $v_x(e, e')$ holds in $\exec$ then $v^{\CC}_x(e, e')$ holds in $\exec$ as well; where $v^{\CC}$ is $\axcc$, the visibility \replaced{formula}{relation} of $\CC$ (\Cref{fig:cc}). 

First, as $v_x(e, e')$ holds in $\exec$, $\writeVarExec{e}{x}{\exec}$ and $\wro_x^{-1}(e) \neq \emptyset$. Moreover, as $v$ is simple, for every $i, 1 \leq i \leq \length{v}$, $\mathsf{Rel}_i^\mathsf{v} \in \{\so, \wro, \rbo\}$. By Property~\ref{def:execution:inclusions} of \Cref{def:execution}, we deduce that $(e, e') \in \rbo^+$. Altogether, we conclude that $v^{\CC}_x(e, e')$ holds in $\exec$.
\end{proof}

\satConsAvailableLWW*

\begin{proof}
\replaced{We define an available implementation of $\sspec^{\CC} = (\CC, \opspec)$.}{For proving (1) $\implies$ (2), we provide an available $\sspec^{\CC}$-implementation; where $\sspec^{\CC} = (\CC, \opspec)$. As $\cmodel$ is arbitration-free w.r.t. $\opspec$, by \Cref{lemma:normal-form:consistency-stronger-than-criterion}, this implies that $\cmodel$ is weaker than $\CC$; so by \Cref{lemma:general-normal-form:consistency-stronger-than-criterion}, any $\sspec^{\CC}$-implementation is also a $\sspec$-implementation. In the following lines, we describe an available implementation $\implementationInstanceCC$ of $\sspec^{\CC}$ and justify its correction.}

\sloppy As discussed in \Cref{sec:lts}, any implementation $\implementationInstanceCC = \implementationCC$ can be characterized by describing its set of states $S_{\mathsf{i}}$, its actions $A_{\mathsf{i}}$, its initial state $ \sigma_0^{\mathsf{i}}$ and its transition function $\Delta_{\mathsf{i}}$.

First, we define $S_{\mathsf{i}}$ as the set of possible values that each object may have; and the declare the initial state any possible state in $S_{\mathsf{i}}$. Next, we define $A_{\mathsf{i}}$ via the synchronized actions $\Events \times (\Vars \times \Events \cup \{\emptyset\})$, as well as the local actions $\esend$ and $\ereceive$. We assume local actions are defined in a similar way to $\Events$, as tuples $a = (\identifier, \replicaID, \operation, \metadataInstance)$, where $\identifier$ is an action identifier, $\replicaID$ is a replica identifier, $\operation$ an operation identifier and $\metadataInstance$ is additional metadata of the action. As for events, we use $\identifierEvent{a}$, $\replicaEvent{a}$, $\operationEvent{a}$ and $\metadataEvent{a}$ for indicating the identifier, replica, operation and metadata of an action $a$.

For describing its transition function, we rely on the definition of $\CC$. As we design $\implementationCC$ to be an available $\sspec^{\CC}$-implementation, we require that any induced abstract execution must be valid w.r.t. $\sspec^{\CC}$. However, \Cref{def:valid} describes validity ``a posteriori'', i.e. validity can only be checked once the event is executed; while transition functions describe validity ``a priori'', i.e. describe a procedure to compute a write-read of a given, not yet added event. For solving this issue, we observe that under $\CC$, that the context of an event $e$ belonging to a synchronized action $a = (e, m)$ only depends on (a) the transitive set of received actions before the last action in its replica and (b) the synchronized actions executed in its own replica. Ensuring transitive communication, i.e. ensuring that every send action on replica $r$ transmits information about all synchronized actions executed or received on replica $r$ before such $\esend$ action suffices to provide $\CC$.

More in detail, for describing the transition function $\Delta_{\mathsf{i}}(t, a)$, we require that (1) $a$ is not present in $t$ and (2) transitive communication is ensured. Also, we require a third condition depending on the type of $a$: 
\begin{itemize}
    \item if $a$ is a synchronized action, we require that (3a) if $a$ represents a read operation, $a = (e, m)$, then $e$ must read from the latest writing event w.r.t. $\ar$ \added{(which coincides with the trace order)} received before $l_r^t$,
    
    \item if $a$ is a $\esend$ action, then (3b) it precedes a synchronized action, and
    \item if $a$ is a $\ereceive$ action, then (3c) there exists a unique preceding $\esend$ action that matches it.
\end{itemize}
where $r = \replicaEvent{a}$ and $l_r^t$ to the last action in trace $t$ whose replica is $r$.

\sloppy On one hand, (1) ensures that $\Delta_{\mathsf{i}}(t, e)$ is well-defined, i.e. in every trace of $\Delta_{\mathsf{i}}$, each action contains each action exactly once. %
On the other hand, (2) and (3a) ensure that $\implementationInstanceCC$ is a $\sspec^{\CC}$-storage implementation while (3b) and (3c) ensure that $\implementationInstanceCC$ is an available storage implementation.

Formally, $\Delta_{\mathsf{i}}(t, a) \downarrow$ if and only if $a \not\in t$ and $\satRspec{t}{a}$ holds; and in such case $\Delta_{\mathsf{i}}(t, a) = t \oplus a$. The predicate $\satRspec{t}{a}$ is described in \Cref{eq:auxiliary-transition-function-cc}.

\begin{equation}
\label{eq:auxiliary-transition-function-cc}
{\arraycolsep=0.1pt
\begin{array}{lll}
\satRspec{t}{a} & = &
\left\{
{\arraycolsep=2pt
\begin{array}{llll}
a = (e, M_t(e)) &  \text{if } \operationEvent{a} & \neq & \esend, \ereceive \\
\sendIfDataSend{t}{a} & \text{if } \operationEvent{a} & = & \esend\\
\; \sendAllData{t}{a} \\
\; \text{and } \maxSend{t}{a} \\
\minRcv{t}{a} &  \text{if } \operationEvent{a} & = & \ereceive\\
\; \text{and } \maxRcv{t}{a} \\
\end{array}}
\right.
\end{array}}
\end{equation}

where $M_t(e)$ is the mapping assigning to the objext $x = \varOf{e}$ the last event that writes on $x$ received by $e$, formally defined using \Cref{eq:context-cc,eq:receive-trace}; and the predicates $\mathsf{sendIfData}$, $\mathsf{sendAllData}$, $\mathsf{maxSend}$, $\mathsf{minRcv}$ and $\mathsf{maxRcv}$ are defined in \Cref{eq:send-if-data,eq:send-all-data,eq:min-send,eq:max-send}.
\begin{equation}
\label{eq:context-cc}
{\arraycolsep=3pt
\begin{array}{lll}
M_t(e) & =&  \left[x \mapsto  \left\{\begin{array}{ll}
   \{\max_{\ar_e^t} E^x_t(e)\} & \text{if } x = \varOf{e} \\
   \emptyset & \text{otherwise}
\end{array}\right.\right]_{x \in \Vars}\\
E^x_t(e)& = & \left\{e' \ \left| \ 
\begin{array}{l}
    e' \in \Events \cap t \ \land \ \writeVarExec{e'}{x}{\execTrace{t}} \ \land\\ 
    (\replicaEvent{e'} = \replicaEvent{e} \lor \mathsf{rec}_t(e', e))
\end{array} \right.\right\} \\
\ar_e^t &= & \ar_{\restriction E^x_t(e) \times E^x_t(e)}
\end{array}
}
\end{equation}

\begin{equation}
\label{eq:receive-trace}
\mathsf{rec}_t(e', e) = \exists r,s \in t \text{ s.t. } \bigwedge
\begin{array}{l}
\operationEvent{r} = \ereceive, 
\replicaEvent{r} = \replicaEvent{e}, \\
\operationEvent{s} = \esend, 
\replicaEvent{s} = \replicaEvent{e'},  \\
\receiveSet{s} = \receiveSet{r}, 
e' <_t s <_t r < e'
\end{array} 
\end{equation}

\begin{flalign}
& {\arraycolsep=1pt
\begin{array}{lll}
\sendIfDataSend{t}{a} & \Coloneqq & \operationEvent{a''} \neq \esend \label{eq:send-if-data}
\end{array}} 
\end{flalign}
\begin{flalign}
& {\arraycolsep=1pt
\begin{array}{lll}
\; \text{where }  a'' & =  & \max_{<_t} \left\{a' \in t \ \left| \ \begin{array}{l}
    \replicaEvent{a'} = \replicaEvent{a} \ \land \ \operationEvent{a'} \neq \ereceive
\end{array}  \right.\right\}
\end{array}} \nonumber
\end{flalign}
\begin{flalign}
& {\arraycolsep=1pt
\begin{array}{lll}
\sendAllData{t}{a} & \Coloneqq & \forall a' \in t. \replicaEvent{a'} = \replicaEvent{a} \land \operationEvent{a'} \neq \esend \\
& & \quad \qquad \implies \mathsf{RV}^x_{a'} \subseteq \receiveSet{a} \\
\end{array}}  \label{eq:send-all-data}
\end{flalign}
\begin{flalign}
\begin{array}{lll}
\; \text{where } \mathsf{RV}^x_{a'} & =  & \left\{ \begin{array}{ll}
    \{e\} & \text{if } \operationEvent{a'} \neq \esend, \ereceive \ \land \\
    &  a' = (e, \_)  \\
    \receiveSet{a'}  & \text{if } \operationEvent{a'} = \ereceive \\
    \emptyset & \text{otherwise} 
\end{array}  \right.
\end{array} \nonumber
\end{flalign}
\begin{flalign}
& 
\begin{array}{ll}
\maxSend{t}{a} \Coloneqq & \nexists a' \in t.\operationEvent{a'} = \esend \land \receiveSet{a} = \receiveSet{a'} \\
\end{array} \label{eq:max-send} \\
&
\begin{array}{ll}
\minRcv{t}{a} \Coloneqq & \exists a' \in t. \operationEvent{a'} = \esend \land  \receiveSet{a} = \receiveSet{a'}
\end{array}  \label{eq:min-send} \\
& {\arraycolsep=1pt
\begin{array}{lll}
\maxRcv{t}{a} \Coloneqq & \nexists a' \in t. &\operationEvent{a'} = \ereceive \ \land \ \replicaEvent{a} = \replicaEvent{a'} \land \receiveSet{a} = \receiveSet{a'} \\
\end{array}} \label{eq:max-rcv} 
\end{flalign}

Note that as $\implementationInstanceCC$ contains $\esend$ and $\ereceive$ actions, as well as events along with their write-read dependencies, $\implementationInstanceCC$ is a storage implementation. 
For proving that $\implementationInstanceCC$ is the searched implementation, we introduce the following notation: for a trace $t$ and an event $e \in t$, $\prefixTraceEvent{t}{e}$ to the trace s.t. $\Delta(\prefixTraceEvent{t}{e}, e)$ is a prefix of $t$.

The rest of the proof, showing that $\implementationInstanceCC$ is an available $\sspec^{\CC}$-implementation, is a consequence of \Cref{lemma:cons-available:consistency,lemma:cons-available:availability,lemma:cons-available:conditionally-finite}.%
\end{proof}

\begin{lemma}
\label{lemma:cons-available:consistency}
The implementation $\implementationInstanceCC$ is an $\sspec^{\CC}$-implementation.
\end{lemma}

\begin{proof}
Let $\programInstance = \program$ be a program. We prove by induction on the length of all traces in $\TracesParallelImplementationCC$ that any trace $t$ is feasible and its induced abstract execution is valid w.r.t. $\sspec^{\CC}$. The base case, when $t = \{\tup{\init_{\programInstance}, \init_{\implementationInstanceCC}}\}$ is immediate as $t$ contains exactly one event that does not read any object. Hence, let us assume that for any trace $t' \in \TracesParallelImplementationCC$ of at most length $k$, $\execTrace{t'}$ is valid w.r.t. $\sspec^{\CC}$; and let us show that for any trace $t$ of length $k+1$, $\execTrace{t}$ is also valid w.r.t. $\sspec^{\CC}$. %
Let $h = \tup{E, \so, \wro}$ and $\exec = \tup{h, \rbo, \ar}$ be respectively the induced history $\historyExecution{t}$ and the induced abstract execution $\execTrace{t}$ \added{where $\ar$ coincides with the trace order}. We denote $\sro$ to the induced order between send-receive actions with the same $\rbo$-$\mathsf{Set}$ on $t$. Before proving that $\exec$ is valid w.r.t. $\sspec^{\CC}$, we show that $t$ is feasible, i.e. $\exec$ satisfies \Cref{def:execution}.

\begin{itemize}
\item \underline{$\rbo = \rbo ; \so^*$:} This is immediate by the definition of induced receive-before.

\item \underline{$\wro \cup \so \subseteq \rbo$:} By definition of $\rbo$, $\so \subseteq \rbo$, so we focus on proving that $\wro \subseteq \rbo$. Let $w,r$ be events and $x$ be an object s.t. $(w, r) \in \wro_x$. In such case, there is a pair of actions $a_r, a_w$ s.t. $r \in a_r$, $w \in a_w$ and $w \in \writesSet{a_r}(x)$. Hence, $\{w\} = \max_{\ar_e^t} E^x_t(e)$. We deduce then that $\mathsf{rec}_t(w, r)$ must hold; which implies that there exists a $\esend$ action $s$ and a $\ereceive$ action $v$ s.t. $\receiveSet{s} = \receiveSet{v}$ and $w <_t s <_t v <_t r$. By $\mathsf{sendAllData}$ predicate, $w \in \receiveSet{s}$. As $\receiveSet{s} = \receiveSet{v}$, $w \in \receiveSet{v}$. By the definition of induced abstract execution, $(w, r) \in \rbo$.

\item \underline{$\rbo \subseteq \ar$:} For proving that $\rbo \subseteq \ar$, as $\rbo$ can be derived by $\sro$ and $\so$,  it suffices to prove that both $\so, \sro \subseteq \ar$. First, as $\so$ is the partial order induced by the total order $<_t$ on actions executed on the same replica, $\so \subseteq \ar$.

Next, for proving that $\sro \subseteq \ar$, let $s$ be a $\esend$ action and let $v$ be a $\ereceive$ action s.t. $(s, v) \in \sro$. Let us consider $p_v^{t} = \prefixTraceEvent{t}{v}$ be the prefix of $t$ before $v$. On one hand, as $p_v^{t}$ is a prefix of $t'$, $\Delta_{\mathsf{i}}(p_v^{t}, v) \downarrow$.
In particular, $\minRcv{p_v^{t}}{v}$ holds; so there is a $\esend$ action $s'$ in $p_v^{t}$ s.t. $\receiveSet{s'} = \receiveSet{v}$. We show that $s' = s$. Otherwise, then w.l.o.g. $s <_t s'$. Note that $\Delta_{\mathsf{i}}(\prefixTraceEvent{t}{s'}, s') \downarrow$ as $\prefixTraceEvent{t}{s'} \oplus s'$ is a prefix of $t'$. In such case, $\maxSend{\prefixTraceEvent{t}{s'}}{s'}$ does not hold; which is impossible as $\Delta_{\mathsf{i}}(\prefixTraceEvent{t}{s'}, s') \downarrow$. Therefore, $s = s'$. As $s' \in p_v^{t}$, $s$ precedes $v$ in $t$; so $(s, v) \in \ar$.

\end{itemize}

After proving that $t$ is feasible, we show that $\exec$ is valid w.r.t. $\sspec^{\CC}$. By \Cref{def:valid}, we need to show that for every event $r$ and object $x$, if $\rspecSimple{r} \uparrow$, $\wro_x^{-1}(r) = \emptyset$, and otherwise, $\wro_x^{-1}(r) = \{\max_{\ar} \context{\CC}{x}{\exec}{r}\}$. Let $r$ be a read event, $x$ be the object it affects and $p = \prefixTraceEvent{t}{r}$. We know by \Cref{eq:context-cc} that $\wro_x^{-1}(r) = \{\max_{\ar_r^p} E^x_p(r)\}$. Observe then that by \Cref{eq:context-cc} and $\rbo$'s definition, $E^x_p(r) = \context{\CC}{x}{t}{r} $. Thus, we conclude that $\wro_x^{-1}(r) = \{\max_{\ar} \context{\CC}{x}{\exec}{r}\}$.
\end{proof}

\begin{lemma}
\label{lemma:cons-available:availability}
For every program $\programInstance$ and every trace $t$ of $\parallelCompositionInstance$, there is no replica waiting in $t$.
\end{lemma}

\begin{proof}
Let $\programInstance = \program$ be a program, $r \in \ReplicaID$ be a replica and $t \in \TracesParallelImplementationCC$ be a reachable trace. Let also be $t_1 \in \Traces[\programInstance]$ and $t_2 \in \Traces[\implementationInstanceCC]$ traces s.t. $t = (t_1, t_2)$. To prove that $r$ is not waiting in $t$, let us suppose that there exists an event $e \in \Events[\programInstance]$ s.t. $\operationEvent{e} \neq \eend$, $\replicaEvent{e} = r$ and $\Delta_{\programInstance}(t_1, e) \downarrow$, and let us prove that there exists a non-$\ereceive$ action $a$ s.t. $\Delta_{\parallelCompositionInstance}(t, a) \downarrow$.

Let $a$ be the action $(e, M_t(e))$; where $M_t(e)$ is described using \Cref{eq:context-cc}. We observe that as $\Delta_{\programInstance}(t_1, e) \downarrow$, $\Delta_{\parallelCompositionInstanceCC}(t, \executedInstance) \downarrow$. Moreover, $\operationEvent{a} \neq \ereceive$. Hence, $r$ is not waiting in $t$; so $\implementationInstanceCC$ is available.
\end{proof}

\begin{lemma}
\label{lemma:cons-available:conditionally-finite}
For every finite program $\programInstance$, the composition $\parallelCompositionInstance$ is also finite.
\end{lemma}

\begin{proof}
Let $\programInstance = \program$ be a finite program. The implementation $\implementationInstanceCC$ is conditionally finite on $\programInstance$ if for every trace $t \in \TracesParallelImplementationCC$ there exists a constant $k_t \in \mathtt{N}$ s.t. $\length{t} \leq k_t$. Let thus $t \in \TracesParallelImplementationCC, t_1 \in \Traces[\programInstance], t_2 \in \Traces[\implementationInstanceCC]$ be traces s.t. $t = (t_1, t_2)$. As $\programInstance$ is finite, the length of $t_1$, $\length{t_1}$, is finite. We show that $k_t \Coloneqq 3 \cdot \length{t_1}$ is the constant we search.

Three cases arise, depending on the type of action we consider. First, by $\mathsf{maxRcv}$ predicate, the number of $\ereceive$ actions coincides with the number of $\ereceive$ actions with distinct metadata; which by $\mathsf{minRcv}$, is bounded by the number of $\esend$ actions in the trace. Then, by $\mathsf{sendIfData}$, the number of $\esend$ actions is bounded by the number of synchronized actions. Finally, by the parallel composition definition, the number of synchronized actions in $t$ and $t_1$ coincide; so such number is bounded by $\length{t_1}$. Altogether, we deduce that $\length{t} \leq 3 \cdot \length{t_1} = k_t$. \qedhere

\end{proof}

\subsection{\replaced{Availability Implies Arbitration-Freeness}{Proof of Lemma~\ref{lemma:cons-available-lww:3-1}}}
\label{ssec:cons-available-lww:3-1}

\added{As explained in \Cref{sec:afc-theorem}, we prove the contrapositive: if $\cmodel$ is not arbitration-free, then no available $\sspec$-implementation exists. Indeed, if $\cmodel$ is not arbitration-free, every normal form $\cmodel'$ of $\cmodel$ contains a simple visibility formula involving $\ar$ (see \Cref{def:arbitration-free-model}). By \Cref{lemma:cons-available-lww:3-1}, such a formula precludes the existence of an available $(\cmodel', \opspec)$-implementation. Consequently, there is no available $(\cmodel, \opspec)$-implementation, since any such implementation would also be an available $(\cmodel', \opspec)$-implementation -- this is an easy observation as $\cmodel$ is equivalent to $\cmodel'$ (see \Cref{th:general-normal-form:existence}).}

\ifbool{diffMode}{\notSatnotConsAvailableLWW*}{}%

\begin{proof}
We assume by contradiction that there is an available implementation $\implementationInstance$ of $\sspec$ 
but \replaced{$\cmodel$ contains a visibility formula $v$ s.t. for some $i, 0 \leq i \leq \length{v}$, $\mathsf{Rel}_i^\mathsf{v} = \ar$}{$\cmodel$ is not arbitration-free w.r.t. $\opspec$}.
We use the latter fact to construct a specific program, which by the assumption, admits a trace (in the composition with this implementation) that contains no $\ereceive$ action. We show that \replaced{any}{the} abstract execution induced by this trace, which is admissible by any available implementation of $\sspec$, is not valid w.r.t. $\sspec$. This contradicts the hypothesis.

\deleted{The key idea is observing that as $\cmodel$ is not arbitration-free w.r.t. $\opspec$, there exists a non-vacuous visibility relation $v$ belonging to $\normalForm{\cmodel}$ that is not arbitration-free; where $\normalForm{\cmodel}$ is the model in normal form $\opspec$-equivalent to $\cmodel$ (\Cref{lemma:normal-form:simple-equivalent-consistency,lemma:general-normal-form:non-vacuous}). Recall that $v$ is a formula defined as in \Cref{eq:visibility-criterion}. Since $v$ is in simple but not arbitration-free, there exists an index $i$ such that $\mathsf{Rel}_{i}^v = \ar$. Let $d_v$ be the largest such index $i$ (last occurrence of $\ar$).}

\replaced{The program $\programInstanceIota$ we construct}{We construct a program $\programInstanceIota$ that} generalizes the litmus program presented in \Cref{fig:motivating-example:prog}. $P$ uses two replicas $r_0, r_1$, two distinguished objects $x_0, x_1$ and a collection of events $e_{i}^{x_l}, 0 \leq i \leq \length{v}, l \in \{0,1\}$. The events are used to ``encode'' two instances $v_{x_0}$ and $v_{x_1}$ \added{of the visibility formula}. 

\added{Let $d_v$ be the largest index $i$ s.t. $\mathsf{Rel}_i^\mathsf{v} = \ar$ (last occurrence of $\ar$).} Then, $v$ is formed of two parts: the path of dependencies from $\event_0$ to $\event_{d_v}$ which is not arbitration-free, and the suffix from $\event_{d_v}$ up to $\event_{\length{v}}$, the arbitration-free part. Thus, $\mathsf{v}$ is of the form:
\begin{align*}
\mathsf{v}_x(\event_0,\event_{\length{v}}) \Coloneqq \exists \event_1, \ldots, \event_{n-1}. & 
\bigwedge_{i=1}^{\length{v}} (\event_{i-1}, \event_{i}) \in \mathsf{Rel}^{\mathsf{v}}_i \ \land  
\writeVar{\event_0}{x} \land \wro_x^{-1}(\event_{\length{v}})\neq  \emptyset %
\end{align*}
where $\mathsf{Rel}_{i}^v \in \{ \so,\wro, \rbo, \ar\}$, for all $i < d_v$, $\mathsf{Rel}_{d_v}^v = \ar$, and $\mathsf{Rel}_{i}^v \in \{\so,\wro, \rbo\}$ for all $i > d_v$.

In the construction, 
we require that replica $r_l$ executes events $e_i^{x_l}$ if $i < d_v$ and events $e_i^{x_{1-l}}$ otherwise -- the replica $r_l$ executes the non arbitration-free part of $v$ for object $x_l$ and the arbitration-free suffix of $v$ for $x_{1-l}$. All objects in replica $r_l$ access (read and/or write) $x_l$ except $e_{\length{v}}^{x_l}$, which access with $x_{1-l}$. We denote by $\tilde{x}_i^{x_l}$ to the unique object that event $e_i^{x_l}$ reads and/or writes.%

More in detail, we construct a set of events, $E^i$, histories, $h^i = \tup{E^i, \so^i, \wro^i}$, and executions, $\exec^i = \tup{h^i, \rbo^i, \ar^i}$, $0 \leq i \leq \length{v}$ inductively, starting from an initial event $\init$, and incorporating at each time a pair of new events, $e_i^{x_0}$ and $e_i^{x_1}$. For simplifying notation, we use the convention $\init = e_{-1}^{x_0} = e_{-1}^{x_1}$. %

For the inductive step, we assume that the abstract execution $\exec^{i-1} = \tup{h^{i-1}, \rbo^{i-1}, \ar^{-1}}$ associated to the history $h^{i-1} = \tup{E^{i-1}, \so^{i-1}, \wro^{i-1}}$ contains events $e_{-1}^{x_0} \ldots e_{i-1}^{x_0}, e_{i-1}^{x_1}$ and is well-defined (satisfies \Cref{def:execution}) and we construct the history $h^i$ and the abstract execution $\exec^i$. 

We distinguish between cases depending on the value $i$: 

\begin{itemize}
    \item \underline{$i = 0$}: In this case, we consider $e_0$ be an event s.t. \replaced{$\wspecSimple{e_0^{x_l}}{\outputEventObj{\init}{\tilde{x}_i^{x_l}}} \downarrow$}{$\wspecSimple{e_0^{x_l}}{\valuewr[\wro^{i-1}]{\init}{\tilde{x}_i^{x_l}}} \downarrow$}. 
    
    \item \underline{$0 < i < \length{v}$, $\mathsf{Rel}_i^\mathsf{v} = \wro$ and $\mathsf{Rel}_{i+1}^\mathsf{v} = \wro$}: In this case, it is easy to see that by \Cref{proposition:general-normal-form:arbitration-well-defined:events-well-defined}, $\opspec$ allows atomic read-write events. We consider $e_i^{x_l}$ be an event s.t. $\rspecSimple{e_i^{x_l}} \downarrow$ and $\wspecSimple{w_i^{x_l}}{\valuewr[\wro^{i-1}]{w_i^{x_l}}{\tilde{x}_i^{x_l}}} \downarrow$.

    \item \underline{$0 < i < \length{v}$ and $\mathsf{Rel}_i^\mathsf{v} \neq \wro$ and $\mathsf{Rel}_{i+1}^\mathsf{v} = \wro$}: In this case, if $\opspec$ allows unconditional writes, then we select $e_i^{x_l}$ as an unconditional write event on object $\tilde{x}_i^{x_l}$. Otherwise, we select event $e_i^{x_l}$ s.t. $\rspecSimple{e_i^{x_l}} \downarrow$ and \replaced{$\wspecSimple{e_i^{x_l}}{\outputEventObj{w_i^{x_l}}{\tilde{x}_i^{x_l}}} \downarrow$}{$\wspecSimple{e_i^{x_l}}{\valuewr[\wro^i]{w_i^{x_l}}{\tilde{x}_i^{x_l}}} \downarrow$}.
    
    \item \sloppy \underline{$0 < i < \length{v}$ and $\mathsf{Rel}_{i+1}^\mathsf{v} \neq \wro$}: In this case, we select $e_i^{x_l}$ to not write $\tilde{x}_i^{x_l}$ unless it is necessary. If $\opspec$ allows read events that are not write events, or if allows conditional atomic read-write events, we select $e_i^{x_l}$ as an event such that $\rspecSimple{e_i^{x_l}} \downarrow$ but \replaced{$\wspecSimple{e_i^{x_l}}{\outputEventObj{w_i^{x_l}}{\tilde{x}_i^{x_l}}} \uparrow$}{$\wspecSimple{e_i^{x_l}}{\valuewr[\wro^{i-1}]{w_i^{x_l}}{\tilde{x}_i^{x_l}}} \uparrow$}. Otherwise, we select event $e_i$ such that $\rspecSimple{e_i^{x_l}} \downarrow$ and \replaced{$\wspecSimple{e_i^{x_l}}{\outputEventObj{w_i^{x_l}}{\tilde{x}_i^{x_l}}} \downarrow$}{$\wspecSimple{e_i^{x_l}}{\valuewr[\wro^{i-1}]{w_i^{x_l}}{\tilde{x}_i^{x_l}}} \downarrow$}.

    \item \underline{$i =  \length{v}$}: In this case, we consider $e_{\length{v}}^{x_l}$ to be an event that reads object $\tilde{x}_i^{x_l}$, i.e. $\rspecSimple{e_{\length{v}}^{x_l}} \downarrow$.
\end{itemize}
where $l \in \{0,1\}$ and \replaced{$w_i^{x_l} = \max_{\ar^{i-1}} \{e \in E^{i-1} \ | \ \wspecSimple{e}{\varOf{e_i^{x_l}}}\downarrow \land \ (e, e_i^{x_l}) \in \so^i\}$}{$w_i^{x_l} = \max_{\ar^{i-1}} \{e \in E^{i-1} \ | \ \valuewr[\wro^{i-1}]{e}{\varOf{e_i^{x_l}}} \downarrow \land (e, e_i^{x_l}) \in \so^i\}$}. We note that as $\init$ writes on every object, $w_i^{x_l}$ is well-defined.

First of all, observe that event $e_i^{x_l}$ is well-defined thanks to \Cref{lemma:cons-available-lww:context-v} and the assumptions on $\opspec$ (\Cref{ssec:validity}). We denote $E^i = E^{i-1} \cup \{e_i^{x_0}, e_i^{x_1}\}$. We observe that w.l.o.g., we can assume that the $\identifierEvent{e_i^{x_0}}$ is bigger than every identifier of an event in $E^{i-1}$ and that $\identifierEvent{e_i^{x_0}} < \identifierEvent{e_i^{x_1}}$.

We conclude the description of $h^i$ and $\exec^i$ by specifying the relations $\so^i, \wro^i, \rbo^i, \ar^i$. We require that $\so^i$ (resp. $\wro^i$, $\rbo^i$, $\ar^i$) contains $\so^{i-1}$ (resp. $\wro^{i-1}$, $\rbo^{i-1}$, $\ar^{i-1}$). Also, we require additional constrains on them due to event $e_i$:
\begin{itemize}
    \item \underline{$\so^i$}: We require that $(e, e_i^{x_l}) \in \so^i$ iff $\replicaEvent{e} = \replicaEvent{e_i^{x_l}}$; as well as $(\init, e_i^{x_l}) \in \so^i$.
    
    \item \underline{$\wro^i$}: If $e_i^{x_l}$ is not a read event, we require that ${\wro_{x_i}^i}^{-1}(e_i^{x_l}) \neq \emptyset$. Otherwise, we require that $(\{w_i^{x_l}\}, e_i^{x_l}) \in \wro^i_{x_i}$.

    \item \underline{$\rbo^i$}: We require that $\rbo^i = \so^i$.

    \item \underline{$\ar^i$}: We impose that for every event $e \in E^i$, $(e, e_i^{x_l}) \in \ar^i$. Also, we impose that $(e_i^{x_0}, e_i^{x_1}) \in \ar^i$.
\end{itemize}

Then, we define $\EventsProgramIota = E^{\length{v}}$ as the set our program employs. The set $\EventsProgramIota$ induces the set of traces $\TracesProgramIota$.%

We define the program $\programInstanceIota = \programIota$, where $\initProgramIota = \init$ and $\Delta_{\mathsf{p}}$ is the transition function defined as follows: for every trace $t \in \TracesProgramIota$ and event $e \in \EventsProgramIota$, $\Delta_{\mathsf{p}}(t, e) \downarrow$ if and only if $e \not\in t$ and every event in $\EventsProgramIota$ whose replica coincide with $e$ and has smaller identifier than $e$ is included in $t$.

Given such a program $\programInstanceIota$, the proof proceeds as follows:
\begin{enumerate}
    \item There exists a finite trace $t$ of $\parallelCompositionInstanceIota$ that contains no receive action (\Cref{lemma:program-a-la-decker:trace-without-receive}): Since $\implementationInstance$ is available, it can always delay receiving messages, and execute other actions instead. Then, as $\programInstanceIota$ is a finite program, such an execution must be finite.
    
    \item The trace $t$ induces a history $h_\mathsf{v} = \tup{E, \so, \wro}$ and an abstract execution $\exec_\mathsf{v} = \tup{h, \rbo, \ar}$ where $\rbo = \so$ \added{($\ar$ is arbitrary as long as $\rbo \subseteq \ar$)}. As $\implementationInstance$ is valid w.r.t. $\sspec$, $\exec_\mathsf{v}$ is valid w.r.t. $\sspec$. Next, we prove that since $\rbo = \so$, events in $\exec_\mathsf{v}$ read the latest value w.r.t. $\so$ written on their associated object in $\exec_\mathsf{v}$ (\Cref{lemma:program-a-la-decker-lww:reads}). In particular, we deduce that all traces of $\programInstanceIota$ without $\ereceive$ events induce the same history and therefore, the induced history does not change when the induced arbitration order changes.
    
    \item \sloppy Since $\ar$ is a total order, either $(e_{d_\mathsf{v}-1}^{x_0}, e_{d_\mathsf{v} - 1}^{x_{1}}) \in \ar$ or $(e_{d_\mathsf{v}-1}^{x_1}, e_{d_\mathsf{v} - 1}^{x_{0}}) \in \ar$. W.l.o.g., assume that $(e_{d_\mathsf{v}-1}^{x_0}, e_{d_\mathsf{v} - 1}^{x_{1}}) \in \ar$. By \Cref{lemma:program-a-la-decker:f0-visible}, we deduce that $e_0^{x_0} \in \context{\cmodel}{x_0}{\exec_\mathsf{v}}{e_{\length{v}}^{x_0}}$. The proof is explained in \Cref{fig:diagram-theorem-proof}: if $(e_{d_\mathsf{v}-1}^{x_0}, e_{d_\mathsf{v} - 1}^{x_{1}}) \in \ar$, then all events $e_i^{x_0}$ form a path in such way that $\mathsf{v}_{x_0}(e_0^{x_0}, \ldots e_{\length{v}}^{x_0})$ holds in $\exec_\mathsf{v}$. 
    
    \item Since $e_{\length{v}}^{x_0}$ is the only event at $r_1$ that reads or writes $x_0$ and events in $\exec_\mathsf{v}$ read the latests values w.r.t. $\so$ in $\exec_\mathsf{v}$, we deduce that $e_{\length{v}}^{x_0}$ reads $x_0$ from $\init$. However, as $e_0^{x_0} \in \context{\cmodel}{x_0}{\exec_\mathsf{v}}{e_{\length{v}}^{x_0}}$ and $\init$ precedes $e_0^{x_0}$ in arbitration order, we deduce that $e_{\length{v}}^{x_0}$ does not read the latest value w.r.t. $\ar$, i.e. $\rspecSimple{e_{\length{v}}^{x_0}} \downarrow$ but $\wro_{x_0}^{-1}(e_{\length{v}}^{x_0}) \neq \{\max_{\ar} \context{\cmodel}{x_0}{\exec_\mathsf{v}}{e_{\length{v}}^{x_0}}\}$. Therefore, $\exec_\mathsf{v}$ is not valid w.r.t. $\sspec$ (see \Cref{def:valid}). This contradicts the hypothesis that $\implementationInstance$ is an implementation of $\sspec$. \qedhere 
\end{enumerate}
\end{proof}

\begin{lemma}
\label{lemma:cons-available-lww:context-v}
Let $\sspec = (\cmodel, \opspec)$ be a storage specification s.t. $\cmodel$ is in normal form w.r.t. $\opspec$. For every visibility \replaced{formula}{relation} $v \in \cmodel$, there exists an abstract execution valid w.r.t. $\sspec$, $\exec$, an object $x$ and events $e_0, \ldots e_{\length{v}}$ s.t. $\rspecSimple{e_{\length{v}}} \downarrow$ and $v_x(e_0, \ldots e_{\length{v}})$ holds in $\exec$.
\end{lemma}

\begin{proof}
Let $v \in \cmodel$ be a visibility \replaced{formula}{relation}. As $\cmodel$ is normal form w.r.t. $\opspec$, $v$ is non-vacuous; so $\cmodel \not\equiv \cmodel \setminus \{v\}$. Hence, there exists an abstract execution valid w.r.t. $\sspec$, $\exec$, an object $x$ and a read event $r$ s.t. $\context{\cmodel}{x}{\exec}{r} \neq \context{\cmodel \setminus \{v\}}{x}{\exec}{r}$. As $\cmodel \setminus \{v\} \preccurlyeq \cmodel$, we conclude that there exists events $e_0, \ldots e_{\length{v}}$ s.t. $r = e_{\length{v}}$ and $v_{x}(e_0, \ldots e_{\length{v}})$ holds in $\exec$.
\end{proof}

\begin{lemma}
\label{lemma:program-a-la-decker:trace-without-receive}
For every available storage implementation, $\implementationInstance$, there exists finite reachable trace $t \in \TracesIotaParallelImplementation$ s.t.
\begin{enumerate}
    \item $t$ does not contain any action $a$ s.t. $\operationEvent{a} = \ereceive$. \label{lemma:program-a-la-decker:trace-without-receive:1}
    \item for every event $e \in \EventsProgramIota$ there exists exactly one action $a \in t$ s.t. $\eventExecuted{a} = e$ and, \label{lemma:program-a-la-decker:trace-without-receive:2}
    \item for every two actions $a, a' \in t$ in the same replica, if $\eventExecuted{a} \downarrow$, $\eventExecuted{a'} \downarrow$ and $\identifierEvent{\eventExecuted{a}} < \identifierEvent{\eventExecuted{a'}}$, then $a <_t a'$\label{lemma:program-a-la-decker:trace-without-receive:3}

\end{enumerate}
\end{lemma}

\begin{proof}
\sloppy

Let $\implementationInstance$ be an available storage implementation. We construct a sequence of traces $\{t^i\}_{i \in \mathbb{N}}$ s.t. for each $i \in \mathbb{N}$ (1) $t^i$ does not contain any $\ereceive$ action, (2a) for every event $e \in \EventsProgramIota$ s.t. $\identifierEvent{e} \leq \identifierEvent{\lastTraceReplica[\replicaEvent{e}]{\pi_1(t^i)}}$ there is exactly one action $a \in t^i$ s.t. $\eventExecuted{a} = e$, (2b) for every event $e \in \EventsProgramIota$ s.t. $\identifierEvent{e} > \identifierEvent{\lastTraceReplica[\replicaEvent{e}]{\pi_1(t^i)}}$ there is no action $a \in t^i$ s.t. $\eventExecuted{a} = e$, and (3) for every two actions $a, a' \in t$, if $\eventExecuted{a} \downarrow$, $\eventExecuted{a'}\downarrow$ and $\identifierEvent{\eventExecuted{a}} < \identifierEvent{\eventExecuted{a'}}$, then $a <_{t^i} a'$.%

Let $t^0 = \init_{\parallelCompositionInstanceIota}$ be the first trace of our sequence. Clearly, $t^0$ satisfy properties (1), (2a), (2b) and (3). Then, let $n \in \mathbb{N}$ and, assuming that the trace $t^n$ satisfy properties (1), (2a), (2b) and (3), we define $t^{n+1}$.
If for every replica $r$ and every event $e \in \EventsProgramIota$, $\Delta_{\programIotaTag}(\pi_1(t^n), e) \uparrow$, we define $t^{n+1} = t^n$. If not, let $r_n$ be a replica and $e_n \in \EventsProgramIota$ be an event s.t. $\Delta_{\programIotaTag}(\pi_1(t^n), e_n) \downarrow$. As $\implementationInstance$ is available, there exists an action $a_n$ s.t. %
$\operationEvent{a_n'} \neq \ereceive$ and $\Delta_{\parallelCompositionInstanceIota}(t^n, a_n) \downarrow$. We then define $t^{n+1} = \Delta_{\parallelCompositionInstanceIota}(t^n, a_n)$.

By induction hypothesis on $t^n$, $t^n$ satisfies properties (1), (2a), (2b) and (3). We show that $t^{n+1}$ also satisfies (1), (2a), (2b) and (3). Without loss of generality, we assume that $t^{n+1} \neq t^n$ as otherwise the result immediately holds. First, as $t^n$ satisfies (1) and $a_n$ is not a $\ereceive$ action, $t^{n+1}$ satisfies property (1). Properties (2a) and (2b) immediately hold from the definition of $\Delta_{\parallelCompositionInstanceIota}$.

Finally, for proving that $t^{n+1}$ satisfies (3), let $a, a' \in t^n$ be distinct actions s.t. $\eventExecuted{a} \downarrow$, $\eventExecuted{a'} \downarrow$ and $\identifierEvent{\eventExecuted{a}} < \identifierEvent{\eventExecuted{a'}}$. If $a, a' \neq a_n$, as $t^n$ satisfies (3), $a <_{t^n} a'$ and therefore, $a <_{t^{n+1}} a'$. Otherwise, note that as $t^n$ satisfies (2b), for every event $e \in \pi_1(t^n)$, $\identifierEvent{e} \leq \identifierEvent{\eventExecuted{a_n}}$. Moreover, as no two events in $\EventsProgramIota$ have identical identifier, traces do not contain the same event twice and $a \neq a'$, we deduce that $a' = a_n$. As $a_n = \lastTraceReplica[r_n]{t^{n+1}}$, we conclude that $a <_{t^{n+1}} a'$.

\ifbool{diffMode}{%
\color{oldVersionColor}%
We define $t^{\infty}$ as the supreme of the traces $\{t^n\}_{n \in \mathbb{N}}$.} By construction, $t^{\infty}$ is a trace in $\Traces[\parallelCompositionInstanceIota]$. As $\programInstanceIota$ is finite and $\implementationInstance$ is available, every trace $t \in \Traces[\parallelCompositionInstanceIota]$ is finite. \deleted{In particular, $t^{\infty}$ must be finite. Therefore, there must exist $k$ s.t. for every $k' > k$, $t^{k'} = t^k = t^{\infty}$. We show that the trace $t^k$ is the searched trace.%
}{%
}%
\color{newVersionColor}
We show by contradiction that there exists some $k\in \mathbb{N}$ s.t. $t^k = t^{k+1}$. Consider the sucession of traces $\{t^n\}_{n \in \mathbb{N}}$ and let us assume that $t^k \neq t^{k+1}$ for any $k \in \mathbb{N}$. In such case, we define $t^{\infty}$ as the limit of such sucession, i.e., the trace obtained by executing events actions $a_i, 0 \leq i \leq \mathbb{N}$ (which are well-defined by construction). Such infinite trace belongs to $\Traces[\parallelCompositionInstanceIota]$. However, as $\programInstanceIota$ is finite and $\implementationInstance$ is available, every trace $t \in \Traces[\parallelCompositionInstanceIota]$ is finite. Thus, $t^{\infty}$ must be finite; which contradicts its construction. Hence, such $k$ exists.
\color{black}

We show that the trace $t^k$ is the searched trace. Clearly, as $t^k$ satisfies (1) and (3), it suffices to prove that it also satisfies (2). On one hand, as $t^k = t^{k+1}$, for every event $e \in \programInstanceIota$, $\Delta_{\programIotaTag}(\pi_1(t^k), e) \uparrow$. Hence, for every replica $r_l, l \in \{0,1\}$, $\lastTraceReplica[r]{\pi_1(t^k)} = e_{\length{v}}^{x_{1-l}}$. By construction of $\EventsProgramIota$, every event $e \in \EventsProgramIota$ with replica $r_l$ has smaller identifier than $e_{\length{v}}^{x_{1-l}}$. Therefore, as $t^k$ satisfies (2a), there is exactly one action $a' \in t^k$ s.t. $\eventExecuted{e'} = e$; so $t^k$ satisfies (2).

\end{proof}

\begin{lemma}
\label{lemma:program-a-la-decker-lww:reads}
\sloppy For every pair of indices $i, -1 \leq i \leq \length{v}$, $l \in \{0,1 \}$,
\begin{itemize}
    \item If $e_i^{x_l}$ is a read event, then $(\{w_i^{x_l}\}, e_i^{x_l}) \in \wro_{\tilde{x}_i^{x_l}}$.

    \item If $e_i^{x_l}$ is a write event s.t. \replaced{$\outputEventObj{e_i^{x_l}}{\tilde{x}_i^{x_l}} \downarrow$}{$\valuewr[\wro^i]{e_i^{x_l}}{\tilde{x}_i^{x_l}} \downarrow$}, then \replaced{$\wspecSimple{e_i^{x_l}}{\outputEventObj{w_i^{x_l}}{\tilde{x}_i^{x_l}}} \downarrow$}{$\valuewr[\wro]{e_i^{x_l}}{\tilde{x}_i^{x_l}} = \valuewr[\wro^i]{e_i^{x_l}}{\tilde{x}_i^{x_l}}$ and $\wspecSimple{e_i^{x_l}}{\valuewr[\wro]{w_i^{x_l}}{\tilde{x}_i^{x_l}}} \downarrow$}.
\end{itemize}
\end{lemma}

\begin{proof}
We prove the result by induction on $i$; where the base case, $i = -1$, trivially holds. For showing the inductive case, let us assume that the result holds for every event $e_{i'}^{x_{l'}}, -1 \leq i' < i, l' \in \{0,1\}$, and let us show it for events $e_i^{x_0}, e_i^{x_1}$. We divide the proof in two blocks, whether $e_i^{x_l}$ is a read event, and $e_i^{x_l}$ is a write event.

For the first part, we note that by construction of $\exec_v$ using \Cref{lemma:program-a-la-decker:trace-without-receive} we know that $\exec_v$ does not contain any $\ereceive$ event, $\rbo = \so$. Hence, as $\exec_v$ is valid w.r.t. $\sspec$, $\wro \subseteq \rbo = \so$. Thus, $e_i^{x_l}$ reads $\tilde{x}_i^{x_l}$ from an event that precedes it in session order. In particular, by \Cref{def:valid}, $\wro_{\tilde{x}_i^{x_l}}^{-1}(e_i^{x_l}) = \{\max_{\ar} \context{\cmodel}{\tilde{x}_i^{x_l}}{\exec_v}{e_i^{x_l}}\}$; so $\wro_{\tilde{x}_i^{x_l}}^{-1}(e_i^{x_l}) = \{w_i^{x_l}\}$.

For the second part, we can assume w.l.o.g. that $e_i^{x_l}$ is a conditional write, as otherwise the result immediately holds. By the choice of $e_i^{x_l}$, in this case, \replaced{we conclude that $\wspecSimple{e_i^{x_l}}{\outputEventObj{w_i^{x_l}}{\tilde{x}_i^{x_l}}} \downarrow$}{$\wspecSimple{e_i^{x_l}}{\valuewr[\wro^i]{w_i^{x_l}}{\tilde{x}_i^{x_l}}} \downarrow$}\deleted{By induction hypothesis, as $w_i^{x_l}$ is an event $e_{i'}^{x_{l'}}, -1 \leq i' < i, l' \in \{0, 1\}$, $\valuewr[\wro]{w_i^{x_l}}{\tilde{x}_i^{x_l}} = \valuewr[\wro^i]{w_i^{x_l}}{\tilde{x}_i^{x_l}}$. Therefore, $\wspecSimple{e_i^{x_l}}{\valuewr[\wro]{w_i^{x_l}}{\tilde{x}_i^{x_l}}} \downarrow$. Proving that $\valuewr[\wro]{e_i^{x_l}}{\tilde{x}_i^{x_l}} = \valuewr[\wro^i]{e_i^{x_l}}{\tilde{x}_i^{x_l}}$ is immediate using \Cref{eq:value-wr} and $\valuewr[\wro]{w_i^{x_l}}{\tilde{x}_i^{x_l}} = \valuewr[\wro^i]{w_i^{x_l}}{\tilde{x}_i^{x_l}}$}.

\end{proof}

\begin{lemma}
\label{lemma:program-a-la-decker:f0-visible}
For every $l \in \{0,1\}$, if $(e_{d_v-1}^{x_l}, e_{d_v - 1}^{x_{1-l}}) \in \ar$, then $e_0^{x_l} \in \context{\cmodel}{x_l}{\exec_v}{e_{\length{v}}^{x_l}}$.
\end{lemma}

\begin{proof}
For proving that $e_0^{x_l} \in \context{\cmodel}{x_l}{\exec_v}{e_{\length{v}}^{x_l}}$, we show that $v_{x_l}(e_0^{x_l}, e_{\length{v}}^{x_l})$ holds in $\exec_v$. Observe that by the choice of events and \Cref{lemma:program-a-la-decker-lww:reads} $\writeVarExec{e_0^{x_l}}{x_l}{\exec_v}$ and $\wro_{x_l}^{-1}(e_{\length{v}}^{x_l}) \neq \emptyset$ holds in $\exec_v$. Therefore, to conclude the result, we prove that for every $i, 1 \leq i \leq \length{v}$, $(e_{i-1}^{x_l}, e_i^{x_l}) \in \mathsf{Rel}_i^{v}$.

For proving it, we observe that $\cmodel$ is in simple form. Thus, for every $i, 1 \leq i \leq \length{v}$, $\mathsf{Rel}_i^\mathsf{v}$ is either $\so, \wro, \rbo$ or $\ar$; which simplify our analysis. First, if $i = d_v$, by definition of $d_v$, $\mathsf{Rel}_i^\mathsf{v} = \ar$. By hypothesis, $(e_{d_v - 1}^{x_l}, e_{d_v - 1}^{x_{1-l}}) \in \ar$. In such case, as $\identifierEvent{e_{d_v - 1}^{x_{1-l}}} < \identifierEvent{e_{d_v}^{x_{l}}}$ and $\replicaEvent{e_{d_v - 1}^{x_{1-l}}} = \replicaEvent{e_{d_v}^{x_l}}$, $(e_{d_v - 1}^{x_{1-l}}, e_{d_v}^{x_l}) \in \so$. Therefore, as $\so \subseteq \ar$ and $\ar$ is a transitive relation, we deduce that $(e_{d_v - 1}^{x_l}, e_{d_v}^{x_{l}}) \in \ar$.

Next, if $i \neq d_v$, we notice that $(e_{i-1}^{x_0}, e_{i}^{x_0}) \in \so \subseteq \rbo \subseteq \ar$. Hence, if $\mathsf{Rel}_i^\mathsf{v}$ is either $\so, \rbo$ or $\ar$, the result immediately holds. Otherwise, if $\mathsf{Rel}_i^{v} = \wro$, we show that $e_i^{x_0}$ is a read event and $e_{i-1}^{x_0} = w_i^{x_0}$; which let us conclude that $(e_{i-1}^{x_0}, e_i^{x_0}) \in \wro$ thanks to \Cref{lemma:program-a-la-decker-lww:reads}.

First, we show that if $i \neq \length{v}$ and $\mathsf{Rel}_i^\mathsf{v} = \wro$, then $w_i^{x_l} = e_{i-1}^{x_l}$. Thanks to the choice of $P$, if $\mathsf{Rel}_i^\mathsf{v} = \wro$, then $e_i^{x_l}$ is a write event s.t. \replaced{$\outputEventObj{e_{i-1}^{x_l}}{\tilde{x}_i^{x_l}} \downarrow$}{$\valuewr[\wro]{e_{i-1}^{x_l}}{\tilde{x}_i^{x_l}} \downarrow$}. By \Cref{lemma:program-a-la-decker-lww:reads}, we deduce that $\writeVar{e_{i-1}^{x_l}}{\tilde{x}_{i-1}^{x_l}}$ in $\exec_v$. As $i \neq \length{v}$, $\tilde{x}_{i-1}^{x_l} = \tilde{x}_i^{x_l}$. Also, as $\mathsf{Rel}_i^\mathsf{v} = \wro$, $\replicaEvent{e_i^{x_l}} = \replicaEvent{e_{i-1}^{x_l}}$. Altogether, we deduce that $e_{i-1}^{x_l}$ is an event writing $\tilde{x}_i^{x_l}$ that is the immediate predecessor of $e_i^{x_l}$ w.r.t. $\so$. Hence, $w_i^{x_l} = e_{i-1}^{x_l}$.

Finally, we show that $\mathsf{Rel}_{\length{v}}^{v} \neq \wro$ and conclude the result. We prove the contrapositive, that if $\mathsf{Rel}_{\length{v}}^\mathsf{v} = \wro$, $v$ is vacuous w.r.t. $\sspec$. If $\mathsf{Rel}_{\length{v}}^\mathsf{v} = \wro$, for every abstract execution $\exec'$ valid w.r.t. $\sspec$, object $x$ and a collection of events $f_0, \ldots f_{\length{v}}$, if $v_x(f_0, \ldots, f_{\length{v}})$ holds in $\exec'$, then $(f_{\length{v}-1}, f_{\length{v}}) \in \wro$. Thus, $\exec'$ is valid w.r.t. $(\cmodel \setminus \{v\}, \opspec)$. Hence, $v$ is vacuous w.r.t. $\opspec$.
\end{proof}

\newpage

\section{Proof of the Arbitration-Free Consistency Theorem \deleted{(\Cref{th:characterization-cons-available})}}
\label{app:main-theorem-general-case-extra-proofs}

\Cref{lemma:cons-available:1-2,lemma:cons-available:3-1} prove the AFC theorem.

\subsection{\replaced{Arbitration-Freeness Implies Availability}{Proof of Lemma~\ref{lemma:cons-available:1-2}}}
\label{ssec:cons-available:1-2}

The proof of (1) $\implies$ (2), essentially coincides with that of \Cref{lemma:cons-available-lww:1-2}: we present an available $\sspec$-implementation that guarantees $\CC$. As \added{in \Cref{lemma:cons-available-lww:1-2},} $\cmodel$ is arbitration-free, \added{so} by \Cref{lemma:saturable:cc-strongest}, this implies that $\cmodel$ is weaker than $\CC$\replaced{. Thanks to}{;so by} \Cref{lemma:general-normal-form:consistency-stronger-than-criterion}, any implementation of $\CC$ also ensures $\cmodel$.

\begin{restatable}[(1)$\implies$(2)]{lemma}{satConsAvailable}
\label{lemma:cons-available:1-2}
\replaced{Let $\opspec$ be a \simpleStorageFullName{} operation specification. There exists an available $(\CC, \opspec)$-implementation.}{Let $\sspec = (\cmodel, \opspec)$ be a storage specification.
If $\cmodel$ is arbitration-free w.r.t. $\opspec$, there exists an available $\opspec$-implementation.}
\end{restatable}

\begin{proof}

The main difference in the construction w.r.t. the implementation shown in \Cref{lemma:cons-available-lww:1-2} corresponds to the transition function, $\Delta_{\mathsf{i}}$. More specifically, the main and only change arise in \Cref{eq:context-cc}, which is substituted by \Cref{eq:general:context-cc}.%

\begin{equation}
\label{eq:general:context-cc}
{\arraycolsep=3pt
\begin{array}{lll}
M_t(e) & =& [x \mapsto \rspecContext{x}{E_t^x(e)}{e}]_{x \in \Vars}\\
E_t^x(e)& = & \left\{e' \ \left| \
\begin{array}{l}
    e' \in \Events \cap t \ \land \ \writeVarExec{e'}{x}{\execTrace{t}} \ \land\\
    (\replicaEvent{e'} = \replicaEvent{e} \lor \mathsf{rec}_t(e', e))
\end{array} \right.\right\} \\
\ar_e^t &= & \ar_{\restriction E_t^x(e) \times E_t^x(e)}
\end{array}
}
\end{equation}

Is immediate to show that $\implementationInstanceCC$ is a storage implementation. Showing that $\implementationInstanceCC$ is an available $\sspec$-implementation is done as in \Cref{lemma:cons-available-lww:1-2}. Observe that \Cref{lemma:cons-available:availability,lemma:cons-available:conditionally-finite} apply to this implementation; so $\implementationCC$ is an available implementation. In \Cref{lemma:general-cons-available:consistency} we show that indeed $\implementationInstanceCC$ is an implementation of $(\CC, \sspec)$, concluding the result.

\end{proof}

\begin{lemma}
\label{lemma:general-cons-available:consistency}
The implementation $\implementationInstanceCC$ is an implementation of $\sspec' = (\CC, \opspec)$.
\end{lemma}

\begin{proof}
Let $\programInstance = \program$ be a program. We prove by induction on the length of all traces in $\TracesParallelImplementationCC$ that any trace $t$ is valid w.r.t. $\sspec'$. The base case, when $t$ contains a single action, is immediate as such action corresponds to the initial event, which does not read any object. Let us assume that for any trace $t' \in \TracesParallelImplementationCC$ of at most length $k$, $\execTrace{t'}$ is valid w.r.t. $\sspec'$; and let us show that for any trace $t$ of length $k+1$, $\execTrace{t}$ is also valid w.r.t. $\sspec'$.

Let $h = \tup{E, \so, \wro}$ and $\exec = \tup{h, \rbo, \ar}$ be respectively the history $\historyExecution{t}$ and the abstract execution $\execTrace{t}$. We denote $\sro$ to the induced order between send-receive actions with the same metadata on $t$. For proving that $\exec$ is valid w.r.t. $\sspec'$, we first prove that $\exec$ is indeed an abstract execution, i.e., $\exec$ satisfies \Cref{def:execution}. In particular, by the construction of $\implementationCC$ (compared with that of \Cref{lemma:cons-available:consistency}), it suffices showing that $\wro \cup \so \subseteq \rbo$.

By definition of $\rbo$, $\so \subseteq \rbo$, so we focus on proving that $\wro \subseteq \rbo$. Let $w,r$ be events and $x$ be an object s.t. $(w, r) \in \wro_x$. In such case, there is a pair of actions $a_r, a_w$ s.t. $r \in a_r$, $w \in a_w$ and $w \in \writesSet{a_r}(x)$. Hence, $w \in \rspecContext{x}{E_t^x(r)}{r}$. We deduce then that $\mathsf{rec}_t(w, r)$ must hold; which implies that there exists a $\esend$ action $s$ and a $\ereceive$ action $v$ s.t. $\receiveSet{s} = \receiveSet{v}$ and $w <_t s <_t v <_t r$. By $\mathsf{sendAllData}$ predicate, $w \in \receiveSet{s}$; so by $\mathsf{minRcv}$, $w \in \receiveSet{v}$. By the induced abstract execution definition, $(w, r) \in \rbo$.

\sloppy Finally, we show that $\exec$ is valid w.r.t. $\sspec'$. By \Cref{def:general-valid}, \replaced{it suffices}{we need} to show that for every event $r$ and object $x$, $\wro_x^{-1}(r) = \rspecContext{x}{\context{\CC}{x}{\exec}{r}}{r}$. Let $r$ be a read event, $x$ be an object and $p = \prefixTraceEvent{t}{r}$. We know by \Cref{eq:general:context-cc} that $\wro_x^{-1}(r) = \rspecContext{x}{E_p^x(r)}{r}$. Observe that by \Cref{eq:general:context-cc} and $\rbo$'s definition, $E_p^x(r)$ coincides with $\context{\CC}{x}{t}{r} $. Thus, so we conclude that $\wro_x^{-1}(r) = \rspecContext{x}{\context{\CC}{x}{t}{r}}{r}$.
\end{proof}

\subsection{\replaced{Availability Implies Arbitration-Freeness}{Proof of Lemma~\ref{lemma:cons-available:3-1}}}
\label{ssec:cons-available:3-1}

\added{The proof of this result mimics that of \Cref{lemma:cons-available-lww:3-1}. We prove the contrapositive: if $\cmodel$ is not arbitration-free, then no available $\sspec$-implementation exists. Indeed, if $\cmodel$ is not arbitration-free, every normal form $\cmodel'$ of $\cmodel$ contains a simple visibility formula involving $\ar$, and such formula precludes the existence of an available $(\cmodel, \opspec)$-implementation (see \Cref{lemma:cons-available:3-1}).}

\notSatnotConsAvailable*

\begin{proof}
\deleted{The proof of this result is inspired by \Cref{lemma:cons-available-lww:3-1}.}

We assume by contradiction that there is an available implementation $\implementationInstance$ of $\sspec$ 
but \replaced{$\cmodel$ contains a visibility formula $v$ non-vacuous w.r.t. $\opspec$ s.t. for some $i, 0 \leq i \leq \length{v}$, $\mathsf{Rel}_i^\mathsf{v} = \ar$.}{$\cmodel$ is not arbitration-free w.r.t. $\opspec$.} We use the latter fact to construct a specific program, which by the assumption, admits a trace (in the composition with this implementation) that contains no $\ereceive$ action. We show that \replaced{any}{the} abstract execution induced by this trace, which is admissible by any available implementation of $\sspec$, is not valid w.r.t. $\sspec$. This contradicts the hypothesis.

\deleted{The key idea is observing that as $\cmodel$ is not arbitration-free w.r.t. $\opspec$, there exists a non-vacuous visibility relation $v \in \generalNormalForm{\cmodel}$ that is not arbitration-free; where $\generalNormalForm{\cmodel}$ is a model in general normal form $\opspec$-equivalent to $\cmodel$ (\Cref{th:general-normal-form:existence}). Recall that $v$ is a formula defined as in \Cref{eq:general-visibility-criterion}. Since $v$ is in simple but not arbitration-free, there exists an index $i$ such that $\mathsf{Rel}_{i}^v = \ar$. Let $d_v$ be the largest such index $i$ (last occurrence of $\ar$). Then, $v$ is formed of two parts: the path of dependencies from $\event_0$ to $\event_{d_v}$ which is not arbitration-free, and the suffix from $\event_{d_v}$ up to $\event_{\length{v}}$, the arbitration-free part.}

\replaced{The program $\programInstanceIota$ we construct}{We construct a program $\programInstanceIota$ that} generalizes the litmus program presented in \Cref{fig:motivating-example:prog}. $\programInstanceIota$ uses two replicas $r_0, r_1$, two distinguished objects $x_0, x_1$ and a collection of events $e_{i}^{x_l}, 0 \leq i \leq \length{v}, l \in \{0,1\}$. The events are used to ``encode'' two instances of $v_{x_0}$ and $v_{x_1}$. 

\added{Let $d_v$ be the largest index $i$ s.t. $\mathsf{Rel}_i^\mathsf{v} = \ar$ (last occurrence of $\ar$). Then, $v$ is formed of two parts: the path of dependencies from $\event_0$ to $\event_{d_v}$ which is not arbitration-free, and the suffix from $\event_{d_v}$ up to $\event_{\length{v}}$, the arbitration-free part.}

For ensuring that $\mathsf{v}_x(e_0^{x_l}, \ldots e_{n}^{x_l})$ holds in an induced abstract execution of a trace without $\ereceive$ actions, we require that if $\mathsf{Rel}_i^{\mathsf{v}} = \wro$, then $e_{i-1}^{x_l}$ is a write event and $e_i^{x_l}$ is a read event. For ensuring that $\writeConstraints{x}{e_0, \ldots e_{\length{v}}}$ holds in such abstract execution, we consider a distinct object $y_E$, also distinct from $x_0, x_1$. These objects represents each different conflict in $v$ in an explicit manner. Intuitively, we require that events $e_i^{x_l}$ write on object $y_E$ (resp. $x_l$) iff $\event_i \in E$.

More formally, we denote by $E_x \in \mathcal{P}(\event_0, \ldots e_{\length{v}})$ to the set s.t. $\writeConflicts[x]{E_x} \in v$. Also, for every $i, 0 \leq i \leq \length{v}, l \in \{0,1\}$, we denote by $X_i^{x_l}$ to the set containing objects $y_E$ (resp. $\hat{x}_i^{x_l}$) iff $E \in \conflictsOfV{v}{i}$ (resp. $E_x \in \conflictsOfV{v}{i}$); where $\hat{x}_i^{x_l} = x_l$ if $i < d_v$ and $x_{1-l}$ otherwise. We denote by $X$ to the union of sets $X_i^{x_l}, 0 \leq i \leq \length{v}, l \in \{0,1\}$.

In the construction, 
we require that replica $r_l$ executes events $e_i^{x_l}$ if $i < d_v$ and events $e_i^{x_{1-l}}$ otherwise -- the replica $r_l$ executes the non arbitration-free part of $v$ for object $x_l$ and the arbitration-free suffix of $v$ for $x_{1-l}$. We denote by $r_i^{x_l}$ to such replica.

More in detail, we construct a set of events, $E^i$, histories, $h^i = \tup{E^i, \so^i, \wro^i}$, and executions, $\exec^i = \tup{h^i, \rbo^i, \ar^i}$, $0 \leq i \leq \length{v}$ inductively, starting from an initial event $\init$, and incorporating at each time a pair of new events, $e_i^{x_0}$ and $e_i^{x_1}$. We use the notation $h^{-1}$ and $\exec^{-1}$ to describe the history and abstract execution containing only $\init$ respectively. For simplifying notation, we use the convention $\init = e_{-1}^{x_0} = e_{-1}^{x_1}$. %

For the inductive step, we assume that the abstract execution $\exec^{i-1} = \tup{h^{i-1}, \rbo^{i-1}, \ar^{-1}}$ associated to the history $h^{i-1} = \tup{E^{i-1}, \so^{i-1}, \wro^{i-1}}$ contains events $e_{-1}^{x_0} \ldots e_{i-1}^{x_0}, e_{i-1}^{x_1}$ and is well-defined (satisfies \Cref{def:execution}) and we construct the history $h^i$ and the abstract execution $\exec^i$. 

The construction of $\exec^i$ follows the structure of that constructed in \Cref{lemma:cons-available-lww:3-1}'s proof, but with the technical details of that used in \Cref{th:general-normal-from:arbitration-well-defined}'s proof.

For the inductive step, we assume that the abstract execution $\exec^{i-1} = \tup{h^{i-1}, \rbo^{i-1}, \ar^{-1}}$ associated to the history $h^{i-1} = \tup{E^{i-1}, \so^{i-1}, \wro^{i-1}}$ contains events $e_{-1}^{x_0} \ldots e_{i-1}^{x_0} e_{i-1}^{x_1}$ and is well-defined (satisfies \Cref{def:execution}) and we construct the history $h^i$ and the abstract execution $\exec^i$. %

In the following, let $l \in \{0,1\}$.

Like in \Cref{th:general-normal-from:arbitration-well-defined}, we define a pair of special objects, $\tilde{x}_i^{x_l}$ and $o_i^{x_l}$. The purpose of object $\tilde{x}_i^{x_l}$ is control the number of events in $\exec$ that write object $x_l$. \Cref{eq:afc:x-i} describes $\tilde{x}_i^{x_l}$; where $\mathsf{choice}$ is a function that deterministically chooses an element from a non-empty set. The object $o_i^{x_l}$ is an object different from objects $x, y_E, E \in \mathcal{P}(\event_0, \ldots \event_{\length{v}})$ and $o_j^{x_{l'}}, -1 \leq j < i, l' \in \{0,1\}$ that we use for ensuring that if $\mathsf{Rel}_i^\mathsf{v} = \wro$, then $(e_{i-1}, e_i) \in \wro$. W.l.o.g., we can assume that $o_i^{x_0} \neq o_i^{x_1}$.

\begin{equation}
\label{eq:afc:x-i}
    \tilde{x}_i^{x_l} = \left\{ \begin{array}{ll}
        \tilde{x}_{i-1}^{x_l} & \text{if } X_i^{x_l} = \emptyset \\
        \hat{x}_i^{x_l} & \text{if } X_i^{x_l} \neq \emptyset \text{ and } \hat{x}_i^{x_l} \in X_i^{x_l} \\
        \choice{X_i} & \text{if } X_i^{x_l} \neq \emptyset \text{ and } \hat{x}_i^{x_l} \not\in X_i^{x_l}
    \end{array}\right.
\end{equation}

We select a domain $D_i^{x_l}$, a set of objects $W_i^{x_l}, W_i^{x_l} \subseteq D_i^{x_l}$ that event $e_i^{x_l}$ must write, and a set of objects $C_i^{x_l} \subseteq D_i^{x_l}$ whose value needs to be corrected for events $e_i^{x_0}, e_i^{x_1}$ in $\exec_{i+1}$ -- in the sense of \Cref{def:execution-corrector}. We distinguishing between several cases:

\begin{itemize}
    \item \underline{$i = 0$ or $0 < i \leq \length{v}$ and $\mathsf{Rel}_i^\mathsf{v} \neq \wro$ and $\conflictsOfV{v}{i} \neq \emptyset$}: In this case, we select $e_i^{x_l}$ to be a write event. If $\opspec$ only allows single-object atomic read-write events, we define $D_i^{x_l} = X_i^{x_l}$; while if not, we consider a domain containing $o_{i-1}^{x_l}, o_i^{x_l}$, every object in $X_i^{x_l}$ but no object from $X \setminus X_i^{x_l}$ nor objects $o_j^{x_{l'}}, 0 \leq j < \length{v}, l' \in \{0,1\} j \neq i-1,i$. Observe that by \Cref{proposition:general-normal-form:arbitration-well-defined:domains}, such domain always exist on $\opspec$. 
    
    If there is an unconditional write event whose domain is $D_i^{x_l}$, we define $W_i^{x_l} = D_i^{x_l}$. Otherwise, we define $W_i^{x_l} = X_i^{x_l} \cup \{o_{i}^{x_l}\}$.%

    \item \underline{$0 < i \leq \length{v}$, $\mathsf{Rel}_i^\mathsf{v} = \wro$ and $\conflictsOfV{v}{i} \neq \emptyset$}: In this case, by \Cref{proposition:general-normal-form:arbitration-well-defined:events-well-defined}, $\opspec$ allows atomic read-write events. If $\opspec$ only allows single-object atomic read-write events, we define $D_i^{x_l} = X_i^{x_l}$; while if not, we consider a domain containing $o_{i-1}, o_i$, every object in $X_i^{x_l}$ but no object from $X \setminus X_i^{x_l}$ nor objects $o_j, 0 \leq j < \length{v}, j \neq i-1,i$. Observe that by \Cref{proposition:general-normal-form:arbitration-well-defined:domains}, such domain always exist on $\opspec$. 
    
    Similarly to the previous case, if there is an unconditional atomic read-write event whose domain is $D_i^{x_l}$, we define $W_i^{x_l} = D_i^{x_l}$. Otherwise, we define $W_i^{x_l} = X_i^{x_l} \cup \{o_{i}^{x_l}\}$.%

    \item \underline{$0 < i \leq \length{v}$ and $\conflictsOfV{v}{i} = \emptyset$}: In this case, by \Cref{proposition:general-normal-form:arbitration-well-defined:events-well-defined}, $\opspec$ allows events that do not unconditionally write. %
    If $\opspec$ allows read events that are not write events, we select $D_i^{x_l}$ to be the domain of any such event and $W_i^{x_l} = \emptyset$. Otherwise, $\opspec$ must allow conditional write events; so we select $D_i^{x_l}$ to be the domain of any such event, $W_i^{x_l} = \emptyset$. %
    Observe that in this case, thanks to the assumptions on $\opspec$ (see \Cref{ssec:assumptions-generalized-specs}), we can assume without loss of generality that whenever $o_{i-1}^{x_l} \in D_{i-1}$, $o_{i-1}^{x_l} \in D_i^{x_l}$ as well; while otherwise, that $\tilde{x}_{i-1}^{x_l} \in D_i^{x_l}$.

\end{itemize}

Finally we describe the event $e_i^{x_l}$ thanks to the sets $D_i^{x_l}$ and $W_i^{x_l}$. If $W_i^{x_l} = D_i^{x_l}$ and $\mathsf{Rel}_i^\mathsf{v} = \wro$, we select an unconditional atomic read-write event whose domain is $D_i^{x_l}$. If $W_i^{x_l} = D_i^{x_l}$ and $\mathsf{Rel}_i^\mathsf{v} \neq \wro$, we select an unconditional write event whose domain is $D_i^{x_l}$. If $W_i^{x_l} = \emptyset$ and $\opspec$ allows read events that are not write events, we select a read event whose domain is $D_i^{x_l}$. Finally, if that is not the case, we select a conditional write event $e_i^{x_l}$ s.t. $\varOf{e_i^{x_l}} = D_i^{x_l}$ and s.t. an execution-corrector exists for $(e_i^{x_l}, W_i^{x_l}, \tilde{x}_i^{x_l}, \exec^{i-1} \oplus e_i^{x_0} \oplus e_i^{x_1})$. Such event always exists by the assumptions on operation specifications (\Cref{ssec:assumptions-generalized-specs}). W.l.o.g. we can assume that $e_i^{x_l}$ happens on replica $r_i^{x_l}$.

For concluding the description of $h^i = \tup{E_i, \so^i, \wro^i}$ and $\exec^i= \tup{h^i, \rbo^i, \ar^i}$, we use an auxiliary history and abstract execution, $h^i_{-1} = \tup{E^i_{-1}, \so^i_{-1}, \wro^i_{-1}}$ and $\exec^i_{-1}= \tup{h^i_{-1}, \rbo^i_{-1}, \ar^i_{-1}}$ respectively. For specifying $\wro^i_{-1}$, we define the context mapping $c^i: \Vars \to \Contexts$ in the same fashion as in \Cref{th:general-normal-from:arbitration-well-defined}:

\begin{equation}
c_i^{x_l}(y) = (F_i^{x_l}(y), \rbo^{i-1}_{\restriction F_i^{x_l}(y) \times F_i^{x_l}(y)}, \ar^{i-1}_{\restriction F_i^{x_l}(y) \times F_i^{x_l}(y)})
\end{equation}
where $F_i^{x_l}(y)$ is the mapping associating each object $y$ with the set of events described below:

\begin{equation}
F_i^{x_l}(y) = \begin{array}{ll}
    \{e \in E^{i-1} \ | \ 
    \begin{array}{l}
    \wspec{\CC}{y}{\exec^{i-1}}{e}\downarrow \text{and} \ (e, e_{i-1}^{x_{1-l}}) \in (\rbo^{i-1})^* 
    \end{array}\} & \text{if } i = d_v \\
    \{e \in E^{i-1} \ | \ 
    \begin{array}{l}
    \wspec{\CC}{y}{\exec^{i-1}}{e}\downarrow \text{and} \ (e, e_{i-1}^{x_l}) \in (\rbo^{i-1})^* 
    \end{array}\} & \text{otherwise} \\
\end{array}\nonumber
\end{equation}

Then, we define $\exec^i_{-1}$ as the abstract execution of the history $h^i_{-1}= \tup{E^i_{-1}, \so^i_{-1}, \wro^i_{-1}}$ obtained by appending $e_i^{x_0}, e_i^{x_1}$ to $h^i_{-1}$ and $\exec^i_{-1}$ as follows: $E^i_{-1}$ contains $E^{i-1}$ and events $e_i^{x_0}, e_i^{x_1}$. First of all, we require that the relations $\so^i_{-1}$, $\wro^i_{-1}$, $\rbo^i_{-1}$ and $\ar^i_{-1}$ contain $\so^{i-1}$, $\wro^{i-1}$, $\rbo^{i-1}$ and $\ar^{i-1}$ respectively. 

With respect to events $e_i^{x_0}, e_i^{x_1}$, we impose that $e_i^{x_l}$ is the maximal event w.r.t. $\so^i_{-1}$ among those on the same replica. Also, $e_i^{x_l}$ is maximal w.r.t. $\wro$ as we define that for every object $z$, $((\wro^i_{-1})_z)^{-1}(e_i^{x_l}) = \rspecContext{z}{c_i^{x_l}(z)}{e_i^{x_l}}$. We also require that $\rbo^i_{-1} = \so^i_{-1}$. With respect to $\ar^i_{-1}$, we impose that $e_i^{x_0}$ succeeds every event in $E^i$ w.r.t. $\ar^i_{-1}$ and that $e_i^{x_1}$ is the maximum event w.r.t. $\ar$ in $\exec^i_{-1}$.

We use $\exec^i_{-1}$ to construct $\exec^i$. If event $e_i^{x_l}$ is not a conditional write event, $\exec^i = \exec^i_{-1}$. Otherwise, if event $e_i^{x_l}$ is a conditional write event, given $W_i^{x_l}$ and object $\tilde{x}_i^{x_l}$, we select an execution-corrector for $e_i^{x_l}$ w.r.t. $(\CC, \opspec)$ and $a_i^{x_l}$. W.l.o.g., we assume that every event mapped by $a_i^{x_l}$ happens on replica $r_i^{x_l}$. Observe that by the choice of sets $D_i^{x_l}$ and $W_i^{x_l}$, and thanks to the assumptions on storages (see \Cref{ssec:assumptions-generalized-specs}), such event(s) are always well-defined.

In addition, we denote by $C_i^{x_l}$ to the set of objects we need to correct for $e_i^{x_l}$. More specifically, if $e_i^{x_l}$ is a conditional write-read, we denote by $C_i^{x_l}$ to the set of objects $y$ s.t. $a_i^{x_l}(y)$ is defined, i.e. $C_i^{x_l} = \{y \in \Vars \ | \ a_i^{x_l}(y) \downarrow\}$. In the case $e_i^{x_l}$ is not a conditional write-read, we use the convention $C_i^{x_l} = \emptyset$. The set of events in $\exec^i$ is the following: $E^i  = E^{i-1} \cup \bigcup_{l \in \{0,1\}}(\{e_i^{x_l}\} \bigcup_{y \in C_i \setminus \{o_{i-1}^{x_l}\}} a_i^{x_l}(y))$. Observe that by the choice of $C_i^{x_l}$, the set $E_i$ is well-defined.

From $\exec^i_{-1}$, we define $\exec^i = \correction{\exec^i_0}{\sequence{a_i}}{e_i}$ as the corrected execution of $\exec$ and $e_i^{x_0}, e_i^{x_1}$ with events $a_i^{x_0}, a_i^{x_1}$. For describing $\exec^i$, we consider $<$ to be a well-founded order over $\Vars$. $\exec^i$ satisfies the following:

\begin{itemize}
    \item \underline{$\so^i$}: Let $y \in C_i^{x_l}$. We require that for every event $e \in E^{i-1}$, $(e, a_i^{x_l}(y)) \in \so^i$ iff $\replicaEvent{e} = r_i^{x_l}, 0 \leq j < i$. We also require that $(\init, a_i^{x_l}(y)) \in \so^i$ and $(a_i^{x_l}(y), e_i^{x_l}) \in \so^i$. Finally, we require that for every objects $y' \in C_i^{x_l}, y' < y$, $(a_i^{x_l}(y'), a_i^{x_l}(y)) \in \so^i$.
    
    \item \sloppy \underline{$\wro^i$}: Let $y$ be an object in $C_i^{x_l}$. For every object $z$, if $z \in C_i^{x_l}$ and $z < y$, we require that $(\wro^i_z)^{-1}(a_i^{x_l}(y)) = \rspecContext{z}{c_i^{x_l}(z) \oplus a_i^{x_l}(z)}{a_i^{x_l}(y)}$; while otherwise, we require that $(\wro^i_z)^{-1}(a_i^{x_l}(y)) = \rspecContext{z}{c_i^{x_l}(z)}{a_i^{x_l}(y)}$. We also require that for every object $z$, if $z \in C_i^{x_l}$, then $(\wro^i_z)^{-1}(e_i^{x_l}) = \rspecContext{z}{c_i^{x_l}(z) \oplus a_i^{x_l}(z)}{e_i^{x_l}}$, while otherwise, $(\wro^i_z)^{-1}(e_i^{x_l}) = \rspecContext{z}{c_i^{x_l}(z)}{e_i^{x_l}}$.

    \item \underline{$\rbo^i$}: Let $y \in C_i^{x_l}$. We require that for every object $y \in  C_i^{x_l}$ and event $e$ s.t. $(e, a_i^{x_l}(y)) \in \so^i$, s.t. $(e, a_i^{x_l}(y)) \in \so^i \cup \wro^i$, $(e, a_i^{x_l}(y)) \in \rbo^i$.   %

    \item \underline{$\ar^i$}: We impose that for every event $e \in E^{i-1}$, $(e, a_i^{x_l}(y)) \in \ar^i, y \in  C_i^{x_l}$. We also require that for every pair of objects $y_1, y_2 \in  C_i$ s.t. $y_1, y_2$, $(a_i^{x_l}(y_1), a_i^{x_l}(y_2)) \in \ar^i$. As a tie-breaker between events associated to $x_0$ and $x_1$, we require that for every pair of events $e \in \{e_i^{x_0}, a_i^{x_0}(y) \ |\ y \in C_i^{x_0}\}$, $e' \in \{e_i^{x_1}, a_i^{x_1}(y) \ |\ y \in C_i^{x_0}\}$, $(e, e') \in \ar^i$.
\end{itemize}

We then define $h^i = \tup{E^i, \so^i, \wro^i}$ and $\exec^i = \tup{h^i, \rbo^i, \ar^i}$. Observe that by construction of $h^i$ and $\exec^i$, as $ \wro^i \subseteq \rbo^i = \so^i$, they satisfy \Cref{def:history,def:execution} respectively; so they are a history and an abstract execution respectively. Also, observe that $\exec^i$ is the corrected abstract  execution of $\exec^i_{-1}$ for events $e_i^{x_0}, e_i^{x_1}$ with events $a_i^{x_0}, a_i^{x_1}$, i.e. $\exec^i = \exec^i_1 = \correction{\exec^i_0}{\sequence{a_i^{x_1}}}{e_i^{x_1}}$, where $\exec^i_0 = \correction{\exec^i_{-1}}{\sequence{a_i^{x_0}}}{e_i^{x_0}}$.

Then, we define $\EventsProgramIota = E^{\length{v}}$ as the set our program employs. The set $\EventsProgramIota$ induces the set of traces $\TracesProgramIota$.%

We define the program $\programInstanceIota = \programIota$, where $\initProgramIota = \init$ and $\Delta_{\mathsf{p}}$ is the transition function defined as follows: for every trace $t \in \TracesProgramIota$ and event $e \in \EventsProgramIota$, $\Delta_{\mathsf{p}}(t, e) \downarrow$ if and only if $e \not\in t$ and every event in $\EventsProgramIota$ whose replica coincide with $e$ and has smaller identifier than $e$ is included in $t$.

The rest of the proof, which proceeds as follows, essentially combines previous results obtained while proving \Cref{lemma:cons-available-lww:3-1} and \Cref{th:general-normal-from:arbitration-well-defined}:
\begin{enumerate}
    \item There exists a finite trace $t$ of $\parallelCompositionInstanceIota$ that contains no receive action (\Cref{lemma:program-a-la-decker:trace-without-receive})
    
    \item The trace $t$ induces a history $h_\mathsf{v} = \tup{E, \so, \wro}$ and an abstract execution $\exec_\mathsf{v} = \tup{h, \rbo, \ar}$ where $\rbo = \so$. As $\implementationInstance$ is valid w.r.t. $\sspec$, $\exec_\mathsf{v}$ is valid w.r.t. $\sspec$. 
        Next, we prove that since $\rbo = \so$, events in $\exec_\mathsf{v}$ read the latests value w.r.t. $\so$ for any object. In particular, we deduce that $\exec_v$ is valid w.r.t. $(\CC, \opspec)$ (\Cref{corollary:afc-theorem:validity-exec-trace}).

    \item \sloppy Since $\ar$ is a total order, either $(e_{d_\mathsf{v}-1}^{x_0}, e_{d_\mathsf{v} - 1}^{x_{1}}) \in \ar$ or $(e_{d_\mathsf{v}-1}^{x_1}, e_{d_\mathsf{v} - 1}^{x_{0}}) \in \ar$. W.l.o.g., assume that $(e_{d_\mathsf{v}-1}^{x_0}, e_{d_\mathsf{v} - 1}^{x_{1}}) \in \ar$. By \Cref{proposition:afc-theorem:visibility}, we deduce that $e_0^{x_0} \in \context{\cmodel}{x_0}{\exec_\mathsf{v}}{e_{\length{\mathsf{v}}}^{x_0}}$. The proof is explained in \Cref{fig:diagram-theorem-proof-general}: if $(e_{d_\mathsf{v}-1}^{x_0}, e_{d_\mathsf{v} - 1}^{x_{1}}) \in \ar$, then all events $e_i^{x_0}$ form a path in such way that $\mathsf{v}_{x_0}(e_0^{x_0}, \ldots e_{\length{\mathsf{v}}}^{x_0})$ holds in $\exec_\mathsf{v}$. If some event $e_i^{x_l}$ is a conditional read-write event, the predicate $\writeConflicts[x]{e_0^{x_0}, \ldots e_{\length{\mathsf{v}}}^{x_0}}$ holds in $\exec_{\mathsf{v}}$ thanks to the corrector events $A_i^{x_l}$.%

    \item As $e_0^{x_0} \in \context{\cmodel}{x_0}{\exec_\mathsf{v}}{e_{\length{v}}^{x_0}}$ but $(e_0^{x_0}, e_{\length{\mathsf{v}}}^{x_0}) \not\in \rbo$ (no message is received), we deduce in \Cref{proposition:general-normal-form:stratified-ar} that $\opspec$ is layered w.r.t. $\ar$. By contrapositive, if $\opspec$ would be layered w.r.t. $\rbo$, as $e_0^{x_0} \in \context{\cmodel}{x_0}{\exec_\mathsf{v}}{e_{\length{\mathsf{v}}}^{x_0}}$, there would exist an event $e$ s.t. $(e_0^{x_0}, e) \in \rbo$ and $e \in \rspec{\cmodel}{x_0}{\exec_\mathsf{v}}{e_{\length{\mathsf{v}}}^{x_0}}$. However, as $\rbo = \so$, $\replicaEvent{e_0^{x_0}} = \replicaEvent{e} = \replicaEvent{e_{\length{\mathsf{v}}}^{x_0}}$ which is false because $\replicaEvent{e_0^{x_0}} = r_0$ and $\replicaEvent{e_{\length{\mathsf{v}}}^{x_0}} = r_1$.

    \item Since $\rspecName$ is maximally layered, we can show that the layer bound of $\rspecName$ is smaller than or equal to the number of arbitration-free suffixes of $\mathsf{v}$ (\Cref{proposition:general-normal-form:k-suffixes}). %
    Observe that an event writes $x_0$ only if it is $\init$ or is an event $e_i^{x_l}$ s.t. $\event_i \in E_x$ and $l = 0$. Any such index $i$ corresponds to a suffix of $\mathsf{v}$. By causal suffix closure, for any arbitration-free suffix $v'$ of $v$ there is a visibility \replaced{formula}{relation} that subsumes $v'$ in $\generalNormalForm{\cmodel}$. As $d_\mathsf{v}$ is the maximum index for which $\mathsf{Rel}_i^\mathsf{v} = \ar$, the number of events writing $x_0$ in replica $r_1$ distinct from $\init$ coincide with the number of arbitration-free suffixes of $\mathsf{v}$. Hence, as $\rspecName$ is layered w.r.t. $\ar$, if its layer bound would be greater than the number of arbitration-free suffixes, $e_{\length{\mathsf{v}}}^{x_0}$ would necessarily read $x_0$ from $\init$ (other events writing $x_0$ are in replica $r_0$ and $e_{\length{\mathsf{v}}}$ only reads from events in $r_1$). However, as $\rspecName$ is maximally-layered and $e_0^{x_0}$ succeeds $\init$ w.r.t. $\ar$ and $\rbo^+$, we would conclude that $e_{\length{\mathsf{v}}}^{x_0}$ would also read $x_0$ from $e_0^{x_0}$. However, this is impossible as $\wro \subseteq \rbo = \so$ but $e_0^{x_0}$ is in replica $r_0$ and $e_{\length{\mathsf{v}}}^{x_0}$ is in replica $r_1$.

    \item Lastly, we show in \Cref{proposition:general-normal-form:layer-bound-implies-v-vacuous} that if the layer bound of $\rspecName$ is smaller than or equal to the number of arbitration-free suffixes of $v$, then $v$ is vacuous w.r.t. $\opspec$, which contradicts the fact that $\mathsf{v}$ is a visibility \replaced{formula}{relation} from the normal form $\generalNormalForm{\cmodel}$.    
    \qedhere
    
\end{enumerate}
\end{proof}

\begin{proposition}
\label{proposition:afc-theorem:contexts}
The abstract execution $\exec^{\length{v}}$ described in \Cref{lemma:cons-available:3-1} satisfies that for every $i, 0 \leq i \leq \length{v}, l \in \{0,1\}$:
\begin{enumerate}

    \item For every object $y \in C_i^{x_l}$, the following conditions hold:
     \begin{enumerate}
        \item For every object $z \in\Vars$, if $z \in C_i^{x_l}$ and $z < y$, $G(a_i^{x_l}(y), z) = F_i^{x_l}(z) \cup \{a_i^{x_l}(z)\}$, while otherwise, $G(a_i^{x_l}(y), z) = F_i^{x_l}(z)$.
        
        \item The execution $\exec^i_l \restriction y$ is valid w.r.t. $(\CC, \opspec)$.
    \end{enumerate}

    \item For the event $e_i^{x_l}$, the following conditions hold:
    \begin{enumerate}
        \item  For every object $z$, if $z \in C_i^{x_l}$, $G(e_i^{x_l}, z) = F_i^{x_l}(z) \cup \{a_i^{x_l}(z)\}$, while otherwise $G(e_i^{x_l}, z) = F_i^{x_l}(z)$.
        \item The execution $\exec^i_l$ is valid w.r.t. $(\CC, \opspec)$.
    \end{enumerate}

\end{enumerate}
where $\context{\CC}{z}{\exec^{\length{v}}}{e} = \tup{G(e, z), \rbo_{\restriction G(e, z) \times G(e, z)}, \ar_{\restriction G(e, z) \times G(e, z)}}$.
\end{proposition}

\begin{proof}
The proof of this result essentially coincides with that of \Cref{proposition:general-normal-form:arbitration-well-defined:contexts}.

We prove the result by induction. In particular, we show that for every $i, -1 \leq i \leq \length{v}$ and object $y$, either (0) $i = -1$ or (1) and (2) hold. The base case, $i = -1$, is immediate as (0) holds; so let us suppose that the result holds for every $j, -1 \leq j < i$, and let us prove it for $i$.

For proving the inductive step, we first prove (1) for $l = 0$, then (2) for $l = 0$, and then (1) and (2) for $l = 1$.  As both (1) and (2) have an identical proof (observe that the role of object $y$ in the former is just to declare that event $a_i^{x_l}(y)$ is well-defined and the role of $l$ is to determine which session must be proven first), we present only the proof of (1) for $l = 0$.

We show (1) by transfinite induction. Let $\alpha$ be an ordinal of cardinality $|\Vars|$. For every $k, 0 \leq k \leq \alpha$, we denote by $V_k$ to the set containing the first $k$ elements in $\Vars$ according to $<$. We show that (1) holds for every $y \in V_k \cap C_i^{x_0}$.

The base, $V_0$ is immediate as $V_0 = \emptyset$. We thus focus on the successor case (i.e., showing that if (1) holds for every object $y \in V_k \cap C_i^{x_0}$ it also holds for $V_{k+1}$), as the limit case is immediate: if $k$ is a limit ordinal, $V_k = \bigcup_{i, i < k} V_i$; so (1) immediately holds. For showing that (1) holds for every object $y \in V_{k+1} \cap C_i^{x_0}$, as by induction hypothesis it holds for every object $y \in V_k \cap C_i^{x_0}$, it suffices to show it for the only object $y \in V_{k+1} \setminus V_i$. W.l.o.g., we can assume that $y \in C_i^{x_0}$; as otherwise the result is immediate. 

We first prove (1a) and then we show (1b). Let $z \in \Vars$ be an object. Two cases arise: $z \in C_i, z < y$ or not. Both cases are identical, so we present the former, i.e., if $z \in C_i^{x_0}, z < y$, then $F_i^{x_0}(z) \cup \{a_i^{x_0}(z)\} = G(a_i^{x_0}(y), z)$.

For proving that $F_i^{x_0}(z) \cup \{a_i^{x_0}(z)\} \subseteq G(a_i^{x_0}(y), z)$, we distinguish whether $i = d_v$ or not. However, the proof essentially coincides in both cases, so we present the case $i = d_v$. We split the proof in two blocks: showing that $F_i^{x_0}(z) \subseteq G(a_i^{x_0}(y), z)$ and showing that $a_i^{x_0}(z) \in G(a_i(y), z)$.

For showing that $F_i^{x_0}(z) \subseteq G(a_i^{x_0}(y), z)$, let $e$ be an event in $F_i^{x_0}(z)$. In such case, $e \in E^{i-1}$, $\wspec{\CC}{z}{\exec^i}{e} \downarrow$ and $(e, e_{i-1}^{x_1}) \in (\rbo^i)^*$. By the construction of $\exec$, it is easy to see that any such event belongs to $E^i$, $\wspec{\CC}{z}{\exec}{e} \downarrow$ and $(e, e_{i-1}^{x_1}) \in (\rbo^{\length{v}})^*$. As $i = d_v$, we deduce that $(e_{i-1}^{x_1}, a_i^{x_0}(y)) \in \rbo^i \subseteq \rbo^{\length{v}}$. Hence, $(e, a_i^{x_0}(y)) \in (\rbo^{\length{v}})^+$; so $e \in G(a_i^{x_0}(y), z)$. This show that $F_i^{x_0}(z) \subseteq G(a_i^{x_0}(y), z)$.

For showing that $a_i^{x_0}(z) \in G(a_i^{x_0}(y), z)$, we observe that $\exec^i_0 = \correction{\exec^i_{-1}}{a_i^{x_0}}{e_i^{x_0}}$. We note that as $z < y$, by induction hypothesis (1b), $\exec^i_0 \restriction z$ is valid w.r.t. $(\CC, \opspec)$. Thus, by Property~\ref{def:execution-corrector:1-context-corrector-writes} of \Cref{def:execution-corrector}, $\wspec{\CC}{z}{\exec^i_0}{a_i(z)} \downarrow$. Hence, $\wspec{\CC}{z}{\exec^i}{a_i^{x_0}(z)} \downarrow$ and $\wspec{\CC}{z}{\exec^{\length{v}}}{a_i^{x_0}(z)} \downarrow$. As $z < y$, $(a_i^{x_0}(z), a_i^{x_0}(y)) \in \so^i \subseteq \so^{\length{v}}$; so we conclude that $a_i^{x_0}(z) \in G(a_i^{x_0}(y), z)$.

We conclude the proof of the inductive step of (1a) by showing the converse i.e. $F_i^{x_0}(z) \cup \{a_i^{x_0}(z)\} \supseteq G(a_i^{x_0}(y), z)$. Let $e \in G(a_i^{x_0}(y), z)$. First of all, by the definition of $\axcc$ visibility \replaced{formula}{relation} (see \Cref{fig:cc}), $e \in G(a_i^{x_0}(y), z)$ iff $\wspec{\CC}{z}{\exec}{e} \downarrow$ and $(e, a_i^{x_0}(y)) \in (\rbo^{\length{v}})^+$. Observe that if $(e, a_i^{x_0}(y)) \in (\rbo^{\length{v}})^+$, by construction of $\exec^{\length{v}}$, such event must belong to $E^i$, $\wspec{\CC}{z}{\exec^i}{e} \downarrow$ and $(e, a_i(y)) \in (\rbo^i)^+$. We prove that if $e \in E^{i-1}$ then $e \in F_i^{x_0}(z)$, while otherwise, if $e \in E^i \setminus E^{i-1}$, then $e = a_i^{x_0}(z)$.

If $e \in E^{i-1}$, as $\wspec{\CC}{z}{\exec^i}{e} \downarrow$, $\wspec{\CC}{z}{\exec^{i-1}}{e} \downarrow$. Also, as $i = d_v$ and $(e, a_i^{x_0}(y)) \in (\rbo^{i})^+$, we deduce that $(e, e_{i-1}^{x_1}) \in (\rbo^{i-1})^*$. In other words, $e \in F_i^{x_0}(z)$. 

Otherwise, if $e \in E^i \setminus E^{i-1}$, we note that by construction of $\exec^{\length{v}}$, the only events in $E^i \setminus E^{i-1}$ s.t. $(e, a_i^{x_0}(y)) \in (\rbo^i)^+$ are events $a_i^{x_0}(w), w \in C_i, w < y$. As $\exec^i_0 = \correction{\exec^i_{-1}}{\sequence{a_i^{x_0}}}{e_i^{x_0}}$ and $z < y$, $\exec^i_0 \restriction z$ is valid w.r.t. $(\CC, \opspec)$. Hence, $\wspec{\CC}{z}{\exec^i}{e} \downarrow$ iff $\wspec{\CC}{z}{\exec^i_0}{e} \downarrow$. Thus, by Property~\ref{def:execution-corrector:1-context-corrector-writes} of \Cref{def:execution-corrector} we conclude that $e = a_i^{x_0}(z)$.

For concluding the inductive step, we show that (1b) holds. This is immediate by the definition of $\wro^i$: for every event $e \in \exec^{i} \restriction y$, by induction hypothesis (1a) or (2a) -- depending on whether $e = e_j^{x_{l'}}$ or $a_j^{x_{l'}}(w)$, where $0 \leq j \leq i, w \in C_i, l' \in \{0,1\}$ -- $(\wro^i)_z^{-1}(e)= \rspec{z}{\CC}{\exec^j_{l'} \restriction y}{e} = \rspec{z}{\CC}{\exec^i_l \restriction y}{e}$. Thus, $\exec^i \restriction y$ is valid w.r.t. $(\CC, \opspec)$.
\end{proof}

A consequence of \Cref{proposition:afc-theorem:contexts} is the following result.

\begin{corollary}
\label{corollary:afc-theorem:validity-exec}
The abstract execution $\exec$ described in \Cref{lemma:cons-available:3-1} is valid w.r.t. $(\CC, \opspec)$. 
\end{corollary}

\Cref{corollary:afc-theorem:validity-exec-trace} is an immediate result from \Cref{corollary:afc-theorem:validity-exec}, obtained by simply observing that $\rbo^{\length{v}} = \so^{\length{v}} = \so = \rbo$.

\begin{corollary}
\label{corollary:afc-theorem:validity-exec-trace}
The abstract execution $\exec_v$ described in \Cref{lemma:cons-available:3-1} is valid w.r.t. $(\CC, \opspec)$. 
\end{corollary}

\begin{proposition}
\label{proposition:afc-theorem:visibility}
For every $l \in \{0,1\}$, if $(e_{d_v-1}^{x_l}, e_{d_v - 1}^{x_{1-l}}) \in \ar$, then the predicate $v_{x_0}(e_0^{x_l}, \ldots e_{\length{v}}^{x_l})$ holds in the abstract execution $\exec =\tup{h, \rbo, \ar}$ described in \Cref{th:general-normal-from:arbitration-well-defined}.
\end{proposition}

\begin{proof}
The proof of this result essentially coincides with that of \Cref{proposition:general-normal-form:arbitration-well-defined:visibility}.

The proof is a simple consequence of $\exec^{\length{v}}$'s construction. To show that $v_{x_0}(e_0^{x_l}, \ldots e_{\length{v}}^{x_l})$ holds in $\exec$, we first show that for every $i, 1 \leq i \leq \length{v}$, $(e_{i-1}^{x_l}, e_i^{x_l}) \in \mathsf{Rel}_i^\mathsf{v}$ and to then prove that $\writeConstraints{x}{e_0^{x_l}, \ldots e_{\length{v}}^{x_l}}$ holds in $\exec$.

We prove that for every $i, 1 \leq i \leq \length{v}$, $(e_{i-1}^{x_l}, e_i^{x_l}) \in \mathsf{Rel}_i^\mathsf{v}$. Four cases arise depending on $\mathsf{Rel}_i^\mathsf{v}$.

\begin{itemize}
    \item \underline{$\mathsf{Rel}_i^\mathsf{v} = \so$}: In this case, by construction of events $e_{i-1}^{x_l}, e_i^{x_l}$, we know that $r_i^{x_l} = r_{i-1}^{x_l}$. Hence, $(e_{i-1}^{x_l}, e_i^{x_l}) \in \so^i \subseteq \so$.
    
    \item \underline{$\mathsf{Rel}_i^\mathsf{v} = \wro$}: In this case, we first show that there is an object $y \in D_i^{x_l} \cap W_{i-1}^{x_l}  \setminus C_i^{x_l}$, and then show that $(e_{i-1}^{x_l}, e_i^{x_l}) \in \wro_y$. For showing the first part, we distinguish between cases depending on whether $o_{i-1}^{x_l} \in D_i^{x_l}$ or not. 
    \begin{itemize}
        \item \underline{$o_{i-1}^{x_l} \in D_i^{x_l}$}: In this sub-case, we show that $y = o_{i-1}^{x_l}$. On one hand, if $\conflictsOfV{v}{i} = \emptyset$, by the choice of event $e_i^{x_l}$, $o_{i-1}^{x_l} \in D_{i-1}^{x_l} \setminus C_i^{x_l}$. On the other hand, if $\conflictsOfV{v}{i} \neq \emptyset$, as $o_{i-1}^{x_l} \in D_{i}^{x_l}$, we deduce that $\opspec$ allows multi-object read-write events. Observe that as $v$ is conflict-maximal, $\conflictsOfV{v}{i-1} \neq \emptyset$. Hence, as $\opspec$ allows multi-object read-write events, we deduce that $o_{i-1}^{x_l} \in D_{i-1}^{x_l} \setminus C_i^{x_l}$. In both cases, as $\conflictsOfV{v}{i-1} \neq \emptyset$ and $o_{i-1}^{x_l} \in D_{i-1}^{x_l}$, by the choice of $W_{i-1}^{x_l}$, we conclude that $o_{i-1}^{x_l} \in W_{i-1}^{x_l}$.

        \item \underline{$o_{i-1}^{x_l} \not\in D_i^{x_l}$}: In this case, we show that $y = \tilde{x}_i^{x_l}$. On one hand, if $\conflictsOfV{v}{i} = \emptyset$, $X_i^{x_l} = \emptyset$; so by the choice of $\tilde{x}_i^{x_l}$ (see \Cref{eq:afc:x-i}), $\tilde{x}_i^{x_l} = \tilde{x}_{i-1}^{x_l}$. By the choice of $D_i^{x_l}$, $\tilde{x}_{i-1}^{x_l} \in D_i^{x_l} \setminus C_i^{x_l}$. Moreover, as $v$ is conflict-maximal, $\conflictsOfV{v}{i-1} \neq \emptyset$; so $\tilde{x}_{i-1}^{x_l} \in X_{i-1}^{x_l}$. By the choice of event $e_{i-1}^{x_l}$, $X_{i-1}^{x_l} \subseteq W_{i-1}^{x_l}$. Altogether, we conclude that $\tilde{x}_i^{x_l} \in W_{i-1}^{x_l}$. 
        
        On the other hand, if $\conflictsOfV{v}{i} \neq \emptyset$, we note that $\tilde{x}_i^{x_l} \in D_i^{x_l} \setminus C_i^{x_l}$. As $o_{i-1}^{x_l} \not\in D_{i}^{x_l}$, we deduce that $\opspec$ only allows single-object read-write events. Thus, $D_i^{x_l} = \{\tilde{x}_i^{x_l}\}$. As $v$ is conflict-maximal, we deduce that $X_i^{x_l} \subseteq X_{i-1}^{x_l}$. As by the choice of $e_{i-1}^{x_l}$, $X_{i-1}^{x_l} \subseteq W_{i-1}^{x_l}$, we conclude that $\tilde{x}_i^{x_l} \in W_{i-1}^{x_l}$.

    \end{itemize}

    We prove now that $(e_{i-1}^{x_l}, e_i^{x_l}) \in \wro_{y}$. First, we show that $\writeVarExec{e_{i-1}^{x_l}}{y}{\exec}$. On one hand, if $e_{i-1}^{x_l}$ is an unconditional write event, $\wspecContext{y}{c_i^{x_l}(y)}{e_{i-1}^{x_l}} \downarrow$. On the other hand, if $e_{i-1}^{x_l}$ is a conditional write event, as $\exec$ is valid w.r.t. $(\CC, \opspec)$ (\Cref{corollary:afc-theorem:validity-exec-trace}) and $y \in W_i^{x_l}$, by Property~\ref{def:execution-corrector:4-e-writes-w} of \Cref{def:execution-corrector}, we deduce that $\wspecContext{y}{c_i^{x_l}(y)}{e_{i-1}^{x_l}} \downarrow$. Then, as $\mathsf{Rel}_i^\mathsf{v} = \wro$, $i \neq d_v$, so $e_{i-1}^{x_l} \in F_i^{x_l}(y)$. Observe that by construction of $\exec$, $e_{i-1}^{x_l}$ is the $so$-maximum event in $c_i^{x_l}(y)$. As every event in $F_i^{x_l}(y)$ is $\so$-related, we deduce that $e_{i-1}^{x_l}$ is the $\ar$-maximum event in $F_i^{x_l}(y)$. We note that as $y \not\in C_i^{x_l}$, by \Cref{proposition:afc-theorem:contexts}, $F_i^{x_l}(y) = G(e_i^{x_l}, y)$. Altogether, $e_{i-1}^{x_l}$ is the $\ar$-maximum event in $\context{\CC}{y}{\exec^{\length{v}}}{e_i^{x_l}}$. As $\rbo^{\length{v}} = \rbo$, we conclude that $e_{i-1}^{x_l}$ is the $\ar$-maximum event in $\context{\CC}{y}{\exec}{e_i^{x_l}}$. As $\rspecName$ is maximally layered, we deduce that $e_{i-1}^{x_l} \in \rspec{\CC}{y}{\exec}{e_i^{x_l}}$. Finally, as $\exec$ is valid w.r.t. $\CC$ (\Cref{corollary:afc-theorem:validity-exec-trace}), we conclude that $(e_{i-1}^{x_l}, e_i) \in \wro_y$.

    \item \underline{$\mathsf{Rel}_i^\mathsf{v} = \rbo$}: In this case, $i \neq d_v$. Then, $\rbo = \so$ and $(e_{i-1}^{x_l}, e_i^{x_l}) \in \so$, we conclude that $(e_{i-1}^{x_l}, e_i^{x_l}) \in \rbo$.

    \item \underline{$\mathsf{Rel}_i^\mathsf{v} = \ar$}: On one hand, if $i = d_v$, by hypothesis, $(e_{i-1}^{x_l}, e_i^{x_l}) \in \ar$. On the other hand, if $i \neq d_v$, $(e_{i-1}^{x_l}, e_i^{x_l}) \in \so$. Thus, $(e_{i-1}^{x_l}, e_i^{x_l}) \in \ar$.

\end{itemize}

For showing that show that $\writeConstraints{x}{e_0, \ldots e_{\length{v}}}$, we show that for every $i, 0 \leq i \leq \length{v}$ and every set $E \in \conflictsOfV{v}{i}$, the event $e_i^{x_l}$ writes on object $y_E$\footnote{For simplifying the proof, we abuse of notation and say that $y_E = x_l$ if $E = E_x$. Observe that $v$ is conflict-maximal, either $\writeConflicts[x]{E_x}$ or $\writeConflicts{E_x}$ do not belong to $v$.}. If $e_i^{x_l}$ is an unconditional write, by the choice of $e_i^{x_l}$, it writes on every object in $D_i^{x_l}$. As $y_E \in D_i^{x_l}$, we conclude that $e_i^{x_l}$ writes on $y_E$. Otherwise, if $e_i^{x_l}$ is a conditional write, we observe that $y_E \in W_i^{x_l}$. Hence, as $\exec^i_0 = \correction{\exec^i_{-1}}{\sequence{a_i^{x_0}}}{e_i^{x_0}}$ and $\exec^i_0$ is valid w.r.t. $(\CC, \opspec)$ (resp. $\exec^i_1 = \correction{\exec^i_{0}}{\sequence{a_i^{x_1}}}{e_i^{x_1}}$ and $\exec^i_1$ is valid w.r.t. $(\CC, \opspec)$) (\Cref{proposition:afc-theorem:contexts}), we deduce using Property~\ref{def:execution-corrector:4-e-writes-w} of \Cref{def:execution-corrector} that $\wspec{\CC}{y_E}{\exec^i}{e_i^{x_l}} \downarrow$. By construction of $\exec$, we conclude that $\wspec{\CC}{y_E}{\exec}{e_i^{x_l}} \downarrow$.
\end{proof}

	}{}%

\end{document}